\documentclass[a4paper,twoside,11pt]{amsbook}
\usepackage[utf8]{inputenc}
\usepackage[ngerman,french,main=english]{babel}
\usepackage{amsmath}
\usepackage{amsthm}
\usepackage{amsfonts,amssymb}
\usepackage[foot]{amsaddr}
\usepackage{mathtools}
\usepackage{mathshortcuts}
\usepackage{enumitem}
\usepackage[hypertexnames=false,backref=page]{hyperref}
\usepackage{cleveref}
\usepackage{autonum}
\usepackage{float}
\usepackage{subcaption}
\usepackage{mleftright}
\usepackage{todonotes}
\usepackage[left=3cm,right=3cm,top=3.5cm,bottom=3.5cm]{geometry}
\usepackage{tikz}
\usetikzlibrary{positioning,shapes.geometric,decorations.markings,arrows,knots,calc,decorations.pathmorphing,decorations.pathreplacing,calc,shapes}
\usepackage{lipsum}
\usepackage{etoolbox}
\usepackage{bm}
\usepackage{blkarray,bigstrut}

\usepackage{chngcntr}
\counterwithout{figure}{chapter}
\counterwithout{table}{chapter}

\usepackage{pgfplots}
\usepgfplotslibrary{colormaps}

\makeindex

\mleftright

\allowdisplaybreaks

\numberwithin{section}{chapter}
\numberwithin{equation}{chapter}

\newcommand{\alpharho}{{{}_\alpha\rho}}
\newcommand{\alphastarrho}{{}_{\alpha^*}\!\rho}
\newcommand{\C}{{\mathcal{C}}}
\newcommand{\Hi}{{\mathcal{H}}}
\newcommand{\Hd}{{\mathcal{H}_3}}
\newcommand{\Obj}{\mathsf{Obj}}
\newcommand{\Hom}{\mathsf{Hom}}

\renewcommand{\id}{\mathsf{id}}
\newcommand{\ev}{\mathsf{ev}}
\newcommand{\coev}{\mathsf{coev}}

\newcommand{\Tr}{\mathsf{Tr}}
\newcommand{\lstar}{\ {}\!\!^*}

\tikzset{->-/.style={decoration={
			markings,
			mark=at position .6 with {\arrow[>=stealth]{>}}},postaction={decorate}}}
\tikzset{-->--/.style={decoration={
			markings,
			mark=at position .52 with {\arrow[>=stealth]{>}}},postaction={decorate}}}
\tikzset{-<-/.style={decoration={
			markings,
			mark=at position .6 with {\arrow[>=stealth]{<}}},postaction={decorate}}}
		
\tikzset{->>-/.style={decoration={
			markings,
			mark=at position .5 with {\arrow[>=stealth]{>}}},postaction={decorate}}}
\tikzset{-<<-/.style={decoration={
			markings,
			mark=at position .5 with {\arrow[>=stealth]{<}}},postaction={decorate}}}
\tikzset{->>>-/.style={decoration={
			markings,
			mark=at position .5 with {\arrow[>=stealth]{>}}},postaction={decorate}}}
\tikzset{-<<<-/.style={decoration={
			markings,
			mark=at position .4 with {\arrow[>=stealth]{<}}},postaction={decorate}}}

\newcommand{\Cfourone}{\begin{tikzpicture}[baseline={([yshift=-3pt]current bounding box.center)}]
	\draw (0,0) to [bend right=60] (0,1);
	\draw (1,0) to [bend left=60] (1,1);
	\end{tikzpicture}}
\newcommand{\Cfourtwo}{\begin{tikzpicture}[baseline={([yshift=-3pt]current bounding box.center)}]
	\draw (0,0) to [bend left=60] (1,0);
	\draw (0,1) to [bend right=60] (1,1);
	\end{tikzpicture}}
\newcommand{\Cfourthree}{\begin{tikzpicture}[baseline={([yshift=-3pt]current bounding box.center)}]
	\draw (0,0) to (0.5,0.3) to (1,0);
	\draw (0,1) to (0.5,0.7) to (1,1);
	\draw (0.5,0.3) to (0.5,0.7);
	\end{tikzpicture}}
\newcommand{\Cfourfour}{\begin{tikzpicture}[baseline={([yshift=-3pt]current bounding box.center)}]
	\draw (0,0) to (0.3,0.5) to (0,1);
	\draw (1,0) to (0.7,0.5) to (1,1);
	\draw (0.3,0.5) to (0.7,0.5);
	\end{tikzpicture}}

\newcommand{\vertex}[3]{\begin{tikzpicture}[scale=0.7,baseline={(0,0.2)}]
	\draw (0,0) -- +(90:1) node[above] {\small $#3$};
	\draw (0,0) -- +(225:1) node[below] {\small $#1$};
	\draw (0,0) -- +(315:1) node[below] {\small $#2$};
	\end{tikzpicture}}

\newcommand{\fuselefttree}[5]{\begin{tikzpicture}[scale = 0.425,baseline={(0,0.4)}]
	\draw (0,0) node[below] {\small $#1$} -- (1.5,1.5) -- (3,0) node[below] {\small $#4$};
	\draw (0.75,0.75) -- (1.5,0) node[below] {\small $#2$};
	\draw (0.7,1.6) node {\small $#3$};
	\draw (1.5,1.5) -- (1.5,2.25) node[above] {\small $#5$};
	\end{tikzpicture}}

\newcommand{\fuserighttree}[5]{\begin{tikzpicture}[font=\scriptsize,scale = 0.425,baseline={(0,0.4)}]
	\draw (0,0) node[below] {\small $#1$} -- (1.5,1.5) -- (3,0) node[below] {\small $#4$};
	\draw (2.25,0.75) -- (1.5,0) node[below] {\small $#2$};
	\draw (2.3,1.6) node {\small $#3$};
	\draw (1.5,1.5) -- (1.5,2.25) node[above] {\small $#5$};
	\end{tikzpicture}}

\newcommand\topstrut[1][1.2ex]{\setlength\bigstrutjot{#1}{\bigstrut[t]}}
\newcommand\botstrut[1][0.9ex]{\setlength\bigstrutjot{#1}{\bigstrut[b]}}

\newcommand{\itemone}{\begin{tikzpicture}[baseline={([yshift=-3pt]current bounding box.center)}]\node[circle,scale=0.6,draw,LinkColor] at (0,0) {$1$};\end{tikzpicture}}
\newcommand{\itemtwo}{\begin{tikzpicture}[baseline={([yshift=-3pt]current bounding box.center)}]\node[circle,scale=0.6,draw,LinkColor] at (0,0) {$2$};\end{tikzpicture}}
\newcommand{\itemthree}{\begin{tikzpicture}[baseline={([yshift=-3pt]current bounding box.center)}]\node[circle,scale=0.6,draw,LinkColor] at (0,0) {$3$};\end{tikzpicture}}
\newcommand{\itemfour}{\begin{tikzpicture}[baseline={([yshift=-3pt]current bounding box.center)}]\node[circle,scale=0.6,draw,LinkColor] at (0,0) {$4$};\end{tikzpicture}}
\newcommand{\itemfive}{\begin{tikzpicture}[baseline={([yshift=-3pt]current bounding box.center)}]\node[circle,scale=0.6,draw,LinkColor] at (0,0) {$5$};\end{tikzpicture}}
\newcommand{\itemsix}{\begin{tikzpicture}[baseline={([yshift=-3pt]current bounding box.center)}]\node[circle,scale=0.6,draw,LinkColor] at (0,0) {$6$};\end{tikzpicture}}

\newcommand{\Fone}{\,\begin{tikzpicture}[baseline={([yshift=-3pt]current bounding box.center)},scale=1]\node[diamond,scale=0.35,draw,black] at (0,0) {\huge $1$};\end{tikzpicture}\,}
\newcommand{\Ftwo}{\,\begin{tikzpicture}[baseline={([yshift=-3pt]current bounding box.center)},scale=1]\node[diamond,scale=0.35,draw,black] at (0,0) {\huge $2$};\end{tikzpicture}\,}
\newcommand{\Fthree}{\,\begin{tikzpicture}[baseline={([yshift=-3pt]current bounding box.center)},scale=1]\node[diamond,scale=0.35,draw,black] at (0,0) {\huge $3$};\end{tikzpicture}\,}
\newcommand{\Ffour}{\,\begin{tikzpicture}[baseline={([yshift=-3pt]current bounding box.center)},scale=1]\node[diamond,scale=0.35,draw,black] at (0,0) {\huge $4$};\end{tikzpicture}\,}
\newcommand{\Fx}{\,\begin{tikzpicture}[baseline={([yshift=-3pt]current bounding box.center)},scale=1]\node[diamond,scale=0.35,draw,black] at (0,0) {\huge $x$};\end{tikzpicture}\,}
\newcommand{\Fy}{\,\begin{tikzpicture}[baseline={([yshift=-3pt]current bounding box.center)},scale=1]\node[diamond,scale=0.35,draw,black] at (0,0) {\huge $y$};\end{tikzpicture}\,}

\newcommand{\VecZ}{\mathbf{Vec}(\mathbb{Z}/p\mathbb{Z})}

\newcommand{\loopi}{\begin{tikzpicture}[baseline={([yshift=-3pt]current bounding box.center)}]\node[circle,scale=0.6,draw] at (0,0) {};\end{tikzpicture}}

\definecolor{LinkColor}{RGB}{221, 6, 0}
\definecolor{LinkColor2}{RGB}{15, 105, 197}
\definecolor{LinkColor3}{RGB}{11, 170, 21}
\definecolor{LinkColor4}{RGB}{238, 143, 0}
\hypersetup{
	colorlinks, 
	linkcolor=black, 
	citecolor=black,
	filecolor=black, 
	urlcolor=black,
}

\definecolor{plotcolor5}{rgb}{0.3455162, 0.39577619999999997, 0.29713619999999996}
\definecolor{plotcolor6}{rgb}{0.737738, 0.790457, 0.29265842999999997}
\definecolor{plotcolor7}{rgb}{0.740646, 0.768237, 0.291835335}
\definecolor{plotcolor8}{rgb}{0.74355452, 0.7460161599999999, 0.29101224}
\definecolor{plotcolor9}{rgb}{0.746463, 0.723795, 0.290189145}
\definecolor{plotcolor10}{rgb}{0.425297, 0.338621, 0.2466408}
\definecolor{plotcolor15}{rgb}{0.539404, 0.328498, 0.22511240000000002}
\definecolor{plotcolor20}{rgb}{0.691761, 0.409259, 0.22008260000000002}
\definecolor{plotcolor25}{rgb}{0.8499, 0.564538, 0.22587000000000002}
\definecolor{plotcolor30}{rgb}{0.944952, 0.717748, 0.24357699999999996}
\definecolor{plotcolor35}{rgb}{0.849015, 0.315572, 0.268106}
\definecolor{plotcolor50}{rgb}{0.844732, 0.300941, 0.48150175000000006}
\definecolor{plotcolor60}{rgb}{0.821381, 0.245522, 0.6544543999999999}

\definecolor{LightGray}{RGB}{220,220,220}

\pgfplotsset{compat=newest}

\theoremstyle{definition}
\newtheorem{hlp}{Helper}[chapter]
\newtheorem{defn}[hlp]{Definition}
\newtheorem{thm}[hlp]{Theorem}

\newtheorem{prop}[hlp]{Proposition}

\newtheorem{exmp}[hlp]{Example}
\newtheorem{rem}[hlp]{Remark}

\crefname{defn}{Definition}{Definitions}
\Crefname{defn}{Definition}{Definitions}
\crefname{thm}{Theorem}{Theorems}
\Crefname{thm}{Theorem}{Theorems}
\crefname{lem}{Lemma}{Lemmas}
\Crefname{lem}{Lemma}{Lemmas}
\crefname{rem}{Remark}{Remarks}
\Crefname{rem}{Remark}{Remarks}
\crefname{prop}{Proposition}{Propositions}
\Crefname{prop}{Proposition}{Propositions}
\crefname{cor}{Corollary}{Corollaries}
\Crefname{cor}{Corollary}{Corollaries}
\crefname{section}{Section}{Sections}
\Crefname{section}{Section}{Sections}
\crefname{equation}{}{}
\Crefname{equation}{}{}
\crefname{figure}{Figure}{Figures}
\Crefname{figure}{Figure}{Figures}
\crefname{table}{Table}{Tables}
\Crefname{table}{Table}{Tables}
\crefname{appendix}{Appendix}{Appendices}
\Crefname{appendix}{Appendix}{Appendices}
\crefname{exmp}{Example}{examples}
\Crefname{exmp}{Example}{examples}
\crefname{chapter}{Chapter}{Chapters}
\Crefname{chapter}{Chapter}{Chapters}

\labelcrefformat{subequation}{#2(#1)#3}
\labelcrefrangeformat{subequation}{#3(#1)#4 to #5(#2)#6}

\title{Thesis}
\author{Ramona Wolf}

\begin{document}


\begin{titlepage}

\begin{center}
	\vspace*{1cm}
	{\LARGE\bfseries
		Microscopic Models for Fusion Categories
		\par}
	\vspace*{10cm}
	\selectlanguage{ngerman}
	{
		Von der Fakultät für Mathematik und Physik\\[2mm]
		der Gottfried Wilhelm Leibniz Universität Hannover\\
		\vspace{8mm}
		zur Erlangung des akademischen Grades\\[2mm]
		{Doktor der Naturwissenschaften}\\[2mm]
		Dr.\ rer.\ nat.\\
		\vspace{8mm}
		genehmigte Dissertation von\par
	}
	\vspace*{1cm}
	{\LARGE
		M.\,Sc. Ramona Wolf
		\par}
	\vspace*{2cm}
	{
		2020
	}
\end{center}

\end{titlepage}

\setcounter{page}{0}

\thispagestyle{empty}

\noindent \textbf{Mitglieder der Prüfungskommission:}\vspace*{0.5\baselineskip}\\
Prof.\ Dr.\ Michèle Heurs (Vorsitzende)\\
Prof.\ Dr.\ Tobias J. Osborne (Betreuer)\\
Prof.\ Dr.\ Reinhard Werner

\vspace*{1.5\baselineskip}

\noindent \textbf{Gutachter:}\vspace*{0.5\baselineskip}\\
Prof.\ Dr.\ Tobias J. Osborne \\
Prof.\ Dr.\ Reinhard Werner\\
Prof.\ Dr.\ Jutho Haegeman

\vspace*{1.5\baselineskip}

\noindent Tag der Promotion: 07.12.2020

\cleardoublepage

\vspace*{8cm}

\thispagestyle{empty}

\begin{flushright}
	\large
	\noindent
	\hspace{40pt}
	\textit{``I almost wish I hadn't gone down that rabbit-hole--and yet--and yet--it's rather curious, you know, this sort of life!''}\\
	\vspace{5pt}
	\normalsize 
	-- Alice
	\newpage
\end{flushright}

\cleardoublepage

\selectlanguage{english}

\newpage

\frontmatter
\chapter*{Abstract}

Besides being a mathematically interesting topic on its own, subfactors have also attracted the attention of physicists, since there is a conjectured correspondence between these and Conformal Field Theories (CFTs). 
Although there is quite a persuasive body of evidence for this conjecture, there are some gaps: there exists a set of exceptional subfactors with no known counterpart CFT. Hence, it is necessary to develop new techniques for building a CFT from a subfactor. Here, it is useful to study the underlying mathematical structure in more detail: The even parts of every subfactor give rise to two Unitary Fusion Categories (UFCs), and it is a promising direction to study quantum spin systems constructed from these categories to find a connection to CFTs. 

The simplest example that requires new techniques for building a CFT is the Haagerup subfactor, since it is the smallest subfactor with index larger than $4$. In this thesis, we investigate the question whether there is a CFT corresponding to the Haagerup subfactor via lattice models in one and two dimensions. The first task here is to find the $F$-symbols of the fusion category since these are crucial ingredients for the construction of a physical model in all of the models we consider in this thesis. We then investigate the following models:
\begin{enumerate}
	\item The golden chain model, which is a one-dimensional spin chain whose ground state can be investigated in order to obtain information about the hypothetical CFT (such as the central charge).
	\item Quantum spin chains (such as the golden chain model) with defects. The construction of the vertices of the defect chain gives insight into a possible way to construct a Unitary Modular Tensor Category (UMTC) via the so-called annular category.
	\item The Levin-Wen model, which is a two-dimensional lattice model with an exactly solvable Hamiltonian that gives rise to a topological quantum field theory. Most interestingly for us, the excitations of the system yield a UMTC.
\end{enumerate}

We find that there is no evidence for a corresponding CFT from the investigation of the UFCs directly and it is necessary to expand these studies to the corresponding UMTC, which can, for instance, be obtained via the excitations of the Levin-Wen model.

\hspace{10pt}

\noindent
\textit{Keywords:} Conformal field theory, anyon chains, fusion categories
\newpage
\chapter*{Kurzzusammenfassung}

Abgesehen davon, dass Subfaktoren ein interessantes mathematisches Gebiet für sich darstellen, haben sie auch die Aufmerksamkeit von Physikern erregt, da vermutet wird, dass es einen Zusammenhang zwischen Subfaktoren und konformen Feldtheorien (CFT) gibt. Obwohl inzwischen eine überzeugende Menge an Hinweisen für die Gültigkeit dieser Vermutung erbracht wurde, existieren immer noch einige Lücken: Es gibt eine Reihe außergewöhnlicher Subfaktoren, für die keine entsprechende CFT bekannt ist. Daher ist es notwendig, neue Techniken für die Konstruktion einer CFT aus einem Subfaktor zu entwickeln. Hier ist es sinnvoll, die zugrunde liegende mathematische Struktur genauer zu untersuchen: Aus den sogenannten ,,even parts'' eines Subfaktors ergeben sich zwei unitäre Fusionskategorien (UFCs). Ein vielversprechender Ansatz ist, aus diesen Kategorien Quantenspinsysteme zu konstruieren und zu untersuchen, um eine Verbindung zu CFTs zu finden.

Das einfachste Beispiel, das neue Techniken zur Konstruktion einer CFT erfordert, ist der Haagerup Subfaktor, da er der kleinste Subfaktor mit einem Index größer als vier ist. In dieser Arbeit untersuchen wir mithilfe von ein- und zweidimensionalen Gittermodellen die Frage, ob eine CFT existiert, die zum Haagerup Subfaktor gehört. Die erste Aufgabe hierbei besteht darin, die $F$-Symbole der Kategorie zu berechnen, da diese bei allen Modellen, die in dieser Arbeit untersucht werden, einen entscheidenden Bestandteil der Konstruktion darstellen. Wir betrachten die folgenden Modelle:
	\begin{enumerate}
		\item Das sogenannte ,,Golden Chain''-Modell, bei dem es sich um eine eindimensionale Spinkette handelt, deren Grundzustand Informationen (wie zum Beispiel den Wert der zentralen Ladung) über die hypothetische CFT enthält.
		\item Quantenspinketten (wie die Golden Chain) mit Defekten. Die Konstruktion der Vertices, die die Kette mit Defekten bilden, liefert Einblicke in eine mögliche Konstruktion einer unitären modularen Tensorkategorie (UMTC) über die sogenannte Annulare Kategorie.
		\item Das Levin-Wen Modell, welches ein zweidimensionales Gittermodell mit exakt lösbarem Hamiltonian ist, das eine topologische Quantenfeldtheorie liefert. Am interessantesten für uns ist, dass die Anregungen des Systems eine UMTC ergeben.
	\end{enumerate}

Wir stellen fest, dass die Untersuchung der UFCs selbst keine Hinweise auf eine Haagerup CFT liefert, und schließen daraus, dass es notwendig ist, diese Untersuchung auf die entsprechende UMTC auszudehnen, die zum Beispiel über die Anregungen des Levin-Wen Modells konstruiert werden kann.

\hspace{10pt}

\noindent
\textit{Schlagwörter:} Konforme Feldtheorie, Anyon-Ketten, Fusionskategorien
\newpage
\chapter*{Acknowledgements}


First, I would like to thank my supervisor Tobias Osborne for introducing me to a fascinating area at the intersection of mathematics and physics, for guiding me through this thesis, and for always sharing valuable advice. I would also like to thank Reinhard Werner for his feedback and advice and for sharing all the interesting stories at lunch at the Mensa. A special thanks goes also to all (past and current) members of the Quantum Information Theory group in Hanover. You have made the day to day process of research a pleasure, especially on days where one of you brought cake to the institute!


During my time as a PhD student, I had the opportunity to visit many great places and to work with a lot of great, inspiring people. I would like to thank Kerstin Beer, Jacob Bridgeman, Dmytro Bondarenko, Cain Edie-Michell, Terry Farrelly, Alexander Hahn, Charles Lim, René Schwonnek, and Deniz Stiegemann for sharing their knowledge and passion with me and for the collaboration on several interesting projects. I would especially like to thank Charles Lim and René Schwonnek for giving me the possibility to visit Singapore and for having me at the CQT. I also especially thank Alexander Hahn for many helpful discussions on all aspects of fusion categories and the joint work on several projects that ended up being a part of this thesis.


There are many people who granted helpful advice and valuable comments on earlier versions of this thesis, who have found several typos and have helped to bring this thesis in its final form: Many thanks to Jacob, Tom, Cain, Alexander H., Alexander K., Ashley, and Christin.


%

Besonderer Dank geht auch an meine Familie, meine Eltern, meinen Bruder und meine beiden Großmütter. Auch wenn ihr keine Ahnung habt, was ich eigentlich mache, habt ihr mich immer unterstützt und jeder von euch hatte in den vergangenen Jahren auf die ein oder andere Weise einen großen Einfluss auf mich. Ohne euch wäre diese Doktorarbeit nicht möglich gewesen. Ich möchte mich auch herzlich bei meiner ,,zweiten Familie'', Sophia und Peter, für die Unterstützung im letzten Jahr bedanken! Und natürlich bei Louis, der süßesten Katze und bestem Kollegen in Homeoffice-Zeiten!

Zuletzt möchte ich mich bei Alex bedanken. Danke für deine Unterstützung, Liebe, und Geduld, und dass du mich daran erinnert hast, auch mal Pause zu machen.
\newpage
\chapter*{Publications}

Publications whose material appears in this thesis:\\

\begin{enumerate}
	\item Alexander Hahn and Ramona Wolf. \emph{Generalized string-net model for unitary fusion categories without tetrahedral symmetry} (2020). 
	\href{https://doi.org/10.1103/PhysRevB.102.115154}{Physical Review B \textbf{102}, 115154},  \href{https://arxiv.org/abs/2004.07045}{\texttt{arXiv:2004.07045}}\vspace{10pt}
	
	\item Jacob C. Bridgeman, Alexander Hahn, Tobias J. Osborne, and Ramona Wolf. \emph{Gauging defects in quantum spin systems: A case study} (2020). \href{https://doi.org/10.1103/PhysRevB.101.134111}{Physical Review B \textbf{101}, 134111}, \href{https://arxiv.org/abs/1910.10619}{\texttt{arXiv:1910.10619}}\vspace{10pt}

	\item Tobias J. Osborne, Deniz E. Stiegemann, and Ramona Wolf. \emph{The F-Symbols for the H3 Fusion Category} (2019). \href{https://arxiv.org/abs/1906.01322}{\texttt{arXiv:1906.01322}}\vspace{10pt}
	
	\item Kerstin Beer, Dmytro Bondarenko, Alexander Hahn, Maria Kalabakov, Nicole Knust, Laura Niermann, Tobias J. Osborne, Christin Schridde, Stefan Seckmeyer, Deniz E. Stiegemann, and Ramona Wolf. \emph{From categories to anyons: a travelogue} (2018). \href{https://arxiv.org/abs/1811.06670}{\texttt{arXiv:1811.06670}}
\end{enumerate}
\vspace{25pt}

Further publications:\\

\begin{enumerate}
	\setcounter{enumi}{4}
	\item René Schwonnek, Koon Tong Goh, Ignatius W. Primaatmaja, Ernest Y.-Z. Tan, Ramona Wolf, Valerio Scarani, and Charles C.-W. Lim. \emph{Robust Device-Independent Quantum Key Distribution} (2020). \href{https://arxiv.org/abs/2005.02691}{\texttt{arXiv:2005.02691}}\vspace{10pt}
	
	\item Kerstin Beer, Dmytro Bondarenko, Terry Farrelly, Tobias J. Osborne, Robert Salzmann, Daniel Scheiermann, and Ramona Wolf. \emph{Training deep quantum neural networks} (2020). \href{https://doi.org/10.1038/s41467-020-14454-2}{Nature Communications \textbf{11}, 808}, \href{https://arxiv.org/abs/1902.10445}{\texttt{arXiv:1902.10445}}\vspace{10pt}
	
	\item Yuan-Yuan Zhao, Guo-Yong Xiang, Xiao-Min Hu, Bi-Heng Liu, Chuan-Feng Li, Guang-Can Guo, René Schwonnek, and Ramona Wolf. \emph{Entanglement Detection by Violations of Noisy Uncertainty Relations: A Proof of Principle} (2019). \href{https://journals.aps.org/prl/abstract/10.1103/PhysRevLett.122.220401}{Physical Review Letters \textbf{122}, 220401}, \href{https://arxiv.org/abs/1810.05588}{\texttt{arXiv:1810.05588}}\vspace{10pt}
\end{enumerate}

\newpage
\tableofcontents
\newpage

\mainmatter

\chapter{Introduction}

Phase transitions are arguably a physical phenomenon that affects our life on a daily basis: Whenever you see ice cream melting in summer (solid to liquid), or steam coming from a tea kettle (liquid to gas), you have just witnessed a phase transition. These kind of phase transitions are thermal transitions, which means that they occur when the temperature $T$ of the system reaches a critical value $T=T_C$. Near this critical point, the correlation length (i.e., the distance over which correlation functions of physical quantities are non-trivial) eventually becomes much larger than the lattice spacing, which implies that the system can be described by a continuous field theory. An infinite correlation length furthermore implies that there is no natural length scale, hence the field theory that describes the system near the critical point $T_C$ is a conformal field theory.

There is another kind of phase transition apart from thermal phase transitions which occurs at $T=0$. When tuning a certain control parameter $g$ (for example, the strength of the magnetic field), phase transitions occur when the parameter reaches a critical point $g=g_C$. This is called a \emph{quantum phase transition} for the following reason: Since a system at zero temperature is always in its lowest-energy state, only quantum fluctuations (as opposed to thermal fluctuations) can be responsible for the transition. Similar to thermal phase transitions, there is no dimensional scale available, hence a system near a quantum phase transition is described by a conformal field theory.

Motivated by the practical applications described above, the study of Conformal Field Theories (CFTs) has also attracted researchers from a purely mathematical background. Mathematical investigations into the theory of quantum fields have led to a conjectured correspondence between subfactors and CFTs by Jones \cite{Jones1990}, which builds on earlier work by Doplicher and Roberts \cite{Doplicher1989}. Evidence for this conjecture was later found by Bischoff \cite{Bischoff2015,Bischoff2016}, who showed that it holds for all subfactors with index less than four, and others \cite{Calegari2010,Xu2017}. However, a general proof of this conjecture is far from being in sight, which implies that it probably requires new techniques to build CFTs from subfactors. The first interesting example here is the Haagerup subfactor \cite{Haagerup1994,asaeda_exotic_1999}, since it is the smallest (finite-depth, irreducible, hyperfinite) subfactor with index more than four. Interestingly, even though it is not known whether there is a corresponding CFT to this subfactor, it is still possible to indirectly study a hypothetical Haagerup CFT \cite{Evans2011}. Hence there is already some knowledge of the properties of the hypothetical CFT even though it is not known whether it exists, which motivates the central question of this thesis:
	\begin{quote}
		\centering
		\textit{Is there a conformal field theory that corresponds to the\\ Haagerup subfactor?}
	\end{quote}

\begin{figure}[t]
	\centering
	\begin{tikzpicture}
	\draw[->,>=stealth] (2.1,0.5) to node [below] {\footnotesize Even parts} (3.9,0.5);
	\draw[->,>=stealth] (6.1,0.5) to node [below] {\footnotesize Levin-Wen} (7.9,0.5);
	\draw[->,>=stealth] (5,1.1) to [bend left=20] node [above] {\footnotesize Anyon chains?} (13,1.1);
	\node at (7,-0.15) {\footnotesize Drinfeld};
	\node at (7,-0.45) {\footnotesize centre};
	\node at (7,-0.8) {\footnotesize Annular};
	\node at (7,-1.1) {\footnotesize category};
	\draw[->,>=stealth] (10.1,0.5) to node [below] {\footnotesize Anyon} (11.9,0.5);
	\node at (11,-0.05) {\footnotesize chains};
	\draw[fill=LightGray,rounded corners] (0,0) rectangle (2,1);
	\node at (1,0.5) {Subfactor};		
	\draw[fill=LightGray,rounded corners] (4,0) rectangle (6,1);
	\node at (5,0.5) {UFC};
	\draw[fill=LightGray,rounded corners] (8,0) rectangle (10,1);
	\node at (9,0.5) {UMTC};
	\draw[fill=LightGray,rounded corners] (12,0) rectangle (14,1);
	\node at (13,0.5) {CFT};
	\end{tikzpicture}
	\caption{\small \label{fig:path}\textbf{The path from subfactors to CFTs.} From a subfactor, one can get two unitary fusion categories (UFCs) via its even parts. Using one of several methods, it is possible to construct the corresponding Unitary Modular Tensor Category (UMTC). Since a UMTC describes anyons, we can build an anyon chain from it and thereby get information about the corresponding Conformal Field Theory (CFT). Building an anyon chain directly from the UFC is simpler than constructing the UMTC first, but it is not guaranteed to succeed.}
\end{figure}
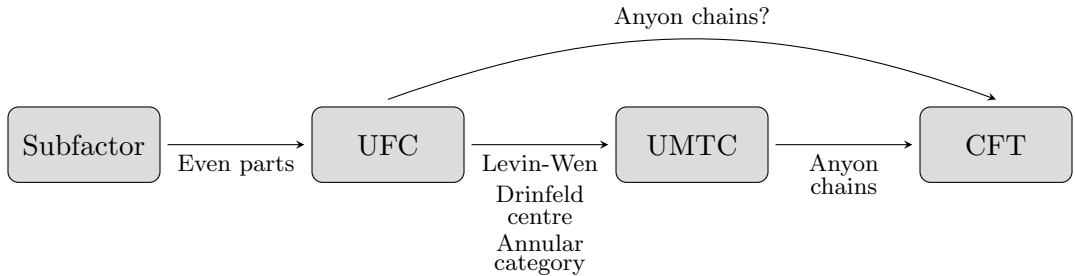

In this thesis, we investigate this question via microscopic models in one and two dimensions. The general approach, as depicted in \cref{fig:path}, is the following: The even parts of a finite-depth subfactor give rise to two Unitary Fusion Categories (UFCs). It is then possible to use one of several techniques to construct a Unitary Modular Tensor Category (UMTC) from the UFCs\footnote{Note that it is not important which of the UFCs we use for the construction since they give rise to the same Drinfeld centre.}, which is called the \emph{Drinfeld centre} or \emph{quantum double} of the subfactor: The direct construction is simply using the definition of the Drinfeld centre, but this approach is tedious and involves complicated and lengthy calculations. Another possibility is constructing a two-dimensional lattice model from one of the UFCs, the so-called Levin-Wen model \cite{Levin2005}. The excitations of the system form a UMTC, which is exactly the Drinfeld centre of the underlying fusion category. A third possibility is a construction via the so-called \emph{tube algebra}, or, more generally, annular category. We encounter these categories in the construction of a one-dimensional chain of particles that has defects in it, i.e., where we allow objects in the chain that do not come from the underlying category. However, this approach is also computationally very challenging.


\begin{figure}[t]
	\begin{tikzpicture}[scale=1]
	\node at (-0.5,2.1) {\textbf{3D}};
	\begin{scope}[scale=0.8]
	\draw[->,>=stealth] (0,0) -- (1,0);
	\draw[->,>=stealth] (0,0) -- (0,1);
	\draw[->,>=stealth] (0,0) -- (225:0.7071);
	\end{scope}
	\draw[->-] (0.75,1.25) arc (180:-180:0.75cm and 0.4cm);
	\draw[->-] (0.75,1.25) arc (180:-180:1.5cm and 0.6cm);
	\shade[ball color = LinkColor!90, opacity = 1] (0.75,1.25) circle (0.25cm);
	\shade[ball color = LinkColor!90, opacity = 1] (3,1.25) circle (0.25cm);
	\node at (4,1.5) {$C_1$};
	\node at (2.45,1.5) {$C_2$};
	\end{tikzpicture}\hspace{15pt}
	\begin{tikzpicture}[scale=1]
	\node at (-0.75,2.25) {\textbf{2D}};
	\draw[fill=LightGray,LightGray] (-0.75,-0.25) -- (4,-0.25) -- (5.5,2.25) -- (0.75,2.25) -- cycle;
	\begin{scope}[scale=0.8,xshift=-0.45cm,yshift=0cm]
	\draw[->,>=stealth] (0,0) -- (1,0);
	\draw[->,>=stealth] (0,0) -- (58:0.7071);
	\end{scope}
	\draw[->-] (0.75,1.25) arc (180:-180:0.75cm and 0.4cm);
	\draw[->-] (0.75,1.25) arc (180:-180:1.5cm and 0.6cm);
	\draw[fill=LinkColor!90,LinkColor!90] (0.75,1.25) ellipse (0.25cm and 0.18cm);
	\draw[fill=LinkColor!90,LinkColor!90] (3,1.25) ellipse (0.25cm and 0.18cm);
	\node at (4,1.5) {$C_1$};
	\node at (2.45,1.5) {$C_2$};
	\end{tikzpicture}
	\caption{\small \label{fig:anyons}\textbf{Topological differences between three- and two-dimensional particle circulation.} In three dimensions, the loop $C_1$ can always be continuously deformed to the path $C_2$, which in turn is contractible to a point. In two dimensions, the paths are topologically distinct: $C_1$ cannot be continuously deformed into $C_2$.}
\end{figure}
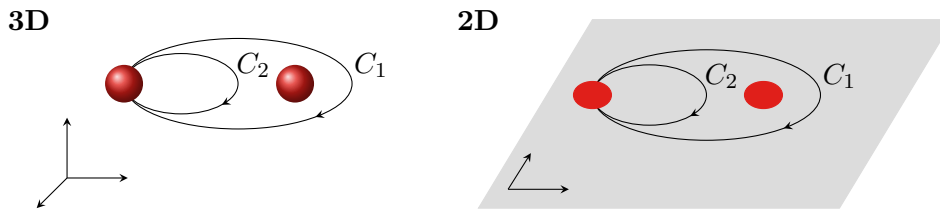

A UMTC is the mathematical model for exotic quasiparticles called \emph{anyons}, which only arise in two-dimensional systems. Their statistical behaviour is much less restricted than those of the commonly known fermions and bosons, which originates from the different topologies of space-time evolutions of point-like particles in two and three dimensions: Consider the process of exchanging two indistinguishable particles as depicted in \cref{fig:anyons}. In three dimensions, the path $C_1$ that describes how one particle encircles the other, can always be continuously deformed to the path $C_2$, which does not encircle the other particle. This path in turn is contractible to a point, denoted $0$, hence the wave function of the system must satisfy
	\begin{equation}
		\Psi(C_1)=\Psi(C_2)=\Psi(0).
	\end{equation}
The process of one particle encircling the other is equivalent to exchanging the two particles twice. Hence, the evolution of the system can be represented by the exchange operator $R$ such that $\Psi(C_1)=R^2\Psi(0)$. Since $C_1$ can be contracted to a point, the exchange operator has to fulfil $R^2=1$. This has only two solutions, $R=+1$ (which corresponds to bosons) and $R=-1$ (which corresponds to fermions). In two dimensions, however, the situation is different: Here, the path $C_1$ cannot be continuously deformed to $C_2$ since it cannot cross the particle. Therefore, the exchange operator $R$ is not required to square to the identity but can be represented by a complex phase or even by a unitary matrix. In the first case, the anyons are called \emph{abelian} and in the second case \emph{non-abelian} (which comes from the fact that matrices, in general, do not commute). These exotic exchange properties can be exploited to implement quantum gates and thus realising topological quantum computation. The fact that the statistics only rely on the topological properties of the system make this approach inherently robust to environmental noise.

There is a remarkable relation between these exotic particles and conformal field theories, which we exploit to draw the final part of the connection between subfactors and CFTs. By building a one-dimensional chain of anyons with nearest-neighbour interactions, it is possible to extract information about the corresponding CFT by studying phase transitions of the system. If the model exhibits a quantum phase transition, we will observe that the correlation length diverges (when the system size goes to infinity), and with it the entanglement entropy of the system. This relation allows us, for example, to get an estimate for the \emph{central charge}, which is an important characteristic of a CFT. This approach was successfully carried out for Fibonacci anyons \cite{feiguin_interacting_2007} and for $SO(5)_2$ anyons \cite{Finch2014}. 

In this thesis, we address several of the stages that are depicted in \cref{fig:path} for the example of the Haagerup subfactor. The first step is easy in this case since the two unitary fusion categories coming from the Haagerup subfactor, denoted $\Hi_1$ and $\Hi_2$, have already been constructed. Moreover, as described in \cref{ch:Haagerup}, all categories in the Morita equivalence class (which is a certain equivalence between categories) are known: Apart from the two UFCs that directly come from the subfactor, there is a third one in the Morita equivalence class, denoted $\Hd$, which is not equivalent to either of the two (only \emph{Morita equivalent}--this is an important difference). Since one characteristic of Morita equivalence is that the Drinfeld centre of the categories are equal, we can do the UMTC construction for any of the three categories. Here, the category $\Hd$ is especially easy for calculations since it is not only a fusion category but also a \emph{trivalent category}. As such, the graphical calculus within the category is particularly simple, as explained in \cref{ch:Trivalent}. However, in order to do graphical calculations within the category it is crucial to calculate the $F$\emph{-symbols}, which is done in \cref{ch:Haagerup}.

The second step, constructing the UMTC from the UFC, is more complicated. As mentioned above, the direct construction of the Drinfeld centre via its definition is an extraordinarily lengthy and complicated calculation (see \cref{sec:Drinfeld}). Another approach via the so-called tube algebra is briefly mentioned in \cref{ch:defects}, but not pursued further since it is computationally very involved. In \cref{ch:defects}, we make a slight detour from our journey from subfactors to CFTs by studying defects in spin chains. Although this does not directly contribute to answering the central question of this thesis, it is a natural generalisation of anyon chains and yields insight into the study of topological phases. Here, we also encounter the so-called \emph{annular category}, whose representations are connected to the Drinfeld centre. For the purpose of finding the UMTC, the most promising approach is the Levin-Wen model, even though it has its own unique challenges: The original model can only be constructed from UFCs that fulfil a certain symmetry condition, the so-called \emph{tetrahedral symmetry}, which is not fulfilled for the Haagerup fusion categories. Hence, the first step here is to find a generalisation of this model that also includes categories without tetrahedral symmetry, which is presented in \cref{ch:LW}. It turns out that in the general Levin-Wen model, finding the excitations is a more complicated task because of the lack of some symmetries. However, there is a promising approach of finding the excitations by constructing a tensor-network representation of the ground state of the model.

Since the possible ways of constructing the Drinfeld centre of the UFCs all involve challenging tasks and time-consuming calculations, it is worth trying a simpler approach first: Although it is not guaranteed to succeed, it is possible to build an anyon chain from the UFC directly instead of using the UMTC. Hence, in \cref{sec:anyons}, we construct the anyon chain from the category $\Hd$ and numerically study the behaviour of the ground state. Unfortunately, we find that using the UFC directly does not give rise to a quantum phase transition, hence it is inevitable that we must construct the UMTC in order to study the connection to the CFT.

This thesis is organized in two parts: In the first part, we introduce the mathematical concepts that are essential for building physical models from fusion categories. Here, we begin with an introduction to the basic concepts of category theory in \cref{ch:cats} and introduce the special case of trivalent categories in \cref{ch:Trivalent}. We then give the basic concepts of the theory of subfactors in \cref{ch:subfactor} with an emphasis on the connection to fusion categories. The final chapter of Part 1, \cref{ch:Haagerup}, discusses the Haagerup subfactor in detail and presents the corresponding fusion categories, especially the category $\Hd$, which is the motivating example of this thesis.

In the second part of this thesis we discuss several microscopic models which are built from fusion categories. We begin with describing anyon chains in \cref{sec:anyons} and present several numerical results for a chain built from the fusion category $\Hd$. In \cref{ch:defects}, we extend this concept by allowing defects to be part of this chain. We study this model for the category $\mathbf{Vec}(\mathbb{Z}/2\mathbb{Z})$. A model that connects fusion categories directly to their quantum double is the Levin-Wen model, which is discussed in \cref{ch:LW}. Here, we present a modification to the original model such that it is applicable to the category $\Hd$. Finally, we summarize the results of this thesis in \cref{ch:conclusion} and discuss potential further directions for future research.

\part{Mathematical tools}

\chapter{Basics of category theory}
\label{ch:cats}

	Category theory takes a fairly general view of mathematical concepts. Instead of concentrating on details, it is the patterns and structures that are brought into focus here. The advantage of this approach is that methods that have proven helpful in one area of mathematics can be transferred to other areas, thus providing new (and possibly easier) proof techniques in these areas. The concept of categories was introduced by Samuel Eilenberg and Saunders Mac Lane in 1945 during their study of algebraic topology, while attempting to understand the processes that preserve mathematical structures \cite{Eilenberg1945}. A detailed treatment of the topic can be found in \cite{MacLane1998} and \cite{Borceux1994}, and more introductory texts are \cite{Leinster2009} and \cite{Awodey2010}. A more physics-oriented explanation of categories can be found in \cite{Baez2010} and \cite{Beer2018}, for example.

	In this thesis, we want to employ category theory to describe complex quantum systems, which is why we restrict ourselves to the basic definitions of category theory here. The systems of interest are \emph{anyons}, which are modelled by a specific type of category, namely Unitary Modular Tensor Categories (UMTCs). These come with the additional structure of a tensor category which allows us to describe composite systems and, moreover, the interaction of particles. Furthermore, they provide a graphical calculus that allows for a simple way of calculating properties of the system. We also introduce another specific type of categories, so-called trivalent categories, which describe some specific examples of anyon models in the next chapter. Although they cannot be used to generally treat anyons, it is worth to study these categories since their graphical calculus is extraordinarily simple. This makes trivalent categories an excellent source of simple examples for constructions that heavily rely on the graphical calculus.
	
	Although we limit ourselves to the basics of category theory here, we cannot avoid presenting a list of definitions, since learning a new mathematical concept is much like learning a new language: first, you need to learn a lot of vocabulary!

\section{Basic definitions}

	\begin{defn}[Category]
		\label{def:cat}
		 A category\index{category} $\C$ consists of:
		 	\begin{itemize}
		 		\item a collection\footnote{We use the term \emph{collection} to indicate that this is not necessarily a set, but is, in general, too large to be a set. Think, for example, of the collection of all sets, which is not a set itself. This avoids running into Russel's paradox.} $\Obj(\C)$ of objects,
		 		\item for each pair of objects $X,Y\in\Obj(\C)$, a collection $\Hom_\C(X,Y)$\footnote{The notation ``Hom'' is of historical origin and stands for homomorphism, which appeared in one of the earliest examples of a category.} of morphisms\index{morphism} (or maps) from $X$ to $Y$,
		 		\item for each $X,Y,Z\in\Obj(\C)$, a function called composition:
		 				\begin{align}
			 				\Hom_\C(Y,Z)\times\Hom_\C(X,Y)&\to\Hom_\C(X,Z)\\
			 				(g,f)&\mapsto g\circ f,
		 				\end{align}
		 		\item for every object $X\in\Obj(\C)$, an element $\id_X\in\Hom_\C(X,X)$ called the identity on $X$.
		 	\end{itemize}
	 	These satisfy the following properties:
	 		\begin{enumerate}
	 			\item associativity: for each $f\in\Hom_\C(W,X)$, $g\in\Hom_\C(X,Y)$, and $h\in\Hom_\C(Y,Z)$ it holds that
	 					\begin{equation}
		 					(h\circ g)\circ f=h\circ(g\circ f),
	 					\end{equation} 
	 			\item identity laws: for every morphism $f\in\Hom_\C(X,Y)$, we have $f\circ\id_X=f=\id_Y\circ f$.
	 		\end{enumerate}
	\end{defn}

	\begin{rem} For convenience and/or clarity, we sometimes switch between different notations throughout this thesis:
		\begin{enumerate}
			\item We often write $X\in\C$ instead of $X\in\Obj(\C)$.
			\item We often write a morphism $f\in\Hom_\C(X,Y)$ as $f:X\to Y$ or depict it as
					\begin{figure}[h!]
					\begin{tikzpicture}
						\node (X) at (0,0) {$X$};
						\node (Y) at (1.5,0) {$Y$};
						\draw[->] (X) to node [above] {$f$} (Y);
					\end{tikzpicture}.
					\end{figure}
			\item We write $\Hom(X,Y)$ instead of $\Hom_\C(X,Y)$ whenever the category the morphisms live in is clear from the context.
		\end{enumerate}
	\end{rem}

	\begin{rem}
		If a morphism $f\in\Hom_\C(X,Y)$, we call $X$ the domain\index{domain} and $Y$ the codomain\index{codomain} of $f$. Every morphism in a category has a definite domain and a definite codomain.
	\end{rem}

	\begin{rem}
		Throughout this thesis, we will often use the terminology that a diagram commutes. This has the following meaning: Whenever there are two paths from an object $X$ to another object $Y$, the map from $X$ to $Y$ that is obtained by composing the maps along one path equals the map obtained by composing maps along the other path. For example, consider the diagram
		\begin{figure}[H]
			\begin{center}
				\begin{tikzpicture}[scale=1.5]
				\node (A) at (0,0) {$A$};
				\node (B) at (2,0) {$B$};
				\node (C) at (0,-1) {$C$};
				\node (D) at (1,-1) {$D$};
				\node (E) at (2,-1) {$E$};
				\draw[->] (A) to node [above] {\small $f$} (B);
				\draw[->] (B) to node [right] {\small $g$} (E);
				\draw[->] (A) to node [left] {\small $h$} (C);
				\draw[->] (C) to node [below] {\small $i$} (D);
				\draw[->] (D) to node [below] {\small $j$} (E);
				\end{tikzpicture}
			\end{center}
		\end{figure}
		\noindent
		This diagram is said to commute if $g\circ f=j\circ i\circ h$.
	\end{rem}

	\begin{defn}[Isomorphism]
		A morphism $f:X\to Y$ in a category $\C$ is an isomorphism\index{isomorphism} if it has an inverse, i.e., there exists a morphism $g:Y\to X$ such that $g\circ f=\id_X$ and $f\circ g=\id_Y$.
	\end{defn}
	This definition of an isomorphism is an example of the abstract way in which notions or concepts can be defined in the language of category theory. It has the advantage that, unlike other common definitions of isomorphisms, it does not use any additional structures than those given by the definition of the category. For example, an isomorphism of sets is usually defined as a bijective function, a definition that makes use of the \textit{elements} of the objects. 
	
	To illustrate the abstract definition, we have a look at some examples that include familiar mathematical objects such as sets and groups:
	
	\begin{exmp}[Category of Sets]
		\label{ex:sets}
		In the category of sets\index{Sets@$\mathbf{Sets}$}, denoted \textbf{Sets}, the objects are all possible sets. This implies that the collection of all objects of \textbf{Sets} itself is \emph{not} a set, but a class. The collection of morphisms from a set $X$ to a set $Y$, $\Hom(X,Y)$, consists of all maps $f:X\to Y$. Composition of morphisms then simply corresponds to the composition of maps. The identity morphism $\id_X:X\to X$ exists for every $X\in\Obj(\mathbf{Sets})$ and is given by the map $\mathbb{I}(a)=a$ for all $a\in X$.
	\end{exmp}

	\begin{exmp}[Group]
		\label{ex:grp}
		A group\index{group} $G$ can be described as a category $\mathcal{G}$ with a single object $\star$ and where all morphisms are isomorphisms. Here, $\Obj(\mathcal{G})=\{\star\}$ and there is only one collection of morphisms, namely $\Hom(\star,\star)$ which is given by the elements $g$ of the group $G$:
			\begin{equation}
				\Hom(\star,\star)=\{g|g\in G\}.
			\end{equation}
		Composition corresponds to the group operation (for example, multiplication) and the identity morphism is given by the unit object of the group. Since in a group every element has a unique inverse, every morphism in $\mathcal{G}$ is an isomorphism. This example illustrates a key guiding principle in category theory, namely, capturing the interesting structure within a category by the morphism sets $\Hom(X,Y)$ rather than in terms of complicated object classes.
	\end{exmp}

	Within a category, we are usually interested in connections, or maps, between objects. We can ask the same question about categories themselves, i.e., if there is a sensible notion of a map between categories. The answer to this question is yes, and the mathematical objects that describe these maps are called \emph{functors}\index{functor}:

	\begin{defn}[Functor]
		Let $\C$ and $\mathcal{D}$ be categories. A functor $F:\C\to\mathcal{D}$ between these categories consists of
			\begin{itemize}
				\item a function that assigns objects in $\C$ to objects in $\mathcal{D}$:
						\begin{align}
							\Obj(\C)&\to\Obj(\mathcal{D})\\
							X&\mapsto F(X),
						\end{align}
				\item for each $X,Y\in\C$, a function that maps morphisms in $\C$ to morphisms in $\mathcal{D}$:
						\begin{align}
							\Hom_C(X,Y)&\to\Hom_\mathcal{D}(F(X),F(Y))\\
							f&\mapsto F(f),
						\end{align}
			\end{itemize}
		which satisfy the following axioms: 
			\begin{enumerate}
				\item $F(g\circ f)=F(g)\circ F(f)$ for all composable morphisms $f,g$,
				\item $F(\id_X)=\id_{F(X)}$ for all $X\in\C$.
			\end{enumerate}
	\end{defn}

	\noindent
	Functors preserve the structure of a category: They preserve domains and codomains, identities, and composition. Hence, a functor $F:\C\to\mathcal{D}$ transforms a diagram in $\C$ into a---probably distorted---diagram in $\mathcal{D}$:
	\begin{figure}[H]
		\begin{center}
			\begin{tikzpicture}[scale=0.85]
				\node (A) at (0,0) {$X$};
				\node (B) at (3,0) {$Y$};
				\node (D) at (3,-2) {$Z$};
				\draw[->] (A) to node [above] {\small $f$} (B);
				\draw[->] (B) to node [right] {\small $g$} (D);
				\draw[->] (A) to node [left] {\small $g\circ f\ \ $} (D);
				\node (B) at (3,-3) {$F(Y)$};
				\node (C) at (0,-5) {$F(X)$};
				\node (D) at (3,-5) {$F(Z)$};
				\draw[->] (B) to node [right] {\small $F(g)$} (D);
				\draw[->] (C) to node [left] {\small $F(f)\ $} (B);
				\draw[->] (C) to node [below] {\small $F(g\circ f)$} (D);
				\node (C1) at (-2,-1) {$\C$};
				\node (C2) at (-2,-4) {$\mathcal{D}$};
				\draw[->] (C1) to node [left] {\small $F$} (C2);
			\end{tikzpicture}
		\end{center}
	\end{figure}

	\begin{defn}[Natural transformation]
		Let $\C$ and $\mathcal{D}$ be categories and $F:\C\to\mathcal{D}$ and $G:\C\to\mathcal{D}$ be functors between them. A natural transformation\index{natural transformation} $\alpha:F\to G$ is a family of morphisms
			\begin{equation}
				\left(F(X)\xrightarrow{\alpha_X}G(X)\right)_{X\in\Obj(\C)}
			\end{equation}
		in $\mathcal{D}$, such that for every morphism $f:X\to X'$ in $\mathcal{C}$, the diagram
			\begin{figure}[H]
				\begin{center}
					\begin{tikzpicture}[scale=1.3]
					\node (A) at (0,0) {$F(X)$};
					\node (B) at (2,0) {$F(X')$};
					\node (C) at (0,-1.5) {$G(X)$};
					\node (D) at (2,-1.5) {$G(X')$};
					\draw[->] (A) to node [above] {\small $F(f)$} (B);
					\draw[->] (B) to node [right] {\small $\alpha_{X'}$} (D);
					\draw[->] (A) to node [left] {\small $\alpha_X$} (C);
					\draw[->] (C) to node [below] {\small $G(f)$} (D);
					\end{tikzpicture}
				\end{center}
			\end{figure}
		\noindent
		commutes. The maps $\alpha_X$ are called the components of $\alpha$.
	\end{defn}

	In the same way that functors can be seen as maps between categories, i.e., as morphisms in the category of categories, natural transformations can be seen as maps between functors. There is indeed a notion of composition of natural transformations: Given two natural transformations $\alpha:F\to G$ and $\beta:G\to H$, with $F,G,H:\C \to\mathcal{D}$, there is a composite natural transformation $\beta\circ \alpha:F\to H$ which is defined by 
		\begin{equation}
			(\beta\circ \alpha)_X=\beta_X\circ \alpha_X
		\end{equation}
	for all $X\in\Obj(\C)$. There is also an identity natural transformation $\id_F:F\to F$, defined by $(\id_F)_X=\id_{F(X)}$ for any functor $F$. Therefore, for any two categories $\C$ and $\mathcal{D}$, one can construct a category whose objects are the functors from $\C$ to $\mathcal{D}$ and whose morphisms are the natural transformations between them. This is called the \emph{functor category}\index{functor category} from $\C$ to $\mathcal{D}$ and denoted $[\C,\mathcal{D}]$.

	\begin{defn}[Natural isomorphism]
		Let $\C$ and $\mathcal{D}$ be categories and $F,G:\C\to\mathcal{D}$ be functors. A natural transformation $\alpha:F\to G$ is a natural isomorphism\index{natural isomorphism} if $\alpha_X:F(X)\to G(X)$ is an isomorphism for all $X\in\Obj(\C)$.
	\end{defn}

	\begin{rem}
		There is an equivalent definition of a natural isomorphism: Given two categories $\C,\mathcal{D}$, a natural isomorphism between functors from $\C$ to $\mathcal{D}$ is an isomorphism in the category $[\C,\mathcal{D}]$.
	\end{rem}

\section{Monoidal categories}

	Monoidal categories provide additional structure beyond the basic definitions of a category, namely the notion of a tensor product, both between objects and between morphisms. A detailed treatment of tensor categories can be found in \cite{Etingof2015}. For a more introductory text we refer to \cite{bakalov_lectures_2001}.
	
	\begin{defn}[Monoidal category]
		A monoidal category\index{monoidal category} is a sextuple $(\C,\otimes,\alpha,\mathbf{1},l,r)$ where $\C$ is a category equipped with a bifunctor
			\begin{equation}
				\otimes:\C\times\C\to\C,
			\end{equation}
		called the tensor product bifunctor. The maps $\alpha$, $l$, and $r$ are natural isomorphisms:
			\begin{align}
				\alpha_{X,Y,Z}:(X\otimes Y)\otimes Z&\to X\otimes(Y\otimes Z),\hspace{10pt}X,Y,Z\in\C,\\
				l_X:\mathbf{1}\otimes X&\to X,\hspace{10pt}X\in\C,\\
				r_X:X\otimes\mathbf{1}&\to X,\hspace{10pt}X\in\C.
			\end{align}
		where $\alpha$ is called the associator\footnote{This notation is not to be confused with the natural transformation, that is also denoted $\alpha$. Since we do not need natural transformations in the remainder of this chapter, we use $\alpha$ to denote the associator from now on.} (or associativity isomorphism), $l$ and $r$ are called left and right unit constraints, and $\mathbf{1}\in\Obj(\C)$ is an object of $\C$ called the unit object, such that the following axioms are fulfilled:
			\begin{enumerate}
				\item \textbf{The pentagon axiom:\index{pentagon axiom}} The following diagram commutes for all objects $W,X,Y,Z\in\Obj(\C)$:
						\begin{figure}[H]
							\centering

						\end{figure}
				\item \textbf{The triangle axiom:\index{triangle axiom}} The following diagram commutes for all objects $X,Y\in\Obj(\C)$:
						\begin{figure}[H]
							\centering
							%
						\end{figure}
				\noindent
				The triangle axiom ensures that adding an removing trivial lines (i.e., units) is consistent with the associativity isomorphism.
			\end{enumerate}
	\end{defn}

	\begin{prop}
		The following diagrams commute for all objects $X,Y\in\Obj(\C)$:
			\begin{figure}[H]
				\centering
				%
			\end{figure}
	\end{prop}
	\begin{proof}
		See \cite[Proposition 2.2.4]{Etingof2015}.
	\end{proof}

\begin{exmp}
	The category \textbf{Sets} introduced in \cref{ex:sets} is a monoidal category where the tensor product is given by the Cartesian product: For two sets $X,Y$ it is defined as
		\begin{equation}
			X\times Y=\{(x,y)|x\in X,y\in Y\}.
		\end{equation}
	The unit object with respect to the tensor product is a one element set.
\end{exmp}

\begin{exmp}
	Consider a group $G$ and let $A$ be an abelian group with operation written multiplicatively. We define the category  $\mathcal{C}_G(A)$ in the following way:
		\begin{itemize}
			\item Objects in $\C_G(A)$ are labelled by the elements of $G$: $\Obj(\C_G(A))=\{\delta_g|g\in G\}$.
			\item Morphisms are defined via the abelian group $A$:
					\begin{equation}
						\Hom\left(\delta_{g_1},\delta_{g_2}\right)=\begin{cases}
							\emptyset, & \mathrm{if\ } g_1\neq g_2\\
							A, & \mathrm{if\ } g_1= g_2.
						\end{cases}
					\end{equation}
		\end{itemize}
	The tensor product of two objects $\delta_{g_1},\delta_{g_2}$ is defined as $\delta_{g_1}\otimes\delta_{g_2}=\delta_{g_1g_2}$. For two morphisms $a,b$, it is given by the multiplication operation in $A$: $a\otimes b=ab$. The unit object with respect to the tensor product is given by the unit element of $G$. This example illustrates that in a monoidal category, the tensor product $X\otimes Y$ does not need to be isomorphic to $Y\otimes X$.
\end{exmp}

In a monoidal category n-fold tensor products can be constructed from any ordered sequence of objects $X_1,\dots,X_n$, namely any parenthesising of the expression $X_1\otimes\dots\otimes X_n$, which are, in general, distinct objects in the category $\C$. However, for $n=3$, there is a canonical identification of the two different parenthesisations of the tensor product, namely $(X_1\otimes X_2)\otimes X_3$ and $X_1\otimes (X_2\otimes X_3)$, given by the associativity isomorphism. A straightforward combinatorial argument then shows that for $n\ge 3$, any two parenthesized products of $X_1,\dots,X_3$ can be identified using a chain of associativity isomorphisms.

It would be convenient if we could say that for this reason, we can safely neglect any parentheses in any monoidal category by identifying all possible parenthesized products with each other. Unfortunately, this leads to the following problem: For $n\ge 4$, there is, in general, more than one possible chain of associativity isomorphisms that maps one parenthesising to another, and it is not clear whether they provide the same identification. For $n=4$, this problem is solved because by the pentagon axiom, we demand that the two possible chains of associativity isomorphisms yield the same identification. But what about the case $n>4$? This is solved by the following theorem of Mac Lane:

	\begin{thm}[Mac Lane's coherence theorem]
		\index{Mac Lane's coherence theorem}
		Let $X_1,\dots,X_n\in\C$. Let $P_1,P_2$ be any two parenthesized products of $X_1,\dots,X_n$ (in this order) with arbitrary insertions of the unit object $\mathbf{1}$. Let 
		$f,g:P_1\to P_2$ be two isomorphisms, obtained by composing associativity and unit isomorphisms and their inverses possibly tensored with identity morphisms. Then $f=g$.
	\end{thm}
	\begin{proof}
		See \cite{MacLane1998} or \cite[Theorem 2.9.2]{Etingof2015}.
	\end{proof}

	\begin{defn}[Equivalence]
		\index{equivalence of categories}
		Two categories $\C$ and $\mathcal{D}$ are called equivalent if there exist functors $F:\C\to\mathcal{D}$ and $G:\mathcal{D}\to\C$ such that
		\begin{align}
		G\circ F&\cong \id_\C\\
		F\circ G&\cong \id_\mathcal{D}.
		\end{align}
		The functors $F$ and $G$ are called equivalences. 
	\end{defn}

	\begin{defn}[Monoidal functor]
		\index{monoidal functor}
		Let $(\C,\otimes,\alpha,\mathbf{1},l,r)$ and $(\C',\otimes',\alpha',\mathbf{1}',l',r')$ be two monoidal categories. A monoidal functor from $\C$ to $\C'$ is a pair $(F,J)$, where $F:\C\to C'$ is a functor, and 
			\begin{equation}
				J_{X,Y}:F(X)\otimes'F(Y)\to F(X\otimes Y)
			\end{equation}
		is a natural isomorphism such that $F(\mathbf{1})$ is isomorphic to $\mathbf{1}'$ and the following diagram commutes for all $X,Y,Z\in\Obj(\C)$:
			\begin{figure}[H]
				\centering
				\begin{tikzpicture}
					\node (1) at (0,0) {$(F(X)\otimes'F(Y))\otimes'F(Z)$};
					\node (2) at (8,0) {$F(X)\otimes'(F(Y)\otimes'F(Z))$};
					\node (3) at (0,-2) {$F(X\otimes Y)\otimes'F(Z)$};
					\node (4) at (8,-2) {$F(X)\otimes'(F(Y\otimes Z))$};
					\node (5) at (0,-4) {$F((X\otimes Y)\otimes Z)$};
					\node (6) at (8,-4) {$F(X\otimes (Y\otimes Z))$};
					\draw[->] (1) to node[above] {\small $\alpha'_{F(X),F(Y),F(Z)}$} (2);
					\draw[->] (1) to node[left] {\small $J_{X,Y}\otimes'\id_{F(Z)}$} (3);
					\draw[->] (3) to node[left] {\small $J_{X\otimes Y,Z}$} (5);
					\draw[->] (2) to node[right] {\small $\id_{F(X)}\otimes' J_{Y,Z}$} (4);
					\draw[->] (4) to node[right] {\small $J_{X,Y\otimes Z}$} (6);
					\draw[->] (5) to node[below] {\small $F(\alpha_{X,Y,Z})$} (6);
				\end{tikzpicture}
			\end{figure}
		A monoidal functor is said to be an equivalence of monoidal categories if it is an equivalence of ordinary categories.
	\end{defn}

	\begin{defn}[Strict]
		A monoidal category is strict\index{strict category} if the associativity isomorphism and the unit constraints are identities, hence for all objects $X,Y,Z\in\Obj(\C)$ there are equalities
			\begin{align}
				(X\otimes Y)\otimes Z&=X\otimes(Y\otimes Z)\\
				X\otimes\mathbf{1}&=X=\mathbf{1}\otimes X.
			\end{align}
	\end{defn}

	\begin{thm}
		\label{thm:strictness}
		Every monoidal category is monoidally equivalent to a strict monoidal category.
	\end{thm}

	\begin{proof}
		See \cite[Theorem 2.8.5]{Etingof2015}.
	\end{proof}

	\begin{rem}
		Note that although \cref{thm:strictness} implies you can always turn your category into a strict one, it is not necessarily useful to do so. For example, in order to make a monoidal category strict one may have to add new objects to it (see \cite[Remark 2.8.6]{Etingof2015}). But if you already have an idea of what the objects in your category should be, you might not want to abandon this to make the category strict, even though it would allow you to work with trivial associativity and unit constraints.
	\end{rem}

	\begin{rem}
		Instead of turning your category into a strict one, it is sometimes more useful to turn it into a \emph{skeletal} one. A skeletal\index{skeletal category} category is one which has only one object in every isomorphism class, i.e., if two objects are isomorphic to each other, they are actually equal. Any category is equivalent to a skeletal one (simply by the axiom of choice), however, a monoidal category does not have to be equivalent to a category that is strict and skeletal at the same time. Although the associators are non-trivial in a skeletal category, these categories are still often particularly nice to work with in actual calculations.
	\end{rem}

\subsection*{Graphical calculus of monoidal categories}

Monoidal categories, often equipped with additional structure, come with a graphical language of \emph{string diagrams}. This often helps to visualise and simplify relations and equations in the category. Such a calculus has first been introduced in \cite{Joyal1991} and has been widely used afterwards. Some introductions with different emphases are \cite{kassel_quantum_1995}, \cite{Selinger2010} and \cite{HV19}. The convention we use throughout this thesis is that diagrams are always read from bottom to top (unless otherwise stated).

Given a category $\C$, a morphism $f:X\to Y$ is depicted as a two-dimensional planar diagram
	\begin{equation}

	\end{equation}
\noindent
Every morphism has an ingoing leg (the domain) and an outgoing leg (the codomain). The identity morphism $\id_X:X\to X$ is represented by a straight line:
	\begin{equation}
		%
	\end{equation}
\noindent
Composition of morphisms corresponds to connecting legs: Given two morphisms $f:X\to Y$ and $g: Y\to Z$, the composed morphism $g\circ f:X\to Z$ is depicted
	\begin{equation}
		%
	\end{equation}
Note that in this graphical calculus associativity of morphisms is implicit. Consider the three morphisms $f:W\to X$, $g:X\to Y$ and $h:Y\to Z$, then both parenthesis structures, $(h\circ g)\circ f$ and $h\circ (g\circ f)$ are depicted
	\begin{figure}[H]
		\centering
		%
	\end{figure}

The previous diagrams are all valid in any category since they just visualise the concepts and properties mentioned in \cref{def:cat}. More specific to a monoidal category is the visualisation of the monoidal product (or tensor product). Given two morphisms $f:X\to X'$ and $g:Y\to Y'$, their tensor product $f\otimes g:X\otimes Y\to X'\otimes Y'$ is the horizontal juxtaposition of the individual diagrams: 
	\begin{equation}
		%
	\end{equation}
The unit object $\mathbf{1}$ is often depicted either as a dashed line or by the absence of any line when it is clear from the context, e.g., the morphism $f:\mathbf{1}\to X\otimes Y$ can be drawn in the following ways:
	\begin{equation}
		%
	\end{equation}

We will use and extend the here presented graphical calculus in the following sections to illustrate different concepts and properties of categories, such as duals of morphisms and braiding structures, and to explain calculations within them, e.g., how to define and evaluate the trace of an object or a morphism graphically.

\section{Fusion categories}
\label{sec:fusioncat}

Fusion categories are monoidal categories that are especially easy to work with since the morphism spaces are linear vector spaces and all properties can be expressed in terms of the simple objects of the category. However, before we state the definition of a fusion category, we need to introduce several adjectives.

	\begin{defn}[Left duals]
		In a monoidal category $(\C,\otimes,\alpha,\mathbf{1},l,r)$, an object $X^*\in\Obj(\C)$ is said to be a left dual\index{left dual} of $X\in\Obj(\C)$ if there exist morphisms
			\begin{align}
				\ev_X&:X^*\otimes X\to\mathbf{1}\\
				\coev_X&:\mathbf{1}\to X\otimes X^*,
			\end{align}
		depicted
			\begin{align}

			\end{align}
		such that the compositions
			\begin{align}
				X\xrightarrow{\coev_X\otimes\,\id_X}(X\otimes X^*)\otimes X&\xrightarrow{a_{X,X^*,X}}X\otimes(X^*\otimes X)\xrightarrow{\id_X\otimes\,\ev_X}X\\
				X^*\xrightarrow{\id_{X^*}\otimes\,\coev_X}X^*\otimes(X\otimes X^*)&\xrightarrow{a_{X^*,X,X^*}^{-1}}(X^*\otimes X)\otimes X^*\xrightarrow{\ev_X\otimes\,\id_{X^*}}X^*
			\end{align}
		are the identity morphisms, which is equivalent to the following equations: 
			\begin{align}
				%
\ .
			\end{align}
		These morphisms are called evaluation\index{evaluation} and coevaluation\index{coevaluation}.
	\end{defn}

	\begin{defn}[Right duals]
		In a monoidal category $(\C,\otimes,\alpha,\mathbf{1},l,r)$, an object $\lstar X\in\Obj(\C)$ is said to be a right dual\index{right dual} of $X\in\Obj(\C)$ if there exist morphisms
		\begin{align}
		\ev_X'&:X\otimes \lstar X\to\mathbf{1}\\
		\coev_X'&:\mathbf{1}\to \lstar X\otimes X,
		\end{align}
		depicted
			\begin{align}
				%
			\end{align}
		such that the compositions
		\begin{align}
		X\xrightarrow{\id_X\otimes\,\coev_X'}X\otimes(\lstar X\otimes X)&\xrightarrow{a_{X,\lstar X,X}^{-1}}(X\otimes{}^*X)\otimes X\xrightarrow{\ev_X'\otimes\,\id_X}X\\
	\lstar X\xrightarrow{\coev_X'\otimes\,\id_{\lstar X}}(\lstar X\otimes X)\otimes \lstar X&\xrightarrow{a_{\lstar X,X,\lstar X}}\lstar X\otimes (X\otimes \lstar X)\xrightarrow{\id_{\lstar X}\otimes\,\ev_{X}'}\lstar X
		\end{align}
		are the identity morphisms, which is equivalent to the following equations:
			\begin{align}
				%
			\end{align}
	\end{defn}

	\begin{rem}
		Note that for an object $X\in\Obj(\C)$, the left (respectively, right) dual object is unique up to a unique isomorphism.
	\end{rem}

	\begin{rem}
		If $X^*\in\C$ is a left dual of an object $X\in\C$, then it is obvious that $X$ is a right dual of $X^*$ with $\ev'_{X^*}=\ev_X$ and $\coev'_{X^*}=\coev_X$, and vice versa. This implies that $\lstar (X^*)\cong X\cong (\lstar X)^*$ for any object in $\C$ with left and right duals. Furthermore, in any monoidal category, it holds that $\mathbf{1}^*=\lstar \mathbf{1}=\mathbf{1}$.
	\end{rem}

	\begin{rem}
		\label{rem:dualmorphs}
		We can use the definitions of evaluation and coevaluation to construct duals of morphisms: If $X$ and $Y$ are objects in a category $\C$ that have left duals $X^*,Y^*$, and $f:X\to Y$ is a morphism between them, one defines the left dual $f^*:Y^*\to X^*$ of $f$ by the composition
			\begin{align}
				f^*\equiv Y^*\xrightarrow{\id_{Y^*}\otimes\coev_X}Y^*\otimes (X\otimes X^*)\xrightarrow{\alpha_{Y^*,X,X^*}^{-1}}(Y^*\otimes X)\otimes &X^*\\
				\xrightarrow{(\id_Y\otimes f)\otimes \id_{X^*}}(Y^*\otimes Y)\otimes X^*\xrightarrow{\ev_Y\otimes\id_{X^*}}&X^*.
			\end{align}
		Using graphical notation, the dual morphism\index{dual morphism} $f^*$ is depicted
			\begin{equation}
			\begin{tikzpicture}[scale=0.85,baseline=(current bounding box.center)]
				\node (X) at (0,0.9) {};
				\node (Y) at (0,4.1) {};
				\node at (-0.25,1.15) {$Y^*\ $};
				\node at (-0.25,3.85) {$X^*\ $};
				\node [rectangle,draw,text width=0.7cm,minimum height=0.7cm,
				text centered,rounded corners, fill=LightGray, name = f] at (0,2.5) {$f^*$};
				\draw (X) -- (f);
				\draw (f) -- (Y);
			\end{tikzpicture}\ \ =\ 
			\begin{tikzpicture}[scale=0.85,baseline=(current bounding box.center)]
				\node at (-0.25,1.75) {$X$};
				\node at (-0.25,3.25) {$Y$};
				\node [rectangle,draw,text width=0.7cm,minimum height=0.7cm,
				text centered,rounded corners, fill=LightGray, name = f] at (0,2.5) {$f$};
				\draw (0,1.5) -- (f);
				\draw (f) -- (0,3.5);
				\draw (0,1.5) arc (-180:0:0.6);
				\draw (0,3.5) arc (0:180:0.6);
				\draw (1.2,1.5) -- (1.2,4.1);
				\draw (-1.2,3.5) -- (-1.2,0.9);
				\node at (-1.45,1.15) {$Y^*\ $};
				\node at (1.45,3.85) {$\ X^*$};
			\end{tikzpicture}
			\end{equation}
		Analogously, if $X,Y\in\Obj(\C)$ have right duals $\lstar X,\lstar Y$ and $f:X\to Y$ is a morphism, one defines the right dual $\lstar f:\lstar Y\to\lstar X$ of $f$ by
			\begin{align}
				\lstar f\equiv\lstar Y\xrightarrow{\coev'_X\otimes\id_{\lstar Y}}(\lstar X\otimes X)\otimes \lstar Y\xrightarrow{\alpha_{\lstar X,X,\lstar Y}}\lstar X\otimes (X\otimes \lstar Y&)\\\xrightarrow{\id_{\lstar X}\otimes(f\otimes\id_{\lstar Y})}\lstar X\otimes(Y\otimes \lstar Y)\xrightarrow{\id_{\lstar X}\otimes\ev'_{Y}}\lstar X&,
			\end{align}
		which is graphically
			\begin{equation}
			\begin{tikzpicture}[scale=0.85,baseline=(current bounding box.center)]
				\node (X) at (0,0.9) {};
				\node (Y) at (0,4.1) {};
				\node at (-0.25,1.15) {$\lstar Y\ $};
				\node at (-0.25,3.85) {$\lstar X\ $};
				\node [rectangle,draw,text width=0.7cm,minimum height=0.7cm,
				text centered,rounded corners, fill=LightGray, name = f] at (0,2.5) {$\lstar f$};
				\draw (X) -- (f);
				\draw (f) -- (Y);
			\end{tikzpicture}\ \ =\ 
			\begin{tikzpicture}[scale=0.85,baseline=(current bounding box.center)]
				\node at (-0.25,1.75) {$X$};
				\node at (-0.25,3.25) {$Y$};
				\node [rectangle,draw,text width=0.7cm,minimum height=0.7cm,
				text centered,rounded corners, fill=LightGray, name = f] at (0,2.5) {$f$};
				\draw (0,1.5) -- (f);
				\draw (f) -- (0,3.5);
				\draw (0,1.5) arc (0:-180:0.6);
				\draw (0,3.5) arc (180:0:0.6);
				\draw (-1.2,1.5) -- (-1.2,4.1);
				\draw (1.2,3.5) -- (1.2,0.9);
				\node at (-1.45,3.85) {$\lstar X\ $};
				\node at (1.45,1.15) {$\ \lstar Y$};
			\end{tikzpicture}
			\end{equation}
	\end{rem}

	\begin{defn}[Rigid]
		An object in a monoidal category is called rigid\index{rigid category} if it has left and right duals. A monoidal category $\C$ is called rigid if every object of $\C$ is rigid.
	\end{defn}
	
	\begin{defn}[$k$-linear]
		Let $k$ be a field. A category $\C$ is said to be $k$-linear\index{k-linear category@$k$-linear category} if for all objects $X,Y\in\Obj(\C)$ the morphisms $\Hom(X,Y)$ form a $k$-vector space and the composition of morphisms in $\C$ is bilinear with respect to $k$.
	\end{defn}
	
	\begin{defn}[Simple and semisimple]
		Let $\C$ be a $k$-linear category. An object $X\in\Obj(\C)$ is called simple\index{simple object} if $\Hom(X,X)=k$. An object is called semisimple\index{semisimple object} if it can be written as a direct sum\footnote{A monoidal category $\C$ is always equipped with a bifunctor $\oplus:\C\times\C\to\C$ that ensures that the direct sum $X_1\oplus X_2$ of two objects in the category is again an object in the category.} of simple objects. A category $\C$ is called semisimple\index{semisimple category} if every object of $\C$ is semisimple.
	\end{defn}
	
	\begin{defn}[Fusion category]
		\label{def:fusioncat}
		A fusion category\index{fusion category} over $\mathbb{C}$ is a $\mathbb{C}$-linear rigid semisimple monoidal category with finitely many simple objects (up to isomorphism) and finite-dimensional morphism spaces such that the identity object is simple.
	\end{defn}

	\begin{rem}
		\label{rem:multiplicities}
		Semisimplicity makes it easy to talk about properties of the category since, generally speaking, everything can be characterized by its action on the simple objects. Furthermore, for a fusion category $\C$, semisimplicity allows us to define multiplicity coefficients in the following way: Label the equivalence classes of simple objects by some set $I$ and choose a representative $X_i$ for each equivalence class $i\in I$. For $i,j\in I$ there are then integers $N_{ij}^k\in\mathbb{Z}_+$ such that
			\begin{equation}
				\label{eq:fusionrule}
				X_i\otimes X_j=\bigoplus_{k\in I} N_{ij}^k X_k.
			\end{equation}
		The notation $N X_k$ denotes the direct sum of $N$ copies of the object $X_k$. We will use these coefficients later in physical applications, where \cref{eq:fusionrule} is called a \emph{fusion rule}\footnote{The name \emph{fusion rule} originates from conformal field theory, where it describes how two excitations can fuse (see \cite{Verlinde1988}).}\index{fusion rule} and the multiplicities $N_{ij}^k$\index{multiplicities} are referred to as \emph{fusion coefficients}\index{fusion coefficients} (see \cref{sec:anyons}). A fusion category is said to be \emph{multiplicity-free}\index{multiplicity-free} if $N_{ij}^k\in\{0,1\}$ for all choices of $i,j,k$.
	\end{rem}

\subsection*{Working in a basis: $F$-symbols for fusion categories}

When studying physical systems in Part $2$ of this thesis, we often choose a basis to do calculations in. This is especially convenient since this means that most of our calculations can be done in terms of matrices and vectors, hence we only need to employ linear algebra. When working in a basis, an important part of the data of a fusion category are $F$-symbols (also called $6j$-symbols). Here, also the notion of fusion coefficients introduced in \cref{rem:multiplicities} comes in handy.

For reasons of clarity, we use a more convenient notation for the morphism space of a trivalent vertex 
 
in the following:
	\begin{equation}
		V_u^{ab}\equiv\Hom(u,a\otimes b),
	\end{equation}
When considering a tensor product of three objects instead of two, there are two ways of decomposing the morphism space $V_u^{abc}$: 
	\begin{align}
		V_u^{abc}&=\bigoplus_p V_p^{ab}\otimes V_u^{pc},\\
		V_u^{abc}&=\bigoplus_q V_u^{aq}\otimes V_q^{bc},
	\end{align}
where the first decomposition corresponds to morphisms given by diagrams of the form
	\begin{equation}
		%
,
	\end{equation}
and the second decomposition of the space corresponds to diagrams of the form
	\begin{equation}
		%
.
	\end{equation}
These two choices are connected via a unitary isomorphism which is given by the so-called $F$-symbol\footnote{We also use the terms $F$-move and $F$-matrix in the following.}\index{F-symbol@$F$-symbol}. This is a map between the two different decompositions of $V_u^{abc}$:
	\begin{equation}
		F_u^{abc}:\bigoplus_p V_p^{ab}\otimes V_u^{pc}\to \bigoplus_q V_u^{aq}\otimes V_q^{bc},
	\end{equation}
whose matrix elements are given as follows:
	\begin{equation}
		\label{eq:Fmove}
		%
.
	\end{equation}
The $F$-matrix $F_u^{abc}$ corresponds to a basis-change matrix of $\Hom(u,a\otimes b\otimes c)$ if the category is strict (i.e., the associator $\alpha_{XYZ}:(X\otimes Y)\otimes Z\to X\otimes(Y\otimes Z)$ is the identity), or to an identification $\Hom(u,(a\otimes b)\otimes c)\cong\Hom(u,a\otimes(b\otimes c))$ if the category is not strict.

If the fusion rules of the underlying category are \emph{not} multiplicity-free, we have to consider a few more indices here. The formula in \cref{eq:Fmove} then reads
	\begin{equation}
	\label{eq:Fmovemult}
		%
.
	\end{equation}
In the following, we restrict our studies to the multiplicity-free case.

\begin{figure}[t]
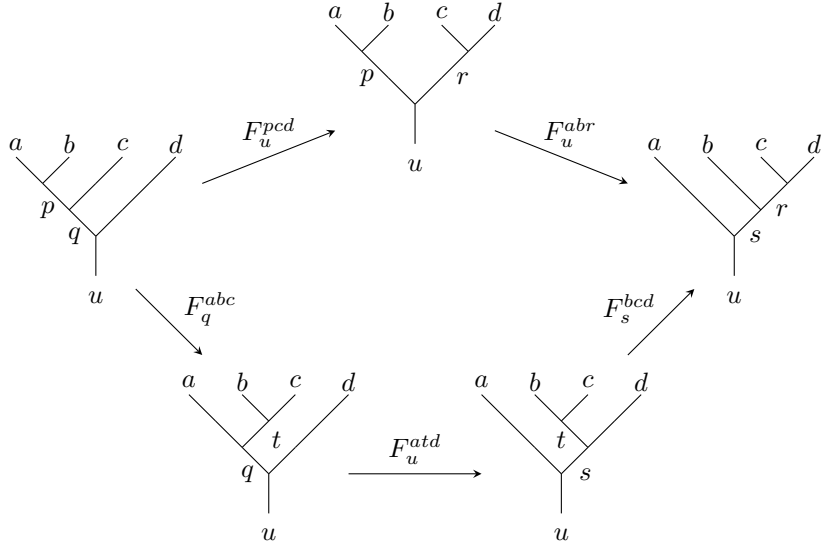

	\centering
	%
	\caption{\small \label{fig:PentagonId}\textbf{Pentagon identity in terms of tree diagrams.} To ensure the consistency of the underlying anyon model, this diagram has to commute.}
\end{figure}

We can continue this train of thought by considering morphism spaces with four objects $\Hom(u,a\otimes b\otimes c\otimes d)$. In this case, there are five different possible decompositions of the morphism space $V_u^{abcd}$. With the same argument as above they have to be connected by a unitary isomorphism. As one can see in \cref{fig:PentagonId}, these isomorphisms are again given by the $F$-moves defined in \cref{eq:Fmove}. However, since now we have more possible decompositions than in the previous case, there are several ways to map from one decomposition to another. For instance, consider the leftmost diagram in \cref{fig:PentagonId}. By employing two $F$-moves (upper path), we can completely reverse the order of splitting and transform the diagram into the rightmost one. However, this is also possible by following the lower path in the diagram that includes three $F$-moves. Therefore, to guarantee the consistency of the fusion category, the two paths have to be identical, or, stated differently, the diagram in \cref{fig:PentagonId} has to commute.

Note that every $F$-move in \cref{fig:PentagonId} only acts on a subtree that is determined by its indices. For instance, the $F$-move that maps from the leftmost tree to the upper middle one acts on the subtree that represents the morphism space $\Hom(u,p\otimes c\otimes d)$, while the subtree representing $\Hom(p,a\otimes b)$ remains unchanged. Hence, there is an additional identity acting on $V_p^{ab}$ that we have omitted here for clarity. This becomes clearer when considering the vector space decompositions where the $F$-moves are applied to as depicted in \cref{fig:pentavec}. Here, we have included the identity maps in the arrow labels. 

\begin{figure}[t]
	\centering
	\begin{tikzpicture}[scale=0.9]
	\node (1) at (0,0) {$\bigoplus_{p,q}V_p^{ab}\otimes V_q^{pc}\otimes V_u^{qd}$};
	\node (2) at (5,2) {$\bigoplus_{p,r}V_p^{ab}\otimes V_u^{pr}\otimes V_r^{cd}$};
	\node (3) at (10,0) {$\bigoplus_{s,r}V_u^{as}\otimes V_s^{br}\otimes V_r^{cd}$};
	\node (4) at (1.25,-2.5) {$\bigoplus_{q,t}V_t^{bd}\otimes V_q^{at} \otimes V_u^{qd}$};
	\node (5) at (8.75,-2.5) {$\bigoplus_{s,t}V_t^{bd}\otimes V_u^{as} \otimes V_s^{td}$};
	\draw[->,>=stealth] (1) to node[above,near start] {\footnotesize $\sum_p F_u^{pcd}\otimes \id_{V_p^{ab}}\hspace{40pt}$} (2);
	\draw[->,>=stealth] (2) to node[above,near end] {\footnotesize $\hspace{40pt}\sum_r F_u^{abr}\otimes \id_{V_r^{cd}}$} (3);
	\draw[->,>=stealth] (1) to node[left] {\footnotesize $\sum_q F_q^{abc}\otimes \id_{V_u^{qd}}$} (4);
	\draw[->,>=stealth] (4) to node[above] {\footnotesize $\sum_t \id_{V_u^{bc}}\otimes F_u^{atd}$} (5);
	\draw[->,>=stealth] (5) to node[right] {\footnotesize $\sum_s \id_{V_u^{as}} \otimes F_s^{bcd}$} (3);
	\end{tikzpicture}
	\caption{\small \label{fig:pentavec}\textbf{Pentagon identity in terms of vector space decompositions.} The $F$-move always acts on parts of the vectors space decomposition while the rest is left unchanged, i.e., the identity acts on it.}
\end{figure}

The requirement that the diagram in \cref{fig:PentagonId} has to commute leads to a condition for the $F$-moves that is stated in terms of the matrix elements, namely the \emph{pentagon equation}\index{pentagon equation}\footnote{Note that in the case with multiplicities, we have to consider additional indices as we did in \cref{eq:Fmovemult}.}:
	\begin{equation}
	\label{eq:pentagon}
		\left(F_u^{abr}\right)_{sp}\left(F_u^{pcd}\right)_{rq}=\sum_t \left(F_s^{bcd}\right)_{rt}\left(F_u^{atd}\right)_{sq}\left(F_q^{abc}\right)_{tp}.
	\end{equation}
The $F$-move has to fulfil this equation for every allowed choice of labels from the set of simple objects in the fusion category that is considered. Hence, the number of variables and equations grows rapidly with the number of simple objects: If $N$ is the number of simple objects in the model, there are, in general, $N^6$ variables for the $F$-moves (since it has six indices) and $N^9$ equations (since the pentagon equation includes nine different labels). This makes it clear that solving the pentagon equation \cref{eq:pentagon} rapidly becomes complicated the more complex the category is. 

\begin{defn}[Unitary fusion category]
	A unitary fusion category\index{unitary fusion category} is a fusion category $\C$ where the $F$-symbols and the left and right unit constraints are unitary.
\end{defn}

In the context of physical systems, we require the $F$-symbols to be unitary to represent physical processes (see \cref{sec:anyons}), hence we usually work with unitary fusion categories.

\section{Braided categories}

	It is possible to equip a monoidal category with the structure of a braiding, which can be thought of as swapping two objects in a category.
	
	\begin{defn}[Braiding]
		\label{def:braiding}
		A braiding\index{braiding} on a monoidal category $\C$ is a natural isomorphism
		\begin{equation}
		c_{X,Y}:X\otimes Y\to Y\otimes X
		\end{equation}
		such that the following diagrams\index{hexagon diagrams} (the \emph{hexagon diagrams}) commute for all $X,Y,Z\in\Obj(\C)$:
		\begin{figure}[H]
			\centering

		\end{figure}
	\end{defn}
	The graphical notation for braiding is the following:
		\begin{equation}
			%
		\end{equation}
	The hexagon equations then take the form (note that associators are not visible in this notation)
		\begin{equation}
			%
		\end{equation}
	When two strings are drawn close to each other it indicates that they are treated as a single composite object for the braiding process. From these diagrams it becomes clear that the hexagon equations impose a quite natural property of the braiding: braiding an object with the tensor product of two objects is the same as braiding it separately with one and with the other afterwards.
	
	\begin{defn}[Braided monoidal category]
		\index{braided category}
		A braided monoidal category is a pair consisting of a monoidal category $\C$ and a braiding.
	\end{defn}
	
	\begin{rem}
		Note that the braiding on a monoidal category is not unique. The same monoidal category can have different structures of a braided category.
	\end{rem}
	
	In a braided monoidal category, the \emph{Yang-Baxter equation}\index{Yang-Baxter equation} holds:
		\begin{equation}
			\begin{tikzpicture}[scale=0.9,baseline=(current bounding box.center)]
				\node at (0,-1.75) {$X$};
				\node at (1,-1.75) {$Y$};
				\node at (2,-1.75) {$Z$};
				\node at (0,3.25) {$Z$};
				\node at (1,3.25) {$Y$};
				\node at (2,3.25) {$X$};
				\begin{knot}[consider self intersections=true,ignore endpoint intersections=false]
					\strand (0,0)to(0,1.5)to[out=90,in=270](1,3);
					\strand (1,0)to[out=90,in=270](2,1.5)to(2,3);
					\strand (2,0)to[out=90,in=270](1,1.5)to[out=90,in=270](0,3);
				\end{knot}
				\begin{knot}[consider self intersections=true,ignore endpoint intersections=false]
					\strand (0,-1.5)to[out=90,in=270](1,0);
					\strand (1,-1.5)to[out=90,in=270](0,0);
					\strand (2,0)to(2,-1.5);
				\end{knot}
			\end{tikzpicture}\ =\begin{tikzpicture}[scale=0.9,baseline=(current bounding box.center)]
				\node at (0,-1.75) {$X$};
				\node at (1,-1.75) {$Y$};
				\node at (2,-1.75) {$Z$};
				\node at (0,3.25) {$Z$};
				\node at (1,3.25) {$Y$};
				\node at (2,3.25) {$X$};
				\begin{knot}[consider self intersections=true,ignore endpoint intersections=false]
					\strand (0,-1.5)to(0,0)to[out=90,in=270](1,1.5)to[out=90,in=270](2,3);
					\strand (1,-1.5)to[out=90,in=270](2,0)to(2,1.5)to[out=90,in=270](1,3);
					\strand (2,-1.5)to[out=90,in=270](1,0)to[out=90,in=270](0,1.5)to(0,3);
				\end{knot}
			\end{tikzpicture}
		\end{equation}
	This equation plays an important role in mathematical knot theory and also relates to problems in condensed matter physics (see \cite{Jimbo1989,Kauffman2001}).

	It is possible to derive a description of the braiding operation when working in a specific basis, similar to how we introduced the $F$-symbols in the previous section. This is done by defining the so-called $R$-matrices. However, since these matrices do not play a role in the rest of Part I of the thesis and will only be needed in the context of the description of anyons in \cref{sec:anyons}, we postpone this discussion to \cref{sec:alganyons}.

\section{Modular tensor categories}
\label{sec:MTC}

We can add further adjectives to braided fusion categories, which will lead to the definition of modular tensor categories. Modular tensor categories are not only fascinating mathematical objects themselves, they are also interesting from a physics perspective, since they are algebraic models of anyon systems \cite{kitaev_anyons_2006,Pachos2009,Wang2010,Rowell2018,Barkeshli2019}. Furthermore, the study of modular tensor categories is closely related to Topological Quantum Field Theories (TQFTs), since the notion of a unitary $(2+1)$ TQFT is essentially the same as that of a unitary modular tensor category (see \cite{Turaev2010,Bartlett2015}). 

Originally, modular tensor categories were invented by Moore and Seiberg \cite{Moore1989} and later equivalently formulated as modular categories in a coordinate-free version by Turaev \cite{Turaev1992}. An accessible introduction to the topic can be found in \cite{Turaev2010} and \cite{bakalov_lectures_2001}.

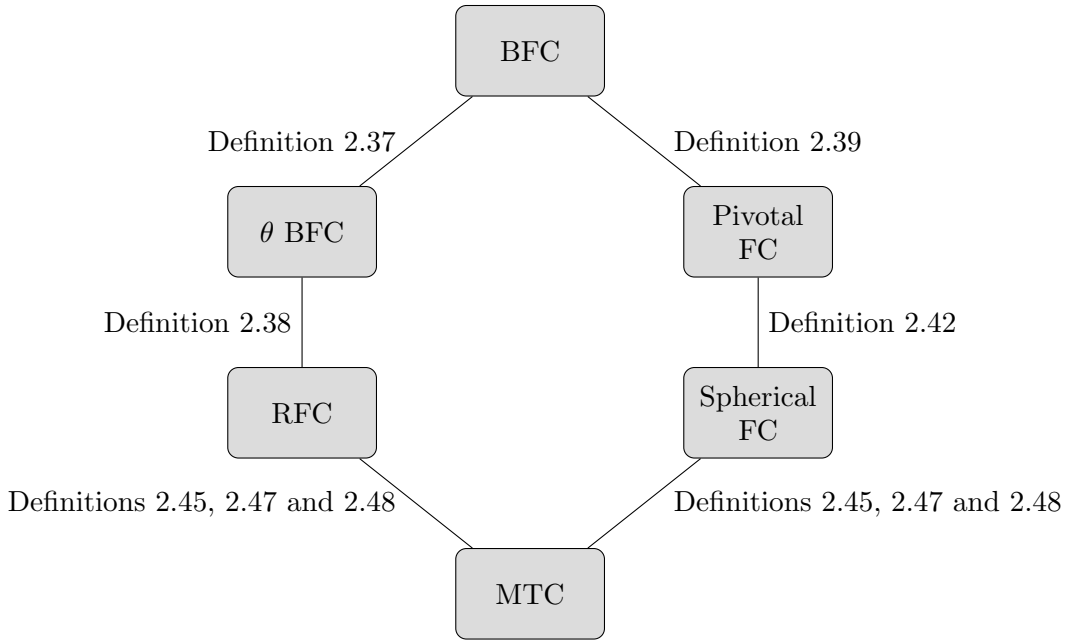
\begin{figure}[t]
		\centering
		\begin{tikzpicture}[scale=1.2]
			\node [rectangle,draw,text width=1.7cm,minimum height=1.2cm,
			text centered,rounded corners, fill=LightGray, name = BFC] at (0,0) {BFC};
			\node [rectangle,draw,text width=1.7cm,minimum height=1.2cm,
			text centered,rounded corners, fill=LightGray, name = tBFC] at (-2.5,-2) {$\theta$ BFC};
			\node [rectangle,draw,text width=1.7cm,minimum height=1.2cm,
			text centered,rounded corners, fill=LightGray, name = RFC] at (-2.5,-4) {RFC};
			\node [rectangle,draw,text width=1.7cm,minimum height=1.2cm,
			text centered,rounded corners, fill=LightGray, name = MTC] at (0,-6) {MTC};
			\node [rectangle,draw,text width=1.7cm,minimum height=1.2cm,
			text centered,rounded corners, fill=LightGray, name = piv] at (2.5,-2) {Pivotal\\ FC};
			\node [rectangle,draw,text width=1.7cm,minimum height=1.2cm,
			text centered,rounded corners, fill=LightGray, name = sph] at (2.5,-4) {Spherical\\ FC};
			\draw (BFC) to node[left] {\cref{def:twist}\ \ } (tBFC);
			\draw (tBFC) to node[left] {\cref{def:ribbon}} (RFC);
			\draw (RFC) to node[left] {\cref{def:preMTC,def:Smatrix,def:MTC}\ \ } (MTC);
			\draw (BFC) to node[right] {\ \ \cref{def:piv}} (piv);
			\draw (piv) to node[right] {\cref{def:sph}} (sph);
			\draw (sph) to node[right] {\ \ \cref{def:preMTC,def:Smatrix,def:MTC}} (MTC);
		\end{tikzpicture}
		\caption{\small \textbf{Two paths to formulate a MTC from a BFC.} One can either equip the Braided Fusion Category (BFC) with a twist $\theta$ that has a ribbon structure (left path), or with a pivotal structure that is spherical (right path).\label{fig:MTC}}
	\end{figure}

There are two different paths to define a Modular Tensor Category (MTC) from a Braided Fusion Category (BFC), both involve a variety of new adjectives: one is via introducing a twist $\theta$ (i.e., a non-zero complex number that is assigned to every object) to formulate the notion of a Ribbon Fusion Category (RFC), the other one is via a pivotal, spherical structure. The two possible ways are sketched in \cref{fig:MTC}. We will first give definitions of all the occurring adjectives and then formulate the definition of a modular tensor category.

	We begin with giving the necessary definitions for the left path, which yields a definition of a modular tensor category via twists and ribbon structures on a braided monoidal category.

	\begin{defn}[Twist]
		\label{def:twist}
		A twist on a braided rigid monoidal category\index{twist} $\C$ is a natural isomorphism $\theta:\id_\C\to\id_\C$ such that it is compatible with the braiding structure $c_{X,Y}$ of the category:
			\begin{equation}
			\label{eq:twist}
				\theta_{X\otimes Y}= c_{Y,X}\circ c_{X,Y}\circ (\theta_X\otimes\theta_Y)
			\end{equation}
		for all $X,Y\in\Obj(\C)$. Graphically, the twist $\theta_X$ of an object $X\in\Obj(\C)$ is depicted as
			\begin{equation}
\ .
			\end{equation}
		\noindent
		The condition \cref{eq:twist} can be depicted as
			\begin{figure}[H]
				%
\ .
			\end{figure}
	\end{defn}
	
	\begin{defn}[Ribbon structure]
		\label{def:ribbon}
		A twist $\theta$ is called a ribbon structure if 	
			\begin{equation}
				\label{eq:ribbon}
				(\theta_X)^*=\theta_{X^*}.
			\end{equation}
		A ribbon tensor category\index{ribbon category} is a braided rigid monoidal category equipped with a ribbon structure. By using the construction of dual morphisms in \cref{rem:dualmorphs}, \cref{eq:ribbon} can be depicted 
			\begin{equation}
				%
\ .
			\end{equation}
	\end{defn}

	We now give the definitions that correspond to the right path in \cref{fig:MTC}, namely those of pivotal and spherical structures.

	\begin{defn}[Pivotal]
		\label{def:piv}
		Let $\C$ be a rigid monoidal category. A pivotal\index{pivotal category} structure on $\C$ is a collection of isomorphisms 
			\begin{equation}
				\phi_X:X\to X^{**}
			\end{equation}
		which is natural in $X$ and satisfies $\phi_{X\otimes Y}=\phi_X\otimes\phi_Y$ for all $X,Y\in\Obj(\C)$. A rigid monoidal category $\C$ equipped with a pivotal structure is said to be pivotal.
	\end{defn}

	\begin{defn}[Trace]
		\label{def:trace}
		Let $\C$ be a rigid monoidal category\index{trace} with a pivotal structure $\phi$, $X\in\Obj(\C)$ and $f\in\Hom_\C(X,X)$. The left canonical (or quantum) trace is then defined
			\begin{equation}
				\Tr^L(f):\mathbf{1}\xrightarrow{\coev_X}X\otimes X^*\xrightarrow{f\otimes\id_{X^*}}X\otimes X^*\xrightarrow{\phi_X\otimes \id_{X^*}}X^{**}\otimes X^*\xrightarrow{\ev_{X^*}}\mathbf{1},
			\end{equation}
		depicted
			\begin{equation}
				\begin{tikzpicture}[scale=0.6,baseline=(current bounding box.center)]
					\node at (0,0.4)[left] {$X$};
					\node at (0,2.5)[left] {$X$};
					\node at (0,4.6)[left] {$X^{**}$};
					\node [rectangle,draw,text width=0.7cm,minimum height=0.7cm,
					text centered,rounded corners, fill=LightGray, name = f] at (0,1.4) {$f$};
					\node [rectangle,draw,text width=0.7cm,minimum height=0.7cm,
					text centered,rounded corners, fill=LightGray, name = ph] at (0,3.6) {$\phi_X$};
					\draw (0,0.5) -- (f);
					\draw (f) -- (ph);
					\draw (ph) -- (0,4.5);
					\draw (0,0.5) arc (-180:0:1.5);
					\draw (0,4.5) arc (180:0:1.5);
					\draw (3,0.5) to node[right] {$X^*$} (3,4.5);
				\end{tikzpicture}
			\end{equation}
		Analogously, the right trace is defined 
			\begin{equation}
				\Tr^R(f):\mathbf{1}\xrightarrow{\coev_{X^*}}X^*\otimes X^{**}\xrightarrow{\id_{X^*}\otimes \phi_X^{-1}}X^*\otimes X\xrightarrow{\id_{X^*}\otimes f}X^*\otimes X\xrightarrow{\ev_{X}}\mathbf{1},
			\end{equation}
		depicted
			\begin{equation}
				\begin{tikzpicture}[scale=0.6,baseline=(current bounding box.center)]
					\node at (0,0.4)[right] {$X^{**}$};
					\node at (0,2.5)[right] {$X$};
					\node at (0,4.6)[right] {$X$};
					\node [rectangle,draw,text width=0.7cm,minimum height=0.7cm,
					text centered,rounded corners, fill=LightGray, name = f] at (0,1.4) {$\phi^{-1}_X$};
					\node [rectangle,draw,text width=0.7cm,minimum height=0.7cm,
					text centered,rounded corners, fill=LightGray, name = ph] at (0,3.6) {$f$};
					\draw (0,0.5) -- (f);
					\draw (f) -- (ph);
					\draw (ph) -- (0,4.5);
					\draw (0,0.5) arc (0:-180:1.5);
					\draw (0,4.5) arc (0:180:1.5);
					\draw (-3,0.5) to node[left] {$X^*$} (-3,4.5);
				\end{tikzpicture}
			\end{equation}
	\end{defn}

	\begin{defn}[Dimension]
		\label{def:ctdim}
		Let $\phi$ be a pivotal structure on a rigid monoidal category $\C$. The dimension\index{dimension} of an object $X\in\Obj(\C)$ with respect to $\phi$ is 
			\begin{equation}
				\dim_\phi(X)=\Tr^L(\id_X)\in\mathsf{End}_\C(\mathbf{1}).
			\end{equation}
		Hence, in a rigid monoidal category over a field $k$ all dimensions are elements of $k$.
	\end{defn}

	\begin{defn}[Spherical]
		\label{def:sph}
		A pivotal structure $\phi$ on a rigid monoidal category $\C$ is spherical if 
			\begin{equation}
				\dim_\phi(X)=\dim_\phi(X^*)
			\end{equation}
		for all objects $X\in\Obj(\C)$. A rigid monoidal category is said to be spherical\index{spherical category} if it is equipped with a spherical structure.
	\end{defn}

	\begin{thm}
		Let $\C$ be a spherical category and $X\in\Obj(\C)$ an object in $\C$. Then for any $f\in\Hom_\C(X,X)$ it holds that	
			\begin{equation}
				\Tr^L(f)=\Tr^R (f).
			\end{equation}
	\end{thm}

	\begin{proof}
		See \cite[Theorem 4.7.15]{Etingof2015}.
	\end{proof}

	\begin{rem}
		\label{rem:trace}
		In a ribbon fusion category $\C$, i.e., a braided fusion category that is equipped with a ribbon structure, we cannot employ the definition of the trace given in \cref{def:trace} since we do not have a pivotal structure. Nevertheless, there is an alternative way to formulate the trace\index{trace} by using braiding and twists. For a morphism $f\in\Hom_\C(X,X)$ it is defined as
			\begin{align}
			\label{eq:tracerib}
				\Tr(f)&=\ev'_{X}\circ(f\otimes\id_{X^*})\circ\coev_X\\
				&=\ev_{X}\circ c_{X,X^*}\circ((\theta_X\circ f)\otimes\id_{X^*})\circ\coev_X,
			\end{align}
		which graphically corresponds to
			\begin{equation}
				\begin{tikzpicture}[scale=0.6,baseline=(current bounding box.center)]
					\node at (0,0.2)[left] {$X$};
					\node at (0,2.8)[left] {$X$};
					\node [rectangle,draw,text width=0.7cm,minimum height=0.7cm,
					text centered,rounded corners, fill=LightGray, name = f] at (0,1.5) {$f$};
					\draw (0,0.5) -- (f);
					\draw (f) -- (0,2.5);
					\draw (0,0.5) arc (-180:0:1.3);
					\draw (0,2.5) arc (180:0:1.3);
					\draw (2.6,0.5) to node[right] {$X^*$} (2.6,2.5);
				\end{tikzpicture}\ =
				\begin{tikzpicture}[scale=0.6,baseline=(current bounding box.center)]
					\node at (0,0.2)[left] {$X$};
					\node at (0,2.8)[left] {$X$};
					\node [rectangle,draw,text width=0.7cm,minimum height=0.7cm,
					text centered,rounded corners, fill=LightGray, name = f] at (0,1.5) {$f$};
					\draw (0,0.5) -- (f);
					\draw (f) -- (0,2.5);
					\begin{knot}[consider self intersections=true,ignore endpoint intersections=false]
						\strand (0,2.5)to[out=90,in=180](0.25,3.25)to[out=0,in=90](0.45,3)to[out=270,in=0](0.25,2.75)to[out=180,in=270](0,3.5);
					\end{knot}
					\begin{scope}[shift={(0,3.5)}]
					\begin{knot}[flip crossing/.list={8},consider self intersections=true,ignore endpoint intersections=false]
						\strand (0,0)to[out=90,in=270](2.1,2);
						\strand (2.6,0)to[out=90,in=270](0.5,2);
					\end{knot}
					\end{scope}
					\draw (0,0.5) arc (-180:0:1.3);
					\draw (0.5,5.5) arc (180:0:0.8);
					\draw (2.6,0.5) to node[right] {$X^*$} (2.6,3.5);
					\node at (2,5.1)[right] {$X$};
					\node at (0.5,6)[left] {$X^*$};
				\end{tikzpicture}
			\end{equation}
		Note that in a ribbon fusion category, left and right traces are equal. As a result, the dimension of an object $X$ is given by $\dim(X)=\Tr(\id_X)$ with the trace given in \cref{eq:tracerib}.
	\end{rem}

	\begin{defn}[Pre-modular category]
		\label{def:preMTC}
		A pre-modular category\index{pre-modular category} can be defined in two equivalent ways:
			\begin{enumerate}
				\item as a ribbon fusion category or
				\item as a braided fusion category equipped with a spherical structure.
			\end{enumerate}
	\end{defn}
	
	\begin{rem}
		From now on, we simply write $\dim(X)$ to denote the dimension of an object $X\in\C$ whenever it is clear from the context whether this is defined via a pivotal structure (as in \cref{def:ctdim}) or a ribbon structure (as in \cref{rem:trace}) or, as it is the case for pre-modular categories, if it can be defined either way.
	\end{rem}
	
	\begin{defn}[$S$-matrix]
		\label{def:Smatrix}
		\index{S-matrix@$S$-matrix}
		Let $\C$ be a pre-modular category. Let $\mathcal{O}(\C)$ denote the set of (isomorphism classes of) simple objects. The $S$-matrix of $\C$ is defined as
			\begin{equation}
				S=(s_{XY})_{X,Y\in\mathcal{O}(\C)},
			\end{equation}
		where
			\begin{equation}
				s_{XY}=\frac{1}{D}\Tr(c_{Y,X}\circ c_{X,Y}),
			\end{equation}
		where $D^2=\sum_{X\in\mathcal{O}(\C)}\dim(X)^2$ is the global quantum dimension.
		Graphically, $s_{XY}$ is depicted
			\begin{equation}\frac{1}{D}\ 
				\begin{tikzpicture}[scale=0.4,baseline=(current bounding box.center)]
					\begin{knot}[flip crossing=2]
					\strand (2,1) circle[radius=2cm];
					\strand (0,1) circle[radius=2cm];
					\end{knot}
					\node at (-1,-1.5) {$X$};
					\node at (3,-1.5) {$Y$};
				\end{tikzpicture}\ .
			\end{equation}
	\end{defn}

	\begin{defn}[Modular tensor category]
		\label{def:MTC}
		A pre-modular category $\C$ is a modular tensor category\index{modular tensor category} if its $S$-matrix is non-degenerate (i.e., invertible).
	\end{defn}

	\begin{rem}
		Analogously to the definition of a unitary fusion category, a modular tensor category is unitary if the $F$-symbols, the left and right unit constraint, and the $R$-matrices are unitary.\index{unitary modular tensor category}
	\end{rem}

	\subsection*{Modular data.} The $S$-matrix is one part of the so-called modular data of a modular tensor category. Modular data\index{modular data} is an invariant of an MTC and is therefore used to classify them. The second part of modular data of an MTC is the so-called $T$-matrix: 
	
	\begin{defn}[$T$-matrix]
		In a modular tensor category $\C$, the $T$-matrix \index{T-matrix@$T$-matrix}is defined as 
			\begin{equation}
				T=(t_{XY})_{X,Y\in\mathcal{O}(\C)},
			\end{equation}
		where 
			\begin{equation}
			\label{eq:tmatrix}
				t_{XX}=\frac{\Tr(c_{X,X})}{d_X}=\frac{1}{d_X}\ 
				\begin{tikzpicture}[baseline=(current bounding box.center)]
				\begin{knot}
				\strand (0,0) to (1,1);
				\strand (1,0) to (0,1);
				\end{knot}
				\draw (1,1) arc (90:-90:5mm);
				\draw (0,0) arc (270:90:5mm);
				\node at (0,-0.4) {$X$};
				\node at (1,-0.4) {$X$};
				\end{tikzpicture}=\theta_X,
			\end{equation}
		and $t_{XY}=0$ for $X\neq Y$ (recall that $\theta_X$ is the twist given in \cref{def:twist}).
	\end{defn}

	Given the modular data $\{S,T\}$ of a modular tensor category $\C$, one can derive several properties of the underlying category from it:
		\begin{enumerate}
			\item There is a formula that expresses the fusion coefficients presented in \cref{rem:multiplicities} in terms of elements of the $S$-matrix, which is the Verlinde formula\index{Verlinde formula}:
				\begin{equation}
					N_{AB}^C=\sum_X \frac{s_{AX} s_{BX} s_{C^* X}}{s_{\mathbf{1} X}}.
				\end{equation}
			\item As shown in \cref{eq:tmatrix}, the twist for the simple objects can be derived from the $T$-matrix:
				\begin{align}
					\theta_X&=t_{XX}.
				\end{align}
			Furthermore, the first row/column of the $S$-matrix gives the dimension of the simple objects.
		\end{enumerate}

\chapter{Trivalent categories}
\label{ch:Trivalent}

In this section, we describe monoidal categories that are generated by a trivalent vertex, which is a rotationally invariant morphism in $\Hom_\C(\mathbf{1},X\otimes X\otimes X)$, where $X$ is a specific simple object in the category. These categories were presented and extensively studied in \cite{Morrison2017} and they can be thought of as fusion categories without finiteness, i.e., there can be an infinite number of simple objects. They are especially nice to work with since their graphical calculus has some additional rules that greatly simplify calculations with string diagrams. We will need a few additional adjectives before we can state the definition. Note that a trivalent category is not necessarily a fusion category or a Modular Tensor Category (MTC), although there are some examples of MTCs and fusion categories which are also trivalent categories. The latter cases are the reason why trivalent categories appear in the context of this thesis and we will discuss some examples of trivalent fusion categories in detail later. In particular, one of the Haagerup fusion categories can be described as a trivalent category, which allows us to exploit its simple graphical calculus to simplify certain calculations in \cref{ch:Haagerup}.

\section{Basic definitions}

Unless otherwise specified, all categories in this chapter are assumed to be monoidal.

	\begin{defn}[Evaluable]
		Let $k$ be a field. A $k$-linear category $\C$ is evaluable\index{evaluable category} if $\dim(\Hom(\mathbf{1},\mathbf{1}))=1$. In fact, $\Hom(\mathbf{1,\mathbf{1}})$ can be identified with the ground field $k$ of the category by sending the empty diagram to $1$.
	\end{defn}

	\begin{defn}[Nondegenerate]
		A pivotal category $\C$ is called nondegenerate\index{nondegenerate category} if for every morphism $f\in\Hom(X,Y)$ there is a morphism $f'\in\Hom(Y,X)$ such that $\Tr(f'\circ f)\neq 0\in\Hom(\mathbf{1,\mathbf{1}})$.
	\end{defn}
	
	\begin{rem}
		For a pivotal category $\C$ and an object $X$ we will use the notation
			\begin{equation}
				\mathfrak{C}_k\equiv\Hom_\C(\mathbf{1},X^{\otimes k}).
			\end{equation}
		We use the convention that $X^{\otimes 0}=\mathbf{1}$.
	\end{rem}
	
	\begin{defn}[Trivalent category]
		A trivalent category\index{trivalent category} is a tuple $(\C,X,\tau)$, where $\C$ is a nondegenerate evaluable pivotal category over $\mathbb{C}$ with an object $X$ with $\dim(\mathfrak{C}_1) = 0$, $\dim(\mathfrak{C}_2) = 1$, and $\dim(\mathfrak{C}_3) = 1$, with a rotationally invariant morphism $\tau\in\mathfrak{C}_3$ called the \emph{trivalent vertex}, such that the category is generated (as a pivotal category) by $\tau$ .
	\end{defn}

	The constraints on the dimensions of certain morphism spaces in the definition of a trivalent category can be interpreted as follows: 
		\begin{enumerate}
			\item Firstly, $\dim\mathfrak{C}_1=0$ implies that there is no map from the object $X$ to the unit. This means that $X$ is not the unit and it is not a sum that contains the unit object.
			\item The second constraint, $\dim\mathfrak{C}_2=1$ says that $X$ is simple, i.e., it is not isomorphic to the unit object and every subobject of $X$ is isomorphic to zero or $X$.
			\item The last constraint, $\dim\mathfrak{C}_3=1$, together with the rotational invariance of the trivalent vertex $\tau\in\mathfrak{C}_3$, implies that we do not need to carry any label that indicates the orientation of the trivalent vertex.
		\end{enumerate}
	
	\begin{rem}
		We can show that the object $X$ is symmetrically self-dual (see \cite[Lemma 2.2]{Morrison2017}), which allows us to omit orientations on strings. Combining these two facts, we can interpret any unoriented planar trivalent string diagram with $n$ boundary points as an element of $\mathfrak{C}_n$.
	\end{rem}

	\begin{rem}
		Any trivalent category is spherical (see \cite[Rem.\ 2.5]{Morrison2017}). Since every object of the category is generated by the simple object $X$, we only need to check that the dimension of $X$ is equal to the dimension of $X^*$, but this is obvious since $X$ is self-dual.
	\end{rem}

\subsection*{Graphical calculus of trivalent categories.} 

	We now explain the additional rules of the graphical calculus that are imposed by the definition of a trivalent category. In the course of this we will also introduce some parameters which play an important role in the classification of these categories. Most importantly, since the category is generated by the translational invariant trivalent vertex $\tau$, we do not need to put an orientation or a label on the strings in the diagrams. Every string is labelled with the object $X$ of the category. Moreover, we are allowed to deform and rotate the diagrams without changing their value. Then, there is basically one additional rule (and one parameter) that simplifies the evaluation of diagrams for each constraint on the dimensions of morphism spaces:
		\begin{enumerate}
			\item Since the category is evaluable, we know that $\dim\mathfrak{C}_0=1$. This implies that each diagram with a loop in it has to be a multiple $d$ of the same diagram without that loop:
				\begin{equation}
				\label{eq:loop}
\ =d.
				\end{equation}
			Here, the parameter $d$ is a characteristic of the category. It has to be non-zero since it is the dimension of the specific simple object $X$ that generates the category. It also follows from the fact that the loop is the pairing of the unique (up to scalar) element of $\mathfrak{C}_2$ with itself, i.e.,
				\begin{equation}
					%
				\end{equation}
			 and that the category is nondegenerate.
			 \item The constraint $\dim\mathfrak{C}_1=0$ implies that there are no diagrams with only one boundary point, i.e, ``lollipops'' are forbidden:
			 	\begin{equation}
			 	%
\ =0.
			 	\end{equation}
			 Whenever some kind of lollipop is part of a diagram, that diagram evaluates to zero.
			 \item The constraint $\dim\mathfrak{C}_2=1$ yields a relation
			 	\begin{equation}
			 	\label{eq:bigon}
				 	%
			 	\end{equation}
			 where the parameter $b$ is again non-zero, because the diagram on the left hand side of the equation has to be non-zero with the same argument as above.
			 \item The constraint $\dim\mathfrak{C}_3=1$ yields another equation:
			 	\begin{equation}
			 	\label{eq:triangle}
				 	%
			 	\end{equation}
			 In this case, the parameter $t$ can be zero.
		\end{enumerate}
	These relations give us three parameters to characterize the underlying category, namely $d$, $b$, and $t$. However, these three parameters are not independent of each other. One can always rescale the trivalent vertex by a constant. Because of this reason, the authors of \cite{Morrison2017} chose the normalisation $b=1$ throughout their paper. We will not choose this normalisation, but a different one due to constraints coming from the physical application later in this thesis, but for now we will just keep all three parameters.
	
	\begin{defn}
		The bilinear inner product of two trivalent graphs $f,g$ with $n$ boundary points is given by	
			\begin{equation}
				\label{eq:innerprod}
				\langle f,g\rangle=\ \begin{tikzpicture}[baseline=(current bounding box.center)]
					\draw (2.1,0) arc(0:180:1.1);
					\draw (2.3,0) arc(0:180:1.3);
					\draw (1.7,0) arc(0:180:0.7);
					\node[rotate=90] at (1,0.9) {$...$};
					\node [circle,draw,minimum size=1.2cm,text centered,fill=LightGray, name = f] at (0,0) {$f$};
					\node [circle,draw,minimum size=1.2cm,text centered,fill=LightGray, name = g] at (2,0) {$g$};		
				\end{tikzpicture}
			\end{equation}
		i.e., the boundary points of the diagrams are connected by $n$ strings. Since this is a diagram in $\Hom(\mathbf{1},\mathbf{1})$, the bilinear inner product is an element of $\mathbb{C}$, the ground field of the category.
	\end{defn}

\begin{exmp}[The Fibonacci category.]
	\label{ex:TrivFib}
	To illustrate the concept of a trivalent category, we consider an example which is also a modular tensor category. The Fibonacci category\index{Fibonacci category}, which we denote $\mathbf{Fib}$, has two simple objects, i.e., $\Obj(\mathbf{Fib})=\{\mathbf{1},X\}$ with the following fusion rule\footnote{Fusing an object $X$ with the unit object $\mathbf{1}$ always yields a trivial fusion rule of the form $X\otimes\mathbf{1}=X$, therefore we usually do not list these kind of fusion rules.}:
		\begin{align}
			X\otimes X&=\mathbf{1}+X.
		\end{align}
	This implies that $X$ is the generating object of the trivalent category and the corresponding trivalent vertex $\tau$ is
		\begin{equation}
			\begin{tikzpicture}[baseline=(current bounding box.center)]
				\draw (0,-0.5) -- (0,-1);
				\draw (0,-1) -- (-0.5,-1.5);
				\draw (0,-1) -- (0.5,-1.5);
				\node at (0,-0.5)[above] {$X$};
				\node at (0.5,-1.5)[below] {$X$};
				\node at (-0.5,-1.5)[below] {$X$};
			\end{tikzpicture}
		\end{equation}
	The characteristic parameters for $\mathbf{Fib}$ are given by
		\begin{equation}
			d^2=d+1,\hspace{30pt}b=1,\hspace{30pt}t=\frac{d-2}{d-1}.
		\end{equation}
	The equation for the parameter $d$ allows two solutions, namely the golden ration $\frac{1+\sqrt{5}}{2}$ and its Galois conjugate, $\frac{1-\sqrt{5}}{2}$. This is one of the reasons why this category is called Fibonacci category. The other reason is because the dimensions of the morphism spaces, $\mathfrak{C}_n$, are given by the Fibonacci series:
		\begin{table}[H]
			\centering
			\begin{tabular}{c|c}
				$n$ & $\dim(\mathfrak{C}_n)$\\ \hline \\[-0.3cm]
				$0$ & $1$\\
				$1$ & $0$\\
				$2$ & $1$\\
				$3$ & $1$\\
				$4$ & $2$\\
				$5$ & $3$\\
				$6$ & $5$\\
				$\dots$ & $\dots$
			\end{tabular}
		\end{table}
	\noindent
	The dimensions of the morphism spaces $\mathfrak{C}_0,\dots,\mathfrak{C}_3$ fulfil the constraints given by the definition of a trivalent category. The dimension of the next morphism space, however, yields an additional equation for the graphical calculus. Note that there are exactly four diagrams in $\mathfrak{C}C_4$ without internal faces, namely 
		\begin{equation}
		\label{eq:C4diags}
			\Cfourone,\hspace{10pt}
			\Cfourtwo,\hspace{10pt}
			\Cfourthree,\hspace{10pt}
			\Cfourfour.
		\end{equation}
	Since the dimension of the morphism space $\mathfrak{C}_4$ is $2$, we find the following two relations between these diagrams (see \cite{Morrison2017}):
		\begin{equation}
		\label{eq:dimC43}
			\Cfourfour-\Cfourthree+\frac{1}{d+1}\ \Cfourone+\frac{1}{d-1}\ \Cfourtwo=0,
		\end{equation}
		\begin{equation}
		\label{eq:dimC42}
			\Cfourone-\frac{1}{d}\ \Cfourtwo=\Cfourthree.
		\end{equation}
	The first equation is in fact true for all trivalent categories with $\dim(\mathfrak{C}_4)\le 3$, while the second equation is a consequence of $\dim(\mathfrak{C}_4)\le 2$ which only holds for the Fibonacci category.
\end{exmp}

\section{Cubic trivalent categories} 
\label{sec:cubictriv}

We will now look into more detail at the classification of so-called \emph{cubic} categories. These are categories where, in addition to the properties of a trivalent category, $\dim(\mathfrak{C}_4)=4$. This implies that there are no relations of the form \cref{eq:dimC43} or \cref{eq:dimC42}, since every diagram listed in \cref{eq:C4diags} is a basis element.

	\begin{defn}[Cubic category]
		A trivalent category $(\C,X,\tau)$ is called cubic\index{cubic category} if $\dim(\mathfrak{C}_4)=4$.
	\end{defn}

In a cubic category, the elements listed in \cref{eq:C4diags} form a basis of $\mathfrak{C}_4$ (this is shown in \cite[Proposition 4.16]{Morrison2017}). However, these are not the only diagrams in $\mathfrak{C}_4$. It is also possible to have diagrams with internal faces, for example
	\begin{equation}
	\label{eq:square}
		\begin{tikzpicture}[baseline=(current bounding box.center)]
			\draw (0,0) rectangle (1,1);
			\draw (0,0) -- (-0.3,-0.3);
			\draw (0,1) -- (-0.3,1.3);
			\draw (1,0) -- (1.3,-0.3);
			\draw (1,1) -- (1.3,1.3);
		\end{tikzpicture}\ .
	\end{equation}
With the rules we have listed so far, we are unable to evaluate these kinds of diagrams, but since it is not a basis element there has to be a decomposition of the above diagram in terms of the basis elements. In \cite{Morrison2017}, this calculation was only done in the case $b=1$. We will need the general case $b\neq 1$ later in this thesis, therefore we calculate the general expression in the following. 

To calculate relations between diagrams in $\mathfrak{C}_4$, we need an orthonormal basis of this space. We know that the diagrams in \cref{eq:C4diags} form a basis of $\mathfrak{C}_4$, hence we can use them do do Gram-Schmidt orthonormalisation. We label the diagrams in the following way:
	\begin{alignat}{2}
		w_1&=\Cfourone\hspace{30pt} w_2&&=\Cfourtwo\\
		w_3&=\Cfourthree\hspace{30pt} w_4&&=\Cfourfour
	\end{alignat}
Using the Gram-Schmidt orthonormalisation algorithm we can find a matrix $\Theta$ such that the orthonormal basis vectors $v_i$ are of the form
	\begin{equation}
		v_i=\sum_k\Theta_{ik} w_k.
	\end{equation}
The matrix $\Theta$ is of the form
	\begin{equation}
		\Theta=
		\begin{pmatrix}
		\frac{1}{|d|} & 0 & 0 & 0 \\
		-\frac{1}{d\sqrt{d^2-1}} & \frac{1}{\sqrt{d^2-1}} & 0 & 0 \\
		-\frac{\mathrm{sgn}(b) \sqrt{d}}{\sqrt{(d^2-a)(d^2-d-1)}} & \frac{\mathrm{sgn}(b)}{\sqrt{(d^2-a)(d^2-d-1)}} & \sqrt{\frac{d^2-1}{b^2d(d^2-d-1)}} & 0 \\
		\frac{b+dt}{d^2-d-1} C_\Theta & \frac{b-bd-t}{d^2-d-1} C_\Theta & \frac{-d^2t+t-b}{b(d^2-d-1)} C_\Theta & C_\Theta
		\end{pmatrix}
	\end{equation}
with
	\begin{equation}
		C_\Theta=\sqrt{\frac{d^2-d-1}{d(b^2d(d-2)-2bt-(d^2-1)t^2}}\ .
	\end{equation}
Using this orthonormal basis, we can write the square diagram \cref{eq:square} (which we will denote $w_\mathrm{sq}$) as a linear combination of the $v_i$:
	\begin{align}
		w_\mathrm{sq}&=\sum_i c_i v_i\\
		&= \sum_i \langle  v_i,w_{\mathrm{sq}} \rangle\ v_i
	\end{align}
with the inner product described in \cref{eq:innerprod}. In terms of the (non-orthonormal) basis diagrams given in \cref{eq:C4diags}, this has the form
	\begin{equation}
	\label{eq:squarelincom}
		\begin{tikzpicture}[scale=0.7,baseline={([yshift=-3pt]current bounding box.center)}]
		\draw (0,0) rectangle (1,1);
		\draw (0,0) -- (-0.3,-0.3);
		\draw (0,1) -- (-0.3,1.3);
		\draw (1,0) -- (1.3,-0.3);
		\draw (1,1) -- (1.3,1.3);
		\end{tikzpicture}=\underbrace{\frac{b(b^2+bt-t^2)}{bd+t+dt}}_{\equiv \alpha}\left(\ 
		\Cfourone+
		\Cfourtwo\ \right)
		+\underbrace{\frac{t^2(d+1)-b^2}{bd+t+dt}}_{\equiv\beta}\left(\ 
		\Cfourthree+
		\Cfourfour\ \right).
	\end{equation}
In this fashion, all diagrams in $\mathfrak{C}_4$ with an arbitrary number of internal faces can be expressed as a linear combination of basis diagrams. This especially equips us with the techniques we need to evaluate any diagram in $\mathfrak{C}_0$ that includes arbitrarily many faces with four edges. Consider for example the following diagram:
	\begin{equation}

	\end{equation}
The inner (red) part of the diagram is exactly the diagram we have expressed in terms of basis elements in \cref{eq:squarelincom}, hence we can insert this relation into the diagram and compute its value:
	\begin{align}
		%
\ \right)\\
		&=
		\alpha\left(b^2 d +b^2 d\right)
		+\beta\left(t^2 bd+t^2 bd\right).
	\end{align}
In the second step, we have used the relation for bigons from \cref{eq:bigon} and the one for triangles from \cref{eq:triangle}. In the last step, we have used the loop relation \cref{eq:loop} and again the bigon relation. In this fashion we can assign to every diagram in $\mathfrak{C}_0$ an element of $\mathbb{C}$, as long as there are enough relations among the different diagrams.
	
\section{The trivalent category $\Hd$}
\label{sec:H3triv}

	We will now introduce the most important example of a category for this thesis, the Haagerup $\mathcal{H}_3$ category\index{trivalent H3@trivalent $\Hd$}, in its trivalent form. The trivalent version of the category is only a subcategory of the full $\mathcal{H}_3$ category, but it already covers a lot of the important properties that we will need throughout this thesis. Especially helpful is the graphical calculus of trivalent categories, since it can be used make statements about parts of the full category. We will see how this works in later chapters.
	
	Note that although the trivalent category is only a subcategory, the full category can be recovered from its trivalent version via the Karoubi envelope (also called the idempotent completion, see for example \cite{Morrison2010} for a detailed explanation of this technique).
	
	The trivalent category $\Hd$ is a cubic category which is characterized by the parameters
		\begin{equation}
			d=\frac{3+\sqrt{13}}{2},\hspace{30pt}b=1,\hspace{30pt}t=-\frac{2}{3}d+\frac{5}{3}.
		\end{equation} 
	Moreover, it has morphism space dimensions $\dim(\mathfrak{C}_5)=11$ and $\dim(\mathfrak{C}_6)=37$. In \cref{sec:cubictriv}, we have already seen a basis for $\mathfrak{C}_4$ and have even constructed an orthonormal basis for this space. As the dimensions of the morphism spaces increase strongly as the number of boundary points increases, doing Gram-Schmidt orthonormalisation soon becomes infeasible. However, we will see later that, when considering the full category, there is a canonical choice of orthonormal basis vectors called the \emph{fusion basis} which is much easier to construct. Therefore, we will only describe non-orthonormal bases for $\mathfrak{C}_5$ and $\mathfrak{C}_6$ in the following.
	
	There are exactly $10$ diagrams with five boundary points in $\mathfrak{C}_5$ that do not contain any faces (to distinguish different diagrams more easily we have added a dotted line around individual diagrams):
		\begin{table}[H]
\begin{tabular}{ccccc}
				\begin{tikzpicture}[scale=0.9,baseline=(current bounding box.center)]
				\draw[dotted] (0,0) circle(1cm);
				\draw (90:1cm) to [bend left=10] (162:0.4cm);
				\draw (18:1cm) to [bend right=10] (306:0.4cm);
				\draw (162:0.4cm) to [bend right=90] (306:0.4cm);
				\draw (306:0.4cm) -- (306:1cm);
				\draw (162:0.4cm) -- (162:1cm);
				\draw (234:0.35cm) -- (234:1cm);
				\end{tikzpicture} &
				\begin{tikzpicture}[scale=0.9,baseline=(current bounding box.center)]
				\rotatebox[origin=c]{72}{
				\draw[dotted] (0,0) circle(1cm);
				\draw (90:1cm) to [bend left=10] (162:0.4cm);
				\draw (18:1cm) to [bend right=10] (306:0.4cm);
				\draw (162:0.4cm) to [bend right=90] (306:0.4cm);
				\draw (306:0.4cm) -- (306:1cm);
				\draw (162:0.4cm) -- (162:1cm);
				\draw (234:0.35cm) -- (234:1cm);}
				\end{tikzpicture} &
				\begin{tikzpicture}[scale=0.9,baseline=(current bounding box.center)]
				\rotatebox[origin=c]{144}{
				\draw[dotted] (0,0) circle(1cm);
				\draw (90:1cm) to [bend left=10] (162:0.4cm);
				\draw (18:1cm) to [bend right=10] (306:0.4cm);
				\draw (162:0.4cm) to [bend right=90] (306:0.4cm);
				\draw (306:0.4cm) -- (306:1cm);
				\draw (162:0.4cm) -- (162:1cm);
				\draw (234:0.35cm) -- (234:1cm);}
				\end{tikzpicture} &
				\begin{tikzpicture}[scale=0.9,baseline=(current bounding box.center)]
				\rotatebox[origin=c]{216}{
				\draw[dotted] (0,0) circle(1cm);
				\draw (90:1cm) to [bend left=10] (162:0.4cm);
				\draw (18:1cm) to [bend right=10] (306:0.4cm);
				\draw (162:0.4cm) to [bend right=90] (306:0.4cm);
				\draw (306:0.4cm) -- (306:1cm);
				\draw (162:0.4cm) -- (162:1cm);
				\draw (234:0.35cm) -- (234:1cm);}
				\end{tikzpicture} &
				\begin{tikzpicture}[scale=0.9,baseline=(current bounding box.center)]
				\rotatebox[origin=c]{288}{
				\draw[dotted] (0,0) circle(1cm);
				\draw (90:1cm) to [bend left=10] (162:0.4cm);
				\draw (18:1cm) to [bend right=10] (306:0.4cm);
				\draw (162:0.4cm) to [bend right=90] (306:0.4cm);
				\draw (306:0.4cm) -- (306:1cm);
				\draw (162:0.4cm) -- (162:1cm);
				\draw (234:0.35cm) -- (234:1cm);}
				\end{tikzpicture}\\
\end{tabular}
		\end{table}
	\noindent
	and, additionally, one element with exactly one face of five edges\footnote{Note that we only take into account faces with five or more edges since in a cubic category, every face with four or less edges can be immediately removed by the relations \cref{eq:loop,eq:bigon,eq:triangle,eq:squarelincom}.}:
		\begin{equation}
			%
		\end{equation}
	Since in $\Hd$ we have $\dim(\mathfrak{C}_5)=11$, the diagrams listed above form a (non-orthonormal) basis of $\mathfrak{C}_5$ (see \cite[Proposition 7.5]{Morrison2017}). The fact that this basis is not an orthonormal one (not even an orthogonal one) can easily be seen by calculating inner products of the basis vectors. For instance, consider the inner product of the first and the sixth element of the list\footnote{The evaluation of this diagram only requires the rules of the graphical calculus \cref{eq:triangle}, \cref{eq:bigon}, and \cref{eq:loop} (in this order) and the fact that diagrams in $\mathfrak{C}_0$ can be rotated and deformed without changing their value.}:
		\begin{equation}
=t b d.
		\end{equation}
	
	Finding a basis in $\mathfrak{C}_6$ is a bit more complicated, because we have to take into account a lot more diagrams since $\dim(\mathfrak{C}_6)=37$. We begin as above by finding all diagrams without faces, which are $34$ in total:
		\begin{table}[H]
			\begin{tabular}{ccc}
				\begin{tikzpicture}[scale=0.9,baseline={([yshift=-3pt]current bounding box.center)}]
					\draw[dotted] (0,0) circle(1.cm);
					\draw (90:1cm) to [bend left=60] (150:1cm);
					\draw (210:1cm) to [bend left=60] (270:1cm);
					\draw (330:1cm) to [bend left=60] (30:1cm);
				\end{tikzpicture}\ +\ 1\text{ rotation,} & 
				\begin{tikzpicture}[scale=0.9,baseline={([yshift=-3pt]current bounding box.center)}]
					\draw[dotted] (0,0) circle(1cm);
					\draw (90:1cm) to [bend left=60] (150:1cm);
					\draw (210:1cm) to (30:1cm);
					\draw (270:1cm) to [bend left=60] (330:1cm);
				\end{tikzpicture}\ +\ 2\text{ rotations,} &
				\begin{tikzpicture}[scale=0.9,baseline={([yshift=-3pt]current bounding box.center)}]
					\draw[dotted] (0,0) circle(1cm);
					\draw (90:1cm) to [bend left=60] (150:1cm);
					\draw (270:1cm) to [bend right=10] (260:0.5cm);
					\draw (330:1cm) to [bend left=10] (340:0.5cm);
					\draw (210:1cm) to [bend left=20] (260:0.5cm);
					\draw (30:1cm) to [bend right=20] (340:0.5cm);
					\draw (260:0.5) to [bend right=20] (340:0.5cm);
				\end{tikzpicture}\ +\ 5\text{ rotations,}\\
\end{tabular}
		\end{table}
	\noindent
	Because of $\dim(\mathfrak{C}_6)=37$, these diagrams obviously do not suffice to form a basis. Therefore, we will also take into account diagrams with one face of which there are seven (we will refer to the second diagram and its rotations as \emph{pentaforks}):
		\begin{table}[H]
			\begin{tabular}{cc}
				\begin{tikzpicture}[scale=0.9,baseline={([yshift=-3pt]current bounding box.center)}]
					\draw[dotted] (0,0) circle(1.cm);
					\draw (90:0.5cm) -- (150:0.5cm) -- (210:0.5cm) -- (270:0.5cm) -- (330:0.5cm) -- (30:0.5cm) -- cycle;
					\foreach \a in {30,90,...,330}
						\draw (\a:0.5cm) -- (\a:1cm);
				\end{tikzpicture}\ , &
				\begin{tikzpicture}[scale=0.9,baseline={([yshift=-3pt]current bounding box.center)}]
					\draw[dotted] (0,0) circle(1.cm);
					\draw (210:1cm) to [bend left=60] (270:1cm);
					\draw (240:0.2cm) -- (240:0.61cm);
					\begin{scope}[scale=0.8,shift={(60:0.25)}]
						\draw (240:0.5) -- (312:0.5) -- (24:0.5) -- (96:0.5) -- (168:0.5) -- cycle;
					\end{scope}
					\draw (330:1cm) to ([scale=0.8,shift={(60:0.25)}]312:0.5);
					\draw (30:1cm) to ([scale=0.8,shift={(60:0.25)}]24:0.5);
					\draw (90:1cm) to ([scale=0.8,shift={(60:0.25)}]96:0.5);
					\draw (150:1cm) to ([scale=0.8,shift={(60:0.25)}]168:0.5);
				\end{tikzpicture}\ +\ 5\text{ rotations}
			\end{tabular}
		\end{table}
	\noindent
	We now have $41$ candidate diagrams for a basis of dimension $37$. It turns out that leaving out any four of the six pentaforks yields a linear independent set of diagrams (see \cite[Proof of Proposition 6.4]{Morrison2017}) and we also know that this set of diagrams spans $\mathfrak{C}_6$ (see \cite[Lemma 6.25]{Morrison2017}), hence they form a basis of $\mathfrak{C}_6$.
	
	These are the vector spaces where almost all our calculations will take place. Later, we will see how to construct orthonormal bases for them in a canonical way, but for that purpose we will need the full category and not only its trivalent description.
	
\chapter{Subfactors and fusion categories}
\label{ch:subfactor}

The study of subfactors \cite{araki_subfactors_1994,Jones1997} has always been connected to physics, more precisely, quantum field theory. Mathematical investigations of these fields have led to a conjectured correspondence between subfactors and Conformal Field Theories (CFTs) by Jones \cite{Jones1990,Jones2014}, which builds on original work of Doplicher and Roberts \cite{Doplicher1989} and was later substantiated by Bischoff \cite{Bischoff2015,Bischoff2016}. Evidence for this conjecture was found, for example, in \cite{Calegari2010,Xu2017}. However, a proof of this conjecture is far from being in sight, since there are still numerous gaps. One example of a subfactor with no known counterpart CFT is the Haagerup subfactor \cite{Haagerup1994}, the smallest (finite-depth, irreducible, hyperfinite) subfactor with index above four. Although some evidence for the existence of such a CFT was provided by Evans and Gannon \cite{Evans2011}, the proof of the existence of the CFT is still an open problem.

The Haagerup subfactor provides the motivating example for most of the work that is done in this thesis. Every finite-depth, irreducible, hyperfinite subfactor gives rise to two fusion categories (its \emph{principle even} and \emph{dual even} parts). In case of the Haagerup subfactor, there is a third category that is Morita equivalent, but not equivalent to these two categories. This category, which is discussed in detail in the next chapter, will be the basis of most of the lattice constructions that are done in the second part of this thesis. 

In this chapter we explain how fusion categories can be constructed from a subfactor. We begin with giving the most important definitions to understand subfactors in terms of operator algebras and their classification via the index before we explore the connection to fusion categories. We also explain the connection between the different fusion categories in terms of algebra objects and module categories and show how to find all subfactors related to a given fusion category. We end this chapter by presenting the construction of the \emph{centre} of a fusion category which provides a way to get a modular tensor category from a fusion category.

\section{Fusion categories from subfactors}

In this section, we explain the connection between subfactors and fusion categories. Since we only use the language of fusion categories in the constructions explained later in this thesis, we only briefly talk about subfactors and mostly focus on the corresponding fusion categories. For a detailed introduction to the theory of subfactors, see for example \cite{Jones1997}. In the following, we use some notions and theorems from basic abstract algebra and operator algebra, such as the standard algebraic tensor product or modules over algebras, which are not introduced here in detail since they can be found in any standard textbook on the subject, for example in \cite{Jacobson2009a,Jacobson2009b}. For a more detailed treatment of von Neumann algebras, see the notes by Vaughan Jones \cite{JonesvNA}. In this thesis, we focus on the definitions and explanations of the corresponding categorical notions of these objects.

\begin{defn}[von Neumann algebra]
	A von Neumann algebra\index{von Neumann algebra} is a self-adjoint subalgebra $M$ of $\mathcal{L}(\Hi)$ (the algebra of all bounded operators on a Hilbert space $\Hi$) that satisfies $M=(M')'$, where
		\begin{equation}
			M'=\{x'\in\mathcal{L}(\Hi):x'x=xx'\text{ for all }x\in M\}
		\end{equation}
	is the commutant of $M$.
\end{defn}

\begin{defn}[Factor]
	A von Neumann algebra $M$ with trivial centre (i.e., it only contains multiples of the identity operator) is called a factor\index{factor}. 
\end{defn}

\begin{defn}[Subfactor]
	A subfactor\index{subfactor} of a factor $M$ is a subalgebra $N\subseteq M$ such that $N$ is also a factor and $N$ contains the identity element of $M$. The subfactor $N$ is said to be irreducible\index{irreducible subfactor} if it has \emph{trivial relative commutant}, i.e., if $N'\cap M=\mathbb{C}$.
\end{defn}

To classify factors, the notion of minimal and finite \emph{projections} in a von Neumann algebra is important.

\begin{defn}[Projection]
	\index{projection}
	An operator $p$ in a von Neumann algebra $M$ is called a projection if $p=pp=p^*$. $p$ is called \emph{minimal} if there is no other projection $q$ with $0<q<p$. It is called \emph{finite} if there is no projection $q<p$ that is equivalent to $p$.
\end{defn}

\begin{defn}
	A factor $M$ is said to be finite\index{finite factor} if the multiplicative identity of $M$ (which always exists) is a finite projection in $M$.
\end{defn}

With these definitions at hand, we can introduce the classification of factors in the way it was done by Murray and von Neumann \cite{Murray1936}: A factor $M$\index{classification of factors} is said to be of
	\begin{enumerate}
		\item type $I$ if there exists a non-zero minimal projection in $M$.
		\item type $II$ if it contains non-zero finite projections and is not of type $I$. If the multiplicative identity of $M$ is a finite projection, then $M$ is of type $II_1$, otherwise it is of type $II_\infty$.
		\item type $III$ if no non-zero projection of $M$ is finite.
	\end{enumerate}
It can be shown (see, for example, \cite{Takesaki1979}) that every factor is either of type $I$, type $II_1$, type $II_\infty$, or type $III$.

We are especially interested in type $II_1$ factors since they have the convenient property that they are equipped with a unique trace and, as we will see, they give rise to fusion categories. To classify subfactors of type $II_1$ factors one can define an \emph{index}, and the trace is the key to this definition because it allows us to associate a dimension to vector spaces on which the factor acts. Let $M$ be a type $II_1$ factor with trace $\Tr_M:M\to\mathbb{C}$. For $x,y\in M$, the trace defines an inner product on $M$ via
	\begin{equation}
		\langle x,y\rangle=\Tr_M(y^* x).
	\end{equation}
The completion of $M$ with respect to the inner product defined above then yields a Hilbert space and is denoted $L^2(M)\equiv L^2(M,\Tr_M)$. The Hilbert space $L^2(M)$ is automatically a left-module over $M$ and a right module over $M^\mathrm{op}$ (which is the von Neumann algebra which is defined to be $M$ as complex vector space, but with the opposite multiplication), hence it is an $M$-bimodule (see \cite{Thom2006}). We call a Hilbert space which is at the same time a bimodule over $M$ a \emph{Hilbert bimodule}\index{Hilbert bimodule} (left and right Hilbert modules are defined analogously). The index of a subfactor is then defined as follows:
	\begin{defn}
		For a type $II_1$ factor $M$ and a subfactor $N\subseteq M$ the index\index{index}, denoted $[M:N]$, measures the dimension of $M$ as an $N$-module:
			\begin{equation}
				[M:N]=\dim_N L^2(M).
			\end{equation}
	\end{defn}
The precise definition of the dimension $\dim_N L^2(M)$ requires more mathematical details than are necessary for the purpose of this thesis, therefore we do not state it here but refer to \cite{Jones1997}, or, for a detailed mathematical treatment, \cite{Lueck2002}. Note that the index is equal to one if and only if $N=M$. 

Before we go into more detail about the possible values the index can take and the classification of subfactors that comes with it, we introduce an important example of a type $II_1$ factor: the \emph{hyperfinite} type $II_1$ factor\index{hyperfinite type 21 factor@hyperfinite type $II_1$ factor}, denoted $\mathcal{R}$. This factor is the unique smallest infinite-dimensional factor in the sense that it can be embedded in every other infinite-dimensional factor, and any infinite-dimensional factor contained in $\mathcal{R}$ is isomorphic to $\mathcal{R}$ (see \cite{Connes1976}). We can think of this factor as the inductive limit of inclusions of $n\times n$-matrices $\overline{\lim_{n\to\infty}M_{2^n\times2^n}(\mathbb{C})}$, i.e., the sequence
	\begin{equation}
		M_{2\times 2}\subseteq M_{4\times 4}\subseteq M_{8\times 8}\subseteq\dots\subseteq M_{2^n\times 2^n}\subseteq M_{2^{n+1}\times 2^{n+1}}\subseteq\dots
	\end{equation}
This means that we embed the $2\times 2$ matrices into $4\times 4$ matrices by building a block diagonal $4\times 4$ matrix, then embed the $4\times 4$ matrices by building a block diagonal $8\times 8$ matrix and so on. 

We now come back to the index of a subfactor. An important result by Jones \cite{Jones1983} gives the possible values the index can take:
	\begin{thm}
		Let $N\subseteq M$ be an inclusion of type $II_1$ factors. Then	
		\begin{equation}
			[M:N]\in\left\{4\cos^2\frac{\pi}{n}|n=3,4,\dots\right\}\cup[4,\infty],
		\end{equation}
		which means that the indices less than four accumulate at four and have gaps in between. Conversely, for any $\lambda\in\{4\cos^2\frac{\pi}{n}|n=3,4,\dots\}\cup[4,\infty]$ there exists a type $II_1$ factor $M$ and a subfactor $N$ with $[M:N]=\lambda$.
	\end{thm}
Furthermore, it can be shown that every index shows up as a possible index in the case where $M\cong N\cong\mathcal{R}$ is the hyperfinite type $II_1$ factor. The index can be visualized as follows:
	\begin{figure}[H]
		\centering
		\begin{tikzpicture}
			\draw[->,>=stealth] (0,0) -- (10,0); 
			\draw (0,-0.25) -- (0,0.25);
			\draw (1,-0.25) -- (1,0.25);
			\draw (2,-0.25) -- (2,0.25);
			\draw (3,-0.25) -- (3,0.25);
			\draw (4,-0.25) -- (4,0.25);
			\node at (0,-0.5) {$0$};
			\node at (1,-0.5) {$1$};
			\node at (2,-0.5) {$2$};
			\node at (3,-0.5) {$3$};
			\node at (4,-0.5) {$4$};
			\foreach \a in {0,0.25,0.5,...,4}{
				\draw (\a,0) -- (\a,-0.125);			
			}
			\foreach \a in {1., 2., 2.61803, 3., 3.24698, 3.41421, 3.53209, 3.61803, 3.68251, 3.73205, 3.77091, 3.80194, 3.82709, 3.84776, 3.86494, 3.87939, 3.89163, 3.90211, 3.91115, 3.91899, 3.92583, 3.93185, 3.93717, 3.94188, 3.94609, 3.94986, 3.95324, 3.9563}{
				\node at (\a,0) {\color{LinkColor} $\bullet$};
			}
			\draw[very thick, color=LinkColor] (4.125,-0.25) -- (4,-0.25) -- (4,0.25) -- (4.125,0.25);
			\draw[thick, color=LinkColor] (4,0) -- (9.915,0);
		\end{tikzpicture}
	\end{figure}

For subfactors with index smaller than four, there is a complete classification via the so-called \emph{standard invariant} \cite{Popa1990,Popa1994}. For subfactors with index four and above, i.e., all those who have an index in the continuous part, there is no complete classification yet. The main complication that arises here is that there are many subfactors which satisfy some technical condition called \emph{non-amenability}, where the results that Popa found cannot be applied. In \cite{Jones2013}, amenable subfactors with index between four and five are classified, but the authors point out that a classification beyond index five is still an open question and it is not even clear whether the question itself makes sense beyond this point. The subfactor that gives rise to the fusion categories we study in this thesis is the one with the smallest index between four and five, namely $\frac{5+\sqrt{13}}{2}$, and it is called the \emph{Haagerup subfactor}. But before we present it in more detail, we discuss how fusion categories can arise from subfactors.

\subsection*{The principal graphs.}

An algebra is a bimodule over any of its subalgebras, and bimodules over an algebra have tensor product structure given by the Connes fusion product\index{Connes fusion product} (see \cite{Connes80,Thom2006}). This is constructed as follows: Consider a left $A$-module $K$. We define the space
	\begin{equation}
		\mathcal{D}(K,\Tr_A)=\Hom_{-A}\left(L^2(A,\Tr_A),K\right)
	\end{equation}
of linear operators from $L^2(A,\Tr_A)$ to $K$ that are compatible with the left $A$-action (i.e., which are left $A$-module homomorphisms). The space of operators $\mathcal{D}(K,\Tr_A)$ is naturally a bimodule over $A$ itself (like every space of module homomorphisms), hence we can define the \emph{Connes fusion} of two $A$-modules $K$ and $L$ via the tensor product of corresponding operator spaces:
	\begin{equation}
		\mathcal{D}(K,\Tr_A)\otimes_A \mathcal{D}(L,\Tr_A).
	\end{equation}
Similar to the construction of $L^2(A,\Tr_A)$, we need to define an inner product on this space and take the completion with respect to this inner product in order to get a Hilbert bimodule (and not just a bimodule). For the construction of this inner product, we first define an $A$-valued inner product on $\mathcal{D}(K,\Tr_A)$ in the following way: For two operators $x,y\in\mathcal{D}(K,\Tr_A)$, we define
	\begin{equation}
		(x,y)=y^* x,
	\end{equation}
which is an element in $A$. The inner product on $\mathcal{D}(K,\Tr_A)\otimes_A \mathcal{D}(L,\Tr_A)$ is then defined as
	\begin{equation}
	\label{eq:inner}
		\langle x_1\otimes y_1,x_2\otimes y_2\rangle=\Tr_A\left((x_1,x_2)\triangleright y_1,y_2\right),
	\end{equation}
where $(x_1,x_2)\triangleright y_1$ denotes the left action of the operator $(x_1,x_2)\in A$ on $y_1$ regarded as an $A$-bimodule element. Finally, the Connes fusion 
	\begin{equation}
		K\otimes'_A L
	\end{equation}
of two Hilbert bimodules $K,L$ is defined as first forming the space $\mathcal{D}(K,\Tr_A)\otimes_A \mathcal{D}(L,\Tr_A)$ and then taking the completion with respect to the inner product defined in \cref{eq:inner}.

We can now apply the Connes fusion product to subfactors. For a subfactor $N\subseteq M$, it is natural to consider the tensor product 
	\begin{equation}
		M_k=M\otimes_N M\otimes_N M\dots\otimes_N M
	\end{equation}
with $k$ copies of $M$ in the tensor product. It is now possible to decompose $M_k$ into $N-N$, $M-M$, $N-M$, and $M-N$ bimodules, which allows us to extract finite-dimensional data from the, in general, infinite-dimensional $M_k$. Following the notions of \cite{Ocneanu1989} and \cite{Haagerup1994}, we define the four types of bimodules as follows: The $N-M$ bimodule is denoted
	\begin{equation}
		\alpha={}_NL^2(M)_M,
	\end{equation}
given by 
	\begin{equation}
		\alpha(n,m)x=nxm, \hspace{10pt}n\in N,\ m\in M,\ x\in L^2(M).
	\end{equation}
Similarly, the $M-N$ bimodule $\bar{\alpha}={}_ML^2(M)_N$ is given by
	\begin{equation}
		\bar{\alpha}(m,n)x=mxn, \hspace{10pt}m\in M,\ n\in N,\ x\in L^2(M).
	\end{equation}
Moreover, we denote $\epsilon_N={}_NL^2(M)_N$ and $\epsilon_M={}_ML^2(M)_M$ the trivial $N-N$ and $M-M$ bimodule, respectively. To simplify the notation, we omit the tensor product between bimodules: For two bimodules $\beta$ and $\gamma$ given by
	\begin{equation}
		\beta={}_P H_Q,\hspace{10pt} \gamma={}_Q K_R,\hspace{10pt} P,Q,R\in\{M,N\}
	\end{equation}
let $\beta\gamma$ denote the $P-R$ bimodule $\beta\otimes_Q\gamma$. With this notation at hand, we can now define the following four sets:
	\begin{enumerate}
		\item Principal even part\index{principal even part} $\Gamma_\mathrm{even}=$ the set of irreducible $N-N$ bimodules occurring in the decompositions of $\epsilon_N,\alpha\bar{\alpha},(\alpha\bar{\alpha})^2\dots$
		\item Principal odd part\index{principal odd part} $\Gamma_\mathrm{odd}=$ the set of irreducible $N-M$ bimodules occurring in the decompositions of $\alpha,\alpha\bar{\alpha}\alpha,\alpha(\bar{\alpha}\alpha)^2\dots$
		\item Dual even part\index{dual even part} $\Gamma'_\mathrm{even}=$ the set of irreducible $M-M$ bimodules occurring in the decompositions of $\epsilon_M,\bar{\alpha}\alpha,(\bar{\alpha}\alpha)^2\dots$
		\item Dual odd part\index{dual odd part} $\Gamma'_\mathrm{odd}=$ the set of irreducible $M-N$ bimodules occurring in the decompositions of $\bar{\alpha},\bar{\alpha}\alpha\bar{\alpha},\bar{\alpha}(\alpha\bar{\alpha})^2\dots$
	\end{enumerate}
We use the notation $\star=\epsilon_N$ and $\star'=\epsilon_M$. The principal graph\index{principal graph} of a subfactor is then a bipartite graph with vertices given by 
	\begin{equation}
		\Gamma=\Gamma_\mathrm{even}\cup\Gamma_\mathrm{odd},
	\end{equation}
which means that the vertices with even distance to $\star$ are the elements of $\Gamma_\mathrm{even}$ and the vertices with odd distance to $\star$ are given by the elements of $\Gamma_\mathrm{odd}$. The edges are constructed in the following way: For any fixed simple object $\gamma\in\Gamma_\mathrm{even}$, calculate $\gamma\alpha\in\Gamma_{\mathrm{odd}}$. For each $\delta\in\Gamma_\mathrm{odd}$ that appears in the decomposition of $\gamma\alpha$ into simple objects of $\Gamma_\mathrm{odd}$, we draw an edge between $\gamma$ and $\delta$ in the graph (note that we have to consider multiplicities here, which corresponds to drawing multiple lines between two objects).

In the same fashion, the dual graph\index{dual principal graph} of a subfactor has vertices
	\begin{equation}
		\Gamma'=\Gamma'_\mathrm{even}\cup\Gamma'_\mathrm{odd},
	\end{equation}
where $\Gamma'_\mathrm{even}$ is the set of vertices with even distance to $\star'$ and $\Gamma'_\mathrm{odd}$ is the set of vertices with odd distance to $\star'$. The edges are constructed analogously to those of the principle graph. The principal and dual graphs, as well as the index of a subfactor, serve as invariants of the subfactor, with the index being a weaker invariant than the graphs. We will see an explicit example of the principal and dual graph of a subfactor when studying the example of the Haagerup subfactor. We can now give the definition of finite depth of a subfactor:
	
\begin{defn}
	A subfactor is said to be of finite depth\index{finite-depth subfactor} if its principal graph is finite.
\end{defn}

Note that the principal graph is finite if and only if the dual principle graph is finite. In this case, their depth differs by at most one. The connection to fusion categories is now straightforward: The principal even part $\Gamma_\mathrm{even}$ and the dual even part $\Gamma'_\mathrm{even}$ have the structure of a monoidal category (see for example \cite{grossman_quantum_2012}). Moreover, if the subfactor is of finite depth, then these categories are fusion categories. These two fusion categories are connected by an equivalence relation called \emph{Morita equivalence} (for more details, see \cite{Nikshych2013,Etingof2005}):
	\begin{defn}[Morita equivalence]
		\index{Morita equivalence}
		Two fusion categories $\C_1$ and $\C_2$ are Morita equivalent, denoted $\C_1\approxeq\C_2$, if there is an invertible bimodule category\footnote{Bimodule categories will be defined in the next section.} ${}_{\C_1}\mathcal{M}_{\C_2}$, i.e., there exists another bimodule category ${}_{\C_2}{\mathcal{M}^\mathrm{op}}_{\C_1}$ such that $\C_1$ can be recovered as a $\C_1-\C_1$ bimodule category  (similar for $\C_2$):
			\begin{align}
				{}_{\C_1}\mathcal{M}_{\C_2}\otimes_{\C_2}{}_{\C_2}{\mathcal{M}^\mathrm{op}}_{\C_1}&\cong{}_{\C_1}{\C_1}_{\C_1},\\
				{}_{\C_2}{\mathcal{M}^\mathrm{op}}_{\C_1}\otimes_{\C_1}{}_{\C_1}\mathcal{M}_{\C_2}&\cong{}_{\C_2}{\C_2}_{\C_2}.
			\end{align}
	\end{defn}

\begin{prop}[\cite{muger_subfactors_2003}]
	\label{prop:Morita}
	For two fusion categories $\C_1$ and $\C_2$, the following statements are equivalent:	
		\begin{enumerate}
			\item $\C_1\approxeq\C_2$
			\item The respective Drinfeld centres $\mathcal{Z}(\C_1)$ and $\mathcal{Z}(\C_2)$ are equivalent as braided fusion categories.
			\item There is an algebra object $A\in\C_1$ such that the category of $A$--$A$ bimodules in $\C_1$ is monoidally equivalent to $\C_2$.
		\end{enumerate}
\end{prop}

So far, we have not discussed the objects that occur in the second and third statement in the above proposition, but we will see that these characterizations of Morita equivalence have a more constructive nature and, therefore, can be shown more easily. We discuss module categories and algebra objects in the next section and introduce the Drinfeld centre in \cref{sec:Drinfeld}.

\section{Module categories and algebra objects}

The link between the different fusion categories that arise from a finite-depth subfactor are so-called \emph{algebra objects}, as pointed out in \cref{prop:Morita}. In this section, we explain how one category can be constructed from the other as a \emph{bimodule category} from an algebra object, before studying an explicit example in the next chapter. We begin by introducing the basic notions of module categories.

	\begin{defn}[Module category]
		\label{def:leftmod}
		Let $\C$ be a monoidal category with associator $\alpha$ and left and right units $l$ and $r$. A left module category\index{module category} over $\C$ is a category $\mathcal{M}$ equipped with a left $\C$-action, which is a functor $\triangleright:\C\times\mathcal{M}\to\mathcal{M}$ and natural isomorphisms
			\begin{align}
				L_{X,Y,M}:(X\otimes Y)\triangleright M&\to X\triangleright(Y\triangleright M)\\
				u^l_M:\mathbf{1}\triangleright M&\to M
			\end{align}
		for $X,Y\in\C$ and $M\in\mathcal{M}$, such that the pentagon diagram 
			\begin{figure}[H]
				\centering
				\begin{tikzpicture}[scale=1.25,decoration={markings,mark=at position 1 with {\arrow[scale=1,thick]{>}}}] 
				\node(up) at (0,1) {$(X\otimes Y)\triangleright(Z\triangleright M)$};
				\node(left) at (-3,0) {$((X\otimes Y)\otimes Z)\triangleright M$};
				\node(right) at (3,0) {$X\triangleright (Y\triangleright (Z\triangleright M))$};
				\node(downleft) at (-2,-1.5) {$(X\otimes (Y\otimes Z))\triangleright M$};
				\node(downright) at (2,-1.5) {$X\triangleright ((Y\otimes Z)\triangleright M)$};
				\draw[postaction={decorate}] (left) -- (up) node[pos=0.7,left=0.3cm] {\small $L_{X\otimes Y,Z,M}$};
				\draw[postaction={decorate}] (up) -- (right) node[pos=0.3,right=0.3cm] {\small $L_{X,Y,Z\otimes M}$};
				\draw[postaction={decorate}] (left) -- (downleft) node[midway,left=0.1cm] {\small $\alpha_{X,Y,Z}\otimes\id_M$};
				\draw[postaction={decorate}] (downleft) -- (downright) node[midway,above] {\small $L_{X, Y\otimes Z,M}$};
				\draw[postaction={decorate}] (downright) -- (right) node[midway,right=0.3cm] {\small $\id_X\otimes L_{Y,Z,M}$};
				\end{tikzpicture}
			\end{figure}
		\noindent
		and the triangle diagram
			\begin{figure}[H]
				\centering
				\begin{tikzpicture}[scale=1.6,decoration={markings,mark=at position 1 with {\arrow[scale=1,thick]{>}}}] 
				\node(up) at (0,-2) {$X\triangleright M$};
				\node(left) at (-1.5,-1) {$(X\otimes\mathbf{1})\triangleright M$};
				\node(right) at (1.5,-1) {$X\triangleright(\mathbf{1}\triangleright M)$};
				\draw[postaction={decorate}] (left)--(up) node[midway,left] {\small $r_X\otimes\id_M\ $};
				\draw[postaction={decorate}] (right)--(up) node[midway,right] {\small $\ \ \id_X\otimes u^l_M$};
				\draw[postaction={decorate}] (left)--(right) node[midway,above] {\small $L_{X,\mathbf{1},M}$};
				\end{tikzpicture}
			\end{figure}
		\noindent
		commute for all objects $X,Y,Z\in\C$ and $M\in\mathcal{M}$.
	\end{defn}
A right module category can be defined analogously with natural isomorphisms $R_{M,X,Y}:M\triangleleft(X\otimes Y)\to (M\triangleleft X)\triangleleft Y$ and $u_M^r:M\triangleleft \mathbf{1}\to M$ and the corresponding commuting pentagon and triangle diagram. If we choose bases for all the vector spaces $\Hom_\mathcal{M}(X\triangleright M,N)$ (similar to when we introduced the $F$-symbols), the map $L_{X,Y,M}$ (the associator) can be written using string diagrams
	\begin{equation}
,
	\end{equation}
where we have coloured the objects from the module category in red for clarity. For a right module, the associator can be written as	
	\begin{equation}
		%
.
	\end{equation}
	
Furthermore, we can define the notion of a \emph{bimodule category} over over a pair of monoidal categories:
	\begin{defn}[Bimodule category]
		Let $\C,\mathcal{D}$ be monoidal categories. A $(\C,\mathcal{D})$-bimodule category\index{bimodule category} is a category $\mathcal{M}$ (also written $\C\curvearrowright\mathcal{M}\curvearrowleft\mathcal{D}$) that has left $\C$-module and right $\mathcal{D}$-module category structures with associativity constraints $L_{X,Y,M}:(X\otimes Y)\triangleright M\to X\triangleright(Y\triangleright M)$ and $u_M^l:\mathbf{1}\triangleright M\to M$, and $R_{M,X,Y}:M\triangleleft(X\otimes Y)\to (M\triangleleft X)\triangleleft Y$ and $u_M^r:M\triangleleft \mathbf{1}\to M$, respectively, which are compatible by a collection of natural isomorphisms 
			\begin{equation}
				C_{X,M,Y}:(X\triangleright M)\triangleleft Y\to X\triangleright (M\triangleleft Y),
			\end{equation}
		called the \emph{middle associativity constraint}, such that the following diagrams commute for all $X,Y\in\C$, $Z,W\in\mathcal{D}$, and $M\in\mathcal{M}$:
		\begin{figure}[H]
			\centering

		\end{figure}
	\end{defn}

As for left and right modules, we can also choose bases for the vector space involved in the definition of the middle associativity constraint of a bimodule category, hence $C_{X,M,Y}$ can be depicted as
	\begin{equation}
		%
.
	\end{equation}

\begin{rem}
	Note that any category $\C$ is a left (respectively, right) module category over itself: The associator $L$ (respectively, $R$) and the unit constraint $u^l$ (respectively, $u^r$) are simply given by the associator $\alpha$ and the unit constraint $l$ (respectively, $r$) of the category $\C$ itself.
\end{rem}

\begin{exmp}[$\VecZ$ bimodules]
	\label{ex:Vecbimod}
	\index{Vec(Z/pZ) category@$\VecZ$ category}
	We now illustrate the concept of bimodules with a concrete category, namely $\VecZ$. We need these bimodules in \cref{ch:defects} to describe defects in lattice systems. This data is taken from \cite{barter_domain_2019}, and we follow the notation used in there. In general, the simple objects in $\VecZ$ are $\{0,1,2,\dots,p\}$, where we use $0$ to denote the vacuum instead of $\mathbf{1}$ since it makes notation easier in this case. The fusion rules are simply given by addition modulo $p$ and the associator is always trivial.
	
	There are several $\VecZ-\VecZ$ bimodules, both invertible and non-invertible ones, but only one family of these bimodules has a non-trivial associator. These bimodules are denoted $F_q$ with $q\in\mathbb{Z}/p\mathbb{Z}$. For a fixed $p$, the category only has one object, denoted by \textasteriskcentered. Left and right action are given by
		\begin{equation}
			\begin{tikzpicture}[scale=1.25,baseline=(current bounding box.center)]
				\draw[very thick, color=LinkColor] (0,0) -- (0,1);
				\draw (-0.5,0) to [bend left] (0,0.5);
				\node at (-0.5,-0.15) {\small $a$};
				\node at (0,-0.15) {\small \color{LinkColor} \textasteriskcentered};
				\node at (0,1.15) {\small \color{LinkColor} \textasteriskcentered};
			\end{tikzpicture},\hspace{30pt}
			\begin{tikzpicture}[xscale=-1.25,yscale=1.25,baseline=(current bounding box.center)]
				\draw[very thick, color=LinkColor] (0,0) -- (0,1);
				\draw (-0.5,0) to [bend left] (0,0.5);
				\node at (-0.5,-0.15) {\small $a$};
				\node at (0,-0.15) {\small \color{LinkColor} \textasteriskcentered};
				\node at (0,1.15) {\small \color{LinkColor} \textasteriskcentered};
			\end{tikzpicture}
		\end{equation}
	and the associator is
		\begin{equation}
			\begin{tikzpicture}[scale=1.25,baseline=(current bounding box.center)]
				\draw[very thick, color=LinkColor] (0,0) to node[left] {\small \textasteriskcentered} (0,1.5);
				\draw (-0.5,0) to [bend left] (0,0.5);
				\draw (0.5,0) to [bend right] (0,1);
				\node at (0.5,-0.25) {\small $b$};
				\node at (-0.5,-0.25) {\small $a$};
				\node at (0,-0.25) {\small \color{LinkColor} \textasteriskcentered};
				\node at (0,1.75) {\small \color{LinkColor} \textasteriskcentered};
			\end{tikzpicture}=e^{\frac{2\pi i}{p}qab}
			\begin{tikzpicture}[scale=1.25,baseline=(current bounding box.center)]
				\draw[very thick, color=LinkColor] (0,0) to node[right] {\small \textasteriskcentered} (0,1.5);
				\draw (-0.5,0) to [bend left] (0,1);
				\draw (0.5,0) to [bend right] (0,0.5);
				\node at (0.5,-0.25) {\small $b$};
				\node at (-0.5,-0.25) {\small $a$};
				\node at (0,-0.25) {\small \color{LinkColor} \textasteriskcentered};
				\node at (0,1.75) {\small \color{LinkColor} \textasteriskcentered};
			\end{tikzpicture}
		\end{equation}
	for $a,b\in\VecZ$. Among the bimodules $F_q$, the case $q=0$ is special: $F_0$ is the only one of these bimodules that is not invertible and, furthermore, it has a trivial associator since $e^{\frac{2\pi i}{p}qab}=1$ for $q=0$. In \cref{ch:defects} we are especially interested in $\mathbf{Vec}(\mathbb{Z}/2\mathbb{Z})-\mathbf{Vec}(\mathbb{Z}/2\mathbb{Z})$ bimodules. Here, the family $F_q$ consists of two bimodules: the non-invertible bimodule $F_0$ with trivial associator, and the invertible bimodule $F_1$ with non-trivial associator given by 
		\begin{equation}
			\begin{tikzpicture}[scale=1.25,baseline=(current bounding box.center)]
				\draw[very thick, color=LinkColor] (0,0) to node[left] {\small \textasteriskcentered} (0,1.5);
				\draw (-0.5,0) to [bend left] (0,0.5);
				\draw (0.5,0) to [bend right] (0,1);
				\node at (0.5,-0.25) {\small $b$};
				\node at (-0.5,-0.25) {\small $a$};
				\node at (0,-0.25) {\small \color{LinkColor} \textasteriskcentered};
				\node at (0,1.75) {\small \color{LinkColor} \textasteriskcentered};
			\end{tikzpicture}=(-1)^{ab}
			\begin{tikzpicture}[scale=1.25,baseline=(current bounding box.center)]
				\draw[very thick, color=LinkColor] (0,0) to node[right] {\small \textasteriskcentered} (0,1.5);
				\draw (-0.5,0) to [bend left] (0,1);
				\draw (0.5,0) to [bend right] (0,0.5);
				\node at (0.5,-0.25) {\small $b$};
				\node at (-0.5,-0.25) {\small $a$};
				\node at (0,-0.25) {\small \color{LinkColor} \textasteriskcentered};
				\node at (0,1.75) {\small \color{LinkColor} \textasteriskcentered};
			\end{tikzpicture}.
		\end{equation}
\end{exmp}

It is possible to define a tensor product on bimodules via the \emph{relative tensor product}\index{relative tensor product}. We will need this later in this thesis when we study spin chains that are constructed from fusion categories, where bimodules are used to add defects to the chain.

\begin{defn}[Relative tensor product]
	For an $(\mathcal{A},\mathcal{B})$ bimodule category $\mathcal{M}$ and a $(\mathcal{B},\mathcal{C})$ bimodule $\mathcal{N}$ the relative tensor product\index{relative tensor product} $\mathcal{M}\otimes_\mathcal{B}\mathcal{N}$ consists of objects $(m,n)\in\mathcal{M}\otimes\mathcal{N}$ along with isomorphisms
		\begin{equation}
			\beta:(m\triangleleft b,n)\to(m,b\triangleright n)
		\end{equation}
	such that these isomorphisms are compatible with the respective module structures, for example, the following diagram commutes:
		\begin{figure}[H]
			\centering
			\begin{tikzpicture}[scale=1.25,decoration={markings,mark=at position 1 with {\arrow[scale=1,thick]{>}}}] 
			\node(up) at (0,1) {$(m\triangleleft(b_1\otimes b_2),n)$};
			\node(left) at (-3,0) {$((m\triangleleft b_1)\triangleleft b_2,n)$};
			\node(right) at (3,0) {$(m,(b_1\otimes b_2)\triangleright n)$};
			\node(downleft) at (-2,-1.5) {$(m\triangleleft b_1,b_2\triangleright n)$};
			\node(downright) at (2,-1.5) {$(m,b_1\triangleright(b_2\triangleright n))$};
			\draw[postaction={decorate}] (up) to node[above, pos=0.6] {\small $R_{m,b_1,b_2}\ \ $} (left) ;
			\draw[postaction={decorate}] (up) to node[above, pos=0.6] {\small $\beta$} (right) ;
			\draw[postaction={decorate}] (left) to node[left,pos=0.6] {\small $\beta\ $} (downleft) ;
			\draw[postaction={decorate}] (downleft) to node[above] {\small $\beta\ $} (downright) ;
			\draw[postaction={decorate}] (right) to node[right,pos=0.6] {\small $\ L_{b_1,b_2,n}$} (downright) ;
			\end{tikzpicture}
		\end{figure}
	\noindent
	where $L$ denotes the associator for the left action in $\mathcal{N}$ and $R$ denotes the associator for the right action in $\mathcal{M}$. Morphisms in $\mathcal{M}\otimes_\mathcal{B}\mathcal{N}$ are morphisms in $\mathcal{M}\otimes\mathcal{N}$ that are compatible with $\beta$.
\end{defn} 
	
	The above definition appears in \cite{Douglas2019} (see also \cite{barter_domain_2019}). Note that $\mathcal{M}\otimes_\mathcal{B}\mathcal{N}$ as defined above is an $(\mathcal{A},\mathcal{C})$ bimodule, and it can be decomposed into simple bimodules $\mathcal{P}$:
		\begin{equation}
			\mathcal{M}\otimes_\mathcal{B}\mathcal{N}\cong\bigoplus_\mathcal{P}N_{\mathcal{M},\mathcal{N}}^\mathcal{P}\ \mathcal{P},
		\end{equation}
	where $N_{\mathcal{M},\mathcal{N}}^\mathcal{P}$ are the coefficients in the decomposition. More details can be found in \cite{Douglas2019}. This decomposition is similar to the fusion rules within a fusion category, therefore it is natural to introduce bimodule trivalent vertices if a given bimodule $\mathcal{P}$ appears in the decomposition of $\mathcal{M}\otimes_\mathcal{B}\mathcal{N}$:
		\begin{equation}
			\begin{tikzpicture}
				\draw[very thick,color=LinkColor] (0,0) -- (90:1);
				\draw[very thick,color=LinkColor2] (0,0) -- (220:1);
				\draw[very thick,color=LinkColor3] (0,0) -- (320:1); 
				\node at (0,1.3) {$\mathcal{P}$};
				\node at (220:1.3) {$\mathcal{M}$};
				\node at (320:1.3) {$\mathcal{N}$};
			\end{tikzpicture}
		\end{equation}

Given a monoidal category $\C$, how can we construct a left (or right) module category or a bimodule category over $\C$ from it? One possible construction exploits so-called \emph{algebra objects} of the monoidal category.
In the following, we present this construction in the case of strict fusion categories, since this is the scenario in which we will apply it later. Remember that every fusion category is equivalent to a strict one (see \cref{thm:strictness}). In general, algebra objects can be defined for any monoidal category. A general and detailed mathematical treatment of this topic can be found in \cite{Etingof2015}.

\begin{defn}[Algebra object]
	\label{def:algebraobject}
	Let $\C$ be a strict fusion category. An object $A\in\C$ is called an algebra object\index{algebra object} if there is a multiplication morphism $m:A\otimes A\to A$ and a unit morphism $i:\mathbf{1}\to A$, satisfying the following constraints:
		\begin{align}
			&m\circ (\id_A\otimes m)=m\circ(m\otimes\id_A)\hspace{10pt}\text{ as maps }A\otimes A\otimes A\to A \label{eq:AO1},\\
			&m\circ(i\otimes \id_A)=\id_A=m\circ(\id_A\otimes i)\hspace{10pt}\text{ as maps }A\to A.\label{eq:AO2}
		\end{align}
\end{defn}

\begin{rem}
	In any fusion category $\C$, the unit object $\mathbf{1}\in\C$ is an algebra object.
\end{rem}

Using the graphical notation that was introduced in the previous chapters, we can represent the maps $m$ and $i$ by a trivalent and a univalent vertex (thereby using the convention that we do not draw the trivial object $\mathbf{1}$), respectively:
	\begin{equation}
.
	\end{equation}
The constraint \cref{eq:AO1} given in the definition above can then be depicted as
	\begin{equation}
	\label{eq:mcondpic}
		%
,
	\end{equation}
and the constraint in \cref{eq:AO2} is depicted
	\begin{equation}
		%
.
	\end{equation}
	
To turn the problem of finding all algebra objects in a fusion category into a finite problem, there is a theorem that bounds the size of possible algebra objects (for a proof, see \cite[Lemma 3.8]{grossman_quantum_2012}):
	\begin{thm}
		\label{thm:algobjdim}
		For a simple algebra object $A$ in a fusion category $\C$, i.e., an algebra object that decomposes into simple objects $X_i\in\Obj(\C)$, it holds that 
			\begin{equation}
				\#X_i\in A\le\dim(X_i).
			\end{equation}
	\end{thm}
	
\begin{exmp}[Fibonacci category]
	\label{ex:AOFib}
	We illustrate the procedure of finding all algebra objects in a fusion category with a simple example, namely the Fibonacci category \textbf{Fib}\index{Fibonacci category} that was introduced in \cref{ex:TrivFib}. Recall that it has two simple objects $\Obj(\mathbf{Fib})=\{\mathbf{1},\tau\}$\footnote{We have changed the notation here: In \cref{ex:TrivFib}, we called the non-trivial object $X$ in order to avoid confusion with the trivalent vertex $\tau$. However, since $\tau$ is the standard notation for this object in \textbf{Fib}, we use it from here on.} and the only non-trivial fusion rule is
		\begin{equation}
			\tau\otimes\tau=\mathbf{1}+\tau.
		\end{equation}
	In \textbf{Fib}, all simple algebra objects are of the form
		\begin{equation}
			a_\mathbf{1}\cdot\mathbf{1}+a_\tau\cdot\tau,
		\end{equation}
	where, according to \cref{thm:algobjdim}, $a_\mathbf{1}\le\mathbf{1}$ and $a_\tau\le\dim(\tau)=\phi\approx 1.618$. Therefore, there are three possible algebra objects: $\mathbf{1}$, $\tau$, and $\mathbf{1}+\tau$. 
		\begin{enumerate}
			\item The vacuum object $\mathbf{1}$ is trivially an algebra object since the conditions in \cref{def:algebraobject} are fulfilled by setting
				\begin{equation}
					\begin{tikzpicture}[scale=0.85,baseline=(current bounding box.center)]
						\draw (0,0) to (90:1);
						\draw (0,0) to (230:1);
						\draw (0,0) to (310:1);	
						\draw[fill=LinkColor2!50!white] (0,0) circle(0.35cm);	
						\node (c) at (0,0) {\small $m_\mathbf{1}$};
						\node at (90:1.25) {\small $\mathbf{1}$};	
						\node at (230:1.25) {\small $\mathbf{1}$};
						\node at (310:1.25) {\small $\mathbf{1}$};	
					\end{tikzpicture}\equiv
					\begin{tikzpicture}[scale=0.7,baseline=(current bounding box.center)]
						\draw (0,0) to (90:1);
						\draw (0,0) to (230:1);
						\draw (0,0) to (310:1);	
						\node at (90:1.25) {\small $\mathbf{1}$};	
						\node at (230:1.25) {\small $\mathbf{1}$};
						\node at (310:1.25) {\small $\mathbf{1}$};
					\end{tikzpicture}.
				\end{equation}
			\item To check whether the object $\tau$ is an algebra object, we need to find the multiplication morphism
				\begin{equation}
					\begin{tikzpicture}[scale=0.85,baseline=(current bounding box.center)]
						\draw (0,0) to (90:1);
						\draw (0,0) to (230:1);
						\draw (0,0) to (310:1);	
						\draw[fill=LinkColor2!50!white] (0,0) circle(0.35cm);	
						\node (c) at (0,0) {\small $m_\tau$};
						\node at (90:1.25) {\small $\tau$};	
						\node at (230:1.25) {\small $\tau$};
						\node at (310:1.25) {\small $\tau$};	
					\end{tikzpicture}= c_{\tau\tau}^{\tau}
					\begin{tikzpicture}[scale=0.7,baseline=(current bounding box.center)]
						\draw (0,0) to (90:1);
						\draw (0,0) to (230:1);
						\draw (0,0) to (310:1);	
						\node at (90:1.25) {\small $\tau$};	
						\node at (230:1.25) {\small $\tau$};
						\node at (310:1.25) {\small $\tau$};
					\end{tikzpicture},
				\end{equation}
				hence we need to determine the coefficient $c_{\tau\tau}^{\tau}$. This is done by checking the condition \cref{eq:mcondpic}:
					\begin{equation}
						{c_{\tau\tau}^\tau}^2\fuselefttree{\tau}{\tau}{\tau}{\tau}{\tau}={c_{\tau\tau}^\tau}^2\fuserighttree{\tau}{\tau}{\tau}{\tau}{\tau}.
					\end{equation}
				On the other hand, applying an $F$-symbol\footnote{The calculation of the $F$-symbols of the Fibonacci category are explained in detail in \cref{ex:FibAnyon}.} to the left hand side of the equation yields
					\begin{equation}
						{c_{\tau\tau}^\tau}^2\fuselefttree{\tau}{\tau}{\tau}{\tau}{\tau}={c_{\tau\tau}^\tau}^2\left(F_\tau^{\tau\tau\tau}\right)_{\mathbf{1}\tau}\fuserighttree{\tau}{\tau}{\mathbf{1}}{\tau}{\tau}+{c_{\tau\tau}^\tau}^2\left(F_\tau^{\tau\tau\tau}\right)_{\tau\tau}\fuserighttree{\tau}{\tau}{\tau}{\tau}{\tau}.
					\end{equation}
				Comparing the two equations yields the system of conditions
					\begin{align}
						0&={c_{\tau\tau}^\tau}^2\left(F_\tau^{\tau\tau\tau}\right)_{\mathbf{1}\tau}={c_{\tau\tau}^\tau}^2\phi^{-\frac{1}{2}}\\
						{c_{\tau\tau}^\tau}^2&={c_{\tau\tau}^\tau}^2\left(F_\tau^{\tau\tau\tau}\right)_{\tau\tau}=-{c_{\tau\tau}^\tau}^2\phi^{-1},
					\end{align}
				which does not have a solution. Therefore, $\tau$ is not an algebra object in \textbf{Fib}. A second (and, in this case, simpler) way to see that $\tau$ is not an algebra object is to check whether the unit morphism $i:\mathbf{1}\to\tau$ exists. Since $\Hom(\mathbf{1},\tau)=\emptyset$, there is no unit morphism for $\tau$, hence it is not an algebra object.
			\item The last candidate is $\mathbf{1}+\tau$. The corresponding multiplication morphism is
				\begin{equation}
					\hspace{30pt}
,
				\end{equation}
				so we need to find five coefficients. We can already conclude that $c_{\mathbf{1}\mathbf{1}}^\mathbf{1}=c_{\tau\mathbf{1}}^\tau=c_{\mathbf{1}\tau}^\tau=1$ from \cref{eq:AO2}, which states that
				\begin{equation}
					%
.
				\end{equation}
				The remaining coefficients can be determined by inserting the expression for $m_{\mathbf{1}+\tau}$ given above into \cref{eq:mcondpic}, which yields
					\begin{equation}
						\hspace{30pt}%
\hspace{-10pt}={c_{\mathbf{1}\mathbf{1}}^\mathbf{1}}^2\fuselefttree{\mathbf{1}}{\mathbf{1}}{\mathbf{1}}{\mathbf{1}}{\mathbf{1}}+{c_{\tau\mathbf{1}}^\tau}^2\fuselefttree{\tau}{\mathbf{1}}{\tau}{\mathbf{1}}{\tau}+c_{\mathbf{1}\tau}^\tau c_{\tau\mathbf{1}}^\tau\fuselefttree{\mathbf{1}}{\tau}{\tau}{\mathbf{1}}{\tau}+\dots
					\end{equation}
				with $13$ terms in total and a similar expression for the right hand side of the equation. Consider those terms with diagrams in $\Hom(\tau\otimes\tau\otimes\tau,\tau)$ and apply $F$-moves to them:
				\begin{align}
					\hspace{10pt}&c_{\tau\tau}^\mathbf{1}c_{\mathbf{1}\tau}^\tau\fuselefttree{\tau}{\tau}{\mathbf{1}}{\tau}{\tau}+{c_{\tau\tau}^\tau}^2\fuselefttree{\tau}{\tau}{\tau}{\tau}{\tau}\\
					&\hspace{20pt}=\sum_xc_{\tau\tau}^\mathbf{1}c_{\mathbf{1}\tau}^\tau\left(F_\tau^{\tau\tau\tau}\right)_{x\mathbf{1}}\fuserighttree{\tau}{\tau}{x}{\tau}{\tau}+\sum_y{c_{\tau\tau}^\tau}^2\left(F_\tau^{\tau\tau\tau}\right)_{y\tau}\fuserighttree{\tau}{\tau}{y}{\tau}{\tau}.
				\end{align}
				Comparing this to the corresponding terms on the right hand side of the equation and setting $c_{\mathbf{1}\mathbf{1}}^\mathbf{1}=c_{\tau\mathbf{1}}^\tau=c_{\mathbf{1}\tau}^\tau=1$ yields the following system of equations:
				\begin{align}
					c_{\tau\tau}^\mathbf{1}&=c_{\tau\tau}^\mathbf{1}\left(F_\tau^{\tau\tau\tau}\right)_{\mathbf{1}\mathbf{1}}+{c_{\tau\tau}^\tau}^2\left(F_\tau^{\tau\tau\tau}\right)_{\mathbf{1}\tau}\\
					{c_{\tau\tau}^\tau}^2&=c_{\tau\tau}^\mathbf{1}\left(F_\tau^{\tau\tau\tau}\right)_{\tau\mathbf{1}}+{c_{\tau\tau}^\tau}^2\left(F_\tau^{\tau\tau\tau}\right)_{\tau\tau},
				\end{align}
				which has the following solution:
				\begin{align}
					c_{\tau\tau}^\mathbf{1}=\sqrt{2+\sqrt{5}}\ {c_{\tau\tau}^\tau}^2,
				\end{align}
				where $c_{\tau\tau}^\tau$ remains a free parameter since it is also not determined by checking the remaining equations. Hence, $\mathbf{1}+\tau$ is an algebra object in \textbf{Fib} with the multiplication morphism determined by the coefficients calculated above.
		\end{enumerate}	
\end{exmp}

Given an algebra object $A$ in a fusion category $\C$, one can build a \emph{module category} over $\C$ from it. For this purpose, we give an alternative definition of a module category in terms of the algebra object:
	\begin{defn}[Module over algebra object]
		\label{def:Amod}
		A left module over an algebra object $A$ in a category $\C$ (or, simply, a left $A$-module)\index{A-module@$A$-module} is a pair $(M,l)$ consisting of an object $M\in\C$ and a morphism $l:A\otimes M\to M$, depicted
			\begin{equation}
,
			\end{equation}
		such that it is compatible with the multiplication morphism $m$ of the algebra object, i.e., the following condition is fulfilled:
			\begin{align}
				l\circ(m\otimes\id_M)=l\circ(\id_A\otimes l)\hspace{10pt}\text{ as maps }A\otimes A\otimes M\to M \label{eq:MO1}.
			\end{align}
	\end{defn}

The constraint stated in \cref{eq:MO1} can be depicted as
\begin{equation}
	%
.
\end{equation}

Analogously, one can define a right module over an algebra object $A\in\C$ via a morphism $r:M\otimes A\to M$ and the corresponding compatibility constraint.

	\begin{defn}[Module category over algebra object]
		The left $A$-modules in $\C$ form a category $\mathsf{Mod}_\C(A)$ whose objects consist of pairs $(M,l)$ given in \cref{def:Amod}\index{module category}. The morphisms in this category (also called $A$-left morphisms) are maps between module objects $(M_1,l_1)\to(M_2,l_2)$ that are compatible with the corresponding morphisms between $M_1,M_2$ in $\C$. More precisely, the following condition has to be fulfilled for $f\in\Hom_\C(M_1,M_2)$:
			\begin{equation}
			\label{eq:morphMod}
				f\circ l_1=l_2\circ(\id_A\otimes f)\hspace{10pt}\text{ as maps }A\otimes M_1\to M_2,
			\end{equation}
		depicted
			\begin{equation}
				\begin{tikzpicture}[scale=0.85,baseline=(current bounding box.center)]
					\draw (0,0) to (90:1);
					\draw (0,0) to (230:1);
					\draw (0,0) to (310:1);	
					\draw[fill=LinkColor!50!white] (0,0) circle(0.325cm);	
					\node (c) at (0,0) {\small $l_1$};
					\node at (90:1.25) {\small $M_1$};	
					\node at (230:1.25) {\small $A$};
					\node at (310:1.3) {\small $M_1$};
					\draw (0,1.5) -- (0,3);
					\draw[fill=LightGray] (0,2.25) circle(0.325cm);
					\node (d) at (0,2.25) {\small $f$};
					\node at (0,3.25) {$M_2$};
				\end{tikzpicture}=
				\begin{tikzpicture}[scale=0.85,baseline=(current bounding box.center)]
					\draw (0,0) to (90:1);
					\draw (0,0) to (230:1);
					\draw (0,0) to (310:1);	
					\draw[fill=LinkColor!50!white] (0,0) circle(0.325cm);	
					\node (c) at (0,0) {\small $l_2$};
					\node at (90:1.25) {\small $M_2$};	
					\node at (230:1.25) {\small $A$};
					\node at (310:1.3) {\small $M_2$};
					\draw (0.77,-2.75) -- (0.77,-1.25);
					\draw[fill=LightGray] (0.77,-2) circle(0.325cm);	
					\node (c) at (0.77,-2) {\small $f$};
					\draw (-0.77,-1.25) -- (-0.77,-2.75);
					\node at (-0.77,-3) {\small $A$};
					\node at (0.77,-3) {\small $M_1$};
				\end{tikzpicture}.
			\end{equation}
	\end{defn}

\begin{rem}
	\label{rem:Caction}
	Note that for any left $A$-module $(M,l)$ and any object $X\in\C$, we can define a functor $\circlearrowleft:\mathsf{Mod}_\C(A)\times \C\to\mathsf{Mod}_\C(A)$ via
		\begin{equation}
			(M,l)\circlearrowleft X=(M\otimes X,l\otimes \id_X).
		\end{equation}
	The object $M\otimes X$ has again the structure of a left $A$-module given by the composition
		\begin{equation}
			A\otimes(M\otimes A)=(A\otimes M)\otimes X\xrightarrow{l\otimes\id_X}M\otimes X.
		\end{equation}
	Hence, the category $\mathsf{Mod}_\C(A)$ is a right $\C$-module category in the sense of \cref{def:leftmod} with the right $\C$-action given by the functor $\circlearrowleft$. With the same argument, the category of right $A$-modules over $\C$ is a left $\C$-module category.
\end{rem}

In this way, we get a general construction of module categories from algebra objects in the category $\C$. An important question that immediately arises is whether every module category over $\C$ can be constructed in this way, and the answer is no. Consider the category $\mathsf{Vec}$ of all finite-dimensional vector spaces. The module category of \emph{all} vector spaces (which includes infinite dimensional ones) is not of the form $\mathsf{Mod}_{\C}(A)$. However, for $\C=\mathsf{Vec}$ any \emph{finite} module category is of the form $\mathsf{Mod}_{\mathsf{Vec}}(A)$. In general, it can be shown that all finite module categories over a finite monoidal category $\C$ are of the form $\mathsf{Mod}_{\mathsf{Vec}}(A)$ for a suitable algebra object $A$ (see \cite[Thm.\ 7.10.1]{Etingof2015} and \cite{Ostrik2003}).

With the definition of an $A$-module at hand we can rephrase the definition of bimodules in terms of algebra objects:
	\begin{defn}[Bimodule]
		Let $A,B$ be two algebra objects in a fusion category $\C$. An $A$--$B$ bimodule\index{bimodule} in $\C$ is a triple $(M,l_A,r_B)$ where $M\in\C$ and $l_A:A\otimes M\to M$, $r_B:M\otimes B\to M$ such that
			\begin{enumerate}
				\item The pair $(M,l_A)$ is a left $A$-module in $\C$.
				\item The pair $(M,r_B)$ is a right $B$-module in $\C$.
				\item The following condition is fulfilled:
						\begin{equation}
							r_B\circ(l_A\otimes\id_B)=l_A\circ(\id_A\otimes r_B),
						\end{equation}
					depicted
					\begin{equation}
.
					\end{equation}
			\end{enumerate}
	\end{defn}
	
	\begin{rem}
		The $A$--$B$ bimodules in a category $\C$ form a category that we denote $\mathsf{Bimod}_\C(A,B)$. A morphism $f$ in this category is both an $A$-left morphism and a $B$-right morphism.
	\end{rem}

\begin{exmp}[Fibonacci category]
	\label{ex:FibMod}
	As shown in \cref{ex:AOFib}, the Fibonacci category \textbf{Fib}\index{Fibonacci category} has only two algebra objects: The trivial one $\mathbf{1}$ and $\mathbf{1}+\tau$ with multiplication morphism $m_{\mathbf{1}+\tau}$ given by
		\begin{equation}
			%
.
		\end{equation}
	We now show how to construct the category $\mathsf{Mod}_{\mathbf{Fib}}(\mathbf{1}+\tau)$ of left module objects over the algebra object $\mathbf{1}+\tau$. The technique we use here is described in \cite{grossman_quantum_2012}. We begin by determining all simple module objects with the following procedure: For this purpose, we calculate the fusion product of the algebra object $\mathbf{1}+\tau$ with all the simple objects in \textbf{Fib}, i.e., 
		\begin{equation}
			\label{eq:AfuseFib}
			A\otimes X_i
		\end{equation}
	for $X_i\in\{\mathbf{1},\tau\}$. We then build a matrix $T$ in the following way: Index each row and each column by a simple object $X_i\in\mathbf{Fib}$. A row of $T$ is given by the outcome of the fusion product \cref{eq:AfuseFib}, where the coefficients in the decomposition into simple objects are the entries in the row corresponding to the object $X_i$. For our example $A=\mathbf{1}+\tau$ we need to calculate
		\begin{align}
			\left(\mathbf{1}\otimes\tau\right)\otimes \mathbf{1}&=\mathbf{1}+\tau\\
			\left(\mathbf{1}\otimes\tau\right)\otimes \tau&=\tau+(\mathbf{1}+\tau)=\mathbf{1}+2\tau.
		\end{align}
	Hence, the matrix $T$ is of the form
		\begin{equation}
			T=\begin{blockarray}{ccc}
			& \mathbf{1} & \tau \\
			\begin{block}{c (cc)}
			\mathbf{1} & 1 & 1 \topstrut \\
			\tau & 1 & 2 \botstrut \\
			\end{block}
			\end{blockarray}\ .
		\end{equation} 
	It can be rewritten as
		\begin{equation}
			T=VV^\mathrm{T},
		\end{equation}
	where $V$ is the matrix that represents the action defined in \cref{rem:Caction}. Note that it is only possible to write $T$ in the form above if it is positive semidefinite, hence if this property is not fulfilled we can directly conclude that the corresponding object is not an algebra object. This matrix is the connection to the category of left modules: For a general fusion category $\C$ and a left module category $\mathsf{Mod}_\C(A)$ constructed from an algebra object $A$, the rows of $V$ are indexed by simple objects $X_i\in\C$ and the columns are indexed by simple objects $M_j\in\mathsf{Mod}_\C(A)$, hence $V$ is of the form
		\begin{equation}
		\label{eq:matrixV}
			V= \begin{blockarray}{ccccc}
			& M_1 & M_2 & M_3 & ...\\
			\begin{block}{c (cccc)}
			X_1 & ... & ... & ... & ... \topstrut\\
			X_2 & ... & ... & ... & ... \\
			X_3 & ... & ... & ... & ... \\
			... & ... & ... & ... & ... \botstrut\\
			\end{block}
			\end{blockarray}\ .
		\end{equation}
	This makes it possible to read off the decomposition of the simple module objects $M_j\in\mathsf{Mod}_\C(A)$ into simple objects $X_i\in\C$ from the columns of $V$. Therefore, we can determine all simple objects in $\mathsf{Mod}_\C(A)$ solely from the fusion outcomes of \cref{eq:AfuseFib}, i.e., from fusion rules in $\C$.
	
	For the algebra object $A=\mathbf{1}+\tau$ in \textbf{Fib} the matrix $V$ is of the form
		\begin{equation}
			V=\begin{pmatrix}
			1 & 0 \\
			1 & 1
			\end{pmatrix}.
		\end{equation}
	We can directly conclude that there are two simple module objects in the category $\mathrm{Mod}_{\mathbf{Fib}}(\mathbf{1}+\tau)$, which are
		\begin{align}
			M_1&=\mathbf{1}+\tau\\
			M_2&=\tau.
		\end{align}
	Note that the algebra object itself is always an object in the category of module objects, since it takes the place of the unit object here. The corresponding morphisms $l_{M_i}$ can be determined from the conditions given in \cref{def:leftmod} in a similar fashion to the procedure we used to determine the multiplication morphism $m_{\mathbf{1}+\tau}$ of the algebra object. The result for $M_1$ is simply the multiplication morphism $m_{\mathbf{1}+\tau}$, while $l_\tau$ is given by
		\begin{equation}
			\begin{tikzpicture}[scale=0.9,baseline=(current bounding box.center)]
				\draw (0,0) to (90:1);
				\draw (0,0) to (230:1);
				\draw (0,0) to (310:1);	
				\draw[fill=LinkColor!50!white] (0,0) circle(0.35cm);	
				\node (c) at (0,0) {\small $l_\tau$};
				\node at (90:1.25) {\small $\tau$};	
				\node at (230:1.25) {\small $\mathbf{1}+\tau$};
				\node at (310:1.25) {\small $\tau$};	
			\end{tikzpicture}=
			\begin{tikzpicture}[scale=0.7,baseline=(current bounding box.center)]
				\draw (0,0) to (90:1);
				\draw (0,0) to (230:1);
				\draw (0,0) to (310:1);	
				\node at (90:1.25) {\small $\tau$};	
				\node at (230:1.25) {\small $\mathbf{1}$};
				\node at (310:1.25) {\small $\tau$};
			\end{tikzpicture}-\frac{1+\sqrt{5}}{2}\ c_{\tau\tau}^\tau
			\begin{tikzpicture}[scale=0.7,baseline=(current bounding box.center)]
				\draw (0,0) to (90:1);
				\draw (0,0) to (230:1);
				\draw (0,0) to (310:1);	
				\node at (90:1.25) {\small $\tau$};	
				\node at (230:1.25) {\small $\tau$};
				\node at (310:1.25) {\small $\tau$};
			\end{tikzpicture}.
		\end{equation}
	Therefore, the category $\mathsf{Mod}_{\mathbf{Fib}}(\mathbf{1}+\tau)$ of left modules has two simple objects: $(\mathbf{1}+\tau,l_{\mathbf{1}+\tau})$ and $(\tau,l_\tau)$. Morphisms between these objects can be calculated by exploiting the condition \cref{eq:morphMod}. Analogously, the category of \emph{right} modules can be constructed from the algebra object $\mathbf{1}+\tau$ by building a matrix $T'$ using the fusion with the algebra object from the right side: $X_i\otimes A$. For the Fibonacci category this does not change the objects of the module category since the fusion rules are symmetric. The corresponding morphisms can be determined using the constraints from the definition of a right $A$-module category.
\end{exmp}

After explaining the construction of module categories via algebra objects, we can now understand the third point in \cref{prop:Morita}: If two categories $\C_1$ and $\C_2$ are Morita equivalent, then we can find an algebra object $A\in\C_1$ such that the category of $A$--$A$ bimodules constructed in the way described earlier is isomorphic to $\C_2$. 

As stated in the previous section, the two fusion categories arising from a finite depth subfactor are always Morita equivalent. However, it is possible that there exist other categories which are also Morita equivalent but not equivalent to these two. Hence, in general, there is a whole \emph{Morita equivalence class} of categories. In fact, the Morita equivalence class of the categories coming from the Haagerup subfactor contains a third category. This category, which we will encounter in \cref{ch:Haagerup}, is connected to the trivalent category $\Hd$ that was introduced in \cref{sec:H3triv} and hence has several nice properties.

\section{From algebra objects to principal graphs}
\label{sec:AOtographs}

So far, we have discussed how to construct fusion categories from a given subfactor. It is also possible to go in the reverse direction: Beginning with a fusion category, one can construct all possible principal graphs by finding all algebra objects $A$ and the corresponding right $A$-module categories. The procedure goes as follows:
	\begin{enumerate}
		\item[1.] Find all algebra objects $A_i$ in a category $\C$.
		\item[2.] For each algebra object $A_i$, construct the category of right $A$-modules $\mathsf{Mod}_\C(A_i)$.
		\item[3.] For a fixed algebra object $A_i$, fix a simple module object $M_j$ in the module category $\mathsf{Mod}_\C(A_i)$.
		\item[4.] Calculate the action of each simple object $X_k\in\C$ on the fixed module object $M_j$, i.e., calculate $X_k\circlearrowright M_j$.
		\item[5.] Construct a graph in the following way: Write down all simple objects $X_k\in\C$ and all simple module objects $M_l\in\mathsf{Mod}_\C(A_i)$. For every time that a simple object $M_l$ appears in the decomposition of $X_k\circlearrowright M_j$ draw a line between $M_l$ and $X_k$.
	\end{enumerate}

Executing this procedure for all algebra objects and all simple module objects yields the collection of all principle graphs associated to the category $\C$.

\begin{exmp}[Fibonacci category]
	To illustrate the technique described above, consider once more the Fibonacci category \textbf{Fib}\index{Fibonacci category} and recall we have already found out about it: In \cref{ex:AOFib} we have calculated the algebra objects of \textbf{Fib}, which turned out to be $\mathbf{1}$ and $\mathbf{1}+\tau$ (step 1). In \cref{ex:FibMod} we have constructed the category of left $\mathbf{1}+\tau$-modules and pointed out that the category of right $\mathbf{1}+\tau$-modules has the same simple objects. Since for the construction of the principal graphs we are only interested in the objects and not the morphisms, we have completed step $2$ for the algebra object $\mathbf{1}+\tau$. The two simple module objects in $\textsf{Mod}_\mathbf{Fib}(\mathbf{1}+\tau)$ are $M_1=\mathbf{1}+\tau$ and $M_2=\tau$. We fix the module object $\mathbf{1}+\tau$ for this example (step 3). The action of the category \textbf{Fib} on this object (step 4) is then given by
		\begin{align}
			\mathbf{1}\circlearrowright(\mathbf{1}+\tau)&=\mathbf{1}+\tau=M_1\\
			\tau\circlearrowright(\mathbf{1}+\tau)&=\mathbf{1}+2\tau=M_1+M_2.
		\end{align}
	Following the construction explained in step $5$ yields the following principal graph: 
		\begin{equation}
		\label{eq:GraphFib}
			\begin{tikzpicture}[baseline=(current bounding box.center),scale=2.5]
				\draw (2,0) -- (3.5,0);
				\node at (2,0) {\color{LinkColor}\Large$\bullet$};
				\node at (3,0) {\color{LinkColor}\Large$\bullet$};
				\node at (2.5,0) {\color{LinkColor2}\Large$\bullet$};
				\node at (3.5,0) {\color{LinkColor2}\Large$\bullet$};
				\node at (2,-0.15) {\small $\mathbf{1}$};
				\node at (2.5,-0.15) {\small $M_1$};
				\node at (3,-0.15) {\small $\tau$};
				\node at (3.5,-0.15) {\small $M_2$};
			\end{tikzpicture}.
		\end{equation}
	This is the principal graph of the $A_4$ subfactor, which can be obtained from applying a construction by Popa \cite{Popa1990} to the module category described above. The $A_4$ subfactor has index $\dim(\mathbf{1}+\tau)=\frac{3+\sqrt{5}}{2}$ (see \cite{Jones2013} and references therein).
	
	Note that the adjacency matrix of the principal graph is also directly given by the matrix $V$ described in \cref{eq:matrixV}, since we fixed the module object in step (3) to be the algebra object itself. Hence, in case we are only interested in the principal graph that corresponds to the algebra object, but not in the other possible principle graphs (or the module category, as in \cref{ex:FibMod}), it suffices to construct the matrix $V$ and read off the graph from it directly. For the category \textbf{Fib} with the algebra object $\mathbf{1}+\tau$, we know from \cref{ex:FibMod} that
		\begin{equation}
			V=\begin{pmatrix}
				1 & 0 \\
				1 & 1
			\end{pmatrix},
		\end{equation}
	which is exactly the adjacency matrix of the graph given in \cref{eq:GraphFib}.
	
	The second possible principle graph (which cannot be read off from the matrix $V$ directly) is constructed by fixing the module object $\tau$ in step (3). The action of the simple objects in \textbf{Fib} is given by
		\begin{align}
			\mathbf{1}\circlearrowright\tau&=\tau=M_2\\
			\tau\circlearrowright\tau&=\mathbf{1}+\tau=M_1,
		\end{align}
	which yields the following graph:
		\begin{equation}
			\begin{tikzpicture}[baseline=(current bounding box.center),scale=2.5]
				\draw (2,0) -- (2.5,0);
				\draw (3,0) -- (3.5,0);
				\node at (2,0) {\color{LinkColor}\Large$\bullet$};
				\node at (3,0) {\color{LinkColor}\Large$\bullet$};
				\node at (2.5,0) {\color{LinkColor2}\Large$\bullet$};
				\node at (3.5,0) {\color{LinkColor2}\Large$\bullet$};
				\node at (2,-0.15) {\small $\mathbf{1}$};
				\node at (2.5,-0.15) {\small $M_2$};
				\node at (3,-0.15) {\small $\tau$};
				\node at (3.5,-0.15) {\small $M_1$};
			\end{tikzpicture}.
		\end{equation}
\end{exmp}

\section{The centre construction}
\label{sec:Drinfeld}

Given a monoidal category, one can construct the \emph{centre}\index{center} (also called \emph{Drinfeld centre}\index{Drinfeld centre}) of this category, which is the categorification of the centre of a ring. More detail on this construction can be found in \cite{muger_subfactors_2003,muger_subfactors_2003-1} and \cite{Etingof2015}. In the following, let $\C$ be a monoidal category with associator
	\begin{equation}
		\alpha_{X,Y,Z}:(X\otimes Y)\otimes Z\to X\otimes (Y\otimes Z).
	\end{equation}

	\begin{defn}
		The centre of a monoidal category $\C$ is the category $\mathcal{Z}(\C)$ defined as follows: The objects of $\mathcal{Z}(\C)$ are pairs $(Z,\gamma)$ where $Z\in\C$ and 
			\begin{equation}
				\gamma_X:X\otimes Z\to Z\otimes X,\hspace{10pt} X\in\C
			\end{equation}
		is a natural isomorphism called \emph{half-braiding}\index{half-braiding} such that the following diagram commutes for all $X,Y\in\C$:
			\begin{equation}
			\label{eq:halfb}

			\end{equation}
		A morphism from $(Z,\gamma)$ to $(Z',\gamma')$ is a morphism $f\in\Hom_\C(Z,Z')$ such that for each $X\in\C$ the following condition is fulfilled:
			\begin{equation}
				\label{eq:centermorph}
				(f\otimes \id_X)\circ\gamma_X=\gamma_X'\circ(\id_X\otimes f),
			\end{equation}
		which can be depicted
		\begin{equation}
			%
\ .
		\end{equation}
	\end{defn}
	
	The centre of a monoidal category is again a monoidal category. The tensor product in this category is given as follows: For two objects $(Z,\gamma)$, $(Z',\gamma')$ in $\mathcal{Z}(\C)$ it is defined as
		\begin{equation}
			(Z,\gamma)\otimes (Z',\gamma')=(Z\otimes Z',\tilde{\gamma})
		\end{equation}
	with
		\begin{equation}
			\tilde{\gamma}_X:X\otimes(Z\otimes Z')\to(Z\otimes Z')\otimes X,\hspace{10pt}X\in\C
		\end{equation}
	given by the following commuting diagram:
		\begin{equation}
			%
		\end{equation}
	The unit object in $\mathcal{Z}(\C)$ is $(\mathbf{1},r^{-1}l)$, where $r$ and $l$ are the right and left unit constraints of the underlying monoidal category $\C$.
	
	\begin{rem}
		If the category $\C$ is strict, condition \cref{eq:halfb} simplifies to 
			\begin{equation}
				\gamma_{X\otimes Y}=(\gamma_X\otimes\id_Y)\circ(\id_X\otimes \gamma_Y)
			\end{equation}
		and the half-braiding $\tilde{\gamma}_X$ of the tensor product is given by
			\begin{equation}
				\tilde{\gamma}_X=(\id_Z\otimes\gamma_X')\circ(\gamma_X\otimes\id_{Z'}).
			\end{equation}
	\end{rem}

The main reason why the centre construction is of interest to the work described in this thesis is because of the following theorem which was proved in \cite{muger_subfactors_2003-1}:
	\begin{thm}
		The centre $\mathcal{Z}(\C)$ of a spherical fusion category $\C$ is a modular tensor category.
	\end{thm}

The above theorem provides a way to construct a modular tensor category from a spherical fusion category, which is especially interesting with regard to the fusion categories associated to the Haagerup subfactor since these categories are not modular themselves. It is therefore a powerful tool in our search for a conformal field theory that corresponds to the Haagerup subfactor: From the modular tensor category, we can construct an anyon chain and investigate whether is has any critical points, which themselves are possibly described by a conformal field theory. This is described in more detail in \cref{sec:anyons}.
\chapter{The Haagerup subfactor and its fusion categories}
\label{ch:Haagerup}

In the study of exotic subfactors, it is natural to ask what the subfactor with the smallest possible index above four is. In \cite{Haagerup1994}, Haagerup investigated this problem by analysing the principal graphs of subfactors with small indices above four. More precisely, he asked the question which non-trivial principal graphs with index near four (in fact, with index in the range $(4,3+\sqrt{3})$) are possible, i.e., have a subfactor associated to them. As a result, he found that there cannot be a subfactor with index in the interval $(4,\frac{5+\sqrt{13}}{2})$, and provides a list with possible pairs of principal graphs\footnote{Note that the pair $(\Gamma,\Gamma')$ of principal graphs is unordered since any subfactor $N\subseteq M$ has a dual with the same index but where the principal and dual principal graph are switched.} in the range $[\frac{5+\sqrt{13}}{2},3+\sqrt{3})$:
{{\setlength{\tabcolsep}{7pt}
		\renewcommand{\arraystretch}{2.5}
		\begin{table}[H]
			\centering
\begin{tabular}{c|c|c}
				& principal graph $\Gamma$& Dual principal graph $\Gamma'$\\\hline
				$\mathcal{H}_n$ & 
				\begin{tikzpicture}[baseline=(current bounding box.center),scale=1.1]
				\draw (0,0) -- (1.75,0);
				\node at (2.25,0) {$\dots$};
				\draw (2.75,0) -- (3.5,0);
				\draw (3.5,0) -- (4.5,0.5);
				\draw (4,0.25) -- (4.5,0);
				\draw (3.5,0) -- (4,-0.25);
				\foreach \a in {0,0.5,1,1.5,3,3.5}{
					\node at (\a,0) {$\bullet$};
				}
				\node at (4,0.25) {$\bullet$};
				\node at (4.5,0.5) {$\bullet$};
				\node at (4.5,0) {$\bullet$};
				\node at (4,-0.25) {$\bullet$};
				\draw[decorate,decoration={brace,amplitude=7pt}] (3.6,-0.15) -- (-0.1,-0.15);
				\node at (1.75,-0.6) {\footnotesize $4n+3$ edges};
				\node at (0,0.25) {$\star$};
				\end{tikzpicture}& \begin{tikzpicture}[baseline=(current bounding box.center),scale=1.1]
				\draw (0,0) -- (1.75,0);
				\node at (2.25,0) {$\dots$};
				\draw (2.75,0) -- (3.5,0);
				\draw (3.5,0) -- (5,0.75);
				\draw (3.5,0) -- (5,-0.75);
				\foreach \a in {0,0.5,1,1.5,3,3.5}{
					\node at (\a,0) {$\bullet$};
				}
				\node at (4,0.25) {$\bullet$};
				\node at (4.5,0.5) {$\bullet$};
				\node at (5,0.75) {$\bullet$};
				\node at (4,-0.25) {$\bullet$};
				\node at (4.5,-0.5) {$\bullet$};
				\node at (5,-0.75) {$\bullet$};
				\draw[decorate,decoration={brace,amplitude=7pt}] (3.6,-0.15) -- (-0.1,-0.15);
				\node at (1.75,-0.6) {\footnotesize $4n+3$ edges};
				\node at (0,0.25) {$\star$};
				\end{tikzpicture}
				\\
\end{tabular}
	\end{table}}
	\vspace{8pt}
	Here, $\mathcal{H}_n$ and $\mathcal{B}_n$ are infinite families of potential subfactors, where $n\in\mathbb{N}$. The indices of the subfactors that correspond to these graphs range from $\mathrm{index}(\mathcal{H}_0)=\frac{5+\sqrt{13}}{2}\approx4.30278$ to $\lim_{n\to\infty}\mathrm{index}(\mathcal{B}_n)\approx 4.65897$. 
	
	However, even though these principal graphs are \emph{possible} graphs, it does not automatically mean that they are realized by a subfactor. While Haagerup's original result did not specify which graphs are actually realized, this question was investigated by various scientists in the subsequent years. It was shown by Haagerup and Asaeda \cite{asaeda_exotic_1999} that the graph $\mathcal{H}_0$ (the \emph{Haagerup subfactor}\index{Haagerup subfactor}) as well as the graph $\mathcal{AH}$ (the \emph{Asaeda-Haagerup subfactor})\index{Asaeda-Haagerup subfactor} are realized by subfactors. An alternative construction of the Haagerup subfactor via a system of endomorphisms of a certain Cuntz algebra by Izumi is given in \cite{izumi_structure_2001}. Additionally, a construction of the Haagerup subfactor via so-called \emph{planar algebras} was presented by Peters in \cite{Peters2010}. Furthermore, it was shown by Bisch \cite{Bisch1998} that none of the graphs $\mathcal{B}_n$ can be the principal graphs of a subfactor since they yield inconsistent fusion rules, and Asaeda and Yasuda \cite{Asaeda2007,Asaeda2008} proved that $\mathcal{H}_n$ is not the principal graph of a subfactor for $n\ge 2$, which only left the existence of a subfactor with principal graph $\mathcal{H}_1$ as an open problem. The classification of subfactors up to index $3+\sqrt{3}$ was completed when it was shown in \cite{bigelow_constructing_2012} that $\mathcal{H}_1$ is realized by the so-called \emph{extended Haagerup subfactor}\index{extended Haagerup subfactor}, using a construction via subfactor planar algebras.
	
	In this thesis, we focus on the subfactor that corresponds to $\mathcal{H}_0$, which is the Haagerup subfactor, and the fusion categories that can be constructed from it. For this purpose, let us analyse the principal and dual principal graph in terms of their respective even and odd parts. We begin with the principal graph $\Gamma$. Remember that $\Gamma_\mathrm{even}$ (respectively, $\Gamma_\mathrm{odd}$) is the set of vertices with even (respectively, odd) distance to the object $\star$. To emphasize this, the elements of $\Gamma_\mathrm{even}$ are coloured red and the elements of $\Gamma_\mathrm{odd}$ are coloured blue in the following figure:
	\begin{figure}[H]
		\centering
		\begin{tikzpicture}[baseline=(current bounding box.center),scale=2.5]
			\draw (2,0) -- (3.5,0);
			\draw (3.5,0) -- (4.5,0.5);
			\draw (4,0.25) -- (4.5,0);
			\draw (3.5,0) -- (4,-0.25);
			\foreach \a in {2,3}{
				\node at (\a,0) {\color{LinkColor}\Large$\bullet$};
			}
			\node at (2.5,0) {\color{LinkColor2}\Large$\bullet$};
			\node at (3.5,0) {\color{LinkColor2}\Large$\bullet$};
			\node at (4,0.25) {\color{LinkColor}\Large$\bullet$};
			\node at (4.5,0.5) {\color{LinkColor2}\Large$\bullet$};
			\node at (4.5,0) {\color{LinkColor2}\Large$\bullet$};
			\node at (4,-0.25) {\color{LinkColor}\Large$\bullet$};
			\node at (2,-0.15) {$\mathbf{1}$};
			\node at (3,-0.15) {$\eta$};
			\node at (4,-0.4) {$\mu$};
			\node at (4,0.4) {$\nu$};
		\end{tikzpicture}
	\end{figure}
	The elements of $\Gamma_\mathrm{even}$ are equipped with labels that denote the object in the corresponding fusion category, which we denote $\mathcal{H}_1$\footnote{Be aware that this notation has nothing to do with the $\mathcal{H}_n$ that denoted the possible principal graphs in Haagerup's list. From now on, we use $\mathcal{H}_i$ with $i\in\{1,2,3\}$ to denote the three different categories in the Morita equivalence class of categories associated to the Haagerup subfactor.}. According to the fusion graph, this category has four simple objects: $\Obj(\mathcal{H}_1)=\{\mathbf{1},\eta,\nu,\mu\}$ with quantum dimensions 
	\begin{equation}
	\dim(\mathbf{1})=1,\hspace{10pt}\dim(\eta)=\frac{3+\sqrt{13}}{2},\hspace{10pt}\dim(\nu)=\frac{1+\sqrt{13}}{2},\hspace{10pt}\dim(\mu)=\frac{5+\sqrt{13}}{2}.
	\end{equation}
	The corresponding fusion rules $X\otimes Y=\sum_Z N_{XY}^Z Z$ for $X,Y,Z\in\Obj(\mathcal{H}_1)$ are listed in \cref{tab:fusionH1}:
	{{\setlength{\tabcolsep}{10pt}
		\renewcommand{\arraystretch}{1.25}
		\begin{table}[H]
			\centering
\begin{tabular}{c||c|c|c|c}
				& $\mathbf{1}$ & $\nu$ & $\eta$ & $\mu$  \\ \hline \hline
				$\mathbf{1}$ & $\mathbf{1}$ & $\nu$ & $\eta$ & $\mu$\\ \hline
				$\nu$ & $\nu$ & $\mathbf{1}+2\nu+2\eta+\mu$ & $2\nu+\eta+\mu$ & $\nu+\eta+\mu$  \\ \hline
				$\eta$ & $\eta$ & $2\nu+\eta+\mu$ & $\mathbf{1}+\nu+\eta+\mu$ & $\nu+\eta$  \\ \hline
				$\mu$ & $\mu$ & $\nu+\eta+\mu$ & $\nu+\eta$ & $\mathbf{1}+\nu$\vspace{8pt}
			\end{tabular}
			\caption{\small \label{tab:fusionH1}Fusion rules for the $\mathcal{H}_1$ fusion category.}
		\end{table}}
		
		In the same fashion we can analyse the dual principal graph $\Gamma'$ whose even part $\Gamma_\mathrm{even}'$ gives rise to the fusion category $\mathcal{H}_2$. As above, even vertices are coloured red and odd vertices are coloured blue, and we add labels to those vertices that correspond to objects in the fusion category.
		\begin{figure}[H]
			\centering
			%
		\end{figure}
		
		Hence, the fusion category $\mathcal{H}_2$ has six simple objects: $\Obj(\mathcal{H}_2)=\{\mathbf{1},\alpha,\alpha^*,\rho,{}_\alpha\rho,\alphastarrho\}$ with the following quantum dimensions:
		\begin{align}
		\dim(\mathbf{1})=\dim(\alpha)=\dim(\alpha^*)=1,\hspace{10pt}\dim(\rho)=\dim({}_\alpha\rho)=\dim(\alphastarrho)=\frac{3+\sqrt{13}}{2}.
		\end{align}
		and fusion rules given in \cref{tab:fusionH2}:
		{{\setlength{\tabcolsep}{10pt}
			\renewcommand{\arraystretch}{1.25}
			\begin{table}[H]
				\centering
				\begin{tabular}{c||c|c|c|c|c|c}
					& $\mathbf{1}$ & $\alpha$ & $\alpha^*$ & $\rho$ & ${}_\alpha
					\rho$ & $\alphastarrho$ \\ \hline \hline
					$\mathbf{1}$ & $\mathbf{1}$ & $\alpha$ & $\alpha^*$ & $\rho$ & ${}_\alpha \rho$ & $\alphastarrho$ \\ \hline
					$\alpha$ & $\alpha$ & $\alpha^*$ & $\mathbf{1}$ & ${}_\alpha \rho$ & $\alphastarrho$ & $\rho$ \\ \hline
					$\alpha^*$ & $\alpha^*$ & $\mathbf{1}$ & $\alpha$ & $\alphastarrho$ & $\rho$ & ${}_\alpha \rho$ \\ \hline
					$\rho$ & $\rho$ & $\alphastarrho$ & ${}_\alpha \rho$ & $\mathbf{1}+ Z$ & $\alpha^*+Z$ & $\alpha+Z$ \\ \hline
					${}_\alpha\rho$ & ${}_\alpha\rho$ & $\rho$ & $\alphastarrho$ & $\alpha+Z$ & $\mathbf{1}+Z$ & $\alpha^*+Z$ \\ \hline
					$\alphastarrho$ & $\alphastarrho$ & ${}_\alpha\rho$ & $\rho$ & $\alpha^*+Z$ & $\alpha+Z$ & $\mathbf{1}+Z$\vspace{8pt}
				\end{tabular}
				\caption{\small \label{tab:fusionH2}Fusion rules for the $\mathcal{H}_2$ and $\mathcal{H}_3$ fusion category, with $Z=\rho+{}_\alpha\rho+\alphastarrho$.}
			\end{table}}
			
			Since these two categories arise as the principal even and the dual even part of the same subfactor, they are Morita equivalent. This implies that there is an algebra object $A_1\in\mathcal{H}_1$ such that $\mathcal{H}_2$ is isomorphic to the category of $A_1$--$A_1$ bimodules over $\mathcal{H}_1$. On the other hand, there is an algebra object $A_2\in\mathcal{H}_2$ such that $\mathcal{H}_1$ is isomorphic to the category of $A_2$--$A_2$ bimodules over $\mathcal{H}_2$. In \cite{grossman_quantum_2012}, this connection was studied in detail and found that the algebra object $A_1\in\mathcal{H}_1$ is $\mathbf{1}+\eta$, and the algebra object $A_2\in\mathcal{H}_2$ is $\mathbf{1}+\rho$.
			
			Moreover, the authors of \cite{grossman_quantum_2012} exploited this technique to determine all fusion categories in the Morita equivalence class of $\mathcal{H}_1$ and $\mathcal{H}_2$ and found that there is exactly one additional category in this class, denoted $\mathcal{H}_3$. It has the same simple objects and fusion rules as $\mathcal{H}_2$ (see \cref{tab:fusionH2}). It can be constructed as the category of $(\mathbf{1}+\alpha+\alpha^*)-(\mathbf{1}+\alpha+\alpha^*)$ bimodules over $\mathcal{H}_2$ or as the category of $(\mathbf{1}+\mu+\nu)-(\mathbf{1}+\mu+\nu)$ bimodules over $\mathcal{H}_1$. The Morita equivalence class is summarized in \cref{fig:MoritaH}.
			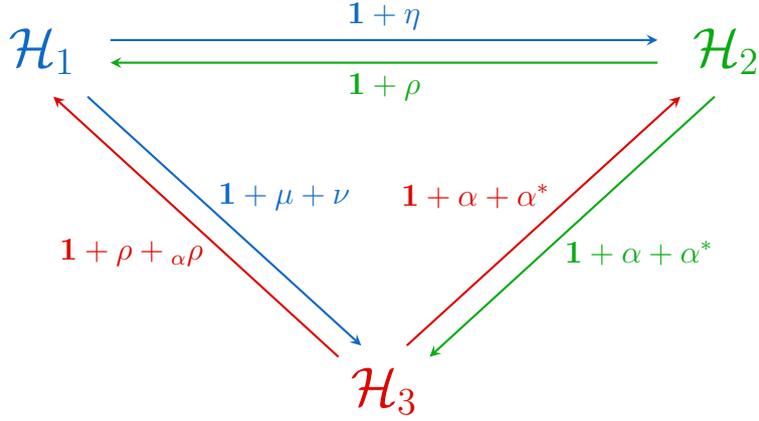
\begin{figure}[H]
				\centering
				\begin{tikzpicture}[scale=3]
				\node (H1) at (0,0) {\color{LinkColor2}\huge$\mathcal{H}_1$};
				\node (H2) at (3,0) {\color{LinkColor3}\huge$\mathcal{H}_2$};
				\node (H3) at (1.5,-1.5) {\color{LinkColor}\huge$\mathcal{H}_3$};
				\draw[->,>=stealth,thick,color=LinkColor2] (0.3,0.05) to node[above] {\large$\mathbf{1}+\eta$} (2.7,0.05);
				\draw[->,>=stealth,thick,color=LinkColor3] (2.7,-0.05) to node[below] {\large$\mathbf{1}+\rho$} (0.3,-0.05);
				\draw[->,>=stealth,thick,color=LinkColor] (1.6,-1.3) to node[left,pos=0.6] {\large$\mathbf{1}+\alpha+\alpha^*\ $} (2.8,-0.2);
				\draw[->,>=stealth,thick,color=LinkColor3] (2.95,-0.2) to node[right,pos=0.6] {\large$\ \mathbf{1}+\alpha+\alpha^*$} (1.7,-1.35);
				\draw[->,>=stealth,thick,color=LinkColor2] (0.2,-0.2) to node[right,pos=0.4] {\large$\ \mathbf{1}+\mu+\nu$} (1.4,-1.3);
				\draw[->,>=stealth,thick,color=LinkColor] (1.3,-1.35) to node[left,pos=0.4] {\large$\mathbf{1}+\rho+{}_\alpha\rho\ $} (0.05,-0.2);
				\end{tikzpicture}
				\caption{\small \label{fig:MoritaH}The Morita equivalence class of fusion categories coming from the Haagerup subfactor. The object on the arrow from $\mathcal{H}_i$ to $\mathcal{H}_j$ indicates the algebra object $A$ from which $\mathcal{H}_j$ can be built as the category of $A$-$A$ bimodules over $\mathcal{H}_i$.}
			\end{figure}
			
	\begin{exmp}
		Using the technique described in \cref{sec:AOtographs} we can recover the principal and dual graphs of the Haagerup subfactor from its fusion categories. The principal graph can be constructed via the module category $\mathsf{Mod}_{\mathcal{H}_1}(\mathbf{1}+\eta)$ and choosing $\mathbf{1}+\eta$ as the fixed simple module object therein. The dual graph can analogously be constructed via $\mathsf{Mod}_{\mathcal{H}_2}(\mathbf{1}+\rho)$ and choosing $\mathbf{1}+\rho$ as the fixed simple module object. We go through both constructions in detail, beginning with the principal graph. 
			
		To construct the principal graph of the Haagerup subfactor, we start with the algebra object $\mathbf{1}+\eta$ in $\Hi_1$. Recall that we first construct the matrix $T$ whose rows are determined by the decomposition of $ X_i\otimes(\mathbf{1}+\eta)$. The result is
			\begin{equation}
				T=\begin{blockarray}{ccccc}
				& \mathbf{1} & \nu & \eta & \mu \\
				\begin{block}{c (cccc)}
				\mathbf{1} & 1 & 0 & 1 & 0 \topstrut \\
				\nu & 0 & 3 & 1 & 1 \\
				\eta & 1 & 1 & 2 & 1 \\
				\mu & 0 & 1 & 1 & 1 \botstrut \\
				\end{block}
				\end{blockarray}\ .
			\end{equation}
		The matrix $V$ that fulfils $T=VV^\mathrm{T}$ is given by
			\begin{equation}
				V=\begin{pmatrix}
					1 & 0 & 0 & 0\\
					0 & 1 & 1 & 1\\
					1 & 1 & 0 & 0\\
					0 & 1 & 0 & 0
				\end{pmatrix}\ .
			\end{equation}
		Interpreting $V$ as an adjacency matrix yields the following graph
			\begin{equation}
				\begin{tikzpicture}[baseline=(current bounding box.center),scale=2]
				\draw (2,0) -- (3.5,0);
				\draw (3.5,0) -- (4.5,0.5);
				\draw (4,0.25) -- (4.5,0);
				\draw (3.5,0) -- (4,-0.25);
				\foreach \a in {2,3}{
					\node at (\a,0) {\color{LinkColor}\Large$\bullet$};
				}
				\node at (2.5,0) {\color{LinkColor2}\Large$\bullet$};
				\node at (3.5,0) {\color{LinkColor2}\Large$\bullet$};
				\node at (4,0.25) {\color{LinkColor}\Large$\bullet$};
				\node at (4.5,0.5) {\color{LinkColor2}\Large$\bullet$};
				\node at (4.5,0) {\color{LinkColor2}\Large$\bullet$};
				\node at (4,-0.25) {\color{LinkColor}\Large$\bullet$};
				\node at (2,-0.15) {\small $\mathbf{1}$};
				\node at (3,-0.15) {\small $\eta$};
				\node at (4,-0.4) {\small $\mu$};
				\node at (4,0.4) {\small $\nu$};
				\end{tikzpicture},
			\end{equation}
		which is exactly the principal graph of the Haagerup subfactor. For the calculation of the dual graph, we begin with the algebra object $\mathbf{1}+\rho$ in $\Hi_2$. The matrix $T$ constructed from the fusion outcomes of $X_i\otimes(\mathbf{1}+\rho)$ is given by
			\begin{equation}
				T=\begin{blockarray}{ccccccc}
				& \mathbf{1} & \alpha & \alpha^* & \rho & {}_\alpha\rho & \alphastarrho\\
				\begin{block}{c (cccccc)}
				\mathbf{1} & 1 & 0 & 0 & 1 & 0 & 0 \topstrut \\
				\alpha & 0 & 1 & 0 & 0 & 1 & 0\\
				\alpha^* & 0 & 0 & 1 & 0 & 0 & 1\\
				\rho & 1 & 0 & 0 & 2 & 1 & 1\\
				{}_\alpha\rho & 0 & 1 & 0 & 1 & 2 & 1\\
				\alphastarrho & 0 & 0 & 1 & 1 & 1 & 2\botstrut \\
				\end{block}
				\end{blockarray}\ ,
			\end{equation}
		and, therefore, $V$ is
			\begin{equation}
				V=\begin{pmatrix}
					1 & 0 & 0 & 0 \\
					0 & 1 & 0 & 0 \\
					0 & 0 & 1 & 0 \\
					1 & 0 & 0 & 1 \\
					0 & 1 & 0 & 1 \\
					0 & 0 & 1 & 1
				\end{pmatrix}.
			\end{equation}
		Note that, apart from interpreting $V$ directly as the adjacency matrix of the principal graph we can also read off the decomposition of the objects $M_i$ in the module category from this matrix:
			\begin{align}
				M_1&=\mathbf{1}+\rho\\
				M_2&=\alpha+\alpharho\\
				M_3&=\alpha^*+\alphastarrho\\
				M_4&=\rho+\alpharho+\alphastarrho.
			\end{align}
		The action $X_i\circlearrowright(\mathbf{1}+\rho)$ for $X_k\in\Obj(\mathcal{H}_2)$ is given by
			\begin{align}
				&\mathbf{1}\circlearrowright(\mathbf{1}+\rho)=\mathbf{1}+\rho=M_1\\
				&\alpha\circlearrowright(\mathbf{1}+\rho)=\alpha+\alpharho=M_2\\
				&\alpha^*\circlearrowright(\mathbf{1}+\rho)=\alpha^*+\alphastarrho=M_3\\
				&\rho\circlearrowright(\mathbf{1}+\rho)=\mathbf{1}+2\rho+\alpharho+\alphastarrho=M_1+M_4\\
				&\alpharho\circlearrowright(\mathbf{1}+\rho)=\alpha+\rho+2\alpharho+\alphastarrho=M_2+M_4\\
				&\alphastarrho\circlearrowright(\mathbf{1}+\rho)=\alpha^*+\rho+\alpharho+2\alphastarrho=M_3+M_4,
			\end{align} 
		which yields the following graph (where we can see that $V$ is indeed the adjacency matrix of this graph) which is the dual principal graph of the Haagerup subfactor:
			\begin{equation}
				\begin{tikzpicture}[baseline=(current bounding box.center),scale=2]
					\draw (2,0) -- (3.5,0);
					\draw (3.5,0) -- (5,0.75);
					\draw (3.5,0) -- (5,-0.75);
					\node at (2,0) {\color{LinkColor}\Large$\bullet$};
					\node at (2.5,0) {\color{LinkColor2}\Large$\bullet$};
					\node at (3,0) {\color{LinkColor}\Large$\bullet$};
					\node at (3.5,0) {\color{LinkColor2}\Large$\bullet$};
					\node at (4,0.25) {\color{LinkColor}\Large$\bullet$};
					\node at (4.5,0.5) {\color{LinkColor2}\Large$\bullet$};
					\node at (5,0.75) {\color{LinkColor}\Large$\bullet$};
					\node at (4,-0.25) {\color{LinkColor}\Large$\bullet$};
					\node at (4.5,-0.5) {\color{LinkColor2}\Large$\bullet$};
					\node at (5,-0.75) {\color{LinkColor}\Large$\bullet$};
					\node at (2,-0.15) {\small $\mathbf{1}$};
					\node at (3,-0.15) {\small $\rho$};
					\node at (4,-0.4) {\small $\alphastarrho$}; 
					\node at (5,-0.9) {\small $\alpha^*$}; 
					\node at (4,0.4) {\small $\alpharho$}; 
					\node at (5,0.9) {\small $\alpha$}; 
					\node at (2.5,-0.15) {\small $M_1$};
					\node at (3.5,-0.15) {\small $M_4$};
					\node at (4.5,0.65) {\small $M_2$};
					\node at (4.5,-0.65) {\small $M_3$};
				\end{tikzpicture}\ .
			\end{equation}
	\end{exmp}

\section{The $F$-symbols for the category $\Hd$}

	The $F$-symbols of a fusion category are important for any problem that involves working in a specific basis, for example the construction of module categories via algebra objects explained above. Moreover, they are a crucial ingredient in the construction of physical models from fusion categories such as one-dimensional spin chains or two-dimensional lattice models (presented in Part $2$ of this thesis). Especially for lattice models, we need the matrix representation of the $F$-symbols. Hence, even though the $F$-symbols are, in principal, obtained by Izumi's construction in \cite{izumi_structure_2001}, some work has to be done to obtain the corresponding matrix representation.
	
	A straightforward way to obtain the $F$-symbols of a given fusion category is to solve the pentagon equation \eqref{eq:pentagon}:
			\begin{equation}
			\label{eq:penta2}
				\left(F_u^{abr}\right)_{sp}\left(F_u^{pcd}\right)_{rq}=\sum_t \left(F_s^{bcd}\right)_{rt}\left(F_u^{atd}\right)_{sq}\left(F_q^{abc}\right)_{tp}.
			\end{equation}
	This equation has to be fulfiled for any combination of labels from the set of simple objects. Hence, the objective here is to solve a set of multivariate equations of polynomials up to third order, where the number of variables and equations is growing with the number of simple objects in the category. This task quickly becomes complicated if the number of simples in the category is too high, and explicit solutions are only known for a handful of cases (a list of many explicit solutions can, for example, be found in \cite{Bond2007}).
	
	In case of the fusion category $\mathcal{H}_3$ there are six simple objects, hence it presents no shortage of challenges: After eliminating trivial equations (i.e., those that contain non-valid $F$-symbols or those that are trivially true), there are $41391$ equations and $1431$ unknowns, resulting in a task that is at the limit of what is easily computable with current state-of-the-art technology and algorithms. To solve this problem, we use the following strategy: We first collect ``seed'' data about the $F$-symbols that can then be input to standard solvers in order to simplify the problem (the detailed description of the procedure can be found in \cite{Osborne2019}). This data comes from three different sources:
		\begin{enumerate}
			\item \textbf{Unitarity.} Since the $F$-symbols are transformations between orthonormal bases, we require them to be unitary. This allows us to use the unitarity condition
				\begin{equation}
					U^\dagger U=UU^\dagger=\mathbb{I}
				\end{equation}
			to get additional (and possibly simpler) equations for the $F$-symbols. This is especially helpful in cases where the majority of variables of a matrix has already been determined: using the unitarity condition, it is likely that we can determine the remaining variables. For a one-dimensional matrix $F_u^{abc}$ this condition becomes
				\begin{equation}
					\|F_u^{abc}\|^2=1.
				\end{equation}
			\item \textbf{Gauge freedom.} To every distinct vertex, there is a gauge freedom assigned that amounts to the choice of basis vertex. Since the pentagon equation is invariant under a gauge transformation, we can use this freedom to fix some of the variables to a certain value, thus reducing the number of variables. We go into more detail about this technique below.
			\item \textbf{The trivalent category $\Hd$.} As described in \cref{ch:Trivalent}, the category $\Hd$ has a description as a trivalent category. As such, it allows for a simple diagrammatic calculus. This can be exploited to obtain some of the variables directly and also additional (possibly simpler) equations. More detail about this technique can be found below.
		\end{enumerate}
	Exploiting these techniques to generate seed data, it is possible to solve the set of equations generated by the pentagon equation. A complete list of the resulting $F$-symbols for the fusion category $\Hd$ can be found in \cref{app:Fsymbols}.
	
\subsection*{Gauge freedom}
	\index{gauge freedom}
	For every distinct vertex in a fusion diagram there is a gauge freedom assigned that accounts for the choice of basis. Suppose we have a vector $\psi\in V_c^{ab}$, where $V_c^{ab}=\Hom(c,a\otimes b)$. Another vector $\psi'\in V_c^{ab}$ can be obtained from $\psi$ via the invertible change of basis transformation $u_c^{ab}$:
		\begin{equation}
			\psi'=u_c^{ab}\psi.
		\end{equation}
	It follows that for an element of an $F$-matrix, we have the transformation
		\begin{equation}
		\label{eq:gauge}
			\left(F_d^{abc}\right)'_{fe}=\frac{u_d^{af} u_f^{bc}}{u_e^{ab} u_d^{ec}}\left(F_d^{abc}\right)_{fe}.
		\end{equation}
	The corresponding transformation for the adjoint of the $F$-matrix\footnote{Note that we work in a unitary setting here, i.e., $\left(F_d^{abc}\right)_{fe}^{-1}=\left(F_d^{abc}\right)^\dagger_{fe}=\overline{\left(F_d^{abc}\right)}_{ef}$.} is
		\begin{equation}
			\left(\left[F_d^{abc}\right]^\dagger\right)_{fe}'=\frac{u_e^{ab} u_d^{ec}}{u_d^{af} u_f^{bc}}\left(F_d^{abc}\right)^\dagger_{fe}.
		\end{equation}
	By fixing one of these ratios we can set a variable to a certain value. This can be exploited to simplify the set (or a subset of the set) of equations. We are allowed to fix these ratios because the pentagon equation is invariant under the transformation \cref{eq:gauge}:
		\begin{thm}
			The pentagon equation \cref{eq:penta2} is invariant under the gauge transformation \cref{eq:gauge}.
		\end{thm}
		\begin{proof}
			Under a gauge transformation, the pentagon equation becomes
			\begin{align}
				\frac{u_u^{xd}u_d^{yc}}{u_a^{xy}u_u^{ac}}\left(F_u^{xyc}\right)_{da}\frac{u_u^{ac}u_c^{zw}}{u_b^{az}u_u^{bw}}\left(F_u^{azw}\right)_{cb}&=\sum_e\frac{u_d^{yc}u_c^{zw}}{u_e^{yz}u_d^{ew}}\left(F_d^{yzw}\right)_{ce}\frac{u_u^{xd}u_d^{bw}}{u_b^{xe}u_u^{bw}}\left(F_u^{xew}\right)_{db}\frac{u_b^{xe}u_e^{yz}}{u_a^{xy}u_b^{az}}\left(F_b^{xyz}\right)_{ea}\\
				\frac{u_u^{xd}u_d^{yc}u_u^{ac}u_c^{zw}}{u_a^{xy}u_u^{ac}u_b^{az}u_u^{bw}}\left(F_u^{xyc}\right)_{da}\left(F_u^{azw}\right)_{cb}&=\frac{u_d^{yc}u_c^{zw}u_u^{xd}u_d^{bw}}{u_b^{xe}u_u^{bw}u_a^{xy}u_b^{az}}\ \sum_e\left(F_d^{yzw}\right)_{ce}\left(F_u^{xew}\right)_{db}\left(F_b^{xyz}\right)_{ea}\\
				\left(F_u^{xyc}\right)_{da}\left(F_u^{azw}\right)_{cb}&=\sum_e\left(F_d^{yzw}\right)_{ce}\left(F_u^{xew}\right)_{db}\left(F_b^{xyz}\right)_{ea},
			\end{align}
			therefore it is gauge invariant.
		\end{proof}

\subsection*{The trivalent category $\Hd$}

	We can exploit the fact that the fusion category $\Hd$ also has a description as a trivalent category to obtain further information about the $F$-symbols. Precisely, this is done by interpreting diagrammatic relations that are properties of the trivalent category as transformations caused by applying an $F$-matrix. In the following, we give an example of how this technique works.
	
	\begin{thm}
		In the fusion category $\mathcal{H}_3$, the following equations hold:
		\begin{align}
			\left(F_\rho^{\rho\rho\rho}\right)_{{}_\alpha\rho \rho}^\dagger\left(F_{{}_\alpha\rho}^{\rho\rho\rho}\right)_{\rho\rho}\sqrt{d_\rho}&=\alpha\left(F_\rho^{\rho\rho\rho}\right)^\dagger_{{}_\alpha\rho \mathbf{1}}+\beta\left(F_\rho^{\rho\rho\rho}\right)^\dagger_{{}_\alpha\rho \rho}\\
			\left(F_\rho^{\rho\rho\rho}\right)^\dagger_{\alphastarrho \rho}\left(F_{\alphastarrho}^{\rho\rho\rho}\right)_{\rho\rho}\sqrt{d_\rho}&=\alpha\left(F_\rho^{\rho\rho\rho}\right)^\dagger_{\alphastarrho \mathbf{1}}+\beta\left(F_\rho^{\rho\rho\rho}\right)^\dagger_{\alphastarrho \rho},
		\end{align}
		where $\alpha=\frac{1}{18}\left(\sqrt{13}+7\right)$ and $\beta=\frac{\sqrt{\sqrt{13}-2}}{3}$.
	\end{thm}
	\begin{proof}
		This statement follows from interpreting the diagram
			\begin{equation}

			\end{equation}
		as an element in $\Hom(\rho,\rho\otimes\rho\otimes\rho)$ in the regular category $\Hd$ and applying $F$-moves to simplify it on the one hand, and comparing this to the corresponding relation \cref{eq:squarelincom} within the trivalent category (note that in the trivalent category, we can rotate strings and vertices without changing the value of a diagram). In the first case, we obtain the relation 
			\begin{equation}
				%
.
			\end{equation}
		Rewriting \cref{eq:squarelincom} as an equation in $\Hom(\rho,\rho\otimes\rho\otimes\rho)$ and bringing it into the same form as the equation above yields
			\begin{align}
				%
			\right).
			\end{align}
		Comparing the different linear combinations of basis vectors for the diagram in question for $x={}_\alpha\rho$ and $x=\alphastarrho$ yields the statement of the proof.
	\end{proof}

	It is possible to exploit various other relations within the trivalent category to gain information about the $F$-symbols. A detailed list of these relations can be found in \cite{Osborne2019}. 
	
	The seed data generated by the methods described above provides enough information such that the set of equations becomes simple enough to be solvable with standard techniques. The $F$-symbols of the $\Hi_3$ category have also independently been obtained by Matthew Titsworth \cite{PrivateTitsworth}.

\section{Algebra objects in $\Hd$}

The $F$-symbols of a category can be exploited to find algebra objects and modules of the category. In this section, we describe how this procedure works by calculating an explicit example. As depicted in \cref{fig:MoritaH}, $\mathcal{H}_1$ can be constructed as the category $\mathsf{Bimod}_{\Hd}(A,A)$ with $A=\mathbf{1}+\rho+{}_\alpha\rho$. First, we show that $\mathbf{1}+\rho+{}_\alpha\rho$ is an algebra object in $\Hd$ by constructing the corresponding multiplication morphism.

The multiplication morphism $m:A\otimes A\to A$ is an element of the morphism space $\Hom(A\otimes A,A)$. To find out the dimension of this space, note that we can write $A=\mathbf{1}+\rho+{}_\alpha\rho$ as
	\begin{equation}
		A=\bigoplus_{X\in\Obj(\Hd)}a_XX,
	\end{equation}
where $\Obj(\Hd)=\{\mathbf{1},\alpha,\alpha^*,\rho,{}_\alpha\rho,\alphastarrho\}$ and $a_X=0$ for $X\in\{\alpha,\alpha^*,\alphastarrho\}$ and $a_X=1$ for $X\in\{\mathbf{1},\rho,{}_\alpha\rho\}$. Hence, the morphism space $\Hom(A\otimes A,A)$ can be written as a combination of morphism spaces only including simple objects:
	\begin{equation}
		\Hom(A\otimes A,A)=\bigoplus_{X,Y,Z\in\Obj(\Hd)}a_Xa_Ya_Z\Hom(X\otimes Y,Z).
	\end{equation}
To obtain the dimension of this space, we simply calculate
	\begin{equation}
		\dim(\Hom(A\otimes A,A))=\sum_{X,Y,Z\in\Obj(\Hd)} a_Xa_Ya_Z\dim(\Hom(X\otimes Y,Z))=15.
	\end{equation}
Therefore, an element $m\in\Hom(A\otimes A,A)$ is a linear combination of the $15$ basis diagrams of this space:
	\begin{align}
		m &= c_{\mathbf{1}\mathbf{1}}^\mathbf{1}\vertex{\mathbf{1}}{\mathbf{1}}{\mathbf{1}} + c_{\mathbf{1}\rho}^{\rho}\vertex{\mathbf{1}}{\rho}{\rho} + c_{\mathbf{1}{}_\alpha\rho}^{{}_\alpha\rho}\vertex{\mathbf{1}}{{}_\alpha\rho}{{}_\alpha\rho} + c_{\rho\mathbf{1}}^{\rho}\vertex{\rho}{\mathbf{1}}{\rho}+	c_{\rho\rho}^{\mathbf{1}}\vertex{\rho}{\rho}{\mathbf{1}}\\ \label{eq:mdecomp} &\hspace{10pt} + c_{\rho\rho}^{\rho}\vertex{\rho}{\rho}{\rho} + c_{\rho\rho}^{{}_\alpha\rho}\vertex{\rho}{\rho}{{}_\alpha\rho} + c_{\rho{}_\alpha\rho}^{\rho}\vertex{\rho}{{}_\alpha\rho}{\rho}+c_{\rho{}_\alpha\rho}^{{}_\alpha\rho}\vertex{\rho}{{}_\alpha\rho}{{}_\alpha\rho}  + c_{{}_\alpha\rho \mathbf{1}}^{ {}_\alpha\rho}\vertex{{}_\alpha\rho}{\mathbf{1}}{{}_\alpha\rho}\\&\hspace{10pt}+ c_{{}_\alpha\rho\rho}^{\rho}\vertex{{}_\alpha\rho}{\rho}{\rho} + c_{{}_\alpha\rho\rho}^{ {}_\alpha\rho}\vertex{{}_\alpha\rho}{\rho}{{}_\alpha\rho}+  
		c_{{}_\alpha\rho  {}_\alpha\rho}^{ \mathbf{1}}\vertex{{}_\alpha\rho}{{}_\alpha\rho}{\mathbf{1}} + c_{{}_\alpha\rho  {}_\alpha\rho}^{\rho}\vertex{{}_\alpha\rho}{{}_\alpha\rho}{\rho} + c_{{}_\alpha\rho  {}_\alpha\rho}^{ {}_\alpha\rho}\vertex{{}_\alpha\rho}{{}_\alpha\rho}{{}_\alpha\rho}.
	\end{align}
In order to find the multiplication morphism $m$ for the algebra object, all $15$ coefficients have to be determined. To solve this problem, we can make use of the different conditions that this morphism has to fulfil (see \cref{def:algebraobject}). Condition \cref{eq:AO2} already fixes five of the coefficients:
	\begin{equation}
		c_{\mathbf{1}\mathbf{1}}^\mathbf{1}=c_{\mathbf{1}\rho}^{\rho}=c_{\mathbf{1}{}_\alpha\rho}^{{}_\alpha\rho}=c_{\rho\mathbf{1}}^\rho=c_{{}_\alpha\rho\mathbf{1}}^{{}_\alpha\rho}=1.
	\end{equation}
This leaves us with $10$ unknowns. Additional equations that help us to determine these coefficients come from condition \cref{eq:AO1} of the definition: Inserting the decomposition \cref{eq:mdecomp} into \cref{eq:mcondpic} yields a linear combination of $81$ terms for the left hand side of the equation:
	\begin{align}
		\begin{tikzpicture}[scale=0.75,baseline=(current bounding box.center)]
			\draw (0,0) to (90:1);
			\draw (0,0) to (230:1);
			\draw (0,0) to (310:1);	
			\draw[fill=LinkColor2!50!white] (0,0) circle(0.35cm);	
			\node (c) at (0,0) {\small $m$};
			\node at (90:1.25) {\small $A$};	
			\node at (230:1.25) {\small $A$};
			\node at (310:1.25) {\small $A$};	
			\begin{scope}[yshift=2.2cm, xshift=0.9cm]
			\draw (0,0) to (90:1);
			\draw (0,0) to (230:1);
			\draw (0,0) to (295:3.225);	
			\draw[fill=LinkColor2!50!white] (0,0) circle(0.35cm);	
			\node (c) at (0,0) {\small $m$};
			\node at (90:1.25) {\small $A$};	
			\node at (295:3.475) {\small $A$};	
			\end{scope}
		\end{tikzpicture}&={c_{\mathbf{1}\mathbf{1}}^\mathbf{1}}^2\fuselefttree{\mathbf{1}}{\mathbf{1}}{\mathbf{1}}{\mathbf{1}}{\mathbf{1}}+c_{\mathbf{1}\mathbf{1}}^\mathbf{1}c_{\mathbf{1}\rho}^\rho\fuselefttree{\mathbf{1}}{\mathbf{1}}{\mathbf{1}}{\rho}{\rho}+c_{\mathbf{1}\mathbf{1}}^\mathbf{1}c_{\mathbf{1}{}_\alpha\rho}^{{}_\alpha\rho}\fuselefttree{\mathbf{1}}{\mathbf{1}}{\mathbf{1}}{{}_\alpha\rho}{{}_\alpha\rho}\\
		&\hspace{10pt}+c_{\mathbf{1}\rho}^\rho c_{\rho\rho}^\mathbf{1}\fuselefttree{\mathbf{1}}{\rho}{\rho}{\rho}{\mathbf{1}} +c_{\mathbf{1}\rho}^\rho c_{\rho \mathbf{1}}^\rho\fuselefttree{\mathbf{1}}{\rho}{\rho}{\mathbf{1}}{\rho}+c_{\mathbf{1}\rho}^\rho c_{\rho\rho}^\rho\fuselefttree{\mathbf{1}}{\rho}{\rho}{\rho}{\rho}+\dots
	\end{align}
and the same number of terms for the right hand side. We can now work out a general form for the equations that can be deduced from this condition: Note that the general form of the terms on the left hand side of the equation is
	\begin{equation}
		c_{xy}^\alpha c_{\alpha z}^w\fuselefttree{x}{y}{\alpha}{z}{w},		
	\end{equation}
where we call this kind of diagram a \emph{left fusion tree}. If the morphism space $\Hom(x\otimes y\otimes z,w)$ is one-dimensional, making a basis transformation by applying the corresponding $F$-move\footnote{Note that these fusion trees are mirrored horizontally compared to the diagrams that appear in the construction of $F$-symbols. Hence, in general, we have to use the complex conjugate of the entries of the $F$-matrix here (see \cref{sec:alganyons}). However, since the $F$-symbols are real for the $\Hi$ category (see \cref{app:Fsymbols}), we can use the standard $F$-symbols here.} yields
	\begin{equation}
	\label{eq:mmorphcond}
		c_{xy}^\alpha c_{\alpha z}^w\fuselefttree{x}{y}{\alpha}{z}{w}=c_{xy}^\alpha c_{\alpha z}^w\left(F_w^{xyz}\right)_{\beta\alpha}\fuserighttree{x}{y}{\beta}{z}{w}
	\end{equation}
On the right hand side of the equation, there will be a term of the form
	\begin{equation}
		c_{x\beta}^w c_{y z}^\beta\fuserighttree{x}{y}{\beta}{z}{w}
	\end{equation}
(a \emph{right fusion tree}) since there is one term for each basis diagram of the morphism space. Comparing the coefficients of the two terms yields the condition
	\begin{equation}
	\label{eq:condm}
		c_{xy}^\alpha c_{\alpha z}^w\left(F_w^{xyz}\right)_{\beta\alpha}=c_{x\beta}^w c_{y z}^\beta.
	\end{equation}
This procedure can easily be generalized to higher-dimensional morphism spaces. In this case, there will be a sum over the label $\beta$ on the right hand side of \cref{eq:mmorphcond}, i.e., a sum over right fusion trees. This implies that every right fusion tree can come from more than one of the left fusion trees. Hence, we have to take sum over all left fusion trees which can give rise to a right fusion tree with label $\beta$ and the general form of condition \cref{eq:condm} is
	\begin{equation}
 		\sum_\alpha c_{xy}^\alpha c_{\alpha z}^w\left(F_w^{xyz}\right)_{\beta\alpha}=c_{x\beta}^w c_{y z}^\beta.
	\end{equation}
Solving this set of equations gives the following solution:
	\begin{align}
		c_{\rho {}_\alpha\rho}^{\rho}&=c_{\alpharho\rho}^\rho\\
		c_{{}_\alpha\rho \rho}^{{}_\alpha\rho}&=c_{\rho\alpharho}^\alpharho\\
		c_{\rho\rho}^\mathbf{1}&=\sqrt{\frac{1}{2} \left(3-\sqrt{2 \sqrt{13}-5}\right)}\ {c_{\rho{}_\alpha\rho}^{{}_\alpha\rho}}^2\\
		c_{{}_\alpha\rho {}_\alpha\rho}^\mathbf{1}&=-\sqrt{\frac{1}{2} \left(3+\sqrt{2 \sqrt{13}-5}\right)}\ p_1\  {c_{\alpharho\rho}^\rho}^2\\
		c_{{}_\alpha\rho {}_\alpha\rho}^\rho&=-\frac{1}{4}\left(1+\sqrt{13}+\sqrt{2 \left(\sqrt{13}-1\right)}\right)\ p_1\ \frac{{c_{\alpharho \rho}^{\rho}}^2}{c_{\rho\alpharho}^\alpharho}\\
		c_{\rho \rho}^{{}_\alpha\rho}&=-\frac{1}{4}\left(1+\sqrt{13}+\sqrt{2 \left(\sqrt{13}-1\right)}\right)\ p_1\ \frac{{c_{\rho \alpharho}^{\alpharho}}^2}{c_{\alpharho\rho}^\rho}\\
		c_{\rho \rho}^\rho&=\frac{1}{4} \left(3-\sqrt{13}-\sqrt{2 \left(\sqrt{13}-1\right)}\right)\ c_{\rho {}_\alpha\rho}^\alpharho\\
		c_{{}_\alpha\rho {}_\alpha\rho}^{{}_\alpha\rho}&=\frac{1}{4} \left(3-\sqrt{13}+\sqrt{2 \left(\sqrt{13}-1\right)}\right)c_{\rho {}_\alpha\rho}^\rho,
	\end{align}
where $p_1\in\{-1,+1\}$ is a parameter coming from the solution of the $F$-symbols in \cref{app:Fsymbols}. Note that two of the coefficients, $c_{\rho\alpharho}^\alpharho$ and $c_{\alpharho\rho}^\rho$ are not determined by the conditions.

\section{Module categories over $\Hd$}

In the previous chapter, we have shown how to construct module categories from algebra objects. Starting from the algebra object $A=\mathbf{1}+\rho+\alpharho$ we now show how to practically determine the category of left modules over $A$ in $\Hd$, following the same techniques we described in \cref{ex:FibMod}. We begin by determining the simple module objects via calculating the matrix $T$. Remember that a row of $T$ is given by the outcome of the fusion product $A\otimes X_i$ for $X_i\in\{\mathbf{1},\alpha,\alpha^*,\rho,\alpharho,\alphastarrho\}$, where the coefficients in the decomposition into simple objects are the entries in the row corresponding to the object $X_i$. The resulting matrix is
	\begin{equation}
		T=\begin{blockarray}{ccccccc}
		& 1 & \alpha & \alpha^* & \rho & {}_\alpha\rho & \alphastarrho\\
		\begin{block}{c (cccccc)}
		1 & 1 & 0 & 0 & 1 & 1 & 0 \topstrut \\
		\alpha & 0 & 1 & 0 & 1 & 0 & 1\\
		\alpha^* & 0 & 0 & 1 & 0 & 1 & 1\\
		\rho & 1 & 1 & 0 & 3 & 2 & 2\\
		{}_\alpha\rho & 1 & 0 & 1 & 2 & 3 & 2\\
		\alphastarrho & 0 & 1 & 1 & 2 & 2 & 3\botstrut \\
		\end{block}
		\end{blockarray}\ .
	\end{equation} 
Recall that $T$ can be rewritten as $T=VV^T$. Hence, $V$ is given by 
	\begin{equation}
		V=\begin{pmatrix}
		1 & 0 & 0 & 0\\
		0 & 1 & 0 & 0\\
		0 & 0 & 1 & 0\\
		1 & 1 & 0 & 1\\
		1 & 0 & 1 & 1\\
		0 & 1 & 1 & 1
		\end{pmatrix}.
	\end{equation}
We can directly conclude that there are four simple module objects in the category $\mathrm{Mod}_{\mathcal{H}_3}(A)$, which are
	\begin{align}
		M_1&=1+\rho+{}_\alpha\rho\\
		M_2&=\alpha+\rho+\alphastarrho\\
		M_3&=\alpha^*+{}_\alpha\rho+\alphastarrho\\
		M_4&=\rho+{}_\alpha\rho+\alphastarrho.
	\end{align}
The corresponding morphisms $l_{M_i}$ can be determined from the conditions given in \cref{def:leftmod} analogously to the calculation of the multiplication morphism $m$ of the algebra object.

\part{Microscopic models}

%
\chapter{Anyon chains}
\label{sec:anyons}

In this chapter, we introduce the concept of \emph{anyon chains}, which is the most important tool in our search for a Conformal Field Theory (CFT) that corresponds to the Haagerup subfactor. Anyons are exotic quasiparticles that can occur in two-dimensional systems, whose statistical behaviour is much less restricted than those of bosons and fermions. Mathematically, they are described by Unitary Modular Tensor Categories (UMTC). 

Anyons are highly interesting for several reasons: Firstly, they can be employed to perform fault-tolerant quantum computation in the context of \emph{topological quantum computation} \cite{Nayak2008,Pachos2009,Wang2010,kitaev_anyons_2006}. Since most of the quasiparticle details are not relevant for the description of anyons (in fact, only topological properties matter), computations using these particles are resistant against errors in the control of the quasiparticles. For instance, in \cite{Kitaev2003} Kitaev proposed a topological quantum error correcting code, the \emph{toric code}, defined on a two-dimensional spin lattice with periodic boundary conditions. In this model, the information is encoded in the ground state space of the Hamiltonian such that errors appear as excitations that can be detected due to an energy gap above the ground states. Another interesting example are computational models coming from the Fibonacci category, since they allow for \emph{universal} quantum computation: any quantum circuit acting on $n$ qubits can be realised using $4n$ physical Fibonacci anyons (see \cite{Freedman2002} and \cite{Preskill2004}).

Anyonic statistics can be observed in the fractional quantum Hall effect \cite{Arovas1984,Halperin1984,Jeon2003,CAS13}. However, it has taken a long time for non-abelian anyons (see below for an explanation of the term \emph{non-abelian}) to actually be observed in an experiment. While there was some evidence for non-abelian anyons reported in \cite{Willett2013}, it has been pointed out in \cite{Keyserlingk2015} that the observed phenomena might come from certain Coulomb effects instead of properties of non-abelian anyons. Recent works show new evidence for the existence of anyonic quasiparticles: While in \cite{Bartolomei2020} anyonic statistics were indirectly observed via noise correlation measurements, the authors of \cite{Nakamura2020} have been able to directly observe these statistics.

Apart from being a promising candidate for topological quantum computation, anyons are an important tool in our search for a Haagerup CFT: As it was investigated in \cite{feiguin_interacting_2007} and \cite{Finch2014}, among others, they are connected to CFTs via a one-dimensional chain constructed from the underlying UMTC. This can be thought of as an analogue to spin chains\index{spin chain}: Consider a chain of spin-$\frac{1}{2}$ particles. Two neighbouring spin-$\frac{1}{2}$ degrees of freedom will either combine to a spin-$0$ state, i.e., a singlet state $|\mathbf{0}\rangle$, or a spin-$1$ state, which is a triplet state $|\mathbf{1}\rangle$. Hence, we can picture this chain with spin-$\frac{1}{2}$ particles on the lattice sites and singlet and triplet states living on the bonds of this chain:
	\begin{figure}[H]
		\centering
		\begin{tikzpicture}[scale=1.5]
			\node at (0,0) {\textbullet};
			\node at (1,0) {\textbullet};
			\node at (2,0) {\textbullet};
			\node at (3,0) {\textbullet};
			\node at (4,0) {\textbullet};
			\node at (5,0) {\textbullet};
			\draw (0.15,0) -- (0.85,0);
			\draw (1.15,0) -- (1.85,0);
			\draw (2.15,0) -- (2.85,0);
			\draw (3.15,0) -- (3.85,0);
			\draw (4.15,0) -- (4.85,0);
			\node at (0,+0.35) {$\frac{1}{2}$};
			\node at (1,+0.35) {$\frac{1}{2}$};
			\node at (2,+0.35) {$\frac{1}{2}$};
			\node at (3,+0.35) {$\frac{1}{2}$};
			\node at (4,+0.35) {$\frac{1}{2}$};
			\node at (5,+0.35) {$\frac{1}{2}$};
			\node at (0.5,-0.25) {\footnotesize $|\mathbf{0}\rangle/|\mathbf{1}\rangle$};
			\node at (1.5,-0.25) {\footnotesize $|\mathbf{0}\rangle/|\mathbf{1}\rangle$};
			\node at (2.5,-0.25) {\footnotesize $|\mathbf{0}\rangle/|\mathbf{1}\rangle$};
			\node at (3.5,-0.25) {\footnotesize $|\mathbf{0}\rangle/|\mathbf{1}\rangle$};
			\node at (4.5,-0.25) {\footnotesize $|\mathbf{0}\rangle/|\mathbf{1}\rangle$};
		\end{tikzpicture}
	\end{figure}

Our goal is to find an analogous way of constructing an anyon chain starting from a UMTC. We begin this chapter by describing anyonic statistics and the underlying algebraic theory in detail, thereby focussing on how the physical properties of anyons arise from the underlying structure of the UMTC. Then we present the constriction of chain models for anyons and explain how a numerical investigation of the ground state yields information about the corresponding conformal field theory. Here, we make use of tensor network methods (more precisely, matrix product states) to be able to efficiently study the model numerically. Finally, we apply these methods to the Haagerup fusion category $\Hd$ in order to investigate whether there is a conformal field theory corresponding to the Haagerup subfactor.

\section{Anyonic statistics}

\index{anyons}
For a moment, we will leave the mathematical concepts we have discussed in the previous chapters and focus on some physical phenomena. As quantum mechanics tells us, indistinguishable particles in three spatial dimensions come in two species: bosons and fermions. Consider now a system of two identical particles. One particle circulates around the other particle via the path $C_1$ as shown in \cref{fig:particleloop}\textbf{(a)}. In three spatial dimensions, the path $C_1$ can be continuously deformed into the path $C_2$ simply by lifting the loop above the second particle. We are only interested in topological properties, hence the particular geometry of the path does not have any effect. $C_2$ can further be deformed into the trivial path (which we simply denote $0$), which is the path that leaves the particle at its original position at all times. 

In two spatial dimensions, the situation is a little different. As shown in \cref{fig:particleloop}\textbf{(b)}, it is now impossible to lift the path $C_2$ above the second particle to continuously deform it into the path $C_1$. We would instead have to cut the path, pass it over the particle and glue it together again, but this would change the topological properties of the path. Hence, in two dimensions, the paths $C_2$ and $C_1$ are topologically different. Despite this difference, note that $C_2$ can still be deformed to the trivial path. 

	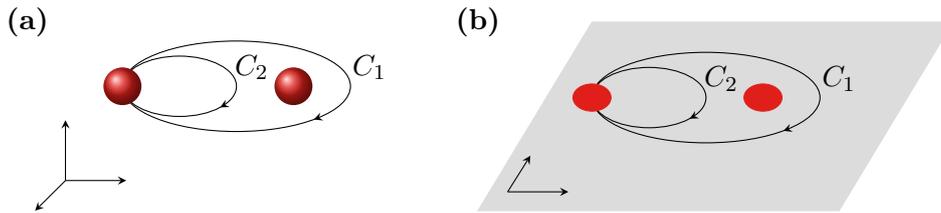
\begin{figure}[t]
		\begin{tikzpicture}[scale=1]
			\node at (-0.5,2.1) {\textbf{(a)}};
			\begin{scope}[scale=0.8]
				\draw[->,>=stealth] (0,0) -- (1,0);
				\draw[->,>=stealth] (0,0) -- (0,1);
				\draw[->,>=stealth] (0,0) -- (225:0.7071);
			\end{scope}
			\draw[->-] (0.75,1.25) arc (180:-180:0.75cm and 0.4cm);
			\draw[->-] (0.75,1.25) arc (180:-180:1.5cm and 0.6cm);
			\shade[ball color = LinkColor!90, opacity = 1] (0.75,1.25) circle (0.25cm);
			\shade[ball color = LinkColor!90, opacity = 1] (3,1.25) circle (0.25cm);
			\node at (4,1.5) {$C_1$};
			\node at (2.45,1.5) {$C_2$};
		\end{tikzpicture}\hspace{15pt}
		\begin{tikzpicture}[scale=1]
			\node at (-0.75,2.25) {\textbf{(b)}};
			\draw[fill=LightGray,LightGray] (-0.75,-0.25) -- (4,-0.25) -- (5.5,2.25) -- (0.75,2.25) -- cycle;
			\begin{scope}[scale=0.8,xshift=-0.45cm,yshift=0cm]
			\draw[->,>=stealth] (0,0) -- (1,0);
			\draw[->,>=stealth] (0,0) -- (58:0.7071);
			\end{scope}
			\draw[->-] (0.75,1.25) arc (180:-180:0.75cm and 0.4cm);
			\draw[->-] (0.75,1.25) arc (180:-180:1.5cm and 0.6cm);
			\draw[fill=LinkColor!90,LinkColor!90] (0.75,1.25) ellipse (0.25cm and 0.18cm);
			\draw[fill=LinkColor!90,LinkColor!90] (3,1.25) ellipse (0.25cm and 0.18cm);
			\node at (4,1.5) {$C_1$};
			\node at (2.45,1.5) {$C_2$};
		\end{tikzpicture}
	\caption{\small \label{fig:particleloop}\textbf{Topological differences between three- and two-dimensional particle circulation.} \textbf{(a)} A particle spans a loop around another particle (path $C_1$). In three dimensions, it is always possible to continuously deform the path $C_1$ into the path $C_2$, which is equivalent to a trivial path. \textbf{(b)} In two dimensions, the paths are topologically distinct: $C_1$ cannot be continuously deformed into $C_2$.}
	\end{figure}

What does this observation tell us about the statistics of these particles? In the three-dimensional case, the wave function $\Psi$ of the system after the circulation $C_1$ has to be exactly the same as the original wave function:
	\begin{equation}
	\label{eq:waveequiv}
		\Psi(C_1)=\Psi(C_2)=\Psi(0).
	\end{equation}
	
Furthermore, note that the circulation of one particle around another is equivalent to performing two exchanges of these particles (plus a spatial translation which is irrelevant because we are only interested in topological properties here), see \cref{fig:particleexchange}. Hence, the single exchange of two particles can yield a phase factor $e^{i\varphi}$ which has to square to one to fulfil \cref{eq:waveequiv}. This has exactly two possible solutions, $\varphi_b=0$ and $\varphi_f=\pi$, which corresponds to the bosonic\index{boson} and fermionic\index{fermion} statistics, respectively.

	\begin{figure}[t]
		\begin{tikzpicture}[scale=1.2]
			\begin{scope}[scale=0.7,xshift=0cm,yshift=0.5cm]
			\draw[->,>=stealth] (0,0) -- (1,0);
			\draw[->,>=stealth] (0,0) -- (0,1);
			\draw[->,>=stealth] (0,0) -- (225:0.7071);
			\end{scope}
			\draw[-->--] (0.75,1.25) to [bend left=30] (3,1.25);
			\draw[-->--] (3,1.25) to [bend left=30] (0.75,1.25);
			\shade[ball color = LinkColor!90, opacity = 1] (0.75,1.25) circle (0.25cm);
			\shade[ball color = LinkColor!90, opacity = 1] (3,1.25) circle (0.25cm);
		\end{tikzpicture}
		\caption{\small \label{fig:particleexchange}\textbf{A single exchange of two particles.} Two successive exchanges of two particles are equivalent to circulating one particle around the other (plus a spatial translation).}
	\end{figure}
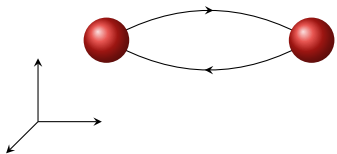

In the two-dimensional case the situation is different. Since the path $C_1$ cannot continuously be deformed into the trivial path, there are a variety of possible statistical evolutions. We can now assign an arbitrary phase factor to the evolution that corresponds to the path $C_1$ (or, equivalently, two successive exchanges of particles) or even a unitary matrix. In the case where we assign an arbitrary phase factor to the evolution, the particles are called \emph{abelian anyons}\index{abelian anyon}. The factor can take any value between the bosonic case $e^{i\varphi_b}=1$ and the fermionic case $e^{i\varphi_f}=-1$\footnote{The fact that these particles can have any statistics has led to the name \emph{anyon} \cite{Wilczek1982}.}. 

Beyond this, it is possible to have more complex statistics that are described by a higher-dimensional unitary matrix. Anyons that have this kind of statistics are called \emph{non-abelian}\index{non-abelian anyon}\footnote{This name is motivated by the fact that matrices do not, in general, commute, in contrast to factors.}. For this kind of evolution to emerge, the wave function that describes the system has to be within a degenerate subspace of states. The matrix evolution then transforms between different states in this subspace, thereby leaving the energy of the system unchanged. Note that the initial considerations we made require the different states in this subspace to be \emph{indistinguishable}. This means that there are no local measurements that can detect the exchange of these particles.

\section{Algebraic theory of anyons}
\label{sec:alganyons}

Mathematically, anyons are described by Unitary Modular Tensor Categories (UMTCs) which have been defined in \cref{sec:MTC}. However, in this chapter we will take a more physical point of view and focus on the \emph{data} we need to describe a system of anyons. We then connect this data with the adjectives that appear in the definition of a UMTC. \Cref{tab:MTCanyons} provides a dictionary between the notation used in the description of anyon chains (that mostly originates from physics) and the categorical terms.

\subsection*{Particle types and fusion rules.}
To describe a system of anyons we first have to specify what kind of particles we allow, i.e., a set
	\begin{equation}
		\mathcal{M}=\{\mathbf{1},a,b,c,\dots\}.
	\end{equation}
Here, $\mathbf{1}$ denotes the (unique) vacuum, while $a,b,c,\dots$ correspond to a \emph{finite} series of different particle types\footnote{In the following, we will use the terminologies \emph{particle} and \emph{anyon} interchangeably.}. In the language of UMTCs, this set of particle types corresponds to the isomorphism classes of simple objects of which there are only finitely many (see \cref{def:fusioncat}).

The fusion rules\index{fusion rule} of an anyon model describe what happens if we bring two anyons close together and treat them as one composite object, hence describing how they statistically behave together. These rules are given in the form
	\begin{equation}
	\label{eq:fusionrules}
		a\otimes b=\sum_c N_{ab}^c\ c,
	\end{equation}
where $a,b,c$ are anyon types and the sum runs over all possible types. Since there might several distinct ways to produce particle $c$ when fusing $a$ and $b$, we have an integer factor $N_{ab}^c$ which is usually referred to as \emph{fusion multiplicity}\index{multiplicities}. 

	\begin{defn}[Multiplicity-free]
		\index{multiplicity-free}
		A fusion rule is multiplicity-free if $N_{ab}^c\in\{0,1\}$ for any choice of particle types $a,b,c$.
	\end{defn}

\begin{table}[t]
	\begin{tabular}{c|c}
		Anyonic system & Unitary MTC\\ \hline\\[-8pt]
		anyon & simple object\\
		fusion & tensor product\\
		fusion rules & fusion rules\\
		antiparticle & dual object\\
		quantum dimension & object dimension\\
		particle exchange & braiding
		\vspace{10pt}
	\end{tabular}
	\caption{\small \label{tab:MTCanyons}Dictionary of terms relating anyonic systems to unitary modular tensor categories (UMTCs).}
\end{table}

The ordering of the particles $a$ and $b$ in the fusion rules \cref{eq:fusionrules} is not important: it holds that
	\begin{equation}
		a\otimes b=b\otimes a.
	\end{equation}	
Fusing a particle with the vacuum is always trivial:
	\begin{align}
		a\otimes\mathbf{1}&=a\\
		\mathbf{1}\otimes b&=b.
	\end{align}
This property can also be expressed as $N_{a\mathbf{1}}^c=\delta_{ac}$ and $N_{\mathbf{1}b}^c=\delta_{bc}$. Each particle type $a$ has a unique antiparticle type $a^*$ such that $N_{ab}^\mathbf{1}=\delta_{ba^*}$. Furthermore, it holds that $\mathbf{1}^*=\mathbf{1}$ and $\left(a^*\right)^*=a$. The fusion multiplicities obey the following relation (see \cite{Bond2007}):
	\begin{align}
		\sum_e N_{ab}^e N_{ec}^d&=\sum_f N_{af}^d N_{bc}^f.
	\end{align}

The fusion rules also tell us whether the anyon model is abelian or non-abelian: In the abelian case, the anyons only have one fusion channel, i.e., every fusion rule is of the form 
	\begin{equation}
	\label{eq:abelian}
		a\otimes b=c.
	\end{equation}
In the non-abelian case there can be multiple fusion outcomes, which corresponds to
	\begin{equation}
	\label{eq:nonabelian}
		\sum_c N_{ab}^c>1.
	\end{equation}

Another important quantity is the so-called \emph{quantum dimension}\index{quantum dimension} of a particle. For a particle $a$ it is defined via the $F$-symbols of the category, which have been introduced in \cref{sec:fusioncat}:
	\begin{equation}
	\label{eq:qdimF}
		d_a=\left|\left(F_a^{aa^* a}\right)_{\mathbf{1}\mathbf{1}}\right|^{-1}.
	\end{equation}
Since the anyon model is unitary, it follows that $d_a\ge 1$ with equality if and only if $a$ is abelian (i.e., fusion with any other particle has exactly one fusion channel as in \cref{eq:abelian}). The total quantum dimension\index{total quantum dimension} of the model is defined as
	\begin{equation}
		\mathcal{D}=\sqrt{\sum_a d_a^2}.
	\end{equation}

Regarding the relation of anyons and modular tensor categories, the fusion operation of anyons corresponds to the tensor product in a UMTC. We have already seen how fusion multiplicities naturally emerge in a fusion category in \cref{rem:multiplicities}. Furthermore, the quantum dimension of a particle is directly related to the dimension of an object in a UMTC as defined in \cref{def:ctdim}. This will become more clear when we introduce the graphical depiction of the quantum dimension in \cref{eq:qdimgraph}.

Recall that the goal of this chapter is to build a one-dimensional chain of anyons similar to how spin chains are constructed. These anyon chains are of the form 
	\begin{equation}
		\begin{tikzpicture}[scale=0.8,baseline=(current bounding box.center)]
			\draw (0,0) -- (-2.5,2.5);
			\draw (-0.5,0.5) -- (1.5,2.5);
			\draw (-1.5,1.5) -- (-0.5,2.5);
			\draw (-2,2) -- (-1.5,2.5);
			\node at (0.25,-0.25) {\small $a_{N+1}$};
			\node at (-2.5,2.75) {\small $a_1$};
			\node at (-1.5,2.75) {\small $a_2$};
			\node at (-0.5,2.75) {\small $a_3$};
			\node at (1.5,2.75) {\small $a_N$};
			\node at (0.5,2.75) {\small $\dots$};
			\node at (-1.9,1.6) {\small $x_1\ $};
			\node at (-1.4,1.1) {\small $x_2\ $};
			\node at (-1,0.6) {\rotatebox{-45}{\small $\dots$}};
		\end{tikzpicture},
	\end{equation}
where every vertex corresponds to the splitting of one anyon into two. Hence, to be able to describe the Hilbert space and the dynamics of this system we need to translate the properties of the underlying fusion category to vector spaces and Hamiltonians. We begin by building vector spaces for the fusion and splitting of anyons.	

Given the fusion rules, we can assign vector spaces to fusion processes\index{fusion space}. For instance, consider two particles $a$ and $b$ that fuse to a particle $c$ with $N_{ab}^c=1$. This implies that there is exactly one way to fuse the particles $a$ and $b$ to $c$. Hence, the vector space that corresponds to this fusion process has to be one-dimensional. We will denote the corresponding fusion space $V_{ab}^c$ and vectors in this space are of the form
	\begin{equation}
		\begin{tikzpicture}[scale=0.5,baseline=(current bounding box.center)]
			\draw (0,0) -- (0,-1);
			\draw (0,-1) -- (-1,-2);
			\draw (0,-1) -- (1,-2);
			\node at (0,0.4) {$c$};
			\node at (-1.1,-2.4) {$a$};
			\node at (1.1,-2.4) {$b$};
		\end{tikzpicture}
	\end{equation}

There is a dual space to $V_{ab}^c$, namely the \emph{splitting space}\index{splitting space} $V_c^{ab}$. Their relation is the following: If $\psi\in V_c^{ab}$ is a splitting vector, then $\psi^\dagger\in V_{ab}^c$ is the dual vector in the fusion space:
	\begin{equation}
		\begin{tikzpicture}[scale=0.5,baseline=(current bounding box.center),yscale=-1]
		\node at (-0.55,-0.65) {$\psi$};
		\draw (0,0) -- (0,-1);
		\draw (0,-1) -- (-1,-2);
		\draw (0,-1) -- (1,-2);
		\node at (0,0.4) {$c$};
		\node at (-1.1,-2.4) {$a$};
		\node at (1.1,-2.4) {$b$};
		\end{tikzpicture},\hspace{20pt}\begin{tikzpicture}[scale=0.5,baseline=(current bounding box.center)]
		\node at (-0.55,-0.65) {$\psi^\dagger$};
		\draw (0,0) -- (0,-1);
		\draw (0,-1) -- (-1,-2);
		\draw (0,-1) -- (1,-2);
		\node at (0,0.4) {$c$};
		\node at (-1.1,-2.4) {$a$};
		\node at (1.1,-2.4) {$b$};
		\end{tikzpicture}.
	\end{equation}
Since these vector spaces are the smallest possible fusion/splitting spaces, we will refer to them as the \emph{minimal} fusion/splitting spaces.

\begin{rem}
	If the fusion rules are multiplicity-free we can omit the Greek letter at the vertex of a fusion/splitting tree, since the space is one-dimensional in this case. In the following, we will usually consider the multiplicity-free case to avoid messy equations and diagrams, hence we will mostly use diagrams without Greek letters. However, we sometimes state an equation in both versions or make a comment to show what changes if we consider multiplicities. It should always be clear from the context which case we are considering since Greek letters occur if and only if the model is not multiplicity-free.
\end{rem}

\subsection*{The standard basis and $F$-matrices.}

When we fuse (or split) several anyons, there is a priori no preferred order of fusion (or splitting, respectively). We can choose several orders of fusion/splitting, which correspond to different choices of bases (we have already encountered this phenomenon when introducing $F$-symbols in \cref{ch:cats}). Therefore, we can choose a \emph{standard basis}\index{standard basis}, by which we mean the decomposition of a bigger fusion/splitting space into minimal fusion/splitting spaces. If we consider the splitting of a particle $u$ into $N$ particles $a_1,a_2,\dots,a_N$, we choose the corresponding standard basis to be of the form
	\begin{equation}
		V_u^{a_1,a_2,\dots,a_N}=\bigoplus_{e_2,e_3,\dots,e_{N-1}}V_{e_2}^{a_1a_2}\otimes V_{e_3}^{e_2a_3}\otimes V_{e_4}^{e_3a_4}\otimes\dots\otimes V_u^{e_{N-1}a_N}.
	\end{equation}
In terms of the previously introduced tree diagrams, the elements of this basis correspond to tree diagrams of the form
	\begin{equation}
.
	\end{equation}
This can analogously be defined for the fusion basis. We can extend the definition of a standard basis to involve both fusion and splitting spaces. The most general form is a vector space $V_{a_1',a_2',\dots,a_M'}^{a_1,a_2,\dots,a_N}$, where $M$ particles $a_1',\dots,a_M'$ are transformed into $N$ particles $a_1,\dots,a_N$. This can be performed in two steps: First, fusing the $M$ ingoing particles into one and then splitting them into the $N$ outgoing particles. In terms of the standard basis, the corresponding tree diagrams are of the form
	\begin{equation}
		%
.
	\end{equation}
We usually use the normalisation convention
	\begin{equation}
	\label{eq:vertexnorm}
		\left(\frac{d_c}{d_a d_b}\right)^\frac{1}{4}
		%
\in V_{ab}^c
	\end{equation}
for trivalent vertices throughout this thesis. Using this convention, the identity operation on two particles $a$ and $b$ can be decomposed in the following way:
	\begin{equation}
	\label{eq:identity}
		%
.
	\end{equation}
The sum over $\alpha$ can be neglected if the fusion rules are multiplicity-free. This relation is also called the \emph{completeness relation}\index{completeness relation}.

Furthermore, the inner product of two vectors is diagrammatically formed by stacking vertices on top of each other such that the fusion/splitting lines connect. The simplest case is the so-called \emph{bigon relation}\index{bigon relation}:
	\begin{equation}
	\label{eq:bigon2}
		%
.
	\end{equation}
This generalizes to more complicated diagrams. From this formula, we can directly see that tadpole diagrams (``lollipops''), i.e., diagrams of the form
	\begin{equation}
		%
	\end{equation}
are forbidden, since these kind of diagrams correspond to the case $c'=\mathbf{1}$ and therefore
	\begin{equation}
		%
,
	\end{equation}
where the right hand side is only non-zero for $c=\mathbf{1}$.

In the following, we will focus on describing properties of splitting spaces since these are the ones that will mainly appear in this thesis. Note that all statements can analogously be made for fusion spaces.

Categorically, to be able to work with the tree diagrams as they are introduced above, we need to choose bases for the involved morphism spaces. Furthermore, as described in \cref{ch:cats}, transforming between different bases is done by applying the correct $F$-matrices\index{F-symbol@$F$-symbol}. Recall that these matrices are defined via the equation
	\begin{equation}
		\begin{tikzpicture}[baseline=(current bounding box.center),yscale=0.9,xscale=0.9]
			\draw (0,0.5) -- (0,1);
			\draw (0,1) -- (-0.5,1.5);
			\draw (-0.5,1.5) -- (-1,2);
			\draw (-0.5,1.5) -- (0,2);
			\draw (0,1) -- (0.5,1.5);
			\draw (0.5,1.5) -- (1,2);
			\node at (0,0.25) {\small $u$};
			\node at (-1,2.25) {\small $a$};
			\node at (0,2.25) {\small $b$};
			\node at (1,2.25) {\small $c$};
			\node at (-0.5,1.1) {\small $p$};
		\end{tikzpicture}=\sum_q \left(F_u^{abc}\right)_{qp}
		\begin{tikzpicture}[baseline=(current bounding box.center),yscale=0.9,xscale=0.9]
			\draw (0,0.5) -- (0,1);
			\draw (0,1) -- (-0.5,1.5);
			\draw (-0.5,1.5) -- (-1,2);
			\draw (0.5,1.5) -- (0,2);
			\draw (0,1) -- (0.5,1.5);
			\draw (0.5,1.5) -- (1,2);
			\node at (0,0.25) {\small $u$};
			\node at (-1,2.25) {\small $a$};
			\node at (0,2.25) {\small $b$};
			\node at (1,2.25) {\small $c$};
			\node at (0.5,1.1) {\small $q$};
		\end{tikzpicture}.
	\end{equation}
Here, the diagram on the left-hand side is in the standard form we have chosen for tree diagrams. In the context of anyons, the above relation can also be motivated from a different point of view than the purely mathematical one: The two different bases both correspond to the splitting of a particle $u$ into the three particles $a,b,c$. Since by fusion we do not mean a physical process, but rather a grouping of particles into one system whose statistical behaviour we are interested in, these two choices are physically the same and are therefore connected via a unitary isomorphism, which is given by the $F$-matrices. 

Recall that the $F$-matrices have to fulfil a consistency condition, the so-called pentagon equation \cref{eq:pentagon}. Furthermore, in a unitary fusion category there are some additional relations between diagrams in terms of $F$-symbols. First, in a unitary fusion category the inverse of an $F$-symbol is given by the hermitian conjugate:
	\begin{equation}
		\left(F_u^{abc}\right)_{qp}^{-1}=\left(F_u^{abc}\right)_{qp}^\dagger=\overline{\Big(F_u^{abc}\Big)}_{pq}.
	\end{equation}
In terms of string diagrams, this means that
	\begin{equation}
.
	\end{equation}
Another consequence of unitarity is that the category is automatically spherical\index{spherical category} (see \cite[Proposition 8.23]{Etingof2005}), which yields a so-called \emph{mirror symmetry}\index{mirror symmetry} (see \cite{Hong2009}). This means that we can find relations between horizontally mirrored tree diagrams in terms of the previously defined $F$-symbols:
	\begin{align}
		%
.
	\end{align}
Furthermore, in any spherical category $\C$ it holds for the quantum dimension of any object $a\in\Obj(\C)$ that $d_{a^*}=d_a$.

\begin{rem}
	\label{rem:Felements}
	Because the $F$-symbol is a transformation between two different bases, the graphical calculus of fusion categories allows us to calculate matrix elements $\left(F_u^{abc}\right)_{qp}$ via the scalar product of elements of the respective orthonormal bases. For this purpose, the orthonormal bases we use are 
		\begin{equation}
			\left\{\frac{1}{\left(d_ad_bd_cd_u\right)^{\frac{1}{4}}}%
\right\}.
		\end{equation}
	The normalisation factor $\frac{1}{\left(d_ad_bd_cd_u\right)^{\frac{1}{4}}}$ comes from the normalisation of vertices we chose in \cref{eq:vertexnorm} together with an additional factor of $\frac{1}{\sqrt{d_u}}$ in order to make the diagram normalised with respect to the trace defined in \cref{def:trace}. The calculation of the matrix element then works via the scalar product of two basis elements:	
		\begin{align}
			\left(F_u^{abc}\right)_{qp}&=\frac{1}{\sqrt{d_a d_b d_c d_u}}
			\Bigg\langle
			%
\\
			&=\frac{1}{\sqrt{d_a d_b d_c d_u}}\left(F_u^{abc}\right)_{qp}\sqrt{\frac{d_b d_c}{d_q}} \sqrt{\frac{d_a d_q}{d_u}}\ d_u\\
			&=\left(F_u^{abc}\right)_{qp},
		\end{align}
	where we have applied the bigon relation \cref{eq:bigon2} twice and also used that the loop value is the quantum dimension, which will be explained in \cref{eq:qdimgraph}.
\end{rem}

In the context of physical models, apart from ensuring that the $F$-moves themselves are consistent, it is also crucial that adding and removing vacuum lines is consistent with the $F$-moves. From a physical perspective, splitting from or fusing with the vacuum is trivial. This is the reason why we can always add an remove vacuum lines in tree diagrams. To guarantee the consistency of this procedure with the model (especially with the $F$-moves), we require that the vector spaces $V_{a}^{a\mathbf{1}}$ and $V_a^{\mathbf{1}a}$ are one-dimensional and, furthermore, that they are canonically isomorphic to $\mathbb{C}$. In other words, these vector spaces have fixed unit vectors $\alpha_a\in V_a^{a\mathbf{1}}$ and $\beta_a\in V_a^{\mathbf{1}a}$. The canonical isomorphisms are simply given by
	\begin{alignat}{3}
		\alpha_a:\mathbb{C}&\to V_a^{a\mathbf{1}},\hspace{20pt}\beta_a:\ &\mathbb{C}\,&\to V_a^{\mathbf{1}a}\\
		z&\mapsto z\alpha_a, &z\,&\mapsto z\beta_a
	\end{alignat}
These isomorphisms have to fulfil the so-called \emph{triangle equation}\index{triangle equation} depicted in \cref{fig:triangle_eq}\,\textbf{(a)} to ensure consistency with the $F$-moves. There are two corollaries of the triangle equation, depicted in \cref{fig:triangle_eq}\,\textbf{(b)} and \textbf{(c)}, which automatically follow from the first triangle equation and the pentagon equation.

	\begin{figure}[t]
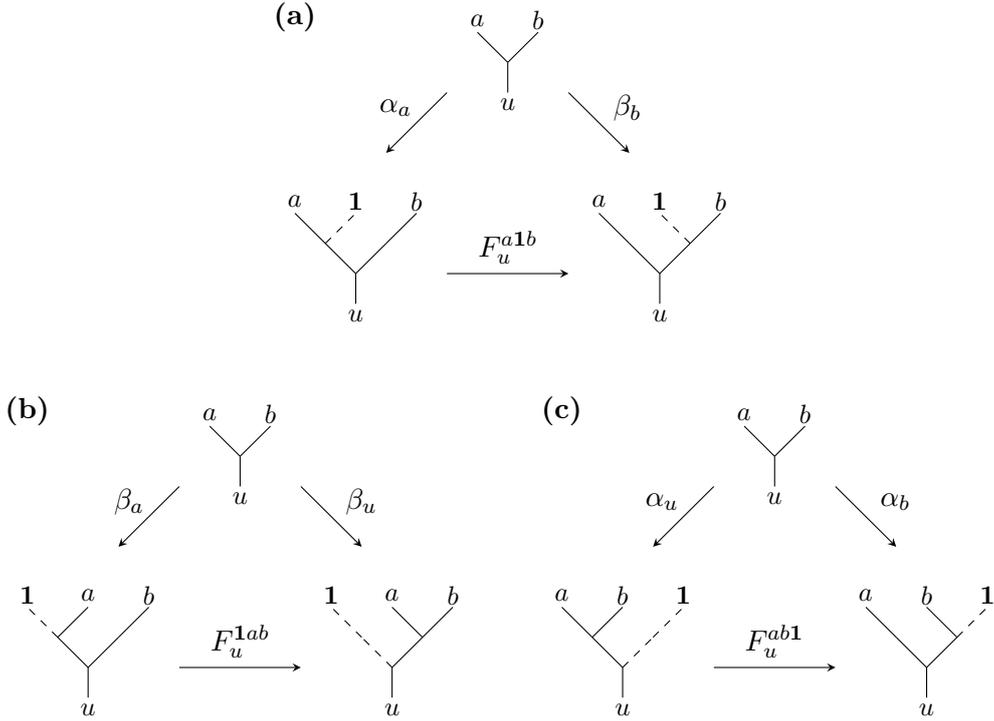

		\centering
		
		\caption{\small \textbf{Diagrammatic depiction of the triangle equation.} \textbf{(a)} The triangle equation ensures that fusion with/splitting from the vacuum lines is compatible with the $F$-move. \textbf{(b)} and \textbf{(c)} are corollaries of the triangle equation that are automatically implied by equation \textbf{(a)} and the pentagon equation.}
		\label{fig:triangle_eq}
	\end{figure}

\subsection*{Bending and tracing.}

We often want to introduce and remove bends in a line. Horizontal bending of lines (such that the line always goes upwards) is trivial:
	\begin{equation}
.
	\end{equation}
In contrast, one has to be very careful when bending lines vertically, for example transformations such as the following:
	\begin{equation}
		%
.
	\end{equation}
Recall that we are free to bend lines horizontally (as explained above) and, furthermore, are allowed to add and remove vacuum lines. Therefore we can modify the above diagram such that we can apply $F$-moves in order to evaluate the effect of bending lines vertically:
	\begin{equation}
	\label{eq:FrobSchur}
		%
,
	\end{equation}
where we have used \cref{eq:bigon2} twice in the last step. Here, the quantity
	\begin{equation}
	\label{eq:FrobSchur2}
		\left(F_a^{aa^*a}\right)_{\mathbf{1}\mathbf{1}}=\frac{\kappa_a}{d_a}
	\end{equation}
has, in general, a non-trivial phase $\kappa_a=\kappa_{a^*}^*$. It is always possible to find a gauge choice such that $\kappa_a=1$ for all self-dual particle types $a$ (i.e., for particles with $a=a^*$), but for those particle types that are not self-dual, $\kappa_a=\pm 1$ is the \emph{Frobenius-Schur indicator}\index{Frobenius-Schur indicator}, which is a gauge-invariant quantity. This is the reason why bending lines vertically is more intricate than bending them horizontally. 

One possible way to account for this problem is by introducing \emph{flags} (following the notation in \cite{kitaev_anyons_2006} and \cite{Bond2007}): When removing a vacuum line from the top of a fusion vertex or from the bottom of a splitting vertex, we indicate this by drawing a right-directed flag:
	\begin{align}
	\label{eq:evaluationcat}

	\end{align}
The cups and caps with left-directed flags are related to those with right-directed flags via the factor $\kappa_a$ (or $\kappa_a^*$, respectively). These cups and caps are already familiar to us as evaluation\index{evaluation} and coevaluation\index{coevaluation} morphisms. They appeared when we introduced left and right duals of objects in \cref{sec:fusioncat}. Using this notation, we can rewrite \cref{eq:FrobSchur} as
	\begin{equation}
		%
.
	\end{equation}
To ensure isotopy invariance when bending lines vertically, one needs to pair up cups and caps with opposing flags since the factors cancel out in this case:
	\begin{equation}
		%
.
	\end{equation}
There are analogous identities for $a^*$:
	\begin{equation}
		%
.
	\end{equation}
Whenever cups and caps are paired up with opposing flags so that the factor cancels out, the flags can be left implicit. This is the convention that we will use from now on and in most cases, the flags are paired up properly so they usually do not show up explicitly.

By defining the same kind of vectors for $a^*$ instead of $a$ and exploiting their relations, one can find several useful identities for the quantum dimension $d_a$ and the factor $\kappa_a$ (see \cite[App.\ E]{kitaev_anyons_2006}):
	\begin{align}
		d_a=d_{a^*},\hspace{20pt}\kappa_{a^*}=\kappa_a^*,\hspace{20pt}|\kappa_a|=1.
	\end{align}
Using this notation and \cref{eq:bigon2}, we can express the quantum dimension of a particle $a$ as 	
	\begin{equation}
	\label{eq:qdimgraph}

		=d_a.
	\end{equation}

These cups and caps can also be used to bend lines down, for example if you want to rotate a splitting vertex into a fusion vertex. Analogously, we can bend lines up to rotate a fusion vertex to become a splitting vertex. The four possible cases are
	\begin{align}
		%
,
	\end{align}
where
	\begin{align}
		A_c^{ab}&=\sqrt{\frac{d_a d_b}{d_c}}\overline{\left(F_b^{a^*a b}\right)}_{c\mathbf{1}} \label{eq:A}\\
		A_{ab}^c&=\overline{A_c^{ab}} \label{eq:As}\\
		B_c^{ab}&=\sqrt{\frac{d_a d_b}{d_c}}\left(F_a^{a b b^*}\right)_{\mathbf{1}c} \label{eq:B}\\
		B_{ab}^c&=\overline{B_c^{ab}}.\label{eq:Bs}
	\end{align}
These formulas are derived by exploiting the completeness relation \cref{eq:identity}. We show how this works by performing the computation of $B_c^{ab}$ step-by-step:
	\begin{align}
		%
,
	\end{align}
where in the last step we have used the fact that $d_b=d_{b^*}$ and that $N_{a\mathbf{1}}^{a'}=\delta_{a,a'}$. The derivations of the other operators use the same techniques. The operators $A_c^{ab}$ and $B_c^{ab}$ are unitary (see \cite[Thm.\ E.6]{kitaev_anyons_2006}), which implies the following identities:	
	\begin{equation}
	\label{eq:Nidentities}
		N_{ab}^c=N_{a^*c}^b=N_{b c^*}^{a^*}=N_{b^*a^*}^{c^*}=N_{c^*a}^{b^*}=N_{cb^*}^a.
	\end{equation}
From the unitarity of these operators we can directly derive the following identities for F-matrices:
	\begin{align}
		\left|\left(F_b^{a^*ab}\right)_{c\mathbf{1}}\right|^2&=\frac{d_c}{d_a d_b},\\
		\left|\left(F_a^{abb^*}\right)_{\mathbf{1}c}\right|^2&=\frac{d_c}{d_a d_b}.\label{eq:Fidentities}
	\end{align}
Furthermore, the fact that the $F$-matrices themselves are unitary matrices implies
	\begin{equation}
	\label{eq:Funitary}
		\left|\left(F_u^{abc}\right)_{ef}\right|^2=1,
	\end{equation}
whenever $e$ and $f$ are the \emph{unique} labels allowed by the fusion rules for $u,a,b,c$, since in this case the $F$-move is a unitary map between one-dimensional vector spaces.

Additionally, we can use the operators $A$ and $B$ to prove an important identity of the quantum dimensions of anyonic particles: 
\begin{thm}
	\label{thm:qdimrelation}
	The quantum dimension\index{quantum dimension} fulfils the key identity
	\begin{equation}
	d_a d_b=\sum_c N_{ab}^c d_c.
	\end{equation}
\end{thm}
\begin{proof}
	This can be shown via the following calculation:
		\begin{align}
			d_a d_b&=
\\
			&=\sum_cN_{ab}^cB_c^{ba}\overline{B^{ba}_c}\overline{A^{ca^*}_b} A_b^{ca^*}B_{a^*}^{c^*b}\overline{B^{c^*b}_{a^*}} d_c\\
			&=\sum_cN_{ab}^c d_c,
		\end{align}
	where the last equality follows from the fact that the operators $A_c^{ab}$ and $B_c^{ab}$ are unitary.
\end{proof}

Another important notion we have to define is the trace of an operator. Similar to the categorical definition of trace given in \cref{def:trace} and \cref{rem:trace}, we take the trace of a diagram by closing it with loops that connect the outgoing lines with the respective incoming lines at the same position. For instance, consider an operator $X$ that maps $n$ incoming particles $a_1,a_2,\dots,a_n$ to $n$ outgoing particles $a_1',a_2',\dots,a_n'$. We can represent this graphically as
	\begin{equation}
		\begin{tikzpicture}[scale=0.7,baseline=(current bounding box.center)]
			\draw (0,0) rectangle (2.5,1.5);
			\node at (1.25,0.75) {$X$};
			\draw (0.25,-0.75) -- (0.25,0);
			\draw (0.75,-0.75) -- (0.75,0);
			\draw (2.25,-0.75) -- (2.25,0);
			\node at (1.5,-0.375) {$\dots$};
			\draw (0.25,2.25) -- (0.25,1.5);
			\draw (0.75,2.25) -- (0.75,1.5);
			\draw (2.25,2.25) -- (2.25,1.5);
			\node at (1.5,1.875) {$\dots$};
			\node at (0.25,-1.1) {\small $a_1$};
			\node at (0.75,-1.1) {\small $a_2$};
			\node at (2.25,-1.1) {\small $a_n$};
			\node at (0.25,2.6) {\small $a'_1$};
			\node at (0.75,2.6) {\small $a'_2$};
			\node at (2.25,2.6) {\small $a'_n$};
		\end{tikzpicture}\ .
	\end{equation}
To calculate the trace\index{trace} of $X$ we connect every outgoing leg with the corresponding ingoing leg:
	\begin{equation}
	\Tr(X)=
		\begin{tikzpicture}[scale=0.7,baseline=(current bounding box.center)]
			\draw (0,0) rectangle (2.5,1.5);
			\node at (1.25,0.75) {$X$};
			\draw (2.25,-0.75) -- (2.25,0);
			\node at (1.5,-0.375) {$\dots$};
			\draw (2.25,2.25) -- (2.25,1.5);
			\node at (1.5,1.875) {$\dots$};
			\node at (-0.1,-0.85) {\small $a_1$};
			\node at (1.9,-0.85) {\small $a_n$};
			\node at (-0.1,2.35) {\small $a'_1$};
			\node at (1.9,2.35) {\small $a'_n$};
			\draw (2.25,2.25) arc (180:0:0.5cm);
			\draw (2.25,-0.75) arc (-180:0:0.5cm);
			\draw (3.25,2.25) -- (3.25,-0.75);
			\node at (4,0.75) {$\dots$};
			\draw[rounded corners=7mm] (0.75,1.5) -- (0.75,3) -- (4.75,3) -- (4.75,-1.5) -- (0.75,-1.5) -- (0.75,0);
			\draw[rounded corners=9mm] (0.25,1.5) -- (0.25,3.5) -- (5.25,3.5) -- (5.25,-2) -- (0.25,-2) -- (0.25,0);
		\end{tikzpicture}\ .
	\end{equation}

\subsection*{Exchange properties.}

Apart from fusing or splitting anyons it is also possible to \emph{braid} them. As we have seen above, the statistics of an anyon model is manifested in the evolution of the wave function of a system when two particles are exchanged\index{braiding}. In two spatial dimensions particles are allowed to exhibit any statistical behaviour as the restrictions that hold in three dimensions do not apply to two-dimensional systems.

When exchanging two particles $a$ and $b$ in the plane, there are two possible ways to do this which are topologically inequivalent:
	\begin{equation}
		\begin{tikzpicture}[scale=0.8]
			\node at (1,-0.25) {$a$};
			\node at (2,-0.25) {$b$};
			\node at (1,1.75) {$b$};
			\node at (2,1.75) {$a$};
			\begin{knot}[consider self intersections=true,ignore endpoint intersections=false]
			\strand (1,0)to[out=90,in=270](2,1.5);
			\strand (2,0)to[out=90,in=270](1,1.5);
			\end{knot}
		\end{tikzpicture}\hspace{50pt}
		\begin{tikzpicture}[scale=0.8]
			\node at (1,-0.25) {$a$};
			\node at (2,-0.25) {$b$};
			\node at (1,1.75) {$b$};
			\node at (2,1.75) {$a$};
			\begin{knot}[consider self intersections=true,ignore endpoint intersections=false]
			\strand (2,0)to[out=90,in=270](1,1.5);
			\strand (1,0)to[out=90,in=270](2,1.5);
			\end{knot}
		\end{tikzpicture}
	\end{equation}
We refer to the left diagram as right-handed braiding, denoted $R_{ab}$, and to the right diagram as left-handed braiding, denoted $R_{ab}^{-1}$. This is analogous to the braiding in the categorical sense as defined in \cref{def:braiding}.

To describe the braiding operator, consider the vector space $V_c^{ab}$ of a particle $c$ splitting into two particles $a$ and $b$. Suppose this vector space is one-dimensional and let $e_c^{ab}$ be its basis vector in terms of fusion trees. When the particles $a$ and $b$ are braided by $R_{ab}$, the vector $e_c^{ab}$ in $V_c^{ab}$ is transformed into a vector $R_{ab}e_c^{ab}$ in the vector space $V_c^{ba}$. The vector space $V_c^{ba}$ also has a corresponding basis vector $e_c^{ba}$. Since both state, $R_{ab}e_c^{ab}$ and $e_c^{ba}$, are non-zero vectors in the one-dimensional vector space $V_c^{ba}$, they are equal up to a phase. This phase is denoted $R_c^{ab}$, i.e., $R_{ab}e_c^{ab}=R_c^{ab}e_c^{ba}$:
	\begin{equation}
	\label{eq:braiding}
		R_{ab}
.
	\end{equation}	
From this relation, we can show that 
	\begin{equation}
		%
	\end{equation}
which tells us about the action of $R_{ab}^{-1}$:
	\begin{equation}
		R_{ab}^{-1}
		%
.
	\end{equation}
Hence, we have two ways of braiding two anyons around each other. Here, $R_c^{ab}$ is a phase factor because the corresponding vector spaces are one-dimensional in the multiplicity-free case. In general, the vector spaces can be higher dimensional, in which case it is a unitary matrix called the $R$-matrix\index{R-matrix@$R$-matrix}. Equation \cref{eq:braiding} then is of the form
	\begin{equation}
		R_{ab}
		%
.
	\end{equation}
Note that unitarity of the $R$-matrix implies that $N_{ab}^c=N_{ba}^c$. It furthermore implies $\left(R_c^{ab}\right)_{\nu\mu}^{-1}=\left(R_c^{ab}\right)_{\nu\mu}^{\dagger}=\overline{\left(R_c^{ba}\right)}_{\mu\nu}$.	

We have to ensure that the braiding operation is compatible with the fusion rules, i.e., $R_{c}^{ba}\neq 0$ if and only if $N_{ab}^c\neq 0$. Additionally, braiding has to be consistent with the $F$-moves. This yields the two so-called \emph{hexagon identities}\index{hexagon equation} that are depicted in \cref{fig:hex}:
	\begin{align}
	\label{eq:hex}
		R_p^{ca} \left(F_u^{acb}\right)_{pq} R_q^{cb}&=\sum_r \left(F_u^{cab}\right)_{rp} R_u^{cr}\left(F_u^{abc}\right)_{qr}\\
	\label{eq:hex2}
		\left(R_p^{ac}\right)^{-1} \left(F_u^{acb}\right)_{pq} \left(R_q^{bc}\right)^{-1}&=\sum_r \left(F_u^{cab}\right)_{rp} \left(R_u^{rc}\right)^{-1}\left(F_u^{abc}\right)_{qr}
	\end{align}

\begin{exmp}[Fibonacci anyons]
	\label{ex:FibAnyon}
	\index{Fibonacci anyons}
	We have already encountered the Fibonacci category $\mathbf{Fib}$ as a trivalent category in \cref{ex:TrivFib} and numerous times afterwards, for example in \cref{ex:AOFib} and \cref{ex:FibMod}. Here, we study how we can build an anyon model from this category. First, recall that the Fibonacci category has two simple objects, namely $\mathbf{1}$ and $\tau$ and that the fusion rules are given by
		\begin{align}
			\tau\otimes\mathbf{1}&=\mathbf{1}\otimes\tau=\tau\\
			\tau\otimes\tau&=\mathbf{1}+\tau.
		\end{align}
	We can immediately see that the fusion rules are multiplicity-free and, moreover, that the anyon model is non-abelian since $\sum_c N_{\tau\tau}^c=2$ (compare to \cref{eq:nonabelian}). From the fusion rules we can also see that $\tau$ is its own antiparticle (for $\mathbf{1}$ this is always the case). 
	
	\begin{figure}[H]
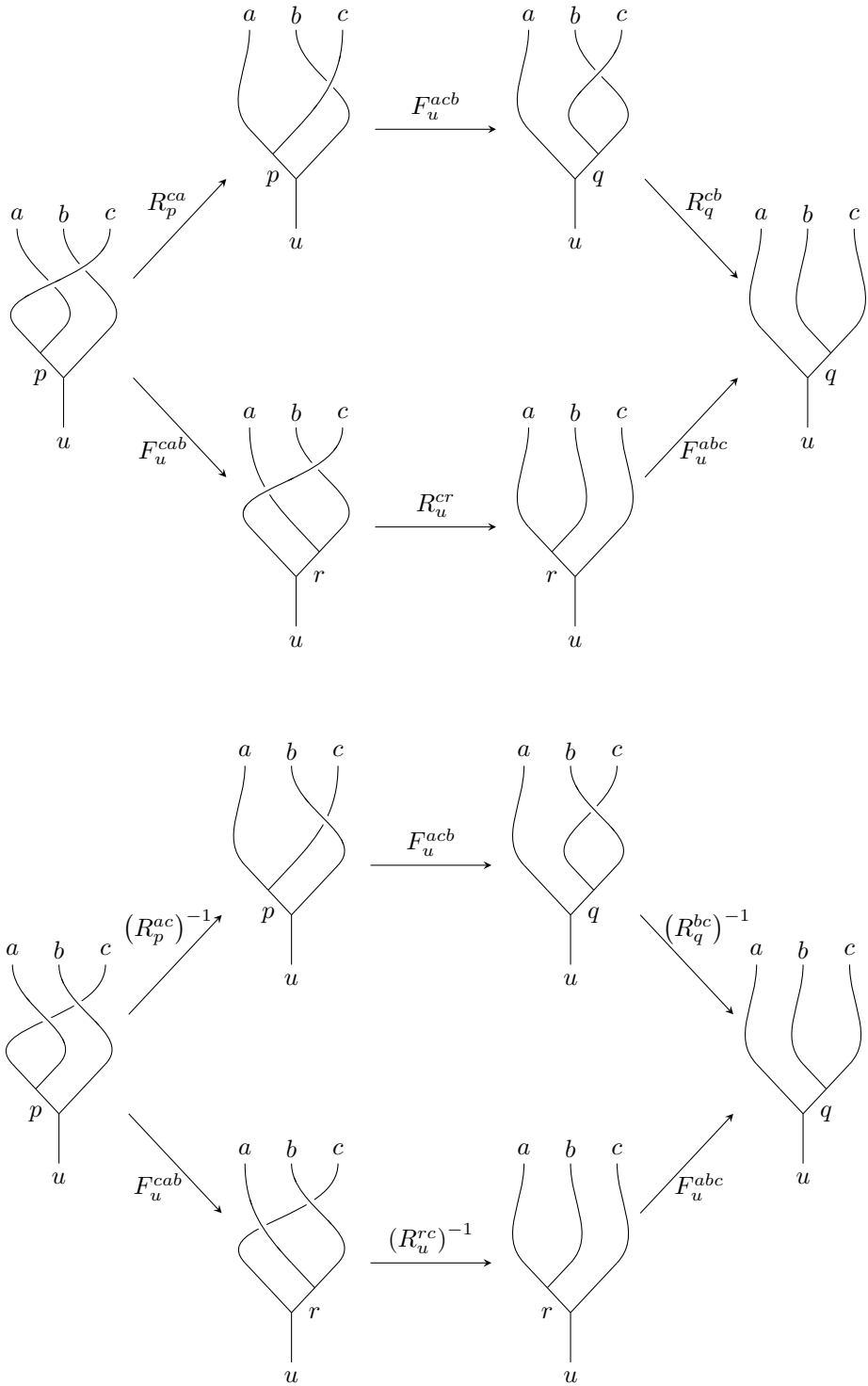

	\vspace{20pt}

	\vspace{20pt}
	\caption{\small \label{fig:hex}\textbf{Hexagon identities.} These diagrams have to commute in order to ensure that the braiding operation is consistent with fusion.}
\end{figure}
	
	\noindent	
	To obtain the $F$-symbols for the Fibonacci model, we have to solve the pentagon equation \cref{eq:pentagon}. In general, this is a difficult task that involves solving a possibly huge number of polynomial equations up to third order in multiple variables. However, we can use several simplifications: For instance, it is always possible to choose a gauging such that $F_u^{abc}=1$ whenever at least one of the labels $a,b,c$ is the vacuum label $\mathbf{1}$. Furthermore, in the Fibonacci example we can also assume that $F_\mathbf{1}^{abc}=1$, which can be explained with the properties of the trivalent form of the category (see \cref{ex:TrivFib}): The only non-trivial case here is when $a=b=c=\tau$. The corresponding tree diagram is equal to the trivalent vertex of the trivalent category with all legs rotated upwards. Since the trivalent vertex is rotationally invariant, the $F$-symbol $F_\mathbf{1}^{\tau\tau\tau}$ is trivial. Therefore, the only non-trivial $F$-matrix is the one with all $\tau$ labels, $F_\tau^{\tau\tau\tau}$. To obtain this matrix, we use a subset of the pentagon equations, namely those that arise when all outer labels in \cref{fig:PentagonId} are set to $\tau$, which is
		\begin{equation}
			\left(F_\tau^{\tau\tau r}\right)_{sp}\left(F_\tau^{p\tau\tau}\right)_{rq}=\sum_{t\in\{\mathbf{1},\tau\}}\left(F_s^{\tau\tau\tau}\right)_{rt}\left(F_\tau^{\tau t\tau}\right)_{sq}\left(F_q^{\tau\tau\tau}\right)_{tp}.
		\end{equation}
		
	This equation has to be fulfilled for all possible choices of labels, hence we can simply set the labels so that we get exactly those equations that help us determine $F_\tau^{\tau\tau\tau}$. A convenient choice is setting $r=q=\mathbf{1}$ and $s=p=\tau$. Note that in this case the sum on the right hand side has only one term: $t$ can only be set to $\tau$, since $t=\mathbf{1}$ would lead to $F$-symbols that correspond to zero-dimensional vector spaces. Therefore, the remaining equation is
		\begin{equation}
			\left(F_\tau^{\tau\tau\mathbf{1}}\right)_{\tau\tau}\left(F_\tau^{\tau\tau\tau}\right)_{\mathbf{1}\mathbf{1}}=\left(F_\tau^{\tau\tau\tau}\right)_{\mathbf{1}\tau}\left(F_\tau^{\tau\mathbf{1}\tau}\right)_{\tau\mathbf{1}}\left(F_\mathbf{1}^{\tau\tau\tau}\right)_{\tau\tau}.
		\end{equation}
	Using the fact that $F_u^{abc}=1$ whenever one of the labels $a,b,c,u$ is $\mathbf{1}$ further simplifies the equation:
		\begin{equation}
			\left(F_\tau^{\tau\tau\tau}\right)_{\mathbf{1}\mathbf{1}}=\left(F_\tau^{\tau\tau\tau}\right)_{\mathbf{1}\tau}\left(F_\tau^{\tau\tau\tau}\right)_{\tau\mathbf{1}}.
		\end{equation}
	Combining this equation with the condition that the matrix is unitary enables us to determine the matrix (up to arbitrary global phases) to be
		\begin{equation}
		\label{eq:Ftau}
			F_\tau^{\tau\tau\tau}=\begin{pmatrix}
				\phi^{-1} & \phi^{-\frac{1}{2}}\\
				\phi^{-\frac{1}{2}} & -\phi^{-1}
			\end{pmatrix},
		\end{equation}
	where $\phi=\frac{1+\sqrt{5}}{2}$ is the golden ratio.

	With the $F$-symbols at hand, we can directly determine the quantum dimensions according to \cref{eq:qdimF}. It is clear that $d_\mathbf{1}=1$. Since $\tau$ is its own antiparticle, the quantum dimension of $\tau$ is given by
		\begin{equation}
			d_\tau=\left|\left(F_\tau^{\tau\tau\tau}\right)_{\mathbf{1}\mathbf{1}}\right|^{-1}=\phi=\frac{1+\sqrt{5}}{2}.
		\end{equation}
	
	To complete the description of the anyon model we also need to know how the braiding of two particles works. To achieve this goal, we need to solve the hexagon identities \cref{eq:hex,eq:hex2}. Here, we pursue the same strategy as before: We first note that braiding a particle with the vacuum is trivial:
		\begin{equation}
			R_\tau^{\tau\mathbf{1}}=1,\hspace{20pt}R_\tau^{\mathbf{1}\tau}=1.
		\end{equation}
	Next, from the multitude of equations we choose those that help us to find a solution for the two non-trivial braidings $R_\mathbf{\mathbf{1}}^{\tau\tau}$ and $R_\tau^{\tau\tau}$. Similar to the pentagon equation, we use the equations where the outer labels in the upper diagram in \cref{fig:hex} are all set to $\tau$, which yields
		\begin{equation}
			R_p^{\tau\tau}\left(F_\tau^{\tau\tau\tau}\right)_{pq}R_q^{\tau\tau}=\sum_r\left(F_\tau^{\tau\tau\tau}\right)_{rp}R_{\tau}^{\tau r}\left(F_\tau^{\tau\tau\tau}\right)_{qr}.
		\end{equation}
	We are left with two labels that we can set, namely $p$ and $q$, which leads to four different equations. Inserting the values for $F_\tau^{\tau\tau\tau}$ that we have computed above \cref{eq:Ftau} yields the following set of equations:
		\begin{align}
			\phi^{-1} \left(R_\mathbf{1}^{\tau\tau}\right)^2&=\phi^{-2}+\phi^{-1}R_\tau^{\tau\tau},\\
			\phi^{-\frac{1}{2}}R_\mathbf{1}^{\tau\tau}R_\tau^{\tau\tau}&=\phi^{-\frac{3}{2}}+\phi^{-\frac{3}{2}}R_\tau^{\tau\tau},\\
			\phi^{-\frac{1}{2}}R_\mathbf{1}^{\tau\tau}R_\tau^{\tau\tau}&=\phi^{-\frac{3}{2}}+\phi^{-\frac{3}{2}}R_\tau^{\tau\tau},\\
			-\phi^{-1}\left(R_\tau^{\tau\tau}\right)^2&=\phi^{-1}+\phi^{-2}R_\tau^{\tau\tau}.
		\end{align}
	Solving these equations leads to a solution for the $R$-matrices:
		\begin{equation}
			R_\mathbf{1}^{\tau\tau}=e^{\frac{4\pi i}{5}},\hspace{20pt}R_\tau^{\tau\tau}=e^{-\frac{3\pi i}{5}}.
		\end{equation}
	Fibonacci anyons are an attractive candidate for topological quantum computation because any unitary transformation of their state space can be approximated arbitrarily accurately by braiding, which was shown in \cite{Bonesteel2005,Hormozi2007}. Therefore, they can be used to perform universal quantum computation. However, there are caveats: Firstly, approximating even simple gates is not straightforward. The $NOT$-gate, for instance, requires thousands of braidings in a specific order (see \cite{Bonesteel2005,Baraban2010}). Secondly, beside the simplicity of the theoretical description of the anyon model, the microscopic systems that support Fibonacci anyons are hard to access. The most promising candidate here is the so-called Read-Rezayi state, which has been proposed to describe the fractional quantum Hall state at a filling fraction of $12/5$ \cite{Read1999}. However, this state is very fragile, so it is questionable whether it can be realised in a laboratory.
\end{exmp}

\begin{exmp}[Ising anyons]
	\label{ex:Ising}
	Since Fibonacci anyons are difficult to realise experimentally, much research has focused on the search for a simpler anyon model, possibly at the cost of losing universality. Here, Ising anyons\index{Ising anyons} are of particular interest. They can only implement logical phase and $NOT$-gates on single qubits, hence they can only implement the \emph{Clifford group} \cite{Ahlbrecht2009}, which can be efficiently simulated with a classical computer. Despite the fact that Ising anyons are not universal, they can still be useful for testing and developing topological quantum technologies since they are easier to realise in a laboratory.
	
	The Ising model consists of three simple objects: The vacuum $\mathbf{1}$ and two non-trivial particles $\psi$ (fermion) and $\sigma$ (anyon) with dimensions
		\begin{equation}
			d_\mathbf{1}=1,\hspace{20pt}d_\psi=1,\hspace{20pt}d_\sigma=\sqrt{2}.
		\end{equation}
	The fusion rules are the following:
		\begin{align}
			\psi\otimes\psi&=\mathbf{1}\\
			\psi\otimes\sigma&=\sigma\\
			\sigma\otimes\sigma&=\mathbf{1}+\psi.
		\end{align}
	The last fusion rule implies that the model is non-abelian, since fusing two $\sigma$ anyons results in either the vacuum or a type-$\psi$ particle. Similar to Fibonacci anyons, solving the Pentagon equation yields that there is one non-trivial $F$-matrix:
		\begin{equation}
			F_\sigma^{\sigma\sigma\sigma}=\frac{1}{\sqrt{2}}\begin{pmatrix}
				1 & 1 \\
				1 & -1
			\end{pmatrix}.
		\end{equation}
	The braiding operator that describes the braiding of two $\sigma$ anyons is given by
		\begin{equation}
			R_\mathbf{1}^{\sigma\sigma}=e^{-\frac{\pi i}{8}},\hspace{20pt}R_\psi^{\sigma\sigma}=e^\frac{3\pi i}{8},
		\end{equation}
	and braiding the fermion $\psi$ with the anyon $\sigma$ is described by 
		\begin{equation}
			R_\sigma^{\psi\sigma}=R_\sigma^{\sigma\psi}=e^{-\frac{\pi i}{2}},
		\end{equation}
	while braiding the fermion $\psi$ with itself shows the expected behaviour: $R_\mathbf{1}^{\psi\psi}=-1$.
\end{exmp}

\section{Building chains from anyon models}

With the mathematical description of anyons and the corresponding graphical notation at hand, we can begin to talk about physical systems of these particles. We concentrate here on the simplest model that one can build, namely a one-dimensional chain of anyons. First, we describe the Hilbert space of the system before we discuss dynamics.

\subsection*{The Hilbert space}

In quantum mechanics, to study a physical system we first need to define its Hilbert space in order to fix the possible states of the system. Here, we make use of the concepts we have introduced in the previous section, more precisely, fusion trees and the standard basis. Note that from now on, we exclusively talk about multiplicity-free anyon models and do not comment on the general case with multiplicities, although all the constructions can be generalised in a straightforward way.

The vector space that describes a chain of $N+1$ particles\index{anyon chain} is the splitting space $V^{a_1a_2\dots a_N}_{a_{N+1}}$:
	\begin{equation}
		\begin{tikzpicture}[scale=1.1]
			\draw (0,0) -- (-2.5,2.5);
			\draw (-0.5,0.5) -- (1.5,2.5);
			\draw (-1.5,1.5) -- (-0.5,2.5);
			\draw (-2,2) -- (-1.5,2.5);
			\node at (0.25,-0.25) {$a_{N+1}$};
			\node at (-2.5,2.75) {$a_1$};
			\node at (-1.5,2.75) {$a_2$};
			\node at (-0.5,2.75) {$a_3$};
			\node at (1.5,2.75) {$a_N$};
			\node at (0.5,2.75) {$\dots$};
			\node at (-1.9,1.6) {$x_1$};
			\node at (-1.4,1.1) {$x_2$};
			\node at (-1,0.6) {\rotatebox{-45}{$\dots$}};
		\end{tikzpicture}
	\end{equation}
This means that the particle $a_{N+1}$ first splits into particles $x_{N-1}$ and $a_N$, $x_{N-1}$ further splits into $x_{N-2}$ and $a_{N-1}$ and so on. When the outer particle types $a_1,\dots,a_{N+1}$, i.e., the labels of the vector space, are fixed, then an element of this space is defined by a choice of the inner labels $x_1,\dots,x_{N-1}$, where the set of possible choices is determined by the fusion rules. In other words, a valid choice of labels, which we also call a \emph{fusion path}\index{fusion path}, fulfils the constraint that
	\begin{equation}
		N_{x_i,a_{i+2}}^{x_{i+1}}=1
	\end{equation}
for all $i\in\{1,\dots,N-1\}$. The basis of the vector space $V^{a_1a_2\dots a_N}_{a_{N+1}}$ is therefore given by all vectors that represent valid choices of inner labels. We call this basis the \emph{anyon chain basis} or \emph{fusion basis}\index{fusion basis} and denote these vectors $|x_1,x_2,\dots x_{N-1}\rangle$\footnote{Note that it is a convention to fix the leftmost label and the rightmost label to $a_1$ and $a_{N+1}$, respectively. We could instead include them in the basis vector as additional labels $x_0$ and $x_N$, hence the basis vector would then be of the form $|x_0,x_1,\dots,x_{N-1},x_N\rangle$.}. 

Recall the spin chain we have introduced in the beginning of this chapter: The fixed anyons in an anyon chain (i.e., the particles $a_1,\dots,a_{N+1}$) correspond to the spin-$\frac{1}{2}$ lattice sites of the spin chain. The labels $x_1,\dotsm,x_{N-1}$, i.e., the degrees of freedom in the anyon chain correspond roughly to the singlet and triplet states that live on the bonds of the spin chain.

We now make some simplifications to this general case. First, we assume that all outer labels are chosen to be \emph{the same} particle type $a$. Furthermore, we draw the chain in a slightly different way as depicted in \cref{fig:anyonchain}. Here, the rightmost label $a$ corresponds to the subscript of the vector space, i.e., we have rotated the chain around $45\degres$. Although this has the disadvantage that one cannot see which are the incoming and which are the outgoing particles, it coincides with the usual depiction of spin chains like the Heisenberg chain. Together with the convention that a chain depicted in this way always corresponds to a vector space of the form $V^{aa\dots a}_a$ (with $N$ $a$'s as superscript), this removes all ambiguity about the order of fusion.

	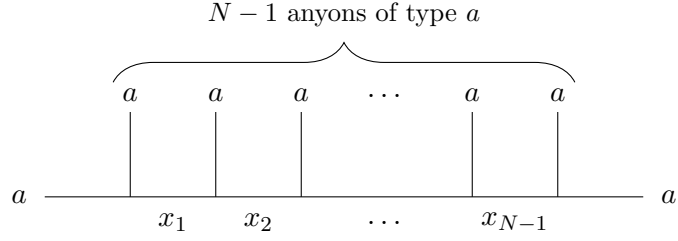
\begin{figure}[t]
		\centering
		\begin{tikzpicture}[scale=2.25]
			\draw (0,0) -- (3.5,0);
			\draw (0.5,0) -- (0.5,0.5);
			\draw (1,0) -- (1,0.5);
			\draw (1.5,0) -- (1.5,0.5);
			\draw (2.5,0) -- (2.5,0.5);
			\draw (3,0) -- (3,0.5);
			\node at (-0.15,0) {$a$};
			\node at (0.5,0.6) {$a$};
			\node at (1,0.6) {$a$};
			\node at (1.5,0.6) {$a$};
			\node at (2,0.6) {$\dots$};
			\node at (2.5,0.6) {$a$};
			\node at (3,0.6) {$a$};
			\node at (3.65,0) {$a$};
			\node at (0.75,-0.15) {$x_1$};
			\node at (1.25,-0.15) {$x_2$};
			\node at (2,-0.15) {$\dots$};
			\node at (2.75,-0.15) {$x_{N-1}$};
			\draw[decorate,decoration={brace,amplitude=15pt}] (0.4,0.675) -- (3.1,0.675);
			\node at (1.75,1.075) {\small $N-1$ anyons of type $a$};
		\end{tikzpicture}
		\caption{\small \label{fig:anyonchain} \textbf{Anyon chain}. A chain corresponding to the vector space $V_a^{aa\dots a}$ with basis vectors $|x_1,x_2,\dots,x_{N-1}\rangle$.}
	\end{figure}

\subsection*{Interactions}

Without any interactions, the state of the anyon chain is simply described by the Hilbert space introduced above. Imposing interactions between the particles splits the degeneracy and we observe a non-trivial collective ground state. The kind of interactions that we impose on the anyon chain are motivated by the Heisenberg interaction for spin-$\frac{1}{2}$ particles (see \cite{Trebst2008}). Recall that two $SU(2)$ spin-$\frac{1}{2}$ particles can either combine to a singlet state $|\mathbf{0}\rangle$ or a triplet state $|\mathbf{1}\rangle$. This can be written, in analogy to the way we write anyonic fusion rules, as
	\begin{equation}
		\frac{1}{2}\otimes\frac{1}{2}=|\mathbf{0}\rangle+|\mathbf{1}\rangle.
	\end{equation}
The usual Heisenberg Hamiltonian\index{Heisenberg chain} is defined as
	\begin{equation}
		H_\mathrm{Heisenberg}^{SU(2)}=J\sum_{(i,j)}\vec{S}_i\cdot\vec{S}_j,
	\end{equation}
where the sum runs over all pairs of spins $i,j$ (which might be restricted to nearest neighbour interactions on a given lattice, depending on the model) and $J$ is a coupling constant. This Hamiltonian can be rewritten using the total spin $\vec{T}_{ij}=\vec{S}_i+\vec{S}_j$ formed by the two spins $\vec{S}_i$ and $\vec{S}_j$:
	\begin{equation}
		H_\mathrm{Heisenberg}^{SU(2)}=\frac{J}{2}\sum_{(i,j)}\left(\vec{T}_{ij}^2-\vec{S}_i^2-\vec{S}_j^2\right).
	\end{equation}
Using the fact that $\vec{T}_{ij}^2|\mathbf{0}\rangle=0$ and $\vec{T}_{ij}^2|\mathbf{1}\rangle=1(1+1)|\mathbf{1}\rangle$ allows us to express the above Hamiltonian in terms of projections onto the pairwise singlet state $\Pi_{ij}^\mathbf{0}$:
	\begin{equation}
		H_\mathrm{Heisenberg}^{SU(2)}=\frac{J}{2}\sum_{(i,j)}\left( \Pi_{ij}^{|\mathbf{0}\rangle}-\frac{3}{2}\right).
	\end{equation}
For $J>0$ (antiferromagnetic coupling), the systems favours an overall singlet state, and for $J<0$ (ferromagnetic coupling) the overall triplet state is favoured.

In a similar fashion, we model local interactions between anyons in an anyon chain by projections onto specific fusion channels. A projection of two neighbouring type-$a$ particles onto a particle of type $e$ at site $i$ of the anyon chain is of the form
	\begin{equation}
		p_i^{(e)}\ \Bigg|
\Bigg\rangle		
	\end{equation}
In this picture, the projector acts on two neighbouring anyons by projecting onto the fusion channel $e$. However, in the standard basis representation of the anyon chain (as depicted in \cref{fig:anyonchain}) this operator locally acts on states $|x_{i-1},x_i,x_{i+1}\rangle$. To be able to project onto a certain fusion channel $e$ we need to rewrite the definition of the projector using $F$-moves:
	\begin{align}
		p_i^{(e)}\ \Bigg|
		%
\Bigg\rangle,
	\end{align}
where $\left(F_{x_{i+1}}^{x_{i-1}aa}\right)^\dagger_{x_i'e}=\overline{\left(F_{x_{i+1}}^{x_{i-1}aa}\right)_{ex_i'}}$ because of unitarity of the $F$-matrices.

The Hamiltonian of the anyon chain system can then be built from these projections. This can be best understood by studying a concrete example.

\begin{exmp}[The golden chain]
	\label{ex:Goldenchain}
	
\index{golden chain} One of the simplest examples of an anyon chain is one built from Fibonacci anyons, which were described in \cref{ex:FibAnyon}. This chain is called the \emph{golden chain} because of the connection between Fibonacci anyons and the golden ratio. It was introduced in \cite{feiguin_interacting_2007} and a more detailed description can be found in \cite{Trebst2008}.

First, we to describe the Hilbert space of the golden chain. Since there is only one non-trivial anyon type in the Fibonacci anyon model, namely $\tau$, the choice of the object $a$ for the anyon chain is straightforward. Hence, the golden chain is of the form
	\begin{figure}[H]
		\centering
		\begin{tikzpicture}[scale=2.25]
			\draw (0,0) -- (3.5,0);
			\draw (0.5,0) -- (0.5,0.5);
			\draw (1,0) -- (1,0.5);
			\draw (1.5,0) -- (1.5,0.5);
			\draw (2.5,0) -- (2.5,0.5);
			\draw (3,0) -- (3,0.5);
			\node at (-0.15,0) {$\tau$};
			\node at (0.5,0.6) {$\tau$};
			\node at (1,0.6) {$\tau$};
			\node at (1.5,0.6) {$\tau$};
			\node at (2,0.6) {$\dots$};
			\node at (2.5,0.6) {$\tau$};
			\node at (3,0.6) {$\tau$};
			\node at (3.65,0) {$\tau$};
			\node at (0.75,-0.15) {$x_1$};
			\node at (1.25,-0.15) {$x_2$};
			\node at (2,-0.15) {$\dots$};
			\node at (2.75,-0.15) {$x_{N-1}$};
		\end{tikzpicture}
	\end{figure}
\noindent
and we need to determine which basis states $|x_1,x_2,\dots, x_{N-1}\rangle$ are allowed by the fusion rules. Remember that there is one non-trivial fusion rule, namely 
	\begin{equation}
		\tau\otimes\tau=\mathbf{1}+\tau.
	\end{equation}
Therefore, successive fusion of two $\tau$-anyons produces either a $\mathbf{1}$ or a $\tau$, so $x_i\in\{\mathbf{1},\tau\}$. Additionally, the constraint that $\mathbf{1}\otimes\tau=\tau$ has to be fulfilled, hence there can never be two $\mathbf{1}$s as two neighbouring intermediate charges. These constraints can also be expressed in terms of a graph\index{fusion graph} whose vertices represent the anyon types of the model, i.e., $\mathbf{1}$ and $\tau$ for the Fibonacci model, and we draw an edge between two labels $a$ and $b$ whenever $N_{a\tau}^b=1$. If $N_{a\tau}^b=N_{b\tau}^a=1$, the edge is undirected, but if one of these multiplicities is zero the edge has a direction. For example, consider a model with particle types $\mathbf{1}$, $a$ and $b$, where fusion with $a$ is given by 
	\begin{align}
		a\otimes a &= \mathbf{1}+b\\
		b\otimes a &= a.
	\end{align}
Hence, we have that $N_{ba}^a=1$, but $N_{aa}^a=0$ and therefore the fusion graph for fusion with $a$ has a directed edge:
	\begin{equation}
		\begin{tikzpicture}
			\draw[fill=LinkColor4!60!white] (-2,0) circle (0.325cm);
			\draw[fill=LinkColor4!60!white] (0,0) circle (0.325cm);
			\draw[fill=LinkColor4!60!white] (2,0) circle (0.325cm);
			\node at (-2,0) {$\mathbf{1}$};
			\node at (0,0) {$a$};
			\node at (2,0) {$b$};
			\draw (-1.65,0) -- (-0.35,0);
			\draw[->,>=stealth] (0.35,0) -- (1.65,0);
			\begin{knot}[consider self intersections=true,ignore endpoint intersections=false]
			\strand[->,>=stealth] (1.9,0.33)to[out=150,in=180](2,1)to[out=0,in=55](2.2,0.4);
			\end{knot}
		\end{tikzpicture}\ .
	\end{equation}

The corresponding graph that describes fusion with $\tau$ in the Fibonacci model is depicted in \cref{fig:fusiongraph}. This graph provides a simple method to find all the basis states of the Hilbert space of a chain of length $N+1$: They are given by the set of all possible paths of length $N-1$ on the graph in \cref{fig:fusiongraph} that start and end at the vertex $\tau$.

	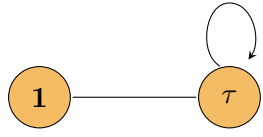
\begin{figure}[t]
		\centering
		\begin{tikzpicture}[scale=1.25]
		\draw[fill=LinkColor4!60!white] (0,0) circle (0.325cm);
		\draw[fill=LinkColor4!60!white] (2,0) circle (0.325cm);
		\node at (0,0) {$\mathbf{1}$};
		\node at (2,0) {$\tau$};
		\draw (0.35,0) -- (1.65,0);
		\begin{knot}[consider self intersections=true,ignore endpoint intersections=false]
		\strand[->,>=stealth] (1.9,0.33)to[out=150,in=180](2,1)to[out=0,in=55](2.2,0.4);
		\end{knot}
		\end{tikzpicture}
		\caption{\small \label{fig:fusiongraph}\textbf{Fusion graph for the golden chain.} A path of length $N-1$ on this graph corresponds to an allowed labelling for a basis state $|x_1,\dots,x_{N-1}\rangle$ of the anyon chain.}
	\end{figure}

The interaction for the golden chain is constructed from projecting onto the possible fusion channels. We use the projectors $p_i^{(e)}$ defined above to construct a local Hamiltonian of the form
	\begin{equation}
		H=\sum_{i=1}^{N-1} H_i
	\end{equation}
in the following way: We assign an energy of $E_\mathbf{1}=-1$ to fusing to the vacuum, and an energy of $E_\tau=0$ to fusing to the $\tau$-anyon, hence fusing to the vacuum is energetically favoured. The local Hamiltonian $H_i$ is then given by
	\begin{align}
	\label{eq:localham}
		H_i=\sum_{e\in\{\mathbf{1},\tau\}} E_e\,p_i^{(e)}=-p_i^{(\mathbf{1})},
	\end{align}
where the projection on the vacuum for two neighbouring $\tau$-anyons is  given by
	\begin{equation}
		p_i^{(\mathbf{1})}|x_{i-1},x_i,x_{i+1}\rangle=\sum_{x_i'}\left(F_{x_{i+1}}^{x_{i-1}\tau\tau}\right)_{\mathbf{1}x_i}\left(F_{x_{i+1}}^{x_{i-1}\tau\tau}\right)^\dagger_{x_i'\mathbf{1}}\,|x_{i-1},x_i',x_{i+1}\rangle.
	\end{equation}
Note that due to the construction of the boundaries of the chain, $x_0=x_N=\tau$.

To obtain a matrix representation of the Hamiltonian, note that we write the action of the local Hamiltonian as
	\begin{equation}
		H_i|x_{i-1},x_i,x_{i+1}\rangle=\sum_{x_i'\in\{\mathbf{1},\tau\}}(H_i)_{x_i',x_i}|x_{i-1},x_i',x_{i+1}\rangle
	\end{equation}
with 
	\begin{equation}
		(H_i)_{x_i',x_i}=-\left(F_{x_{i+1}}^{x_{i-1}\tau\tau}\right)_{\mathbf{1}x_i}\left(F_{x_{i+1}}^{x_{i-1}\tau\tau}\right)^\dagger_{x_i'\mathbf{1}}.
	\end{equation}
The basis for the local states $|x_{i-1},x_i,x_{i+1}\rangle$ the Hamiltonian acts on is given by the allowed labellings according to the fusion rules:
	\begin{equation}
		\{|\mathbf{1}\tau\mathbf{1}\rangle,|\mathbf{1}\tau\tau\rangle,|\tau\tau\mathbf{1}\rangle,|\tau\mathbf{1}\tau\rangle,|\tau\tau\tau\rangle\}.
	\end{equation}
Therefore, the local Hamiltonian $H_i$ is given by the following $5\times 5$ matrix:
	\begin{equation}
		H_i=-\begin{pmatrix}
			1 & 0 & 0 & 0 & 0 \\
			0 & 0 & 0 & 0 & 0 \\
			0 & 0 & 0 & 0 & 0 \\
			0 & 0 & 0 & \varphi^{-2} & \varphi^{-\frac{3}{2}}\\
			0 & 0 & 0 & \varphi^{-\frac{3}{2}} & \varphi^{-1} 
		\end{pmatrix}.
	\end{equation}

Studying the ground space of the chain Hamiltonian $H=\sum_i H_i$ yields the central charge of the corresponding conformal field theory, which is $c=\frac{7}{10}$, see \cite{feiguin_interacting_2007} and \cref{ex:Goldenchainnumerics}. The methods that can be used to study whether an anyon chain gives rise to a conformal field theory are explained in \cref{sec:numerics}.
\end{exmp}

\section{Numerical methods for anyon chains}
\label{sec:numerics}

How can we learn about the connection between fusion categories and conformal field theories by studying anyon chains? The key word here is \emph{criticality}. Systems that undergo a phase transition can exhibit a critical point\index{critical point} if the phase transition is continuous, and the theory that describes the system at this critical point possibly is a Conformal Field Theory (CFT).
In this section, we briefly explain the most important concepts of continuous phase transitions and criticality and refer to \cite{Sachdev2017} for a more detailed treatment of the topic. We concentrate here on those quantities that can be easily investigated numerically, and we also discuss which numerical methods are suitable for this purpose.

We are interested in zero-temperature phase transitions, which, unlike in classical systems, have a rich structure in quantum mechanics because of possible ground state entanglement. Here, we define a phase transitions as a point at which the ground state energy density of the system is a non-analytic function of some control parameter of the Hamiltonian such as the magnetic field strength. Phase transitions are often divided into two classes, namely thermal (which occur at certain temperatures, for example, the transition from solid to liquid) and continuous phase transitions. The latter kind of phase transitions is the one we focus on here.

Continuous phase transitions\index{continuous phase transitions} are characterized by scale-invariant fluctuations. This means that at any length scale significantly larger than the lattice spacing of the system, correlation functions are invariant under spatial scaling transformations
	\begin{equation}
		x\to cx
	\end{equation}
for some scaling factor $c$. In the non-scale-invariant case, correlation functions $C$ typically decay exponentially
	\begin{equation}
		C(x)\approx e^{-\frac{x}{\xi}},
	\end{equation}
where $\xi$ is the \emph{correlation length}\index{correlation length} of the system, which imposes a natural length scale. Such a scale does not exist in the scale-invariant case, hence the correlation length diverges
	\begin{equation}
		\xi\to\infty.
	\end{equation}
This is called \emph{criticality} and points at which this phenomenon appears are called \emph{critical points}. 

Systems at the critical point are often not only scale-invariant but also invariant under \emph{conformal transformations} (which are those transformations that preserve angles locally), hence they have a description as a conformal field theory. A large number of $2D$ CFTs\index{conformal field theory} can be classified by a single number known as the \emph{central charge}, which can be numerically determined by studying properties of the ground state of the system at the critical point.	

There are several indicators for the criticality of a system that can be numerically investigated. First, scale-invariance implies that the gap between the ground state and the first excited state of the system (the \emph{spectral gap}) vanishes in the limit of infinite system size. This is because a non-zero gap imposes a natural energy scale, which would break the scale-invariance. In critical systems the energy spectrum is continuous, hence it is possible to observe excitations with arbitrarily small energies.

\begin{figure}[t]
	\centering
	\begin{tikzpicture}[scale=2.25]
		\draw (0,0) -- (3.5,0);
		\draw (0.5,0) -- (0.5,0.5);
		\draw (1,0) -- (1,0.5);
		\draw (1.5,0) -- (1.5,0.5);
		\draw (2,0) -- (2,0.5);
		\draw (2.5,0) -- (2.5,0.5);
		\draw (3,0) -- (3,0.5);
		\draw[dashed] (1.75,0.6) -- (1.75,-0.3);
		\node at (-0.25,0) {$\dots$};
		\node at (3.75,0) {$\dots$};
		\node at (0.875,-0.25) {$\rho_A$};
		\node at (2.625,-0.25) {$\rho_B$};
	\end{tikzpicture}
	\caption{\small \label{fig:ente}\textbf{Dividing the chain into two subsystems.} The scaling of the entanglement entropy of one of the subsystems reflects the critical behaviour of the system.}
\end{figure}
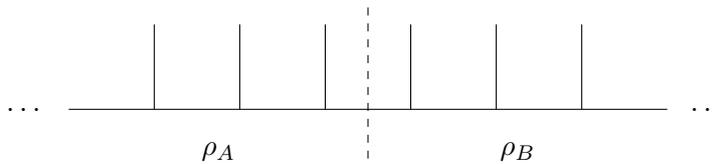

The second indicator of criticality is the behaviour of the \emph{entanglement entropy}\index{entanglement entropy} of the ground state with growing system size, which can also be exploited to calculate an estimate for the central charge of the model. In general, the entanglement entropy describes the entanglement of a subsystem $A$ and is defined by the von Neumann entropy of the state of the subsystem:
	\begin{equation}
		S(\rho_A)=-\Tr\left(\rho_A\log(\rho_A)\right).
	\end{equation}
In the concrete case of an anyon chain, we cut the chain in half (see \cref{fig:ente}) and calculate the entropy of one of the subsystems. Note that in case the overall state of the chain is pure, it does not matter which subsystem we consider since
	\begin{equation}
		S\left(\rho_A\right)=S\left(\rho_B\right)\footnote{This is a well-known result in quantum information theory, hence we do not give any more details here. It can, for example, be found in \cite{Wilde2016}.}.
	\end{equation}
The scaling behaviour of the entanglement entropy when the system grows in size reflects the scaling behaviour of the system in the following way: In general, at the critical point the entanglement entropy diverges, which is the first indicator for a critical system. Furthermore, the relation between the correlation length $\xi$ and the entropy $S$ (see \cite{Holzhey1994} and \cite{Calabrese2009}), given by
	\begin{equation}
	\label{eq:xiS}
		S(\xi)\propto\frac{c}{6}\log\left(\xi\right),
	\end{equation}
allows us to calculate an estimate for the central charge $c$\index{central charge}.

\subsection*{Tensor network methods}

Studying quantum many-body systems is, in general, a difficult and computationally challenging task due to the high-dimensional Hilbert spaces that arise with a growing number of particles. The study of anyon chains is no exception. To circumvent this problem, we use tensor network methods, or, more precisely, matrix product states, to approximately simulate the evolution of the anyon chain of interest. Since introducing the theory of tensor networks and matrix product states in detail goes beyond the scope of this thesis and, furthermore, there is already some excellent introductory literature, we refer to \cite{Bridgeman2017a} for a more in-depth introduction to this concept. Here, we only give the most important definitions that are necessary to understand the numerical investigations done in this thesis.

The theory of tensor networks\index{tensor network} generally uses a convenient graphical language. A general rank-$r$ tensor $T$ of dimensions $d_1\times d_2\times\dots\times d_r$ is an element of $\mathbb{C}^{d_1\times d_2\times\dots\times d_r}$. A single tensor is simply represented by a geometric shape (a circle in this example) with legs sticking out of it, where each leg corresponds to an index. For instance, a rank-$4$ tensor $T$ is depicted
\begin{equation}
T_i^{jkl}=
.
\end{equation}
In this graphical representation scalars, vectors and matrices are simply special cases of a general rank-$r$ tensor:
\begin{enumerate}
	\item A scalar $c\in\mathbb{C}$ is a rank-$0$ tensor and hence has no indices, which means that it has no legs:
	\begin{equation}
	%
	\end{equation}
	\item A matrix $A$ (a rank-$2$ tensor) has two legs:
	\begin{equation}
	%
	\end{equation}
\end{enumerate}

We can build larger networks by stacking several of these objects together, which is done by contracting legs that correspond to shared indices. With this operation we can represent important operations such as matrix-vector multiplication and matrix multiplication:
\begin{enumerate}
	\item Matrix-vector multiplication: $A\vec{v}=\sum_j A_{ij} v_j=\vec{w}$
	\begin{equation}
	%
	\end{equation}
\end{enumerate}
The trace of a matrix is built by simply contracting the two legs:
\begin{equation}
\Tr(A)=\sum_i A_{ii}=
%
\ .
\end{equation} 

A very important concept in tensor network theory is the Singular Value Decomposition (SVD)\index{singular value decomposition}. For a rank-$2$ tensor $T$ it is given by
\begin{equation}
T_{ab}=\sum_{i,j} U_{ai} S_{ij} \bar{V}_{bj},
\end{equation}
where $U$ and $V$ are unitary matrices and $S$ is positive semi-definite and diagonal. $S$ has the singular values of $T$ on its diagonal, typically arranged in non-increasing order. Graphically, this is depicted
\begin{equation}
%
\end{equation}
The singular value decomposition is the key to approximating high-dimensional states. By truncating the singular value matrix $S$ after the highest $k$ values the matrix $T$ can be approximated by
\begin{equation}
T^{(k)}=US^{(k)}V^\dagger.
\end{equation}

After having established the notation and the most important concepts, we want to use this to represent quantum many-body systems. Consider a general state $|\psi\rangle$ of $N$ qudits (i.e., $d$-dimensional quantum systems):
\begin{equation}
\label{eq:qstate}
|\psi\rangle=\sum_{i_1,i_2,\dots,i_N=0}^{d-1} C_{i_i,i_2,\dots,i_N}|i_1\rangle|i_2\rangle\dots |i_N\rangle.
\end{equation}
This could, for example, be the state of an anyon chain of length $N+2$. It is completely determined by the rank-$N$ tensor $C$. Note that it consists of $d^N$ numbers and therefore grows exponentially with the system size $N$, which makes it a computationally inefficient description of the quantum state. We can use the tensor network notation introduced above to replace this big tensor $C$ with a network of tensors with smaller rank. The final tensor network representation then typically depends on a polynomial number of parameters, making it computationally more efficient. 

The tensor network representation of $|\psi\rangle$ is derived as follows: By splitting the first index in \cref{eq:qstate} out from the rest and performing a singular value decomposition we get the Schmidt decomposition
\begin{equation}
|\psi\rangle=\sum_i\lambda_i |L_i\rangle\otimes |R_i\rangle,
\end{equation}
where the $\lambda_i$ are the Schmidt coefficients and $\{|L_i\rangle\}$ and $\{|R_i\rangle\}$ are orthonormal sets of vectors. We can represent it graphically (here for $N=4$) as
\begin{equation}
\label{eq:cut1}
\begin{tikzpicture}[scale=0.9, baseline=(current bounding box.center)]
\draw[fill=LinkColor!50!white] (0,0) rectangle (4,0.75);
\node at (2,0.375) {\small $\psi$};
\draw (0.5,0) -- (0.5,-0.5);
\draw (1.5,0) -- (1.5,-0.5);
\draw (2.5,0) -- (2.5,-0.5);
\draw (3.5,0) -- (3.5,-0.5);
\draw[dashed] (1,1) -- (1,-0.75);
\end{tikzpicture}=
\begin{tikzpicture}[scale=0.9, baseline=(current bounding box.center)]
\draw (0.5,0.375) -- (3.5,0.375);
\draw[fill=LinkColor2!50!white] (0.25,0) rectangle (1,0.75);
\draw[fill=LinkColor4!50!white] (1.75,0.375) circle (0.375cm);
\draw[fill=LinkColor2!50!white] (2.5,0) rectangle (5.5,0.75);
\node at (0.625,0.375) {\small $L$};
\node at (1.75,0.375) {\small $\lambda$};
\node at (4,0.375) {\small $R$};
\draw (0.625,0) -- (0.625,-0.5);
\draw (4,0) -- (4,-0.5);
\draw (5,0) -- (5,-0.5);
\draw (3,0) -- (3,-0.5);
\end{tikzpicture}
\end{equation}
where $\lambda$ is a diagonal matrix with the Schmidt coefficients on the diagonal. The Schmidt coefficients correspond to the singular values in the decomposition \cref{eq:cut1}, hence they capture the entanglement structure along this cut. We can now successively introduce additional cuts to $R$ and perform the singular value decomposition on each of the cuts, ending up with a representation of the state $|\psi\rangle$ in terms of local tensors $M$ and diagonal matrices of singular values $\lambda$ that quantify the entanglement across the respective cut:
\begin{equation}
.
\end{equation}
We can further contract the tensors $\lambda^{(i)}$ that represent the singular values into the local tensors\footnote{It is actually important which tensors the singular values are contracted into. This relates to gauge fixing the matrix product state and is explained in detail in \cite{Bridgeman2017a}.} $M^{(i)}$ to arrive at the form
\begin{equation}
\label{eq:MPS}
|\psi\rangle=
%
.
\end{equation}
The state depicted in \cref{eq:MPS} is a Matrix Product State\index{matrix product state} (MPS). For reasons of convenience, it is often modified to
\begin{equation}
|\psi\rangle=
%
\end{equation}
which corresponds to taking the trace of the tensors $A^{(i)}$:
	\begin{equation}
		|\psi\rangle=\sum_{i_1,i_2,\dots, i_N}\Tr\left(A_{i_1}^{(1)}A_{i_2}^{(2)}\dots A_{i_N}^{(N)}\right) |i_1 i_2\dots i_n\rangle,
	\end{equation}
or, in the translational invariant case,
	\begin{equation}
		\label{eq:TIMPS}
		|\psi\rangle=\sum_{i_1,i_2,\dots, i_N}\Tr\big(A_{i_1}A_{i_2}\dots A_{i_N}\big) |i_1 i_2\dots i_n\rangle.
	\end{equation}

So far, we have not reduced the computational complexity of the representation since the above form is both general and exact. However, the matrix product form of the state makes it easy to find good approximations to the exact state. Consider the translation invariant case, where the matrices $A_i$ all have the same rank $r$. For some value of $r$, say $r^*$, the form in \cref{eq:TIMPS} is the exact representation of $|\psi\rangle$. However, if we truncate the matrices to a rank $D\le r^*$, we only include the largest $D$ Schmidt coefficients in the description of the state. We are left with an approximate form of $|\psi\rangle$ which is much easier to handle computationally because of the decreased number of parameters. We call the rank $D$ the \emph{bond dimension}\index{bond dimension} of the matrix product state.

We can use the tensor network representation of the quantum state to numerically simulate the behaviour of the quantum system we are interested in. Matrix product states provide an efficient description of one-dimensional gapped systems, since in these cases the entanglement entropy of a large enough subsystem will saturate. At the critical point, however, the entanglement entropy diverges, as we have seen above, hence the MPS cannot fully capture the behaviour of the system in the limit of infinite system size for finite bond dimensions. However, it turns out that \emph{how} the MPS approximation truncates the correlations of the system at the critical point allows us to draw conclusions about the scaling behaviour of the entanglement entropy (hence it is called \emph{finite entanglement scaling}), see \cite{Tagliacozzo2008} and \cite{Haegeman2011}.

However, when we try to use matrix product states to simulate anyon chains, we run into a problem: the MPS are formulated on a space that possesses a tensor product structure, which is not necessarily compatible with the fusion rules of the anyon model. How do we solve this problem? First, note that there is a natural way to embed the Hilbert space $\Hi$ of an anyon chain into a larger Hilbert space $\tilde{\mathcal{H}}$ that possesses tensor product structure. The larger Hilbert space is simply spanned by \emph{all} states $|x_1,x_2,\dots\rangle$ regardless of whether the sequence $x_1,x_2,\dots$ is allowed by the fusion rules. The vector space that consists only of the allowed sequences, which we call the \emph{anyon subspace} in this context, is a subspace of this larger Hilbert space.

One possible solution is to modify the Hamiltonian such that it acts on the larger Hilbert space by adding penalty terms that suppress contributions from states outside the anyon subspace. In the case of anyon chains, this means that we add terms which are diagonal on two neighbouring labels for all forbidden combinations: Let $M$ be the adjacency matrix describing the fusion constraints of the anyon model and $\tilde{H}$ the Hamiltonian on the larger Hilbert space $\tilde{\mathcal{H}}$. The modified Hamiltonian is then of the form
	\begin{equation}
	\label{eq:penaltyham}
		\tilde{H}^\prime=\tilde{H}+C\sum_i\tilde{P}_i,
	\end{equation}
with a constant $C$ and the penalty term $\tilde{P}_i$ given by
	\begin{equation}
		\tilde{P}_i=\sum_{x_i,x_{i+1}}\left(1-M_{x_ix_{i+1}}\right)|x_ix_{i+1}\rangle\langle x_ix_{i+1}|.
	\end{equation}

Once we have an MPS representation of the anyon chain, we can use it to calculate properties of the ground state. For instance, we are interested in the behaviour of the energy gap and the entanglement entropy with growing system size. In case we have a translational-invariant MPS as in \cref{eq:TIMPS} (a so-called \emph{uniform} MPS\index{uniform matrix product state}) it is even possible to study infinite chains. Here, we study the behaviour of the entanglement entropy with growing bond dimension instead of growing system size. We can relate the entanglement entropy directly to the bond dimension of the MPS by modifying \cref{eq:xiS}, since the correlation length $\xi$ also depends on the bond dimension (see \cite{Stojevic2015}). The resulting formula is 
	\begin{equation}
	\label{eq:SD}
		S=\frac{1}{\sqrt{\frac{12}{c}}+1}\log_2(D)+k,
	\end{equation}
where $c$ is the central charge of the corresponding CFT. Hence, from the scaling of $S(D)$ vs. $\log_2(D)$ we can get an estimate of the central charge.

To calculate the ground state of the system we have used the package evoMPS \cite{evoMPS} written in the python programming language. It provides algorithms to evaluate the real or imaginary time evolution for MPS of both finite chains with open boundary conditions or uniform MPS (i.e., infinite chains) via the Time-Dependent Variational Principle (TDVP). It is based on algorithms presented in \cite{Haegeman2011} and \cite{Milsted2013}, among others. In this thesis, we use it to calculate ground states for infinite chains, in case of Fibonacci anyons and for the $\Hd$ chain. The application of the TDVP to uniform MPS is described in detail in \cite{Haegeman2011} and \cite{Haegeman2013}. The code used for the calculations of this thesis is based on code written by Marius Lewerenz \cite{PrivateMarius}.

\begin{exmp}[The golden chain]
	\label{ex:Goldenchainnumerics}
	In case of the Fibonacci anyon chain\index{golden chain} presented in \cref{ex:Goldenchain} there is only one possible Hamiltonian we have to study:
	\begin{equation}
	H^\mathbf{Fib}=-\sum_i p_i^{(\mathbf{1})}.
	\end{equation}
	Before we can numerically study the model with tensor network methods, we need to construct the Hamiltonian that acts on the larger Hilbert space that possesses tensor product structure. Instead of the five-dimensional basis, the full Hilbert space has eight basis states:
	\begin{equation}
	\{|\mathbf{1}\mathbf{1}\mathbf{1}\rangle,|\mathbf{1}\mathbf{1}\tau\rangle,|\mathbf{1}\tau\mathbf{1}\rangle,|\mathbf{1}\tau\tau\rangle,|\tau\mathbf{1}\mathbf{1}\rangle,|\tau\mathbf{1}\tau\rangle,|\tau\tau\mathbf{1}\rangle,|\tau\tau\tau\rangle\},
	\end{equation}
	where the subset of allowed sequences is
		\begin{equation}
			\{|\mathbf{1}\tau\mathbf{1}\rangle,|\mathbf{1}\tau\tau\rangle,|\tau\mathbf{1}\tau\rangle,|\tau\tau\mathbf{1}\rangle,|\tau\tau\tau\rangle\},
		\end{equation} 
	hence the penalty matrix $P$ has non-zero entries on the diagonal at positions one, two, and five. As a result, the modified Hamiltonian (that also includes penalty terms) is, according to \cref{eq:penaltyham}, given by 
	\begin{equation}
	\tilde{H}^\prime_i=-\begin{pmatrix}
	0 & 0 & 0 & 0 & 0 & 0 & 0 & 0\\
	0 & 0 & 0 & 0 & 0 & 0 & 0 & 0\\
	0 & 0 & 1 & 0 & 0 & 0 & 0 & 0\\
	0 & 0 & 0 & 0 & 0 & 0 & 0 & 0\\
	0 & 0 & 0 & 0 & 0 & 0 & 0 & 0\\
	0 & 0 & 0 & 0 & 0 & 0 & 0 & 0\\
	0 & 0 & 0 & 0 & 0 & 0 & \varphi^{-2} & \varphi^{-\frac{3}{2}}\\
	0 & 0 & 0 & 0 & 0 & 0 & \varphi^{-\frac{3}{2}} & \varphi^{-1} 
	\end{pmatrix}+C\ \begin{pmatrix}
	1 & 0 & 0 & 0 & 0 & 0 & 0 & 0\\
	0 & 1 & 0 & 0 & 0 & 0 & 0 & 0\\
	0 & 0 & 0 & 0 & 0 & 0 & 0 & 0\\
	0 & 0 & 0 & 0 & 0 & 0 & 0 & 0\\
	0 & 0 & 0 & 0 & 1 & 0 & 0 & 0\\
	0 & 0 & 0 & 0 & 0 & 0 & 0 & 0\\
	0 & 0 & 0 & 0 & 0 & 0 & 0 & 0\\
	0 & 0 & 0 & 0 & 0 & 0 & 0 & 0
	\end{pmatrix}.
	\end{equation}
	
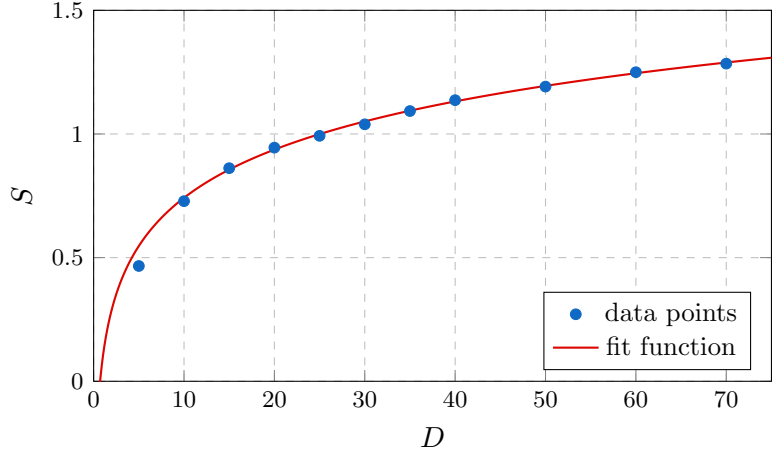
\begin{figure}[t]
	\centering
	\begin{tikzpicture}[scale=1]
	\begin{axis}[xmin=0,xmax=75,
	xlabel=$D$,ylabel=$S$,
	width=10.5cm,height=6.5cm,
	ticklabel style={font=\footnotesize},
	ymin=0,ymax=1.5,
	grid=both,
	grid style=dashed,
	samples=500,
	legend pos=south east]
	\addplot[color=LinkColor2,only marks,mark=*] coordinates {
		(5, 0.46621461440833206)
		(10, 0.7283756853371707)
		(15, 0.8617259091066211)
		(20, 0.9447738593650602)
		(25, 0.9925288284432728)
		(30, 1.0390100589825766)
		(35, 1.0927468750995466)
		(40, 1.136645)
		(50, 1.191610)
		(60, 1.249809)
		(70, 1.2844213667669817)
	};
	\addlegendentry{\small data points};
	\addplot[LinkColor, thick][domain=0:75] {0.0944869 + 0.281158*ln(x)};
	\addlegendentry{\small fit function};
	\end{axis}
	\end{tikzpicture}
	\caption{\small \label{fig:Fibent}Entanglement entropy $S$ for different values of the bond dimension $D$ for $H^\mathbf{Fib}_i=-p_i^{(\mathbf{1})}$.}
\end{figure}
	
	Investigating the behaviour of this system with tensor network methods allows us to study the connection between entanglement entropy and bond dimension. The resulting data points for the entropy $S$ for different bond dimensions $D$ are shown in blue in \cref{fig:Fibent}, together with a fit function (red). Using the formula in \cref{eq:SD}, we get an estimate for the central charge of the corresponding conformal field theory:
		\begin{equation}
		\label{eq:cFib}
			c\approx 0.716751.
		\end{equation}
	For unitary CFTs in $(1+1)$ dimensions, unitarity restricts the possible values of the central charge to
		\begin{equation}
			c=1-\frac{6}{m(m+1)},\hspace{15pt}m=2,3,4,\dots
		\end{equation}
	for $c<1$. For $c>1$, the allowed central charges are continuous and $c$ can take any value (see \cite{Belavin1984} and \cite{Friedan1984}). Hence, the only possible values of $c$ near the value \cref{eq:cFib} are $\frac{1}{2}$, $\frac{7}{10}$, and $\frac{4}{5}$, so we can conclude that the value in \cref{eq:cFib} is consistent with $c=\frac{7}{10}$ (as was done in \cite{feiguin_interacting_2007}\footnote{Here, the authors found a value of $c=0.701\pm0.001$ in their numerical studies, which is even closer to $\frac{7}{10}$ than the value we have found in our investigation.}). This value  corresponds to the classical $2D$ tricritical Ising model known as the RSOS model, see \cite{Andrews1984}.
	
	The Fibonacci chain model has been extended to also include three-anyon interactions in addition to pairwise interaction. In \cite{Trebst2008a}, the authors find that this leads to a rich phase diagram with multiple critical and gapped phases.
\end{exmp}

\section{The $\Hd$ chain}
\label{sec:H3chain}

We now want to apply the methods described above to study a chain that is built from the Haagerup fusion category $\Hd$. As explained earlier, the underlying mathematical theory of anyons is a Unitary Modular Tensor Category (UMTC), not a unitary fusion category (UFC). But what happens if we do the same construction starting from a UFC instead of a UMTC? Does the resulting chain still give rise to a critical model? Since a UMTC has certain properties that a fusion category lacks (for example, a fusion category is not braided), one has to be very careful when constructing an anyon chain from a fusion category. However, the general construction does not explicitly require any properties that are exclusive to a UMTC, hence it is, in principle, possible to construct chains from UFCs\footnote{Strictly speaking, the resulting chain is not an \emph{anyon} chain since anyons are described by UMTCs. In an abuse of notation we still call chains built from UFCs anyon chains, keeping in mind that the underlying category is a UFC instead of a UMTC.}. We can therefore construct a chain from the $\Hd$ fusion category and study whether it gives rise to a conformal field theory.

Evidence for the existence of a conformal field theory that corresponds to the Haagerup subfactor was already found by Evans and Gannon in \cite{Evans2011}. Here, the authors constructed the (hypothetical) CFT from the quantum double of the Haagerup subfactor as a vertex operator algebra and found that it has central charge $c=8$. Hence, we expect that the numerical investigations of a Haagerup anyon chain with matrix product states will reveal that the central charge is $c=8$.

As a first step, we need to specify the Hilbert space of the model. A natural choice for the object $a$ in \cref{fig:anyonchain} that specifies the outer labels of the chain in the $\Hd$ case is $\rho$ because we can get any other simple object of the category by fusing $\rho$'s\footnote{Although $\alpha$ and $\alpha^*$ are not in the decomposition of $\rho\otimes\rho$, they do appear in the decomposition of $\rho\otimes\rho\otimes\rho$.} (Recall the fusion rules of $\Hd$ from \cref{tab:fusionH2}). Hence, the resulting chain is of the form\index{H3 chain@$\Hd$ chain}
\begin{figure}[H]
	\centering

\end{figure}

Similar to the example of the Fibonacci anyon chain, we can get the basis states of the Hilbert space $\mathcal{H}_{N+2}$ from the corresponding fusion graph (see \cref{fig:fusiongraphH3}).

\begin{figure}[t]
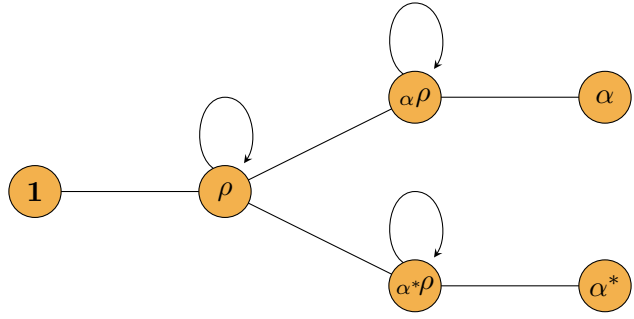

	\centering
	%
	\caption{\small \label{fig:fusiongraphH3}\textbf{Fusion graph for the $\Hd$ chain.} A path of length $N-1$ on this graph corresponds to an allowed labelling for a basis state $|x_1,\dots,x_{N-1}\rangle$ of the Hilbert space $\mathcal{H}_{N+2}$ of the chain.}
\end{figure}

In contrast to the Fibonacci anyon chain, there are more possibilities to build projectors in the $\Hd$ case, which is due to the fact that there are more simple objects in the decomposition of $\rho\otimes\rho$:
\begin{equation}
\rho\otimes\rho=\mathbf{1}+\rho+\alpharho+\alphastarrho.
\end{equation}
As a result, we can define projections for $e\in\{\mathbf{1},\rho,\alpharho,\alphastarrho\}$:
\begin{equation}
p_i^{(e)}|x_{i-1},x_i,x_{i+1}\rangle=\sum_{x_i'}\left(F_{x_{i+1}}^{x_{i-1}\rho\rho}\right)_{ex_i}\left(F_{x_{i+1}}^{x_{i-1}\rho\rho}\right)^\dagger_{x_i'e}\,|x_{i-1},x_i',x_{i+1}\rangle,
\end{equation}
using the $F$-symbols given in \cref{app:Fsymbols}. Hence, the most general form of the local Hamiltonians $H_i$ is
\begin{equation}
H_i=c_\mathbf{1}\ p_i^{(\mathbf{1})}+c_\rho\  p_i^{(\rho)}+c_{\alpharho}\ p_i^{(\alpharho)}+c_{\alphastarrho}\  p_i^{(\alphastarrho)},
\end{equation}
where $c_i\in\mathbb{R}$. Therefore, we have an infinite number of possible Hamiltonian which makes finding the right one (i.e., the one that corresponds to a critical model) an almost impossible task. Fortunately, we can restrict the parameter space by switching to hyper-spherical coordinates:
\begin{equation}
\label{eq:fullham}
H_i=r\cos\psi\ p_i^{(\mathbf{1})}+r\sin\psi\cos\theta\  p_i^{(\rho)}+r\sin\psi\sin\theta\cos\varphi\ p_i^{(\alpharho)}+r\sin\psi\sin\theta\sin\varphi\  p_i^{(\alphastarrho)},
\end{equation}
with $r\in\mathbb{R}$, $\psi,\theta\in [0,\pi]$, and $\varphi\in[0,2\pi]$. Furthermore, we can set $r=1$ since this simply corresponds to rescaling the ground state energy but does not change the behaviour of the model, and especially does not have an effect of the criticality of the model. This leaves us with only three parameters in bounded intervals which we have to search through in order to find a Hamiltonian that gives a critical model.

\subsection*{Numerical results}

Since it is a rather complicated and numerically challenging task (in the sense that there are many parameter constellations to check even after setting $r=1$) to investigate the full Hamiltonian given in \cref{eq:fullham}, we begin our numerical investigation by checking some easy special cases. For instance, a natural guess for a critical model is the Hamiltonian we have already used in the Fibonacci case:
\begin{equation}
\label{eq:ham1}
H=-\sum_i p_i^{(\mathbf{1})}.
\end{equation}
The plot in \cref{fig:Ent1} shows the entanglement entropy for different values of the bond dimension (blue data points) together with a fit function (red) according to \cref{eq:SD}. From the diagram it is not immediately clear whether the entropy diverges or saturates for $D\to\infty$. Since the numerical calculations for this Hamiltonian take a very long time for high values of the bond dimension, it is difficult to study the behaviour of the entanglement entropy in this region. However, investigations of the scaling of the entanglement entropy of a finite chain with growing system size $N$ and the scaling of the ground state energy density by Ashley Milsted \cite{PrivateAsh} have shown that the Hamiltonian does not correspond to a CFT. A second indicator that this Hamiltonian is not a CFT is the scaling of the entanglement entropy vs. the bond dimension. If there is a corresponding CFT, the fit in \cref{fig:Ent1} yields an estimate of the central charge according to \cref{eq:SD}. For the Hamiltonian studied here this estimate is $c=3.11$, which is far from the expected value $c=8$\footnote{Note that it is, in principle, possible that via the investigation of the category $\Hd$ we find a CFT with central charge different than $c=8$ (which was found in \cite{Evans2011}). However, we believe that there is only one CFT associated to a certain subfactor.}.

\begin{figure}[t]
	\centering
	\begin{tikzpicture}[scale=1]
	\begin{axis}[xmin=0,xmax=35,
	xlabel=$D$,ylabel=$S$,
	width=10cm,height=7cm,
	ticklabel style={font=\footnotesize},
	ymin=0,ymax=2.5,
	grid=both,
	grid style=dashed,
	samples=500,
	legend pos=south east]
	\addplot[color=LinkColor2,only marks,mark=*] coordinates {
		(5, 1.61687)
		(10, 1.6431)
		(15, 1.96131)
		(20, 2.14134)
		(25, 2.19271)
		(30, 2.21964)
	};
	\addlegendentry{\small data points};
	\addplot[LinkColor, thick][domain=0:40] {0.691725*ln(x)};
	\addlegendentry{\small fit function};
	\end{axis}
	\end{tikzpicture}
	\caption{\small \label{fig:Ent1}Entanglement entropy $S$ for different values of the bond dimension $D$ for for $H_i=-p_i^{(\mathbf{1})}$, together with a fit according to \cref{eq:SD}.}
\end{figure}

Similarly, we can check if there is a critical model if we only use the projection onto the $\rho$ object, i.e., the Hamiltonian
	\begin{equation}
		\label{eq:hamrho}
		H=-\sum_i p_i^{(\rho)}.
	\end{equation}
The resulting plot for the entanglement entropy for different values of the bond dimension is given in \cref{fig:Entrho}. It is obvious that the entanglement entropy $S$ does not diverge for $D\to\infty$ since already at bond dimensions around $50$ the values saturate. The estimate for the central charge is $c=10.18$, which is also far from the expected value. In the plot shown in \cref{fig:Entrho} it is obvious that the best fit according to \cref{eq:SD} does not represent the data points very well which is due to the fact that the values saturate instead of going to infinity. As a result, we find that neither the Hamiltonian built from the $\mathbf{1}$-projection in \cref{eq:ham1} nor the one built from the $\rho$-projection in \cref{eq:hamrho} correspond to a critical model.

\begin{figure}[t]
	\centering
	\begin{tikzpicture}[scale=1]
	\begin{axis}[xmin=0,xmax=65,
	xlabel=$D$,ylabel=$S$,
	width=10cm,height=7cm,
	ticklabel style={font=\footnotesize},
	ymin=0,ymax=2.5,
	grid=both,
	grid style=dashed,
	samples=500,
	legend pos=south east]
	\addplot[color=LinkColor2,only marks,mark=*] coordinates {
		(5, 0.683731)
		(6, 0.815984)
		(7, 1.20413)
		(8, 1.39829)
		(9, 1.43381)
		(10, 1.47473)
		(15, 1.50059)
		(20, 1.50581)
		(25, 1.51761)
		(50, 1.53847)
		(60, 1.53871)
	};
	\addlegendentry{\small data points};
	\addplot[LinkColor, thick][domain=0:65] {0.486824*ln(x)};
	\addlegendentry{\small fit function};
	\end{axis}
	\end{tikzpicture}
	\caption{\small \label{fig:Entrho}Entanglement entropy $S$ for different values of the bond dimension $D$ for $H_i=-p_i^{(\rho)}$, together with a fit according to \cref{eq:SD}.}
\end{figure}
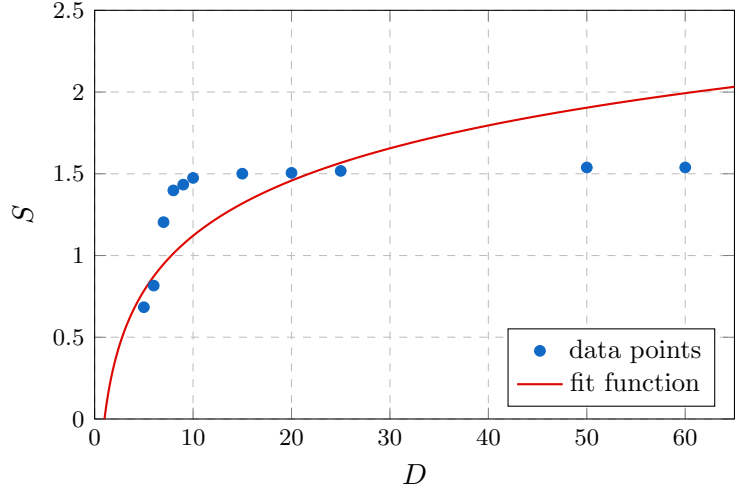

Since the previous approaches did not yield a critical model, we have to make the model a bit more complicated. Instead of studying the two projections separately, we can combine them in a single Hamiltonian:
\begin{equation}
H=-\sum_i c_\mathbf{1}\ p_i^{(\mathbf{1})}+c_\rho \ p_i^{(\rho)}
\end{equation}
with $c_\mathbf{1},c_\rho\in\mathbb{R}$. In this description, the parameter space that we have to check is infinite which makes it difficult to find the correct Hamiltonian. A more practical description arises when we switch to polar coordinates instead of Cartesian ones using $r\in\mathbb{R}$ and $\theta\in[0,2 \pi]$:	
	\begin{equation}
	\label{eq:2paramHam}
		H=-\sum_i r\sin\theta\ p_i^{(\mathbf{1})}+r\cos\theta \ p_i^{(\rho)}
	\end{equation}
with $r\in\mathbb{R}$ and $\theta\in[0,2\pi]$. Finally, we can set $r=1$ since this does not change the general behaviour of the system. Hence, we only have to check different values of $\theta\in[0,2\pi]$ and look for indicators of critical points. As described above, one indicator is the scaling of the entanglement entropy with the bond dimension. 

In \cref{fig:Ent1rho}, the entanglement entropy $S$ is depicted for different bond dimensions (indicated by different colours) and for different values of $\theta$. There are essentially three regions that indicate a possible critical point: $\theta\approx 0$, $\theta\approx\pi$, and $\theta\approx 4.5$. We have already investigated the first two cases since $\theta=0$ corresponds to the Hamiltonian $H_i=-p_i^{(\rho)}$ and $\theta=\pi$ corresponds to $H_i=-p_i^{(\mathbf{1})}$. A similar investigation of the scaling behaviour of the entanglement entropy for \cref{eq:2paramHam} with $\theta\approx 4.5$ shows that the entanglement entropy saturates with growing bond dimension (similar to the other two cases), hence it does not correspond to a critical point.

\begin{figure}[t]
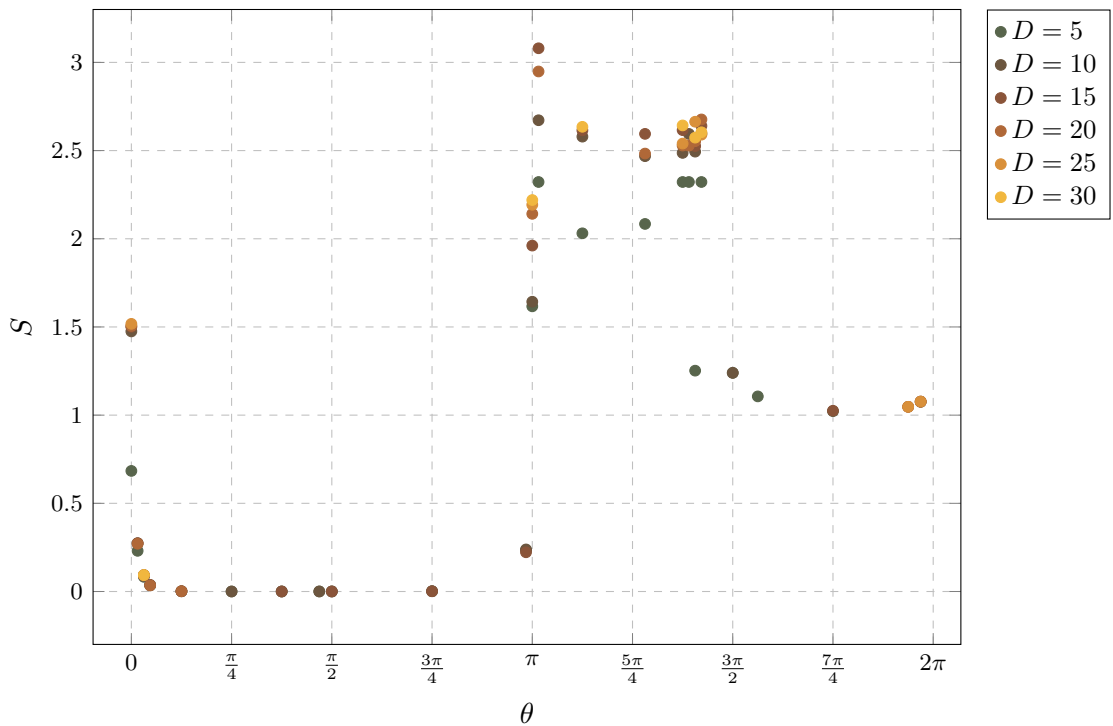

	\centering

	\caption{\small \label{fig:Ent1rho}Entropy landscape for for $H_i=-\big(\sin\theta\ p_i^{(\mathbf{1})}+\cos\theta\  p_i^{(\rho)}\big)$.}
\end{figure}

Since all previous approaches have failed to give rise to a critical model, it remains to consider the full Hamiltonian and scan the whole parameter space for an interesting point, i.e., study the Hamiltonian
	\begin{equation}
		H=-\sum_i \cos\psi\ p_i^{(\mathbf{1})}+\sin\psi\cos\theta\  p_i^{(\rho)}+\sin\psi\sin\theta\cos\varphi\ p_i^{(\alpharho)}+\sin\psi\sin\theta\sin\varphi\  p_i^{(\alphastarrho)}.
	\end{equation}
To identify interesting points, we begin with a rather coarse overview of the maximum value of the entanglement entropy $S$ for different values of $\psi,\theta,\varphi$ without analysing the connection between entropy and bond dimension. We expect that at a critical point, the entropy diverges, therefore it suffices to look for points with an unusually high maximum entropy compared to the surrounding region. For the coarse overview, we split the interval $[0,\pi]$ into $16$ pieces and calculate the entanglement entropy for bond dimensions in the range $D\in[5,10,\dots,25]$ and take the maximum value. The complete list of numerical results can be found in \cref{app:H3chain}. Here, we show an example of the kind of results that we get. 

\begin{figure}[t]
	\begin{tikzpicture}
	\node at (0,0) {\textbf{(a)}};
	\end{tikzpicture}\\
	\includegraphics[width=0.8\textwidth]{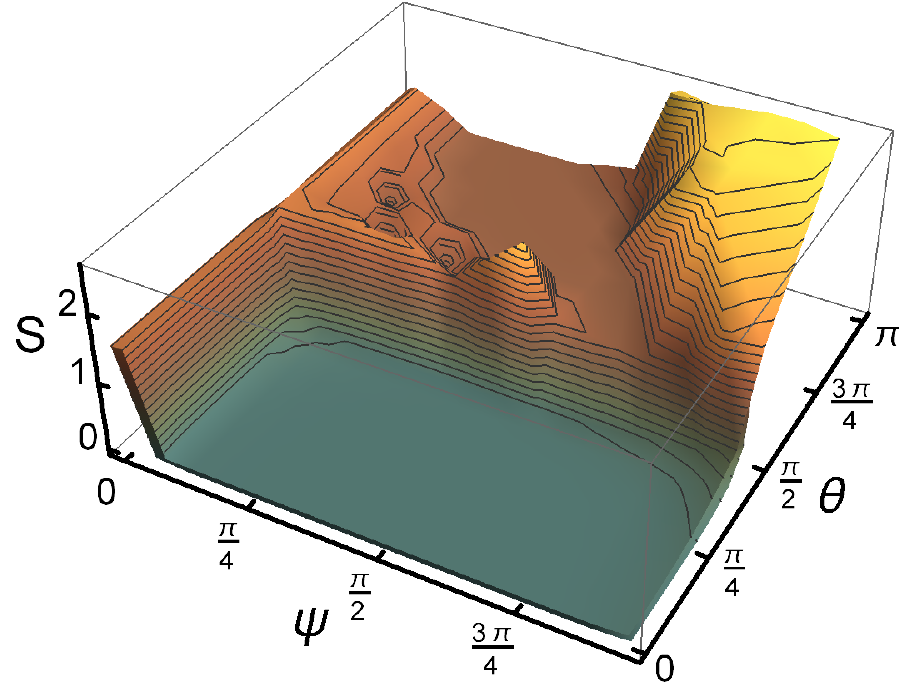}\\
	\vspace{20pt}
	\begin{tikzpicture}
	\node at (0,0) {(\textbf{b)}};
	\end{tikzpicture}\\
	\includegraphics[width=0.8\textwidth]{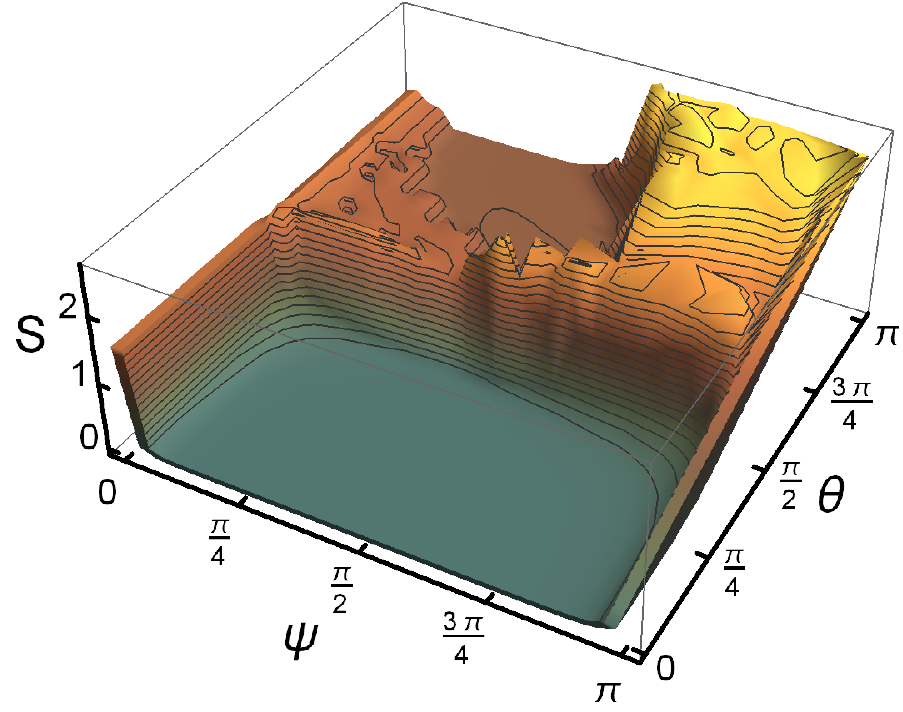}
	\caption{\small \label{fig:fullent}\textbf{Entropy landscape for }$\varphi=\frac{19\pi}{16}$: \textbf{(a)} Coarse diagram: $\psi,\theta\in[0,\pi]$, where the interval is split into $16$ pieces. Here, we observe an isolated peak in the middle of the diagram. \textbf{(b)} Finer diagram: The interval $[0,\pi]$ is split into $32$ pieces. The isolated peak is split into several smaller peaks.}
\end{figure}

In \cref{fig:fullent} the entropy landscape for $\varphi=\frac{19\pi}{16}$ is shown. The upper plot in \cref{fig:fullent}\textbf{(a)} shows the coarse overview, while the lower plot in \cref{fig:fullent}\textbf{(b)} gives a finer landscape where the interval $[0,\pi]$ is split into $32$ pieces instead of $16$. In general, there seem to be roughly three different regions in this landscape: There is one big region (green) where the entropy stays zero regardless of the bond dimension. This region exists in all plots in \cref{app:H3chain} for $\theta\in[0,\frac{\pi}{2}]$ (roughly), regardless of the values of $\varphi$ and $\psi$. Another region, depicted in brown, has a value of $S\approx 1$, and the third region (yellow) consists of points with $S\approx 2$. This is consistent with the two-dimensional entropy landscape depicted in \cref{fig:Ent1rho}. Hence, candidates for critical points are at the borders between these regions. In the coarse overview we are looking for points where the entanglement entropy decreases, as explained above. In most of the plots the transition between the regions does not show an unusual high value for the entropy. 

In one of the plots, which is shown in \cref{fig:fullent}\textbf{(a)}, we observe a rather isolated peak at the boarder of the $S\approx 0$ and the $S\approx 1$ region, hence we take a closer look at this plot. In \cref{fig:fullent}\textbf{(b)}, a finer version of the landscape is shown where we have collected more data points. Here, it becomes clear that the emergence of the isolated peak was merely due to the small number of data points than a reflection of the physical behaviour of the system. Furthermore, a more detailed analysis of the scaling of the entanglement entropy shows that it saturates instead of diverging with growing bond dimension.

In the end, we have not been able to identify any critical points for the anyon chain built from the fusion category $\Hd$.

\section{Further directions}

We conclude from our numerical investigation of anyon chains built from the fusion category $\Hd$ that the system does not show critical behaviour, hence it does not correspond to a conformal field theory. However, this does not mean that there is no conformal field theory associated to the Haagerup subfactor. It is merely an indicator that building the anyon chain directly from the fusion category instead of a modular category is the wrong approach to the problem. A more promising yet also more complicated idea is to start from the quantum double of the fusion category $\Hd$ (which is indeed a unitary modular tensor category) and study the corresponding anyon chain. 

Before following this approach, for completeness we can also check whether the other two categories in the Morita equivalence class, $\Hi_1$ and $\Hi_2$, yields a CFT via the anyon chain construction. Even though we do not expect a positive result here, these calculations are still easier to do than going via the quantum double. Furthermore, since the fusion rules of the category $\Hi_1$ contain multiplicities (which is also the case for the quantum double of these categories), this construction can be helpful to understand the additional challenges that come with multiplicities in anyon chains. In contrast to the quantum double, which has twelve simple objects, $\Hi_1$ only has four simples and is therefore easier to handle.

The approach that uses the UMTC, however, brings new challenges: First, we have to construct the quantum double of the fusion category $\Hd$\footnote{We could also use $\Hi_1$ or $\Hi_2$, for that matter, since they are Morita equivalent and therefore yield the same quantum double.}. Some work has already been done in this direction, for instance the simple objects have been identified and the modular data has been constructed (see \cite{izumi_structure_2001,SEUNGMOON2008,Evans2011,Grossman2015}). However, an important ingredient in the construction of an anyon chain Hamiltonian are the $F$-symbols of the category, which have not yet been determined for the quantum double of the Haagerup subfactor. Hence, before we can numerically investigate the anyon chain that corresponds to the quantum double, we have to find its $F$-symbols. Since the category includes twelve simple objects and, moreover, has fusion rules with multiplicities, finding the $F$-symbols is a much more challenging task than it was for the $\Hd$ category.

\chapter{Defects in quantum spin systems}
\label{ch:defects}

In the previous chapter, we have only considered chains with labels from the set of simple objects of a category. However, it is possible to study this setting in a more general way, namely by allowing the labels to come from outside the category. This corresponds to having \emph{defects}\index{defect} in the chain, indicated in red in \cref{fig:defect}.

	\begin{figure}[H]
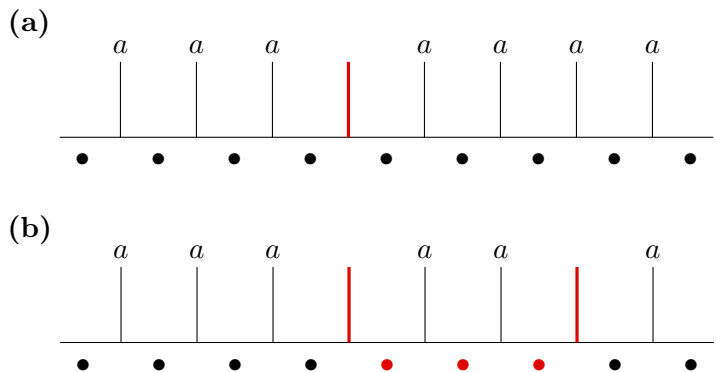

		\centering

		\caption{\small \label{fig:defect} \textbf{Defects (red) in a spin chain.} \textbf{(a)} A single defect in a chain constructed from the object $a$, with objects from the category denoted $\bullet$. \textbf{(b)} Having two defects in a chain, the possibilities for objects in between them (red bullets) can be changed.}
	\end{figure}

In general, defects play an important role in the study of \emph{topological phases}\index{topological phase} \cite{Wen1990a,Wen1990b}. Without considering defects, topological phases are already an interesting area in their own right: Due to their insensitivity to environmental noise, they are a promising candidate for topological quantum computation and quantum error-correction \cite{Dennis2002,Kitaev2003,Nayak2008,Terhal2015,Pastawski2015,Brown2016}. The ground space of a topologically ordered system can be used to store quantum data, and by braiding and fusing the emergent quasi-particle excitations it is possible to manipulate the encoded information in a robust manner. However, in many phases (which includes those most suitable for experimental realization) the computational power is severely limited. Here, it has been found that the introduction of defects to the system can enhance the computational power and a lot of work has been done to understand these defects (see, for example, \cite{Fuchs2013,Barkeshli2013,Cong2016,Bridgeman2017,Cong2017,Kesselring2018}).

In this chapter, we study the effects of defects in physical systems via a microscopic approach in order to make sense of a dynamical theory of defects for quantum spin systems. We also discuss how to tackle certain problems in this context with methods from category theory, thereby expanding the chain model we have introduced in the previous chapter. We first discuss defects in topologically ordered systems without using category theory via the two-dimensional toric code example. Then we go into more detail about how to use methods from category theory to make the study of the ground space of the system easier. 

Here, we consider defects in a one-dimensional chain that is built from a fusion category $\C$ (as described in the previous chapter). Defects in this chain are therefore $\C-\C$ bimodules, i.e., we are fusing an element of the category of bimodules to the chain. To realize this, we need to have a description of the corresponding vertices. While in the previous chapter we have only considered vertices within the category, we now need to know how bimodules and objects of the category fuse together. After having constructed the necessary vertices, it is then possible to define a local Hamiltonian from projectors onto simple objects in the same fashion as we did in the previous chapter. We demonstrate how this works with a detailed example, namely a chain constructed from the category $\mathbf{Vec}(\mathbb{Z}/2\mathbb{Z})$, which is the category of vector spaces over the field $\mathbb{Z}/2\mathbb{Z}$ (see \cite{Bridgeman2020}).

This chapter does not directly contribute to answering the central question of this thesis, namely whether there is a CFT that corresponds to the Haagerup subfactor. However, adding defects to an anyon chain is a natural generalisation of the anyon chains studied in the previous chapter and an interesting topic by itself because of its connection to topological phases, hence we study these systems here. Furthermore, the construction of the vertices of the $\mathbf{Vec}(\mathbb{Z}/2\mathbb{Z})$ chain with defects yields insight into how computations are done within the so-called annular category. This category is interesting with regard to UMTCs since there is a connection between the simple objects of the Drinfeld centre of a category and representations of the annular category, which we elaborate on later in this chapter.

\section{Gauging defects in lattice systems}

Adding defects to a physical system in general changes its dynamical behaviour. Consider a system where each particle is described by the Hilbert space $\mathbb{C}^d$. An example for a two-dimensional system that is often considered is a two-dimensional regular lattice on a torus as depicted in \cref{fig:torus}, which for instance appears in Kitaev's toric code \cite{Kitaev2003}\index{toric code}. In the following, we illustrate the concept of defects by missing spin defects in a lattice on a torus. The Hilbert space of a general lattice system with an unknown number of such particles is given by the \emph{distinguishable Fock space}\index{Fock space}\footnote{At first sight, it seems counter-intuitive that we do not know how many particles we have but still can individually identify and address the particles. A physical example of such a system is a one-dimensional optical lattice in which at each site we either have zero atoms or one two-level atom. We furthermore assume that the atoms are lined up in a contiguous line with no gaps (which can be achieved by imposing some kind of potential gradient). In the resulting lattice we can identify and address the individual atoms via their lattice position, although we do not know the total number of particles.}

\begin{figure}[t]
	\centering
	\begin{tikzpicture}
	\begin{axis}[
	axis equal image,
	hide axis,
	z buffer = sort,
	view = {122}{30},
	colormap={mycolormap}{color=(LightGray) color=(white) color=(LightGray)},
	scale = 1.5
	]
	\addplot3[
	surf,
	shader = faceted interp,
	samples = 25,
	samples y= 50,
	domain = 0:2*pi,
	domain y = 0:2*pi,
	colormap name = mycolormap,
	thin
	] (
	{(3+sin(deg(\x)))*cos(deg(\y))},
	{(3+sin(deg(\x)))*sin(deg(\y))},
	{cos(deg(\x))}
	);
	\end{axis}
	\end{tikzpicture}
	\caption{\small \label{fig:torus}\textbf{Lattice on a torus.} This is an example for a two-dimensional lattice with periodic boundary conditions.}
\end{figure}
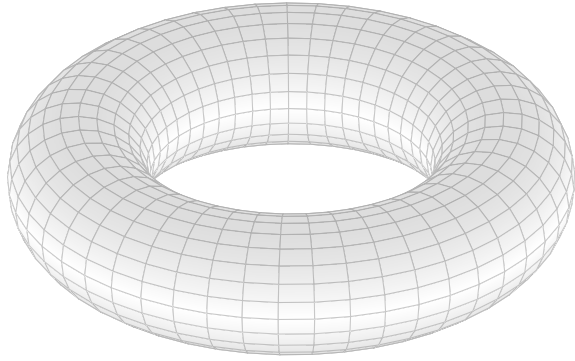

\begin{equation}
\mathfrak{F}\big(\mathbb{C}^d\big)\cong\bigoplus_{N=0}^\infty\Hi_N\cong\bigoplus_{N=0}^\infty\big(\underbrace{\mathbb{C}^d\otimes\mathbb{C}^d\otimes\dots\otimes\mathbb{C}^d}_{N\ \mathrm{factors}}\big)
\end{equation}
with the convention that the space that describes zero particles (i.e., the vacuum), is described by the Hilbert space  $(\mathbb{C}^d)^{\otimes 0}\cong\mathbb{C}$. We can incorporate defects to this system in the following way: For instance, if a system (with a single site) is comprised of either zero quantum spins or one distinguishable quantum spin, the Hilbert space is given by
\begin{equation}
\label{eq:Fockone}
\mathfrak{F}_{\le 1}\big(\mathbb{C}^d\big)\cong\mathbb{C}\oplus\mathbb{C}^d.
\end{equation}
Here, the subscript indicates that we have at most one quantum spin present in the system. 

Suppose now that we have a system with $n$ sites, where on each site there either is a quantum spin present or not. The absence of a spin corresponds to a defect\footnote{Note that choosing the defect to be an absent spin is only one possible choice. The construction works for arbitrary defects and also situations where different kinds of defects are present in the same system, although the description might become more complicated.} in the system. We can write the overall Hilbert space of the system as a tensor products of single-site Hilbert spaces \cref{eq:Fockone}:
\begin{equation}
\label{eq:Fnone}
\mathfrak{F}_{\le n}\big(\mathbb{C}^d\big)\equiv\bigotimes_{j=0}^n\mathfrak{F}_{\le 1}\big(\mathbb{C}^d\big)\cong\big(\mathbb{C}\oplus\mathbb{C}^d\big)^{\otimes n},
\end{equation}
which can be rewritten by expanding out the tensor factors as
\begin{equation}
\label{eq:Fntwo}
\mathfrak{F}_{\le n}\big(\mathbb{C}^d\big)\cong\bigoplus_{j=0}^n\mathbb{C}^{\binom{n}{j}}\otimes\Hi_j,
\end{equation}
where $\Hi_j=(\mathbb{C}^d)^{\otimes j}$ is the Hilbert space of $j$ distinguishable spins. We can understand this representation as follows: The space $\mathbb{C}^{\binom{n}{j}}$ can be interpreted as the configuration space of $j$ identical scalar particles arranged on a system of $n$ sites, i.e., it takes care of identifying which locations are occupied with spins, while $\Hi_j$ represents the actual distinguishable spins at those positions. At first sight, it is not clear that \cref{eq:Fnone} and \cref{eq:Fntwo} yield the same representation. A quick way to convince yourself that they are actually equal is by checking their respective dimensions. In the first case \cref{eq:Fnone}, we have
\begin{equation}
\dim\Big(\mathfrak{F}_{\le n}\big(\mathbb{C}^d\big)\Big)=\dim\Big(\big(\mathbb{C}\oplus\mathbb{C}^d\big)^{\otimes n}\Big)=(d+1)^n.
\end{equation}
In the second case \cref{eq:Fntwo} we get
\begin{equation}
\dim\Big(\mathfrak{F}_{\le n}\big(\mathbb{C}^d\big)\Big)=\sum_{j=0}^n\binom{n}{j}d^j=(d+1)^n,
\end{equation}
where the last equality is a consequence of the binomial theorem. An example for the different defect configurations that can appear in a higher-dimensional lattice system are given in \cref{fig:defect}. As we will see in the example of the toric code, different configurations of the same number of defects can result in different dimensions of the ground space of the system.

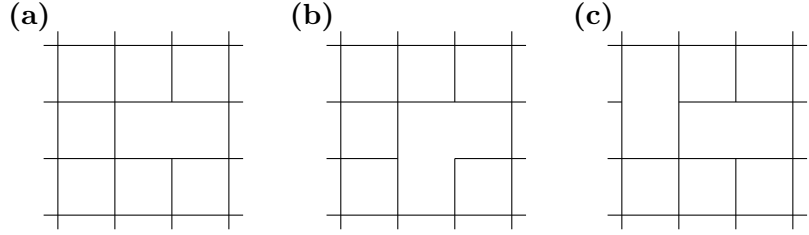
\begin{figure}[t]
	\centering
	\begin{tikzpicture}[scale=0.75]
	\node at (-0.5,3.5) {\textbf{(a)}};
	\draw (-0.25,0) -- (3.25,0);
	\draw (-0.25,1) -- (3.25,1);
	\draw (-0.25,2) -- (3.25,2);
	\draw (-0.25,3) -- (3.25,3);
	\draw (0,-0.25) -- (0,3.25);
	\draw (1,-0.25) -- (1,3.25);
	\draw (2,-0.25) -- (2,1);
	\draw (2,2)-- (2,3.25);
	\draw (3,-0.25) -- (3,3.25);
	\end{tikzpicture}\hspace{10pt}
	\begin{tikzpicture}[scale=0.75]
	\node at (-0.5,3.5) {\textbf{(b)}};
	\draw (-0.25,0) -- (3.25,0);
	\draw (-0.25,1) -- (1,1);
	\draw (2,1) -- (3.25,1);
	\draw (-0.25,2) -- (3.25,2);
	\draw (-0.25,3) -- (3.25,3);
	\draw (0,-0.25) -- (0,3.25);
	\draw (1,-0.25) -- (1,3.25);
	\draw (2,-0.25) -- (2,1);
	\draw (2,2)-- (2,3.25);
	\draw (3,-0.25) -- (3,3.25);
	\end{tikzpicture}\hspace{10pt}
	\begin{tikzpicture}[scale=0.75]
	\node at (-0.5,3.5) {\textbf{(c)}};
	\draw (-0.25,0) -- (3.25,0);
	\draw (-0.25,1) -- (3.25,1);
	\draw (-0.25,2) -- (0,2);
	\draw (1,2) -- (3.25,2);
	\draw (-0.25,3) -- (3.25,3);
	\draw (0,-0.25) -- (0,3.25);
	\draw (1,-0.25) -- (1,3.25);
	\draw (2,-0.25) -- (2,1);
	\draw (2,2)-- (2,3.25);
	\draw (3,-0.25) -- (3,3.25);
	\end{tikzpicture}
	\caption{\small \label{fig:2Ddefects} \textbf{Defect configurations in a two-dimensional lattice.} \textbf{(a)} A single defect. \textbf{(b)} Two adjacent defects. \textbf{(c)} Two distant defects: This configuration gives rise to a different ground space than two adjacent defects.}
\end{figure}

The Hilbert space $\mathfrak{F}_{\le n}\big(\mathbb{C}^d\big)$ provides the kinematical data to describe a system of at most $n$ defects on a lattice with $n$ sites. We now want to incorporate \emph{dynamics} to this system. In quantum mechanics, this is usually done via a Hamiltonian $H$ such that the time evolution of the system is given by
\begin{equation}
\label{eq:timeev}
U_t=e^{-itH}.
\end{equation}
Hence, introducing dynamics on the lattice system described above requires finding a Hermitian matrix that acts on $\mathfrak{F}_{\le n}\big(\mathbb{C}^d\big)$. Although this sounds simple in the abstract, there is a way to introduce dynamical information more indirectly, which is particularly helpful when considering topologically ordered systems such as the toric code.

Instead of specifying the time evolution operator in \cref{eq:timeev} we introduce dynamical information indirectly by describing the ground space $\mathcal{V}\subset\mathcal{H}$ of a specific Hamiltonian $H$. We show how this works with the example of Kitaev's toric code. Here, we have a two-dimensional lattice on a torus as depicted in \cref{fig:torus}. For simplicity we usually do not draw the torus but a simpler, two-dimensional version of the lattice with periodic boundary conditions that is equivalent to the lattice on the torus, for example a $n\times n$ lattice is depicted

\begin{figure}[H]
	\centering	
	\begin{tikzpicture}[scale=0.65]
	\foreach \a in {0,1,...,7}{
		\draw (\a,0) -- (\a,7);	
		\draw (0,\a) -- (7,\a);			
	}
	\draw[LinkColor, very thick] (0,0) -- (7,0);
	\draw[LinkColor, very thick] (0,7) -- (7,7);
	\draw[LinkColor2, very thick] (0,0) -- (0,7);
	\draw[LinkColor2, very thick] (7,0) -- (7,7);
	\node at (0,-0.5) {\small $(1,1)$};
	\node at (6,-0.5) {\small $(n,1)$};
	\node at (-1,6) {\small $(1,n)$};
	\node at (0,0) {$\bullet$};
	\node at (0,6) {$\bullet$};
	\node at (6,0) {$\bullet$};
	\end{tikzpicture}
\end{figure}

Here, we identify the upper and the lower line (red) and the left and the right line (blue), respectively, to get the same boundary conditions that we have on a torus. The particles we consider are spins that can either be in the up state or in the down state, hence the single-site Hilbert space is $\mathbb{C}^2$. The overall Hilbert space of the system is then
\begin{equation}
\mathcal{H}_\mathbb{T} =\bigoplus_{j=1}^{n^2}\mathbb{C}^2.
\end{equation}
The Hamiltonian in this case consists of two kinds of terms: Vertex operators that apply a Pauli-$Z$ operator to each edge adjacent to a vertex $\mathbf{v}$ and plaquette operators that apply a Pauli-$X$ operator to each edge of a plaquette $\mathbf{p}$:
\begin{equation}
\label{eq:TCham}
H_\mathbb{T}=-\sum_\mathbf{v}
\begin{tikzpicture}[scale=1.4,baseline={([yshift=-3pt]current bounding box.center)}]
\draw (0,0.5) -- (1,0.5);
\draw (0.5,-0) -- (0.5,1);
\draw[fill=white,draw=white] (0.25,0.5) circle (0.08cm);
\draw[fill=white,draw=white] (0.5,0.25) circle (0.11cm);
\draw[fill=white,draw=white] (0.75,0.5) circle (0.08cm);
\draw[fill=white,draw=white] (0.5,0.75) circle (0.11cm);
\node at (0.25,0.5) {\footnotesize $Z$};
\node at (0.5,0.25) {\footnotesize $Z$};
\node at (0.75,0.5) {\footnotesize $Z$};
\node at (0.5,0.75) {\footnotesize $Z$};
\end{tikzpicture}
-\sum_\mathbf{p}
\begin{tikzpicture}[baseline={([yshift=-3pt]current bounding box.center)}]
\draw (0,0) -- (0,1) -- (1,1) -- (1,0) -- cycle;
\draw[fill=white,draw=white] (0.5,0) circle (0.1cm);
\draw[fill=white,draw=white] (0,0.5) circle (0.15cm);
\draw[fill=white,draw=white] (0.5,1) circle (0.1cm);
\draw[fill=white,draw=white] (1,0.5) circle (0.15cm);
\node at (0.5,0) {\footnotesize $X$};
\node at (1,0.5) {\footnotesize $X$};
\node at (0.5,1) {\footnotesize $X$};
\node at (0,0.5) {\footnotesize $X$};
\end{tikzpicture}.
\end{equation}
If we consider a regular $n\times n$ lattice on the torus without any defects, the ground eigenspace $\mathcal{V}_\mathbb{T}$ of this Hamiltonian is four-dimensional.

We can now analyse the ground eigenspace of this system when adding defects to it. If we introduce a single defect to the lattice at site $e$ (as depicted in \cref{fig:2Ddefects}\textbf{(a)}) a short combinatorial calculation shows that the new ground eigenspace $\mathcal{V}_{\mathbb{T}\backslash e}$ is still four-dimensional. However, the problem gets more intricate when introducing more defects to the lattice. With two defects, there are already two different situations one has to consider: The two defects can be adjacent to each other, leading to a single bigger puncture in the lattice (depicted in \cref{fig:2Ddefects}\textbf{(b)}). Here, the ground eigenspace is $\mathbb{C}^2\otimes \mathbb{C}^2$, i.e., we still have a four-dimensional ground space. However, in case the two defects are not adjacent (as depicted in \cref{fig:2Ddefects}\textbf{(c)}) we get the eight-dimensional ground space $\mathbb{C}^2\otimes\mathbb{C}^2\otimes\mathbb{C}^2$.

In general, we do not want to specify how many defects there are but find the ground eigenspace of the system for \emph{up to} $n^2$ defects, i.e., we are interested in finding the ground space of $\mathfrak{F}_{\le n^2}\big(\mathbb{C}^2\big)$:
	\begin{equation}
		\mathcal{V}_{\le n^2}^\mathbb{T}\big(\mathbb{C}^2\big)=\bigoplus_{e_1,e_2,\dots,e_{n^2}=0}^1\mathcal{V}_{e_1,e_2,\dots,e_{n^2}},
	\end{equation}
where $e=0$ means that there is a defect at site $e$ and $e=1$ means that there is a spin at site $e$. $\mathcal{V}_{e_1,e_2,\dots,e_{n^2}}$ is the ground eigenspace for the Hamiltonian \cref{eq:TCham} for the corresponding defect configuration. This is a big direct sum: Writing out the first terms of it yields
	\begin{equation}
	\label{eq:groundspace}
		\mathcal{V}_{\le n^2}^\mathbb{T}\big(\mathbb{C}^2\big)=\big(\mathbb{C}^2\otimes\mathbb{C}^2\big)\oplus\mathbb{C}^{n^2}\otimes\big(\mathbb{C}^2\otimes\mathbb{C}^2\big)\oplus\dots,
	\end{equation}
where the first summand corresponds to the ground eigenspace without defects, the second summand corresponds to the $n^2$ different possibilities of having a single defect and so on. It is obvious that writing out the full direct sum is not easy since figuring out the ground eigenspace for all numbers of defects (with their different configurations) is an intricate combinatorial problem. 

This problem becomes even more intricate when we allow the defects to move (i.e., ``gauge'' the defects). This is needed to get a full dynamical theory where the assumption that we actually can know where the defects are located is possibly unjustified. To account for the effects of such a limited detection ability we impose an equivalence relation on $\mathcal{V}_{\le n}^\mathbb{T}\big(\mathbb{C}^2\big)$ that identifies \emph{physically indistinguishable} defect configurations. Since this is an \emph{operational} notion, there are several possible notions of indistinguishability that we can use to define an equivalence relation. The one we use here is as follows: Suppose we have a state $|\phi\rangle$ of the system with defect configuration $e_1,e_2,\dots,e_{n^2}$ and a second state $|\psi\rangle$ for a defect configuration with the same number of defects but at possibly different locations $e_1',e_2',\dots,e_{n^2}'$. We say that the two states $|\phi\rangle$ and $|\psi\rangle$ are \emph{equivalent} if there is a unitary circuit that transforms $|\phi\rangle$ into $|\psi\rangle$. This requires, in particular, that the spaces $\mathcal{V}_{e_1,e_2,\dots,e_{n^2}}$ and $\mathcal{V}_{e_1',e_2',\dots,e_{n^2}'}$ have the same dimension. This equivalence relation collapses many of the terms in \cref{eq:groundspace}:
	\begin{equation}
		\mathcal{V}_{\le n}^\mathbb{T}\big(\mathbb{C}^2\big)/\tilde=\big(\mathbb{C}^2\otimes\mathbb{C}^2\big)\oplus\big(\mathbb{C}^2\otimes\mathbb{C}^2\big)\oplus\dots,
	\end{equation}
where the $n^2$ possibilities of the single-defect location are collapsed to one possibility since the different configuration are all equivalent.

Of course, the above description is not restricted to the toric code model. However, even for this rather simple two-dimensional system the combinatorial problems that arise when writing out the ground eigenspace are highly complicated. This is why it is helpful to apply methods from category theory to model defects in the system. In the following, we discuss this method via a one-dimensional spin chain with $\mathbf{Vec}(\mathbb{Z}/2\mathbb{Z})$ fusion rules.

\section{A spin chain with defects}
\label{sec:Hilbertspace}

The first step when discussing a physical model is to identify the Hilbert space we are working with. For this purpose, we forget about the category formalism for a moment and simply consider a one-dimensional spin chain of $N$ particles. Each of these particles can be in a spin-up or a spin-down state, hence the Hilbert space of a single particle is $\mathbb{C}^2$. The resulting spin chain\index{spin chain} is then of the form
	\begin{equation}
		\begin{tikzpicture}
			\foreach \a in {0,1,...,6}{
				\node at (\a,0) {\large $\bullet$};	
			}
			\node at (7,0) {$\dots$};
			\node at (8,0) {\large $\bullet$};
			\foreach \a in {0,1,...,6}{
				\node at (\a,-0.5) {$\mathbb{C}^2$};	
			}
			\node at (8,-0.5) {$\mathbb{C}^2$};
		\end{tikzpicture}
	\end{equation}
Since the Hilbert space of a composite system is given by the tensor product of the individual Hilbert spaces, the total Hilbert space of a chain of $N$ spins is
	\begin{equation}
		\mathcal{H}_0=\bigotimes_{i=1}^N\mathbb{C}^2.
	\end{equation}

We now introduce defects to the chain, i.e., particles that behave differently than the spin-up/spin-down particles that are represented by the black bullets above. We begin by putting a single defect, indicated in red, at position $j$ in the chain:
	\begin{equation}
		\begin{tikzpicture}
			\foreach \a in {0,1,2,4,5,6}{
				\node at (\a,0) {\large $\bullet$};	
			}
			\node at (7,0) {$\dots$};
			\node at (3,0) {\large \color{LinkColor} $\bullet$};	
			\node at (8,0) {\large $\bullet$};
			\foreach \a in {0,1,2,4,5,6}{
				\node at (\a,-0.5) {$\mathbb{C}^2$};	
			}
			\node at (8,-0.5) {$\mathbb{C}^2$};
		\end{tikzpicture}
	\end{equation}
The defect can be anything that behaves differently than the particles of the chain, for example a particle with no spin. In this example, there is only one state this particle can be in (the ``no spin'' state), hence the corresponding Hilbert space is $\mathbb{C}$. As a result, the total Hilbert space $\mathcal{H}_1$ of a chain with one defect only differs to $\mathcal{H}_0$ at position $j$, where $\mathbb{C}^2$ is replaced by $\mathbb{C}$:
	\begin{equation}
		\mathcal{H}_1^{(j)}=\left(\bigotimes_{i=1}^{j-1}\mathbb{C}^2\right)\otimes\mathbb{C}\otimes\left(\bigotimes_{i=j+1}^{N}\mathbb{C}^2\right),
	\end{equation}
where the subscript describes the number of defects in the chain and the superscript indicates the position of the defect. It is also possible to consider both possibilities together, having no defect and having a defect at position $j$. This is described by the direct sum of the two Hilbert spaces:
	\begin{equation}
		\Hi=\Hi_0\oplus\Hi_1^{(j)}.
 	\end{equation}
We can generalize this model one step further by not specifying the position of the defect, i.e., allowing the defect to move. In this case, we get an additional sum over all positions $j$:
	\begin{equation}
		\Hi=\Hi_0\oplus\left(\bigoplus_{j=1}^N\Hi_1^{(j)}\right).
	\end{equation}
The above Hilbert space is the most general one for a chain with at most one defect. We can go one step further and allow \emph{two} defects in the chain, say at positions $j$ and $k$:
	\begin{equation}
		\begin{tikzpicture}
			\foreach \a in {0,1,2,4,6}{
				\node at (\a,0) {\large $\bullet$};	
			}
			\node at (7,0) {$\dots$};
			\node at (3,0) {\large \color{LinkColor} $\bullet$};	
			\node at (5,0) {\large \color{LinkColor} $\bullet$};
			\node at (8,0) {\large $\bullet$};
			\foreach \a in {0,1,2,4,6}{
				\node at (\a,-0.5) {$\mathbb{C}^2$};	
			}
			\node at (8,-0.5) {$\mathbb{C}^2$};
		\end{tikzpicture}
	\end{equation}
Similar to the description of a single defect, the Hilbert space of a chain with two defects is given by
	\begin{equation}
		\Hi_2^{(j,k)}=\left(\bigotimes_{i=1}^{j-1}\mathbb{C}^2\right)\otimes\mathbb{C}\otimes\left(\bigotimes_{i=j+1}^{k-1}\mathbb{C}^2\right)\otimes\mathbb{C}\otimes\left(\bigotimes_{i=k+1}^{N}\mathbb{C}^2\right).
	\end{equation}
Analogous to the single-defect case we allow the defects to move and, furthermore, we include the case of having either a single defect or no defect at all. This gives the Hilbert space for a spin chain with \emph{at most} two defects:
	\begin{equation}
		\Hi=\Hi_0\oplus\left(\bigoplus_{j=1}^N\Hi_1^{(j)}\right)\oplus\left(\bigoplus_{j=1}^N\bigoplus_{k\neq j}\Hi_2^{(j,k)}\right).
	\end{equation}
This procedure can be expanded to an arbitrary number of defects in the chain until we have a defect at every position. The Hilbert space in this case is simply $\Hi_N=\bigoplus_{i=1}^N\mathbb{C}$. Finally, we can define the Hilbert space of a chain of $n$ particles with an indefinite\footnote{Since the chain has only $N$ particles, the number of defects is naturally limited to a maximum of $N$.} number of defects that are allowed to move: 
	\begin{equation}
	\label{eq:Hilbertall}
		\Hi=\bigoplus_{n=\#\mathrm{ defects}}\Hi_n,
	\end{equation}
where 
	\begin{equation}
		\Hi_n\equiv\bigoplus_{j=1}^N\bigoplus_{k_1\neq j}\bigoplus_{k_2\neq k_1,j}\dots\bigoplus_{k_n\neq k_1,k_2,\dots,j}\Hi_n^{(j,k_1,k_2,\dots,k_n)}.
	\end{equation}
This Hilbert space is, unfortunately, highly complicated and therefore difficult to work with. However, there is a simpler way to describe it: At each site, the particle can be in one of three states: spin up, spin down, or no spin. Hence, we effectively have a three-level system at each site of the chain and the overall Hilbert space described in \cref{eq:Hilbertall} can be written as
	\begin{equation}
	\label{eq:totalHilb}
		\Hi\cong\bigotimes_{j=1}^N\left(\mathbb{C}\oplus\mathbb{C}^2\right)
	\end{equation}
which is much simpler than the previous expression. This description will be helpful in the following chapters when we study an explicit example of a spin chain with defects, namely a chain built from objects of the $\mathbf{Vec}(\mathbb{Z}/2\mathbb{Z})$ fusion category.

\section{A $\mathbf{Vec}(\mathbb{Z}/2\mathbb{Z})$ spin chain}

The spin chain introduced in the previous section and especially the effects of adding defects to it can be described via the fusion category $\mathbf{Vec}(\mathbb{Z}/2\mathbb{Z})$. First, remember that $\mathbf{Vec}(\mathbb{Z}/2\mathbb{Z})$ has two simple objects: $\Obj(\mathbf{Vec}(\mathbb{Z}/2\mathbb{Z}))=\{0,1\}$, where the vacuum is denoted $0$\footnote{Here, we do not follow our convention of denoting the vacuum $\mathbf{1}$ for convenience to be able to state the fusion rules as addition modulo $2$.} and $1$ is the non-trivial simple object in the category. The fusion rules are given by addition modulo 2, which results in the following fusion table:
	{{\setlength{\tabcolsep}{10pt}
			\renewcommand{\arraystretch}{1.25}
	\begin{table}[H]
		\centering
		\begin{tabular}{c||c|c}
			& 0 & 1 \\\hline\hline
			0 & 0 & 1 \\\hline
			1 & 1 & 0
		\end{tabular}
	\end{table}}
\noindent
Furthermore, the $F$-symbols of the category are trivial.

Let us now study the Hilbert space of the spin chain constructed from $\mathbf{Vec}(\mathbb{Z}/2\mathbb{Z})$\index{Vec(Z/2Z) chain@$\mathbf{Vec}(\mathbb{Z}/2\mathbb{Z})$ chain}. The choice of basis vectors here is very restricted. Suppose we set $a=1$. The resulting chain is\footnote{Note that we have not set the leftmost and the rightmost label to $a=1$ here, but included them into the labels of the basis state.} 
	\begin{figure}[H]
		\centering
		\begin{tikzpicture}[scale=2]
			\draw (0.1,0) -- (4.4,0);
			\draw (0.5,0) -- (0.5,0.5);
			\draw (1,0) -- (1,0.5);
			\draw (1.5,0) -- (1.5,0.5);
			\draw (2,0) -- (2,0.5);
			\draw (2.5,0) -- (2.5,0.5);
			\draw (3,0) -- (3,0.5);
			\draw (3.5,0) -- (3.5,0.5);
			\draw (4,0) -- (4,0.5);
			\node at (0.5,0.6) {$1$};
			\node at (1,0.6) {$1$};
			\node at (1.5,0.6) {$1$};
			\node at (2,0.6) {$1$};
			\node at (2.5,0.6) {$1$};
			\node at (3,0.6) {$1$};
			\node at (3.5,0.6) {$1$};
			\node at (4,0.6) {$1$};
			\node at (0.25,-0.15) {$x_0$};
			\node at (0.75,-0.15) {$x_1$};
			\node at (1.25,-0.15) {$x_2$};
			\node at (1.75,-0.15) {$x_3$};
			\node at (2.25,-0.15) {$x_4$};
			\node at (2.75,-0.15) {$x_5$};
			\node at (3.25,-0.15) {$x_6$};
			\node at (3.75,-0.15) {$x_7$};
			\node at (4.25,-0.15) {$x_8$};
		\end{tikzpicture}
	\end{figure}

There are only two possible basis states for the chain depicted above: $|1,0,1,0,1,0,1,0\rangle$ and $|0,1,0,1,0,1,0,1\rangle$. This is in fact true for any chain of this form, independent of its length, because the fusion rules demand that the $x_i$ are alternating labels. The situation is similar if we set $a=0$. The two possible basis state are (independent of the length of the chain) $|0,0,0,0,\dots\rangle$ and $|1,1,1,1,\dots\rangle$. Hence, for a fixed label $a$, we can interpret the Hilbert space of the chain as that of a qubit, which is $\mathbb{C}^2.$

This situation changes once we allow defects to fuse with the chain. Since the objects in the chain are elements of the category $\mathbf{Vec}(\mathbb{Z}/2\mathbb{Z})$, defects are $\mathbf{Vec}(\mathbb{Z}/2\mathbb{Z})$--$\mathbf{Vec}(\mathbb{Z}/2\mathbb{Z})$ bimodules. In the model we describe here we need the vacuum to occur in the fusion of two defects, which requires that the bimodule we use is invertible. 
Therefore, from the bimodules we have presented in \cref{ex:Vecbimod} we choose the invertible bimodule $F_1$ to model the defects. Recall that $F_1$ has only one simple object, so we generally omit writing a label for the bimodule object but indicate it with a red line. If it is helpful to use a label, we denote it \textasteriskcentered. 

Introducing one defect to the chain as depicted in \cref{fig:defect} does not change the dimension of the Hilbert space. The basis states are still determined by the labels on the boundary. However, fusing more than one defect to the chain opens up new possibilities: Consider a chain with two defects:

	\begin{figure}[H]
		\centering

	\end{figure}
\noindent
Here, the bullets between the two defects (indicated in red) are not determined by the outer labels of the chain since, in general, all of the following vertices are possible:
	\begin{figure}[H]
		\centering
		%
	\end{figure}

Note that once we fix one of the objects represented by the red bullets in the above chain, the fusion rules determine the values of the remaining red bullets, independent of the number of bullets. 
Additional to fusing defects to the chain we can also allow objects from the bimodule to live on the horizontal lines of the chain, i.e., the labels $x_i$ are not only simple objects of the category $\mathbf{Vec}(\mathbb{Z}/2\mathbb{Z})$ but also of the bimodule $F_1$. Consider this situation in a chain where \emph{only} defects are fused to the chain, which is of the form
	\begin{figure}[H]
		\centering
,
	\end{equation}
which means that an object living on a blue line can either be from the category $\mathbf{Vec}(\mathbb{Z}/2\mathbb{Z})$ (black line), which is $0$ or $1$, or from the bimodule $F_1$ (red line), which can only be \textasteriskcentered. This implies that, according to the discussion in \cref{sec:Hilbertspace}, the Hilbert space that corresponds to an orange line is $\mathbb{C}^2\oplus\mathbb{C}\cong\mathbb{C}^3$. Hence, valid configurations of the chain are of the form
	\begin{equation}
	\label{eq:defectstates}
		%
	\end{equation}

The vertices that appear in the above chain are part of the data of the bimodule, see \cref{ex:Vecbimod}. However, a new vertex appears when introducing dynamics to the system. The Hamiltonian we consider is of the same form as the one for the golden chain defined in \cref{eq:localham}, i.e., fusion to the vacuum object $0$ is energetically favoured by assigning an energy of $-1$ to it. The full Hamiltonian is then given by
	\begin{equation}
	\label{eq:defectham}
		H=-\sum_i\frac{1}{\sqrt{2}} 
		%
,
	\end{equation}
where the normalization factor results from the fact that $\dim(\mathrm{\textasteriskcentered})=\sqrt{2}$. This Hamiltonian includes the vertex
	\begin{equation}
	\label{eq:bmvertex}
		%
,
	\end{equation}
which is neither defined in the category $\mathbf{Vec}(\mathbb{Z}/2\mathbb{Z})$ nor in the bimodule $F_1$. Hence, in order to study the action of the Hamiltonian in \cref{eq:defectham} on the chain we need to construct this vertex. In other words, our objective is to build the extended category $\mathbf{Vec}(\mathbb{Z}/2\mathbb{Z})\oplus F_1$.

\subsection*{Construction of the extended category}

To compute the vertex in \cref{eq:bmvertex} we need to construct a basis for the corresponding morphism space for each choice of labels on the vertex. Here, we utilize the so-called \emph{annular category}.

	\begin{defn}[Three-string annular category]
		Let $\mathcal{A}$, $\mathcal{B}$, and $\mathcal{C}$ be fusion categories and $\mathcal{M}$ an $(\mathcal{A},\mathcal{B})$ bimodule, $\mathcal{N}$ a $(\mathcal{B},\mathcal{C})$ bimodule, and $\mathcal{P}$ an $(\mathcal{A},\mathcal{C})$ bimodule. The three-string annular category\index{annular category} $\mathbf{Ann}_{\mathcal{M},\mathcal{N};\mathcal{P}}(\mathcal{A},\mathcal{B},\mathcal{C})$ is defined as follows: Simple objects are triplets $(m,n;p)\in\mathcal{M}\times\mathcal{N}\times\mathcal{P}$. The basis for a morphism space $(m,n;p)\to(m',n';p')$ is given by valid diagrams on the annulus (up to isotopy and local relations), i.e., diagrams of the form
			\begin{equation}
			\label{eq:annbasis}
\ .
			\end{equation}
		For two morphisms $f:(m,n;p)\to(m',n';p')$ and $g:(m',n';p')\to(m'',n'';p'')$, the composition $g\circ f$ corresponds to drawing the diagram $g$ outside of the diagram $f$: 
			\begin{equation}
				%
\ ,
			\end{equation}
		after which, isotopy and the $F$-symbols of the respective fusion categories can be used to reduce the diagram to a sum of diagrams of the form \cref{eq:annbasis}.
	\end{defn}

Note that it is possible to define $n$-string annular categories analogously. Two common examples are the one-string annular category with the string labelled by objects from the category itself, which is known as the \emph{tube algebra}\index{tube algebra} \cite{Ocneanu1994,Evans1995,Evans1998}, and the one-string annular category with the string labelled by an invertible bimodule, which is called \emph{dube algebra}\index{dube algebra} \cite{Williamson2017}. However, for the purpose of constructing bimodule vertices we only need the three-string annular category. The objective in our case is to find representations of the three-string annular category with bimodules $\mathcal{M}=\mathcal{N}=F_1$ and $\mathcal{P}=\mathbf{Vec}(\mathbb{Z}/2\mathbb{Z})$. In the following, we use the notation $\mathbf{Vec}_{\mathbb{Z}_2}$ instead of $\mathbf{Vec}(\mathbb{Z}/2\mathbb{Z})$ for clarity and omit drawing the outer circle of the diagram in \cref{eq:annbasis}. 

To construct vertices in the extended category $\mathbf{Vec}_{\mathbb{Z}_2}$ we use the so-called ``inflation trick'' that was developed in \cite{Bridgeman2019,Bridgeman2020a,Bridgeman2020b}. This is a technique to construct representations of the annular category using primitive idempotents of the category. In our example, we need representations of morphisms in the annular category $\mathbf{Ann}_{F_1,F_1;\mathbf{Vec}_{\mathbb{Z}_2}}(\mathbf{Vec}_{\mathbb{Z}_2},\mathbf{Vec}_{\mathbb{Z}_2},\mathbf{Vec}_{\mathbb{Z}_2})$. More precisely, our goal it to find a basis for each vertex in the extended category $\mathbf{Vec}_{\mathbb{Z}_2}\oplus F_1$ (that is not determined either in $\mathbf{Vec}_{\mathbb{Z}_2}$ or in $F_1$) in terms of annular diagrams. In the following, we explicitly show step-by-step how the vertex in \cref{eq:bmvertex} is constructed using this technique.

\subsection*{Step 1: Compute isomorphism classes of simple objects}

The first step is to identify the isomorphism classes of simple objects in the category and pick a representative for each class. For the category $\mathbf{Ann}_{F_1,F_1;\mathbf{Vec}_{\mathbb{Z}_2}}(\mathbf{Vec}_{\mathbb{Z}_2},\mathbf{Vec}_{\mathbb{Z}_2},\mathbf{Vec}_{\mathbb{Z}_2})$, simple objects are of the form $(\mathrm{\textasteriskcentered},\mathrm{\textasteriskcentered},a)$, hence morphisms are given by diagrams of the form
	\begin{equation}
		\begin{tikzpicture}[scale=1.2,baseline={([yshift=-3pt]current bounding box.center)}]
			\draw (0,0) circle (0.5cm);
			\draw[LinkColor, very thick] (-60:0.5) -- (-60:1.6);
			\draw[LinkColor, very thick] (-120:0.5) -- (-120:1.6);
			\draw ([shift=(-60:1cm)]0,0) arc (-60:90:1cm);
			\draw ([shift=(-120:1.15cm)]0,0) arc (-120:-60:1.15cm);
			\draw ([shift=(240:0.85cm)]0,0) arc (240:90:0.85cm);
			\draw (90:0.5) -- (90:1.6);
			\node at (90:.3cm){\small $a$};
			\node at (90:1.8cm){\small $a+x+z$};
			\node at (-60:.3cm){};
			\node at (-60:1.8cm){};
			\node at (-120:.3cm){};
			\node at (-120:1.8cm){};
			\node at (-1,0.3) {\small $x$};
			\node at (1.15,0.3) {\small $z$};
			\node at (0,-1.3) {\small $y$};
		\end{tikzpicture}\ .
	\end{equation}
Since any morphism between simple objects is an isomorphism, the figure above implies that $(\mathrm{\textasteriskcentered},\mathrm{\textasteriskcentered},a)\cong(\mathrm{\textasteriskcentered},\mathrm{\textasteriskcentered},a+x+z)$ for all possible labels $x,y,z\in\mathbf{Vec}_{\mathbb{Z}_2}$. Hence, there is only one isomorphism class of simple objects and we pick $(\mathrm{\textasteriskcentered},\mathrm{\textasteriskcentered},0)$ as its representative.

\subsection*{Step 2: Find primitive idempotents}

To find a representation of the annular category, we need to find the simple objects in the \emph{Karoubi envelope} (see \cite{barter_domain_2019} for an explanation of this technique). This can be done by constructing the primitive idempotents. In the following, we use the notation $T_iT_j\equiv T_i\circ T_j$.
	\begin{defn}
		A set of primitive idempotents\index{primitive idempotent} is a set of morphisms $\{T_i\}$ which fulfil
			\begin{align}
				T_iT_i&=T_i,\\
				T_iT_j&=0,\\
				T_i&\neq \sum_jf_j,
			\end{align}
		where the $f_j$ are (not necessarily primitive) idempotents, i.e., $f_jf_j=f_j$.
	\end{defn}

This means that we need to find morphisms $T: (\mathrm{\textasteriskcentered},\mathrm{\textasteriskcentered},0)\to(\mathrm{\textasteriskcentered},\mathrm{\textasteriskcentered},0)$ which square to themselves and are orthogonal to each other. Since we furthermore look for \emph{primitive} idempotents, it is also essential that they cannot be written as a sum of idempotents. Candidates for these idempotents are those morphisms where $x+z=0$, since otherwise it is not possible to compose $T$ with itself. Therefore, a morphism is labelled by the two labels $x$ and $y$, hence there are four possible candidates $T_{x,y}$ (note that we do not draw lines when $x,y,z$ are the vacuum object $0$):
	\begin{align}
		T_{0,0}=
.
	\end{align}
	
The first morphism, $T_{0,0}$, can be interpreted as the identity morphism and it obviously squares to itself. $T_{0,1}^2$ and $T_{1,0}^2$ are computed as follows\footnote{Remember that the associator of the category $\mathbf{Vec}_{\mathbb{Z}_2}$ is trivial. Furthermore, the bimodule $F_1$ has a trivial left associator $L$ as a left module category and a trivial right associator $R$ as a right module category. Also, the dimension of all simple objects is one, so bubble popping does not introduce any factors.}:
	\begin{align}
		T_{0,1}^2=
		%
.
	\end{align}
so both of them do not square to themselves but to $T_{0,0}$. The calculation for the fourth candidate is a little more intricate, since this is the only one where the non-trivial associator of $F_1$ appears:
	\begin{align}
		%
.
	\end{align}
The first equality here follows from applying the non-trivial associator of the bimodule $F_1$, which introduces a negative sign to the diagram. As the previous candidates, $T_{1,1}$ also squares to $T_{0,0}$. Hence, none of these candidates is a primitive idempotent. In fact, similar computations show that the general multiplication rule for the $T_{i,j}$ is
	\begin{equation}
	\label{eq:multrule}
		T_{a,b}T_{c,d}=T_{a+c,b+d}.
	\end{equation} 
However, since we have four candidates $T_{i,j}$ for the primitive idempotents, we know that the algebra of primitive idempotents is four-dimensional. In an algebra, it is also possible to consider linear combinations of the candidates. To find suitable linear combinations of the candidates that form a primitive idempotent, it is helpful to work with matrices instead of diagrams on the annulus. More precisely, we need to construct a set of four-dimensional matrices that multiplies in the same way as the four candidates. Since we already know the primitive idempotents for four-dimensional matrices, we can then use these matrices to easily decompose them into primitive idempotents. 

The easiest diagram to translate into a matrix is $T_{0,0}$ since it is the identity. Therefore, the corresponding matrix is 
	\begin{equation}
		M_{0,0}=\begin{pmatrix}
			1 & 0 & 0 & 0\\
			0 & 1 & 0 & 0\\
			0 & 0 & 1 & 0\\
			0 & 0 & 0 & 1
		\end{pmatrix}.
	\end{equation}
We have already seen that it is an idempotent (it squares to itself), but it is not a primitive one since it can be written as a sum of matrices that square to themselves:
	\begin{equation}
	\label{eq:primi}
		M_{0,0}=\begin{pmatrix}
		1 & 0 & 0 & 0\\
		0 & 0 & 0 & 0\\
		0 & 0 & 0 & 0\\
		0 & 0 & 0 & 0
		\end{pmatrix}+\begin{pmatrix}
		0 & 0 & 0 & 0\\
		0 & 1 & 0 & 0\\
		0 & 0 & 0 & 0\\
		0 & 0 & 0 & 0
		\end{pmatrix}+\begin{pmatrix}
		0 & 0 & 0 & 0\\
		0 & 0 & 0 & 0\\
		0 & 0 & 1 & 0\\
		0 & 0 & 0 & 0
		\end{pmatrix}+\begin{pmatrix}
		0 & 0 & 0 & 0\\
		0 & 0 & 0 & 0\\
		0 & 0 & 0 & 0\\
		0 & 0 & 0 & 1
		\end{pmatrix}.
	\end{equation}
Since the matrix that represents the second candidate has to fulfil $M_{0,1}^2=M_{0,0}$ a possible candidate is
	\begin{equation}
		M_{0,1}=\begin{pmatrix}
		1 & 0 & 0 & 0\\
		0 & 1 & 0 & 0\\
		0 & 0 & -1 & 0\\
		0 & 0 & 0 & -1
		\end{pmatrix}.
	\end{equation}
Similar matrices can be found for the remaining two candidates:
	\begin{equation}
		M_{1,0}=\begin{pmatrix}
		-1 & 0 & 0 & 0\\
		0 & 1 & 0 & 0\\
		0 & 0 & -1 & 0\\
		0 & 0 & 0 & 1
		\end{pmatrix},\hspace{20pt}
		M_{1,1}=\begin{pmatrix}
		-1 & 0 & 0 & 0\\
		0 & 1 & 0 & 0\\
		0 & 0 & 1 & 0\\
		0 & 0 & 0 & -1
		\end{pmatrix}.
	\end{equation}
It is straightforward to check that the set of matrices $\{M_{i,j}\}$ fulfils the general multiplication rule for the four candidates \cref{eq:multrule}. We now need to express the four primitive idempotents from \cref{eq:primi} as linear combinations of the $M_{i,j}$ to be able to translate them back into annular diagrams. The general formula for a primitive idempotent is given by
	\begin{equation}
	\label{eq:primidem}
		P_{x,y}=\frac{1}{4}\sum_{a,b}(-1)^{ax+by}M_{a,b}.
	\end{equation}
In terms of annular diagrams, primitive idempotents are hence depicted
	\begin{align}
		P_{x,y}&=\frac{1}{4}\sum_{a,b}(-1)^{ax+by}

		\right).
	\end{align}

\subsection*{Step 3: Check for isomorphism classes of primitive idempotents}

In general, the primitive idempotents we have found in the previous step can be isomorphic to each other. In this case, we have to pick one representative for each isomorphism class. Two idempotents are isomorphic to each other if there are matrices within the algebra that can be multiplied to $P_{x,y}$ to get $P_{x',y'}$. For example, we can get $P_{1,0}$ from $P_{0,1}$ via 
	\begin{equation}
	\label{eq:isomP}
		%
=P_{1,0}.
	\end{equation}
However, the matrices we used here are not elements of the matrix algebra since elements of the algebra only have entries on the diagonal. If the matrices were elements of the algebra then they would form an isomorphism between $P_{0,1}$ and $P_{1,0}$, hence we would pick one of them as a representative of the isomorphism class. However, in our case none of the four primitive idempotents are isomorphic to each other so we do not need to pick representatives, hence we find that there are four isomorphism classes.

In general, this step can me much more complicated if the algebra is higher-dimensional. Fortunately, there is a trick that helps us to see how many isomorphism classes there are, namely the \emph{Artin-Wedderburn theorem}\index{Artin-Wedderburn theorem}\index{Artin-Wedderburn theorem} (see for example \cite{Beachy1999}). It states that every semisimple matrix algebra is isomorphic to a finite direct sum of full matrix algebras, i.e.,
	\begin{equation}
		\mathcal{M}\cong \bigoplus\mathcal{M}_d
	\end{equation}
with $\dim\mathcal{M}_d=d^2$. For each full matrix algebra, we get exactly one primitive idempotent: We can, for example, pick the primitive idempotent $\mathrm{diag}(1,0,0,\dots)$. Every other primitive idempotent is isomorphic to this one since equations like \cref{eq:isomP} always exist within a \emph{full} matrix algebra. 

In our case, the algebra $\mathcal{M}$ is four-dimensional. The two possible decompositions into full matrix algebras are $\mathcal{M}\cong\mathbb{C}\oplus\mathbb{C}\oplus\mathbb{C}\oplus\mathbb{C}$ and $\mathcal{M}\cong\mathcal{M}_2(\mathbb{C})$. The first decomposition is the correct one since we have found four non-isomorphic primitive idempotents by the construction above. In case of a five-dimensional algebra $\mathcal{M}'$, for example, the two possible decompositions are $\mathcal{M}'\cong\mathbb{C}\oplus\mathcal{M}_2(\mathbb{C})$ and $\mathcal{M}'=\mathbb{C}\oplus\mathbb{C}\oplus\mathbb{C}\oplus\mathbb{C}\oplus\mathbb{C}$, hence there are either two or five primitive idempotents. These examples illustrate how using the Artin-Wedderburn theorem can make the task of determining the number of primitive idempotents much easier, since it drastically limits the number of possibilities.

\subsection*{Step 4: Build the full representation}

Remember that the goal of the whole calculation is to find a representation for the vertex
	\begin{equation}
		\begin{tikzpicture}[scale=1,baseline={([yshift=-3pt]current bounding box.center)}]
			\draw[very thick, LinkColor] (0,0) -- (-0.5,-0.5);
			\draw[very thick, LinkColor] (0,0) -- (0.5,-0.5);
			\draw (0,0) -- (0,0.6);
			\node at (0,0.75) {\small $x$};
		\end{tikzpicture},
	\end{equation}
with $x\in\{0,1\}$ in terms of annular diagrams. This corresponds to finding a basis for the morphism space $\Hom\big((\mathrm{\textasteriskcentered},\mathrm{\textasteriskcentered},0),(\mathrm{\textasteriskcentered},\mathrm{\textasteriskcentered},x)\big)$ for every possible $x$.

With the set of primitive idempotents at hand we can now build the full representation of the annular category, which is done by putting all possible annuli on the outside of the idempotents. This procedure allows us to find all basis vectors for the morphism spaces, i.e., all possible vectors of the form\footnote{Note that, in general, we would have to do this for every representative of an isomorphism class of simple objects, but in this example wo only have one.}
	\begin{equation}
	\label{eq:annulivec}
\ .
	\end{equation}
The basis vectors are determined by the choice of labels $\alpha,\beta,\gamma$, hence there are $2^3=8$ possible basis vectors for each vertex. However, it is possible that some of these vectors are linearly dependent, as we will see. We begin by putting the general form of the annuli \cref{eq:annulivec} around the primitive idempotents given in \cref{eq:primidem} and simplify the resulting diagram using the respective associators:
	\begin{align}
		\frac{1}{4}\sum_{a,b}(-1)^{ax+by}
		%
,
	\end{align}
where we have set $\alpha'\equiv\alpha+\gamma$, $a'\equiv a+\beta$, and $b'\equiv b+\gamma$ in the last step. This is a vector in the morphism space $\Hom\big((\mathrm{\textasteriskcentered},\mathrm{\textasteriskcentered},0),(\mathrm{\textasteriskcentered},\mathrm{\textasteriskcentered},\alpha')\big)$, which corresponds to the morphism space 	
	\begin{equation}
		%
.
	\end{equation}
of the extended category $\mathbf{Vec}_{\mathbb{Z}_2}\oplus F_1$. Our goal is now to find a basis for this morphism space for every choice of $\alpha'$. For a fixed representation (i.e., fixed $x,y$) this vector is unique up to a multiplicative scalar (with parameters $\beta,\gamma$) for each choice of $\alpha'\in\{0,1\}$:
	\begin{align}
		%
.
	\end{align}
Therefore, the morphism space is one-dimensional for every $\alpha$ and we define the basis vector to be
	\begin{equation}
		%
.
	\end{equation}
According to the ``inflation trick'' developed in \cite{Bridgeman2019,Bridgeman2020a,Bridgeman2020b} we can pick out the representation $P_{0,0}$ and only work with this one from here on. Hence, the basis vectors we use in the following computations are of the form
	\begin{equation}
		%
.
	\end{equation}

\subsection*{Step 5: Find the associator of the extended category}

After we have found a representation of the vertex, we need to calculate the corresponding $F$-symbols to complete the description of the extended category. Note that here we use the following definition of the $F$-symbols since it is more compatible to the definition of the associators of the module and bimodule categories:
	\begin{equation}
	\label{eq:FsymGD}
		%
.
	\end{equation}
Since we are in the setting of a unitary fusion category here, these are related to the originally defined $F$-symbols by complex conjugation (see \cref{sec:alganyons}).

The data of the category $\mathbf{Vec}_{\mathbb{Z}_2}$ and the bimodule $F_1$ already includes some of the $F$-symbols we need: Within $\mathbf{Vec}_{\mathbb{Z}_2}$, we have that
	\begin{equation}
		\left(F_{abc}^{a+b+c}\right)_{b+c,a+b}=1
	\end{equation}
for $a,b,c\in\mathbf{Vec}_{\mathbb{Z}_2}$. Furthermore, since $F_1$ has trivial associators as a left and right module category, it follows that
	\begin{align}
		\left(F_{ab\mathrm{\textasteriskcentered}}^\mathrm{\textasteriskcentered}\right)_{\mathrm{\textasteriskcentered},a+b}&=1\\
		\left(F_{\mathrm{\textasteriskcentered}ab}^\mathrm{\textasteriskcentered}\right)_{a+b,\mathrm{\textasteriskcentered}}&=1.
	\end{align}
Finally, the associator of the bimodule is given by
	\begin{equation}
		\left(F_{a\mathrm{\textasteriskcentered}b}^\mathrm{\textasteriskcentered}\right)_{\mathrm{\textasteriskcentered},\mathrm{\textasteriskcentered}}=(-1)^{ab}.
	\end{equation}
However, there are still some $F$-symbols that are not determined by the present data, namely
	\begin{equation}
		\left(F_{a\mathrm{\textasteriskcentered}\mathrm{\textasteriskcentered}}^{a+b}\right)_{b,\mathrm{\textasteriskcentered}},\hspace{10pt}
		\left(F_{\mathrm{\textasteriskcentered}a\mathrm{\textasteriskcentered}}^b\right)_{\mathrm{\textasteriskcentered},\mathrm{\textasteriskcentered}},\hspace{10pt}
		\left(F_{\mathrm{\textasteriskcentered}\mathrm{\textasteriskcentered}a}^{a+b}\right)_{\mathrm{\textasteriskcentered},b},\hspace{10pt}
		\left(F_{\mathrm{\textasteriskcentered}\mathrm{\textasteriskcentered}\mathrm{\textasteriskcentered}}^\mathrm{\textasteriskcentered}\right)_{b,a}.
	\end{equation}

To compute these matrices we use the vertex normalization introduced in \cref{eq:vertexnorm}, hence we have the relation
	\begin{equation}
.
	\end{equation}
For the black string in our diagrams, i.e., those strings that belong to the category $\mathbf{Vec}_{\mathbb{Z}_2}$, the sum on the right hand side has only one term and the coefficients are given by the dimensions $d_0=d_1=1$, which yields the relation
	\begin{equation}
	\label{eq:normGD}
		%
.
	\end{equation}
Furthermore we also have the following relations including the bimodule:
	\begin{equation}
		%
.
	\end{equation}

The computation of the $F$-symbol $F_{\mathrm{\textasteriskcentered}\mathrm{\textasteriskcentered}a}^{a+b}$ is straightforward: We simply insert the representation of the vertex we just found into the diagram and use those associators that are already known to simplify the diagram:
	\begin{align}
		%
,
	\end{align}
hence the corresponding $F$-matrix is 
	\begin{equation}
		\left(F_{\mathrm{\textasteriskcentered}\mathrm{\textasteriskcentered}a}^{a+b}\right)_{\mathrm{\textasteriskcentered},b}=1.
	\end{equation}
With similar calculations we get the two additional $F$-symbols:
	\begin{align}
		\left(F_{a\mathrm{\textasteriskcentered}\mathrm{\textasteriskcentered}}^{a+b}\right)_{b,\mathrm{\textasteriskcentered}}&=1,\\
		\left(F_{\mathrm{\textasteriskcentered}a\mathrm{\textasteriskcentered}}^{b}\right)_{\mathrm{\textasteriskcentered},\mathrm{\textasteriskcentered}}&=(-1)^{ab}.
	\end{align}
It remains to calculate the matrix elements of $F_{\mathrm{\textasteriskcentered}\mathrm{\textasteriskcentered}\mathrm{\textasteriskcentered}}^\mathrm{\textasteriskcentered}$. From the definition of the $F$-symbols \cref{eq:FsymGD}, we get 
	\begin{equation}
.
	\end{equation}
Using the same approach as before to calculate $F_{\mathrm{\textasteriskcentered}\mathrm{\textasteriskcentered}\mathrm{\textasteriskcentered}}^\mathrm{\textasteriskcentered}$ is circular, so we need a different technique here. Consider applying the four-string annulus given by
	\begin{equation}
		%
	\end{equation}
to both sides of the equation. The left hand side becomes
	\begin{align}
		%
.
	\end{align}
The diagrams on the right hand side can be simplified analogously. The resulting equation is
	\begin{align}
		&(-1)^{(a+x_0)(x_1+x_3)}%
.
	\end{align}
We can now also apply an $F$-move to the left hand side of the above equation, which yields
	\begin{align}
		&(-1)^{(a+x_0)(x_1+x_3)}%
\right).
	\end{align}
Comparing the coefficients of the first diagram on the right hand side in the two equations above yields
	\begin{align}
		(-1)^{(x_1+x_3)x_2}\left(F_{\mathrm{\textasteriskcentered}\mathrm{\textasteriskcentered}\mathrm{\textasteriskcentered}}^\mathrm{\textasteriskcentered}\right)_{0,a}&=(-1)^{(a+x_0)(x_1+x_3)}\left(F_{\mathrm{\textasteriskcentered}\mathrm{\textasteriskcentered}\mathrm{\textasteriskcentered}}^\mathrm{\textasteriskcentered}\right)_{x_1+x_3,a+x_0+x_2}\\
		\Rightarrow \left(F_{\mathrm{\textasteriskcentered}\mathrm{\textasteriskcentered}\mathrm{\textasteriskcentered}}^\mathrm{\textasteriskcentered}\right)_{0,a}&=(-1)^{(a+x_0+x_2)(x_1+x_3)}\left(F_{\mathrm{\textasteriskcentered}\mathrm{\textasteriskcentered}\mathrm{\textasteriskcentered}}^\mathrm{\textasteriskcentered}\right)_{x_1+x_3,a+x_0+x_2}.
	\end{align}
For $a=0$ and by setting $a'\equiv x_0+x_2$ and $b'=x_1+x_3$ we arrive at the expression
	\begin{equation}
		\left(F_{\mathrm{\textasteriskcentered}\mathrm{\textasteriskcentered}\mathrm{\textasteriskcentered}}^\mathrm{\textasteriskcentered}\right)_{0,0}=(-1)^{a'b'}\left(F_{\mathrm{\textasteriskcentered}\mathrm{\textasteriskcentered}\mathrm{\textasteriskcentered}}^\mathrm{\textasteriskcentered}\right)_{b',a'}.
	\end{equation}
The value of $\left(F_{\mathrm{\textasteriskcentered}\mathrm{\textasteriskcentered}\mathrm{\textasteriskcentered}}^\mathrm{\textasteriskcentered}\right)_{0,0}$ is fixed by our choice of normalization \cref{eq:normGD} to (compare to \cref{eq:FrobSchur2})
	\begin{equation}
		\left(F_{\mathrm{\textasteriskcentered}\mathrm{\textasteriskcentered}\mathrm{\textasteriskcentered}}^\mathrm{\textasteriskcentered}\right)_{0,0}=\frac{\kappa_\mathrm{\textasteriskcentered}}{\sqrt{2}},
	\end{equation}
where $\kappa_\mathrm{\textasteriskcentered}=\pm 1$ is the Frobenius-Schur indicator of the object \textasteriskcentered. Hence, the matrix elements of the $F$-symbol in question are given by
	\begin{equation}
		\left(F_{\mathrm{\textasteriskcentered}\mathrm{\textasteriskcentered}\mathrm{\textasteriskcentered}}^\mathrm{\textasteriskcentered}\right)_{b,a}=(-1)^{ab}\frac{\kappa_\mathrm{\textasteriskcentered}}{\sqrt{2}}.
	\end{equation}

With the full set of $F$-symbols at hand, we now have to check whether they fulfil the pentagon equation. As a matter of fact, the $F$-symbols we have found coincide with those of the Ising category (compare to \cref{ex:Ising}) and hence fulfil the pentagon equation.

\subsection*{The Hamiltonian}

After the construction of the extended category $\mathbf{Vec}_{\mathbb{Z}_1}\oplus F_1$ we are now prepared to study the action of the Hamiltonian 
	\begin{equation}
		H=-\sum_i\frac{1}{\sqrt{2}} 

	\end{equation}
on the defect chain. We first rewrite the Hamiltonian using the identity
	\begin{equation}
		%
\ 
		\right),
	\end{align}
where we used the $F$-symbols from \cref{ex:Vecbimod}. Let us now analyse how the Hamiltonian acts locally on three sites. Remember from \cref{sec:Hilbertspace} that the Hilbert space for one site is $\mathbb{C}^2\oplus\mathbb{C}$, hence the Hilbert space for three sites is
	\begin{equation}
		\left(\mathbb{C}^2\oplus\mathbb{C}\right)\otimes\left(\mathbb{C}^2\oplus\mathbb{C}\right)\otimes\left(\mathbb{C}^2\oplus\mathbb{C}\right)=\left(\mathbb{C}^2\otimes\mathbb{C}\otimes\mathbb{C}^2\right)\oplus\left(\mathbb{C}\otimes\mathbb{C}^2\otimes\mathbb{C}\right)\oplus \text{forbidden states},
	\end{equation}
where the forbidden states are those that are not of the form \cref{eq:defectstates} which includes states where all the objects on all three sites are from the bimodule. Hence, there are two valid local configurations of the chain:
	\begin{equation}
		\begin{tikzpicture}[baseline={([yshift=-3pt]current bounding box.center)}]
			\draw[very thick, color=LinkColor] (0,0) -- (1,0);
			\draw (-0.5,0) to node[below] {$a$} (0,0);
			\draw (1,0) to node[below] {$b$} (1.5,0);
			\draw[very thick, color=LinkColor] (0,0) -- (0,1);
			\draw[very thick, color=LinkColor] (1,0) -- (1,1);
		\end{tikzpicture}\hspace{20pt}\mathrm{and}\hspace{20pt}
		\begin{tikzpicture}[baseline={([yshift=-3pt]current bounding box.center)}]
			\draw (0,0) to node[below] {$a$} (1,0);
			\draw[very thick, color=LinkColor] (-0.5,0) -- (0,0);
			\draw[very thick, color=LinkColor] (1,0) -- (1.5,0);
			\draw[very thick, color=LinkColor] (0,0) -- (0,1);
			\draw[very thick, color=LinkColor] (1,0) -- (1,1);
		\end{tikzpicture},
	\end{equation}
where $a,b\in\{0,1\}$. We are interested in a formulation of the local Hamiltonian as an operator acting on a general basis state $|x_{i-1},x_i,x_{i+1}\rangle\in\left(\mathbb{C}^2\oplus\mathbb{C}\right)\otimes\left(\mathbb{C}^2\oplus\mathbb{C}\right)\otimes\left(\mathbb{C}^2\oplus\mathbb{C}\right)$, i.e., a state of the form
	\begin{equation}
		|x_{i-1},x_i,x_{i+1}\rangle=\begin{tikzpicture}[baseline={([yshift=-3pt]current bounding box.center)}]
			\draw[very thick, color=LinkColor] (0,0) -- (0,1);
			\draw[very thick, color=LinkColor] (1,0) -- (1,1);
			\draw[very thick, color=LinkColor2] (-0.5,0) -- (1.5,0);
			\node at (-0.25,-0.25) {\small $x_{i-1}$};
			\node at (0.5,-0.25) {\small $x_{i}$};
			\node at (1.25,-0.25) {\small $x_{i+1}$};
		\end{tikzpicture}
	\end{equation}
with $x_{i-1},x_i,x_{i+1}\in\{0,1,\mathrm{\textasteriskcentered}\}$. The Hamiltonian consists of two terms: One that acts as the identity and a non-trivial term. Therefore, we know that an operator that describes the action of the local Hamiltonian $H_i$ is of the form
	\begin{equation}
		H_i|x_{i-1},x_i,x_{i+1}\rangle=-\frac{1}{\sqrt{2}}\left(\mathbb{I}+A\right)|x_{i-1},x_i,x_{i+1}\rangle
	\end{equation}
with some operator $A$. Note that we do not need to worry about how the operator acts on forbidden states such as $|\mathrm{\textasteriskcentered},\mathrm{\textasteriskcentered},\mathrm{\textasteriskcentered}\rangle$ or $|\mathrm{\textasteriskcentered},\mathrm{\textasteriskcentered},0\rangle$, for example. To get an expression for the operator $A$, we need to study how the second, non-trivial term acts on the two valid configurations:
	\begin{enumerate}
		\item The first configuration includes all states of the form $|a,\mathrm{\textasteriskcentered},b\rangle$. Therefore, the object in the middle, $x_i$, is projected onto the state $|\mathrm{\textasteriskcentered}\rangle\langle\mathrm{\textasteriskcentered}|$. The action of the Hamiltonian on the objects $x_{i-1}=a$ and $x_{i+1}=b$ is as follows:
			\begin{equation}
			\label{eq:action1}
				\begin{tikzpicture}[baseline={([yshift=-3pt]current bounding box.center)}]
					\draw[very thick, color=LinkColor] (0,0) -- (1,0);
					\draw (-0.5,0) to node[below] {$a$} (0,0);
					\draw (1,0) to node[below] {$b$} (1.5,0);
					\draw (0,0.5) to node[above] {$1$} (1,0.5);
					\draw[very thick, color=LinkColor] (0,0) -- (0,1);
					\draw[very thick, color=LinkColor] (1,0) -- (1,1);
				\end{tikzpicture}=\begin{tikzpicture}[baseline={([yshift=-3pt]current bounding box.center)}]
					\draw[very thick, color=LinkColor] (0,0) -- (1,0);
					\draw (-0.5,0) to node[below] {$a$} (0,0);
					\draw (1,0) to node[below] {$b$} (1.5,0);
					\draw (0,0.5) to [bend left] node[above] {$1$} (0.35,0);
					\draw (1,0.5) to [bend right] node[above] {$1$} (0.65,0);
					\draw[very thick, color=LinkColor] (0,0) -- (0,1);
					\draw[very thick, color=LinkColor] (1,0) -- (1,1);
				\end{tikzpicture}=(-1)^{a+b}\begin{tikzpicture}[baseline={([yshift=-3pt]current bounding box.center)}]
					\draw[very thick, color=LinkColor] (0,0) -- (1,0);
					\draw (-0.5,0) to node[below] {$a$} (0,0);
					\draw (1,0) to node[below] {$b$} (1.5,0);
					\draw[very thick, color=LinkColor] (0,0) -- (0,1);
					\draw[very thick, color=LinkColor] (1,0) -- (1,1);
				\end{tikzpicture}\equiv Z_a\otimes Z_b,
			\end{equation}
		where $Z_a,Z_b$ are Pauli-$Z$ operators acting on the $\mathbb{C}^2$ part of the Hilbert space. This does not specify the action of the Hamiltonian on the $\mathbb{C}$ part of the Hilbert space of the two outer labels. However, since states of the form $|\mathrm{\textasteriskcentered},\mathrm{\textasteriskcentered},b\rangle$ and $|a,\mathrm{\textasteriskcentered},\mathrm{\textasteriskcentered}\rangle$ are forbidden anyway, we do not need to specify the action of the Hamiltonian on them and simply use $0$ as a placeholder. As a result, for this configuration the operator $A$ is defined as follows:
			\begin{equation}
				A=\left(Z\oplus 0\right)\otimes|\mathrm{\textasteriskcentered}\rangle\langle\mathrm{\textasteriskcentered}|\otimes \left(Z\oplus 0\right).
			\end{equation}
		\item The situation is similar for the second configuration, which consists of states of the form $|\mathrm{\textasteriskcentered},a,\mathrm{\textasteriskcentered}\rangle$. Hence, the outer sites are projected onto $|\mathrm{\textasteriskcentered}\rangle\langle\mathrm{\textasteriskcentered}|$ while the second term of the Hamiltonian acts as a Pauli-$X$ operator on the middle object:
			\begin{equation}
			\label{eq:action2}
				\begin{tikzpicture}[baseline={([yshift=-3pt]current bounding box.center)}]
					\draw (0,0) to node[below] {$a$} (1,0);
					\draw (0,0.5) to node[above] {$1$} (1,0.5);
					\draw[very thick, color=LinkColor] (-0.5,0) -- (0,0);
					\draw[very thick, color=LinkColor] (1,0) -- (1.5,0);
					\draw[very thick, color=LinkColor] (0,0) -- (0,1);
					\draw[very thick, color=LinkColor] (1,0) -- (1,1);
				\end{tikzpicture}=\begin{tikzpicture}[baseline={([yshift=-3pt]current bounding box.center)}]
					\draw (0,0) to node[below] {$a+1$} (1,0);
					\draw[very thick, color=LinkColor] (-0.5,0) -- (0,0);
					\draw[very thick, color=LinkColor] (1,0) -- (1.5,0);
					\draw[very thick, color=LinkColor] (0,0) -- (0,1);
					\draw[very thick, color=LinkColor] (1,0) -- (1,1);
				\end{tikzpicture}\equiv X_a.
			\end{equation}
		With the same argumentation as above, the resulting operator $A$ for this configuration is
			\begin{equation}
				A=|\mathrm{\textasteriskcentered}\rangle\langle\mathrm{\textasteriskcentered}|\otimes \left(X\oplus 0\right)\otimes |\mathrm{\textasteriskcentered}\rangle\langle\mathrm{\textasteriskcentered}|.
			\end{equation}
	\end{enumerate}

In summary, the local Hamiltonian is given by
	\begin{equation}
		H_i=-\frac{1}{\sqrt{2}}\Big(\mathbb{I}+\left(Z\oplus 0\right)\otimes|\mathrm{\textasteriskcentered}\rangle\langle\mathrm{\textasteriskcentered}|\otimes \left(Z\oplus 0\right)+|\mathrm{\textasteriskcentered}\rangle\langle\mathrm{\textasteriskcentered}|\otimes \left(X\oplus 0\right)\otimes |\mathrm{\textasteriskcentered}\rangle\langle\mathrm{\textasteriskcentered}|\Big).
	\end{equation}
For a chain of a fixed length $N$, there are exactly two possible configurations the states of the chain can be in, which are determined by fixing one of the labels, for example the leftmost one. For instance, a chain with $N=9$ can be in the following configuration:
	\begin{align}
,
	\end{align}
which effectively is a system of five qubits:
	\begin{equation}
		%
	\end{equation}		
On this Hilbert space, i.e., the Hilbert space of five qubits, the effective Hamiltonian is
	\begin{equation}
		H_\mathrm{eff}=-\frac{1}{\sqrt{2}}\sum_{i\in\text{half chain}}\mathbb{I}+Z_iZ_{i+1}+X_i,
	\end{equation}
which is exactly the Hamiltonian that corresponds to the transverse field Ising model\index{transverse field Ising model}. The second possible configuration for this chain is 
	\begin{align}
		%
,
	\end{align}
which corresponds to a chain of four qubits, hence the resulting effective Hamiltonian is the same (with the only difference that it is acting on four instead of five qubits in our example). As a result, if we do not fix any labels we get the direct sum of two copies of the transverse field Ising Hamiltonian. 

In general, an important requirement for the construction of a defect chain is that the category $\C$ has an invertible $\C$--$\C$ bimodule to model the defects. Unfortunately, this is not the case for the Haagerup fusion categories $\Hi_i$: There are only trivial $\Hi_i$--$\Hi_i$ bimodules \cite{grossman_quantum_2012}. These cannot be used for constructing a defect chain since in this case, the resulting defect chain is simply the same as the chain without defects. 

However, constructing the tube algebra $\mathcal{A}$ (i.e., the one-string annular category) of a fusion category $\C$ turns out to be useful in order to construct the Drinfeld centre $\mathcal{Z}(\C)$ of the category. The reason behind this is that finite dimensional irreducible representations of $\mathcal{A}$ are in one-to-one correspondence with simple objects of $\mathcal{Z}(\C)$ (see \cite{izumi_structure_2000} and \cite{muger_subfactors_2003-1} and also \cite{Jones2016}). Hence, the methods for constructing vertices in the annular category presented in this chapter can be used to construct the quantum double of a fusion category. This is even applicable to the Haagerup fusion categories since it is not necessary to include bimodules. For the construction of the tube algebra we only need objects from the category itself. However, this construction can become very tedious if the category is complex, which is the reason why this approach has not yet been pursued for the Haagerup fusion categories.

\chapter{String-net models for unitary fusion categories}
\label{ch:LW}

In the study of microscopic models for fusion categories we are not limited to one-dimensional systems. There is also a prominent example of a two-dimensional lattice model for the study of topologically ordered systems, namely the \emph{Levin-Wen model} \cite{Levin2005}. Although it is possible to define this model for any lattice form, it is usually discussed for the honeycomb lattice, which goes well with the structure of fusion categories since it only consists of trivalent vertices. The Levin-Wen model itself yields a Topological Quantum Field Theory (TQFT), but it is also a powerful tool in our search for a Conformal Field Theory (CFT) corresponding to the Haagerup subfactor.  Firstly, it requires as input a unitary fusion category, which is exactly the kind of category we get from a subfactor. Secondly, the quantum double of the category (which is the most promising candidate for finding a CFT) arises naturally in terms of the excitations of the Hamiltonian of the Levin-Wen model. Hence, calculating the excitations of the model yields a unitary modular tensor category, which, in turn, can then be studied in terms of anyon chains in order to find the corresponding CFT (as explained in \cref{sec:anyons}).

Even though this approach seams promising, there is one caveat: In the original paper by Levin and Wen the category does not only need to be a unitary fusion category, but the $F$-symbols of the category have to fulfil an additional constraint called \emph{tetrahedral symmetry}. However, not every unitary fusion category does fulfil this constraint. Several counterexamples are known (see for example \cite{Hong2009}), most importantly the Haagerup fusion category $\Hd$. However, the tetrahedral symmetry condition is not crucial for the model to work which was pointed out for example in \cite{Hong2009}, hence it is possible to construct a Hamiltonian in the sense of Levin and Wen from a unitary fusion category that does not fulfil tetrahedral symmetry.

In this chapter, we first elaborate on the tetrahedral symmetry condition and the connection of the Levin-Wen model to Turaev-Viro state sums, a mathematical theory that is equivalent to the Levin-Wen model but naturally circumvents the problem of tetrahedral symmetry. We then present the Levin-Wen model, first in its original form where tetrahedral symmetry holds before we show the construction without this constraint in detail. We also prove that the resulting Hamiltonian has the same properties as the original one and, furthermore, that it can be transformed into the original one when imposing tetrahedral symmetry. We then discuss how the excitations of the model can be computed and why this is especially difficult in the general approach where tetrahedral symmetry does not hold. Here, once again tensor networks will come into play as a practical tool for the investigation of physical systems. We show how to express the ground state of the Levin-Wen model as a tensor network for fusion categories without tetrahedral symmetry. In the end, we elaborate on possible generalisations of the model, namely either losing the constraint that the fusion rules are multiplicity-free or the constraint that the fusion category is unitary. While the first one is a true (but tedious) generalisation of the construction, the second one leads to unphysical models.

\newpage
\section{Tetrahedral symmetry}
\label{sec:tetrahedral}

The tetrahedral symmetry condition\index{tetrahedral symmetry} is an additional constraint on the $F$-symbols of the fusion category which was required to be fulfilled in the original paper by Levin and Wen \cite{Levin2005}. It is motivated by the symmetries that arise for classical $6j$-symbols, see for example \cite{Roberts1999}. It is derived by evaluating diagrams that correspond to a tetrahedral string-net configuration. Consider the following diagram:
	\begin{align}
\\
	&=\left(F_k^{j^*i^*l^*}\right)_{nm}\sqrt{\frac{d_j d_k}{d_n}}\sqrt{\frac{d_id_l}{d_n}}d_n\\
	&=\sqrt{d_id_jd_kd_l}\ \left(F_k^{j^*i^*l^*}\right)_{nm}.
	\end{align}
Imagine now that the above diagram lies on a sphere. In this case, topological invariance and parity invariance (two aspects required for the Levin-Wen model) require that the value of the diagram is invariant under all $24$ symmetries (both rotational and reflection symmetries and their combinations) of the tetrahedron
	\begin{equation}
		%
	\caption{\small \label{fig:tetrahedron}\textbf{Symmetries of the tetrahedron.} These configurations are related by tetrahedral symmetry transformations: \textbf{(b)} is obtained from \textbf{(a)} by a reflection at the plane that connects the line $n$ with the centre of $m$. \textbf{(c)} is obtained from \textbf{(a)} by a reflection at the plane that connects the line $m$ with the centre of $n$. \textbf{(d)} is obtained from \textbf{(a)} by a reflection at the plane that connects the line $i$ with the centre of $k$.}
\end{figure}
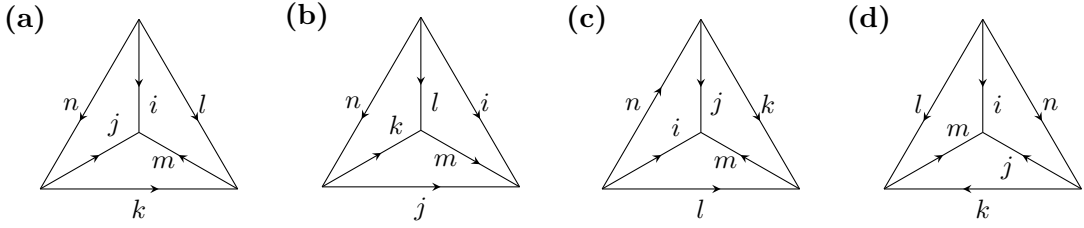

Three results of these symmetry transformations are depicted in \cref{fig:tetrahedron}. The corresponding values of these diagrams are
	\begin{align}
		\mathbf{(b)}&=\sqrt{d_id_jd_kd_l}\left(F_{j^*}^{k^*l^*i^*}\right)_{nm^*}\\
		\mathbf{(c)}&=\sqrt{d_id_jd_kd_l}\left(F_l^{i^*j^*k^*}\right)_{n^*m}\\
		\mathbf{(d)}&=\sqrt{d_id_kd_md_n}\left(F_{k^*}^{m^*i^*n^*}\right)_{lj}.
	\end{align}
Since the value of the diagram is required to be invariant under all these symmetry transformations we get the following equation:
	\begin{equation}
		\left(F_k^{j^*i^*l^*}\right)_{nm}=\left(F_{j^*}^{k^*l^*i^*}\right)_{nm^*}=\left(F_l^{i^*j^*k^*}\right)_{n^*m}=\sqrt{\frac{d_md_n}{d_ld_j}}\left(F_{k^*}^{m^*i^*n^*}\right)_{lj}.
	\end{equation}
With a slight renaming of the labels we arrive at the tetrahedral symmetry condition:
	\begin{equation}
		\label{eq:tetrahedralsymmetry}
		\left(F_i^{jkl}\right)_{nm}=\left(F_{l^*}^{kji^*}\right)_{n^*m}=\left(F_{j^*}^{i^*lk}\right)_{nm^*}=\sqrt{\frac{d_md_n}{d_ld_j}}\left(F_{i^*}^{m^*kn^*}\right)_{l^*j^*}.
	\end{equation}

However, not every Unitary Fusion Category (UFC) fulfils the tetrahedral symmetry condition. One example for a UFC that does not fulfil this symmetry is the Haagerup fusion category $\Hd$: Consider, for example, the following transformation:
	\begin{equation}
		\begin{tikzpicture}[baseline=(current bounding box.center),yscale=0.9,xscale=0.9]
			\draw (0,0.5) -- (0,1);
			\draw (0,1) -- (-0.5,1.5);
			\draw (-0.5,1.5) -- (-1,2);
			\draw (-0.5,1.5) -- (0,2);
			\draw (0,1) -- (0.5,1.5);
			\draw (0.5,1.5) -- (1,2);
			\node at (0,0.25) {\small $\rho$};
			\node at (-1,2.25) {\small $\mathbf{1}$};
			\node at (0,2.25) {\small $\alpha$};
			\node at (1,2.25) {\small $\alphastarrho$};
			\node at (-0.5,1.1) {\small $\alpha$};
		\end{tikzpicture}= \left(F_\rho^{\mathbf{1}\alpha\alphastarrho}\right)_{\rho\alpha}
		\begin{tikzpicture}[baseline=(current bounding box.center),yscale=0.9,xscale=0.9]
			\draw (0,0.5) -- (0,1);
			\draw (0,1) -- (-0.5,1.5);
			\draw (-0.5,1.5) -- (-1,2);
			\draw (0.5,1.5) -- (0,2);
			\draw (0,1) -- (0.5,1.5);
			\draw (0.5,1.5) -- (1,2);
			\node at (0,0.25) {\small $\rho$};
			\node at (-1,2.25) {\small $\mathbf{1}$};
			\node at (0,2.25) {\small $\alpha$};
			\node at (1,2.25) {\small $\alphastarrho$};
			\node at (0.5,1.1) {\small $\rho$};
		\end{tikzpicture}.
	\end{equation}
According to \cref{eq:tetrahedralsymmetry}, we get the following identities:
	\begin{equation}
		\left(F_\rho^{\mathbf{1}\alpha\alphastarrho}\right)_{\rho\alpha}=\left(F_{\alphastarrho}^{\alpha\mathbf{1}\rho}\right)_{\rho\alpha}=\left(F_\mathbf{1}^{\rho\alphastarrho\alpha}\right)_{\rho\alpha^*}=\left(F_\rho^{\alpha^*\alpha\rho}\right)_{\alphastarrho\mathbf{1}}.
	\end{equation}
However, only the first and the last $F$-symbol coincide (they are both equal to one, see \cref{app:Fsymbols}). The second and third $F$-symbol are not valid: $\left(F_{\alphastarrho}^{\alpha\mathbf{1}\rho}\right)_{\rho\alpha}$ corresponds to a transformation between zero-dimensional vector spaces and $\left(F_\mathbf{1}^{\rho\alphastarrho\alpha}\right)_{\rho\alpha^*}$ has the wrong outer labels, although the vector space is one-dimensional. Due to the different dimensions of the vector spaces, it is not possible to find a gauge transformation that can be applied to the $F$-symbols in order to fulfil the tetrahedral symmetry condition.

Therefore, it has been questioned whether the tetrahedral symmetry condition is necessary for the Levin-Wen model, for example in \cite{Hung2012} and \cite{Lin2014}. The string-net construction in the sense of Levin and Wen is furthermore closely related to so-called Turaev-Viro state sums \cite{Turaev1992a}: It was shown, for example, in \cite{Kadar2010,Koenig2010,Kirillov2011} (and revisited more recently in \cite{Barkeshli2019a,Bauer2019}), that the string-net space of the Levin-Wen construction for a unitary fusion category $\C$ is equal to the state space of a \emph{Turaev-Viro} topological quantum field theory for $\C$. This itself is isomorphic to the state space of the \emph{Reshetikin-Turaev} theory for the Drinfeld double of $\C$, see for example \cite{Kirillov2010,Turaev2010a,Balsam2010}. In both of these theories tetrahedral symmetry does not play any role. While we do not go into detail about these theories here, this observation gives a strong hint that tetrahedral symmetry is not a necessary condition for the construction. In fact, in \cite{Hong2009} it was shown that \emph{any} unitary fusion category yields an exactly solvable Hamiltonian via the string-net construction. 

However, the literature of Turaev-Viro state sums does not provide an explicit way to construct the Hamiltonian from a category the way it was done by Levin and Wen. Especially for physicists who do not necessarily have a solid background in category theory or topology, this lack of a constructive way of defining the Hamiltonian is a serious hurdle when using the Levin-Wen approach for arbitrary UFCs. Therefore, although solved in principle, the construction of the Levin-Wen Hamiltonian without tetrahedral symmetry was still noted as an open problem in the physics literature, for example in \cite{Hung2012}. In \cite{Lin2014}, the authors attacked this problem in the case of fusion categories where the objects form an abelian group under the fusion operation. In \cite{Lan2014}, the authors discuss how to calculate the quasiparticle statistics of the excitations of the model, but do not provide a construction of either the Hamiltonian or the excitations themselves.

Therefore, in the following we describe in detail how the Hamiltonian can be constructed without imposing the tetrahedral symmetry condition on the $F$-symbols, hence making it applicable to all unitary fusion categories.

\section{The Levin-Wen model}

\index{Levin-Wen model}
In their seminal paper \cite{Levin2005}, Levin and Wen explain how topological phases emerge from microscopic degrees of freedom of a physical system. In their string-net model, the universal properties of such a phase are described by the ground state wave function, which is determined from local constraints. The underlying mathematical framework of this theory is that of unitary fusion categories. In the original paper, these categories have to fulfil some additional constraints, among them the tetrahedral symmetry condition introduced in the previous section. In this section we present the basics of the string-net model and explain how the Hamiltonian is constructed. 

As the name suggests, sting-net models are networks of strings of different types\index{string-net}. We focus here on networks that are built from trivalent vertices, which means that at each node exactly three strings meet. However, not all combinations of string types are allowed to meet at a node since the model is restricted by the data of the underlying fusion category. In general, one needs to input the following data to define a string-net model:
	\begin{enumerate}
		\item \textbf{String types.} We need to specify which types of strings (i.e., labels) are allowed in the model. They are given by the simple objects of the underlying category. We usually label different string types with integers: $i=0,1,2,\dots,N$, where $0$ represents the vacuum or unit object\footnote{This notation is different to the category theory language, where we always use $\mathbf{1}$ to indicate the unit object. However, in the string-net literature it is common to use $0$ for the vacuum string.}. The number of simple objects of the category also yields the total number of different string types $N+1$.
		\item \textbf{Branching rules.} It is necessary to specify which string types are allowed to meet at a vertex to identify the allowed string-net configurations:
			\begin{equation}
				\begin{tikzpicture}[scale=0.75,baseline=(current bounding box.center)]
					\draw[->-,>=stealth] (90:1) -- (0,0);
					\draw[->-,>=stealth] (210:1) -- (0,0);
					\draw[->-,>=stealth] (330:1) -- (0,0);
					\node at (90:1.25) {\small $i$};
					\node at (210:1.25) {\small $k$};
					\node at (330:1.25) {\small $j$};
				\end{tikzpicture}.
			\end{equation}
		These rules can be obtained from the fusion rules of the underlying UFC: For instance, the above vertex is an allowed configuration if $N_{kj}^{i^*}>0$. To every string-type $i$ we associate a complex number $d_i$ which is given by the quantum dimension of the simple object in the category.
		\item \textbf{String orientations.} Every string has an arrow to indicate its direction. With every string type $i$ we associate a \emph{dual} string type $i^*$ such that $(i^*)^*=i$. This is possible because a UFC is rigid an therefore has left and right duals. A string of type $i^*$ corresponds to a type-$i$ string with opposite direction:
			\begin{equation}
.
			\end{equation}
	\end{enumerate}
After specifying this data, defining the Hilbert space of the model is straightforward: The states are given by linear combinations of different  spatial configurations of string-nets.

\begin{rem}
	Before describing the Hamiltonian of the model, some remarks on the graphical notation are in order: 
	\begin{enumerate}
		\item First, we usually use dashed lines to draw a string that corresponds to the vacuum and omit the label $0$:
			\begin{equation}
				%
\ .
			\end{equation}
		In this case, it is also not necessary to indicate the direction of the string since the unit object is its own dual.
		\item For clarity we sometimes draw diagrams without indicating the direction of the arrows. In this case, the convention is that all arrows point upwards, for example
			\begin{equation}
				%
.
			\end{equation}
		\item We use the same normalisation of vertices as discussed in \cref{eq:vertexnorm}. This implies that in graphical calculations we can use the completeness relation \cref{eq:identity} and the bigon relation \cref{eq:bigon2}.
		\item We do not use horizontal lines in the diagrams in this chapter (in contrast to the original paper \cite{Levin2005}). The reason behind this is that the goal of this chapter is to do a construction of the model without imposing tetrahedral symmetry, and without this condition the meaning of horizontal lines is ambiguous. Hence, throughout this chapter we make an effort to translate all diagrams from the original paper to ones without horizontal lines.
	\end{enumerate}	
\end{rem}

In \cite{Levin2005}, the $F$-symbols of the category are required to fulfil several conditions:
	\begin{align}
		\left(F_i^{ijj^*}\right)_{0k}&=\sqrt{\frac{d_k}{d_i d_j}}\ \delta_{ik}^k \label{eq:LWcond1}\\
		\left(F_i^{jkl}\right)_{nm}=\left(F_{l^*}^{kji^*}\right)_{n^*m}&=\left(F_{j^*}^{i^*lk}\right)_{nm^*}=\sqrt{\frac{d_md_n}{d_ld_j}}\left(F_{i^*}^{m^*kn^*}\right)_{l^*j^*} \label{eq:LWcond2}\\
		\left(F_u^{abr}\right)_{sp}\left(F_u^{pcd}\right)_{rq}&=\sum_t \left(F_s^{bcd}\right)_{rt}\left(F_u^{atd}\right)_{sq}\left(F_q^{abc}\right)_{tp} \label{eq:LWcond3}\\
		\overline{\left(F_i^{jkl}\right)}_{nm}&=\left(F_{i^*}^{j^*k^*l^*}\right)_{n^*m^*} \label{eq:LWcond4},
	\end{align}
where  
	\begin{equation}
	\label{eq:delta}
		\delta_{ij}^k=\begin{cases}
		1, & i\otimes j=k \text{ is an allowed fusion}\\
		0, &\mathrm{otherwise.}
		\end{cases}
	\end{equation}
The first one is a normalisation condition similar to the one we derived in \cref{eq:Fidentities}, but a bit more restrictive. The second one is the tetrahedral symmetry condition we derived in \cref{sec:tetrahedral}. The third one is simply the pentagon equation \cref{eq:pentagon}. The last one is a unitarity condition. For the purpose of introducing the Levin-Wen string-net model in this section we assume all of these conditions to be true. In the next section, we explain how to do the construction without imposing all of these conditions\footnote{Note that the pentagon equation always has to be fulfilled since the input category is a unitary fusion category, hence we also require it in the general case.}.

With the definition of the string-net Hilbert space at hand we can write down a Hamiltonian on this space. In general, a string-net Hamiltonian can be any local operator acting on string-net states. The form of this Hamiltonian is determined by certain local relations on its ground state wave function $\Phi$ which arise from scale-invariance, which means the following: We expect that two states within the same topological phase to look the same at long distances. That is, the two wave functions will only differ in short distance details. For instance, at long length scales, a local bubble is irrelevant and will simply look like a string: 
	\begin{equation}
		\Phi\left(

		\right)
	\end{equation}
However, if $i\neq j$, the configuration on the right hand side is invalid, which yields
	\begin{equation}
		\Phi\left(
		%
		\right)=0.
	\end{equation}
As a result, we conclude that the amplitude vanishes if $i\neq j$. Similar considerations yield the following set of local relations: 
	\begin{align}
		\Phi\left(
		%
		\right). \label{eq:groundstate5}
	\end{align}
The grey rectangles represent the remaining parts of the string-net configuration that is left unchanged by the local relations. Note that we have slightly changed the relation in contrast to the original formulation to avoid horizontal lines. In \cite{Levin2005}, the relations \cref{eq:groundstate4} and \cref{eq:groundstate5} are summarized as
	\begin{equation}
	\label{eq:normalcond}
		\Phi\left(
		%
		\right).
	\end{equation}
The operators $G$ and $H$ in \cref{eq:groundstate4} and \cref{eq:groundstate5} are related to the $F$-symbols of the underlying fusion category. They are computed by applying the completeness relation \cref{eq:identity} to the diagrams on both sides. For instance, the computation of $H$ works as follows:
	\begin{align}
		%
,
	\end{align}
which yields
	\begin{equation}
	\label{eq:Hop}
		\left(H_{ij}^{kl}\right)_{nm}=\sqrt{\frac{d_m d_n}{d_i d_l}}\left(F_n^{kmj}\right)_{li}.
	\end{equation}
A similar calculation for $G$ yields
	\begin{equation}
	\label{eq:Gop}
		\left(G_{ij}^{kl}\right)_{nm}=\sqrt{\frac{d_m d_n}{d_j d_k}}\overline{\left(F_n^{iml}\right)}_{kj}.
	\end{equation}
Note that imposing tetrahedral symmetry on the $F$-symbols implies that the vertices in \cref{eq:groundstate4} and \cref{eq:groundstate5} can be moved up and down arbitrarily, i.e.
	\begin{equation}
		%
.
	\end{equation}
This can also be seen by manipulating the operators $G$ and $H$ using the conditions listed in \cref{eq:LWcond1,eq:LWcond2,eq:LWcond3,eq:LWcond4}:
	\begin{align}
		\left(G_{ij}^{kl}\right)_{nm}&=\sqrt{\frac{d_m d_n}{d_j d_k}}\overline{\left(F_n^{iml}\right)}_{kj}\\
		&=\sqrt{\frac{d_m d_n}{d_j d_k}}\left(F_{n^*}^{i^*m^*l^*}\right)_{k^*j^*}\\
		&=\sqrt{\frac{d_m d_n}{d_j d_k}}\sqrt{\frac{d_j d_k}{d_l d_i}}\left(F_n^{km^* j}\right)_{li}\\
		&=\left(H_{ij}^{kl}\right)_{nm^*}.
	\end{align}
Hence, if we impose tetrahedral symmetry on the $F$-symbols, the operators $G$ and $H$ are equal. Moreover, the conditions \cref{eq:groundstate4} and \cref{eq:groundstate5} are then equal to the original relation \cref{eq:normalcond} in the form
	\begin{equation}
		\Phi\left(
		\begin{tikzpicture}[baseline={([yshift=-3pt]current bounding box.center)}, decoration={markings,mark=at position .6 with {\arrow[>=stealth]{>}}},scale=0.6]
		\draw[{postaction=decorate}] (0.8,0.5) to node[below] {\small $i$} (1.5,1);
		\draw[{postaction=decorate}] (1.5,1) to node[above] {\small $k$} (0.8,1.5);
		\draw[{postaction=decorate}] (3.2,0.5) to node[below] {\small $j$} (2.5,1);
		\draw[{postaction=decorate}] (2.5,1) to node[above] {\small $l$} (3.2,1.5);
		\draw[{postaction=decorate}] (2.5,1) to node[above] {\small $m$} (1.5,1);
		\draw[fill=LightGray,rounded corners,LightGray] (-0.2,0) rectangle (0.8,2);
		\draw[fill=LightGray,rounded corners,LightGray] (3.2,0) rectangle (4.2,2);
		\end{tikzpicture}\right) 
		=\sum_n \left(F_j^{i^*kl}\right)_{nm}\ \Phi\left(
		\begin{tikzpicture}[baseline={([yshift=-3pt]current bounding box.center)}, decoration={markings,mark=at position .6 with {\arrow[>=stealth]{>}}},scale=0.6]
		\draw[{postaction=decorate}] (0.8,0.5) to node[below] {\small $i$} (2,0.5);
		\draw[{postaction=decorate}] (3.2,0.5) to node[below] {\small $j$} (2,0.5);
		\draw[{postaction=decorate}] (2,1.5) to node[above] {\small $k$} (0.8,1.5);
		\draw[{postaction=decorate}] (2,1.5) to node[above] {\small $l$} (3.2,1.5);
		\draw[{postaction=decorate}] (2,0.5) to node[right] {\small $n$} (2,1.5);
		\draw[fill=LightGray,rounded corners,LightGray] (-0.2,0) rectangle (0.8,2);
		\draw[fill=LightGray,rounded corners,LightGray] (3.2,0) rectangle (4.2,2);
		\end{tikzpicture}
		\right).,
	\end{equation}
which can be seen by applying the tetrahedral symmetry condition \cref{eq:LWcond2} twice:
	\begin{align}
		\left(F_j^{i^*kl}\right)_{nm}&=\sqrt{\frac{d_m d_n}{d_l d_i}}\left(F_{j^*}^{m* k n^*}\right)_{l^* i}\\
		&=\sqrt{\frac{d_m d_n}{d_l d_i}} \left(F_n^{km^* j}\right)_{li}\\
		&=\left(H_{ij}^{kl}\right)_{nm^*}.
	\end{align}
As a result, we can say that the conditions \cref{eq:groundstate4} and \cref{eq:groundstate5} are indeed a true generalization of the original condition \cref{eq:normalcond} by Levin and Wen and replacing their condition does not have an effect on the model in case we assume tetrahedral symmetry.

After having specified the local constraints that uniquely specify the ground state we can construct the exactly solvable lattice Hamiltonian that has exactly these states as ground states. We concentrate here on the honeycomb lattice where the degrees of freedom are on the edges, which goes well with the constraint of using only trivalent vertices. The Hamiltonian consists of two types of operators: ones that act on vertices $\mathbf{v}$ and ones that act on plaquettes $\mathbf{p}$. 

The vertex operators\index{vertex operator} (also called \emph{electric charge operators}\index{electric charge operator}) are denoted $Q_\mathbf{v}$. They always act on three degrees of freedom that meet at a vertex (see \cref{fig:honeycomb}) and they ensure that the ground state of the Hamiltonian only consists of configurations that are allowed by the branching rules:
	\begin{equation}
		Q_\mathbf{v}\Bigg|

	\caption{\small \textbf{Hamiltonian on the honeycomb lattice.} A general Hamiltonian consists of operators that act on vertices $Q_\mathbf{v}$ and ones that act on plaquettes $B_\mathbf{p}$.\label{fig:honeycomb}}
\end{figure}

The plaquette operator\index{plaquette operator} $B_\mathbf{p}$ (also called \emph{magnetic flux operator}\index{magnetic flux operator}) imposes dynamics to the system. It acts on all twelve degrees of freedom of a plaquette (see \cref{fig:honeycomb}). It is a linear combination of $N+1$ terms, one term for each string type:
	\begin{equation}
		B_\mathbf{p}=\sum_{s=0}^N a_s B_\mathbf{p}^s,
	\end{equation}
with $a_s=\frac{d_s}{D^2}$, where $D^2=\sum_{i=0}^N d_i^2$ is the total quantum dimension\footnote{In general, the coefficients $a_s$ only have to fulfil the condition $a_{s^*}=a_s^*$ but are otherwise arbitrary. However, the authors of \cite{Levin2005} find that the choice we make here corresponds to a smooth continuum limit where the ground state is topologically invariant. Furthermore, this choice is a requirement for the operators $Q_\mathbf{v}$ and $B_\mathbf{p}$ to be projection operators, hence we adapt it here.}. Each ot the individual terms $B_\mathbf{p}^s$ acts on the plaquette $\mathbf{p}$ by inserting a loop of string type $s$ and fusing this loop into the internal links of the plaquette. Hence, the operator only changes the configuration of internal edges from $g,h,i,j,k,l$ to $g',h',i',j',k',l'$ but leaves the outer strings unchanged: 
	\begin{equation}
	\label{eq:Bcoeff}
		B_\mathbf{p}^s\ \Bigg|

		\Bigg\rangle.
	\end{equation}
As shown in \cite{Levin2005}, the coefficients $\left(B_\mathbf{p}^s\right)_{ghijkl}^{g'h'i'j'k'l'}$ are given by 
	\begin{equation}
	\label{eq:LWoriginal}
		\left(B_\mathbf{p}^s\right)_{ghijkl}^{g'h'i'j'k'l'}=\left(F_{s^*}^{la^*g'^*}\right)_{l'^* g}\left(F_{s^*}^{gb^*h'^*}\right)_{g'^* h}\left(F_{s^*}^{hc^*i'^*}\right)_{h'^* i}\left(F_{s^*}^{id^*j'^*}\right)_{i'^* j}\left(F_{s^*}^{j^*e^*k'^*}\right)_{j'^* k}\left(F_{s^*}^{kf^*l'^*}\right)_{k'^* l}.
	\end{equation}
The exactly solvable Hamiltonian on the honeycomb lattice is then given by the sum of all $Q_\mathbf{v}$ and $B_\mathbf{p}$ operators:
	\begin{equation}
		H=-\sum_\mathbf{v}Q_\mathbf{v}-\sum_\mathbf{p}B_\mathbf{p},
	\end{equation}	
where the negative sign ensures that those string-net configurations that obey the local constraints \cref{eq:groundstate1,eq:groundstate2,eq:groundstate3,eq:groundstate4,eq:groundstate5} are energetically favoured and therefore in the ground state. In this form, the Hamiltonian can be constructed for any unitary fusion category that fulfils the conditions \cref{eq:LWcond1,eq:LWcond2,eq:LWcond3,eq:LWcond4}. However, as shown in \cref{sec:tetrahedral}, the category we are mostly interested in here (the Haagerup fusion category $\Hd$) does not fulfil the tetrahedral symmetry condition. Therefore, it is necessary to derive an expression for the Hamiltonian without using this constraint in order to make the model applicable to \emph{any} unitary fusion category.

\section{Hamiltonian without tetrahedral symmetry}

We now go into detail about the construction of the string-net model without imposing the tetrahedral symmetry condition on the $F$-symbols and show all necessary calculations for the model in detail. More precisely, we do not require that the conditions \cref{eq:LWcond1}, \cref{eq:LWcond2} and \cref{eq:LWcond4} are fulfilled. We follow the construction presented in \cite{Hahn2020}.

We have seen above that the Hamiltonian for the Levin-Wen model consists of two types of terms: The operators $Q_\mathbf{v}$ act on the vertices $\mathbf{v}$ and ensure that the ground state of the model only consists of configurations that obey the fusion rules of the category. These operators do not involve any $F$-symbols, hence they do not change by relaxing the tetrahedral symmetry condition. The plaquette operator $B_\mathbf{p}$, however, does change, since we apply $F$-moves to fuse a loop into the string-net configuration. We go through the construction of this operator in detail. Recall that a plaquette operator $B_\mathbf{p}$ acting on a fixed plaquette $\mathbf{p}$ is of the form:
	\begin{equation}
		B_\mathbf{p}=\sum_{s=0}^Na_s B_\mathbf{p}^s,
	\end{equation}
where we choose the coefficients to be $a_s=\frac{d_s}{D^2}$ with $D^2=\sum_{i=0}^Nd_i^2$. To get an expression for the coefficients $\left(B_\mathbf{p}^s\right)_{ghijkl}^{g'h'i'j'k'l'}$ defined in \cref{eq:Bcoeff} we have to insert a loop into the plaquette and fuse it to the internal edges. The first step in this calculation is to apply the completeness relation \cref{eq:identity} at every internal link:

\begin{align}
	&

	\label{eq:bigdiagram}
\end{align}

We can now simplify each of the six corners individually:
	\begin{enumerate}
		\item[\itemone] To be able to apply an $F$-move to the first diagram, we need to add a vacuum line\footnote{In fact, this vacuum line is present in this diagram anyway since the operation corresponds to the evaluation of $s^*$ and $s$ as given in \cref{eq:evaluationcat}. We usually omit the vacuum lines for convenience.} (indicated by the dashed line) here:
			\begin{equation}
				%
.
			\end{equation}
		Now we can apply a series of $F$-moves and other simplifications such as the bigon relation \cref{eq:bigon2} to the diagram on the right-hand side to simplify it to a trivalent vertex:
			\begin{align}
				%
.
			\end{align}
		\item[\itemtwo] After slightly redrawing the second diagram, we can directly apply a $G$-move to simplify it:
			\begin{align}
				%
.
			\end{align}
		\item[\itemthree] Slightly redrawing the third diagram allows us to apply an $H$-move:
			\begin{align}
				%
.
			\end{align}
		\item[\itemfour] The situation here is similar to the first diagram. To be able to apply an $F$-move, we first need to insert a vacuum line\footnote{Similar to diagram \itemone, the vacuum line is here due to the coevaluation of $s^*$ and $s$ (see \cref{eq:coevaluationcat}).}:
			\begin{equation}
				%
.
			\end{equation}
		Then we apply a series of $F$-moves and other simplifications to simplify the diagram:
			\begin{align}
				%
.
			\end{align}
		\item[\itemfive] Here, we can directly apply a $G$-move to the diagram after slightly redrawing it:
			\begin{align}
				%
.
			\end{align}
		\item[\itemsix] The last diagram can be simplified by applying an $H$-move:
			\begin{align}
				%
.
			\end{align}
	\end{enumerate}

Inserting the above simplification into \cref{eq:bigdiagram} gives an expression for the diagram with the loop in terms of diagrams without a loop:
	\begin{align}
		&%
.
	\end{align}
From this equation we can conclude that the matrix elements of the plaquette operator $B_\mathbf{p}$ are given by
	\begin{align}
		\left(B_{\mathbf{p}}^s\right)_{ghijkl}^{g'h'i'j'k'l'}(abcdef)&=\sqrt{d_s d_{s^*}} \sqrt{\frac{d_{g^*} d_h d_{i'} d_j d_{k^*} d_{l'^*}}{d_{g'^*} d_{h'} d_i d_{j'} d_{k'^*} d_{l^*}}}\left(F_{g^*}^{g^*s^* s}\right)_{0g'^*}\overline{\left(F_{b^*}^{g'^*sh}\right)}_{h' g^*}\overline{\left(F_{i'}^{shc^*}\right)}_{ih'}\\&\hspace{20pt}\left(F_{i'}^{sjd}\right)_{ij'}\left(F_j^{s^*sj}\right)_{j' 0} \overline{\left(F_e^{k^*s^*j'}\right)}_{j k'^*} \overline{\left(F_{l'^*}^{fk^*s^*}\right)}_{k'^*l^*}\left(F_{l'^*}^{a^*g^*s^*}\right)_{g'^*l^*}.\label{eq:Bop}
	\end{align}
This formula is clearly more complicated than the original one, since it has additional factors with quantum dimensions and more $F$-moves. Moreover, the symmetry that was present in the labelling of the $F$-symbols in \cref{eq:LWoriginal} is not present in this formula. However, it can be shown that by imposing the conditions \cref{eq:LWcond1,eq:LWcond2,eq:LWcond3,eq:LWcond4} the formula \cref{eq:Bop} can be transformed into the original one \cref{eq:LWoriginal}, see \cref{sec:originalHam}.

\section{Properties of the Hamiltonian}

The Hamiltonian constructed above has some important properties, similar to the one that has been constructed by Levin and Wen in the original paper \cite{Levin2005}:
	\begin{enumerate}
		\item It is hermitian, which is crucial for the Hamiltonian to describe an actual physical system.
		\item It is a projector for the parameter choice $a_s=\frac{d_s}{D^2}$.
		\item The operators $B_\mathbf{p}$ and $Q_\mathbf{v}$ all commute with each other when applied to two different plaquettes, which makes the Hamiltonian exactly solvable.
	\end{enumerate}
We show each of these statements in the following. Note that the first part of the Hamiltonian, the vertex operator $Q_\mathbf{v}$ trivially fulfils these properties: It is hermitian since the individual operators are one-dimensional real values. It is a projector because applying it twice to a vertex does not change the validity of the vertex, and by the same argument it also commutes with itself. Hence, we only have to prove these statements for the plaquette operator $B_\mathbf{p}$.

\begin{thm}
	The Hamiltonian $H=\sum_\mathbf{v}Q_\mathbf{v}+\sum_\mathbf{p}B_\mathbf{p}$ with $B_\mathbf{p}$ given by \cref{eq:Bop} is hermitian.
\end{thm}
\begin{proof}
	Hermicity is a crucial property for the Hamiltonian in order to describe a physical system. To show that the Hamiltonian constructed above is hermitian, first recall that the operator $B_\mathbf{p}$ maps the string-net configuration $\vec{p}\equiv\{g,h,i,j,k,l\}$ of a plaquette $\mathbf{p}$ to another configuration $\vec{p}\,'\equiv\{g',h',i',j',k',l'\}$ via
		\begin{equation}
			\begin{tikzpicture}[scale=0.6,baseline={([yshift=-0.1cm]current bounding box.center)}]
				\draw[->-] (90:1) -- (150:1);
				\draw[->-] (150:1) -- (210:1);
				\draw[->-] (210:1) -- (270:1);
				\draw[->-] (270:1) -- (330:1);
				\draw[->-] (330:1) -- (30:1);
				\draw[->-] (30:1) -- (90:1);
			\foreach \a in {30,90,150,210,270,330} \draw[-<-] (\a:1) -- (\a:1.5);
			\node at (0,0) {$\vec{p}$};
			\end{tikzpicture}\mapsto\sum_{\vec{p}\,'}C(\vec{p},\vec{p}\,')
			\begin{tikzpicture}[scale=0.6,baseline={([yshift=-0.1cm]current bounding box.center)}]
				\draw[->-] (90:1) -- (150:1);
				\draw[->-] (150:1) -- (210:1);
				\draw[->-] (210:1) -- (270:1);
				\draw[->-] (270:1) -- (330:1);
				\draw[->-] (330:1) -- (30:1);
				\draw[->-] (30:1) -- (90:1);
				\foreach \a in {30,90,150,210,270,330} \draw[-<-] (\a:1) -- (\a:1.5);
				\node at (0,0) {$\vec{p}\,'$};
			\end{tikzpicture}.
		\end{equation}
	Here, $C(\vec{p},\vec{p}\,')$ denotes the matrix elements of $B_\mathbf{p}$ derived in \cref{eq:Bcoeff}. Note that this is a function of \emph{all} $12$ labels of each plaquette, but we only indicate those that are changed by the transformation ($\vec{p}$ and $\vec{p}\,'$) since the outer labels are not affected by the transformation. Hence, in order to show that the operator $B_\mathbf{p}$ is hermitian we need to show that
		\begin{equation}
			C(\vec{p},\vec{p}\,')=\overline{C(\vec{p}\,',\vec{p})}.
		\end{equation}
	This equality can be shown using the graphical calculus of fusion categories. As described in \cref{rem:Felements}, the matrix element of a transformation between diagrams that belong to two different orthonormal bases can be calculated by computing the inner product of the initial string diagram with the final string diagram. In the case of hexagonal plaquettes, an orthonormal basis is given by
		\begin{equation}
			\left\{\frac{1}{N_{\vec{p}}}\ 

			\right\},
		\end{equation}
	where $N_{\vec{p}}$ is a normalisation factor that ensures that the diagrams are normalised with respect to the trace. In the following we use the notation $|\vec{p}\rangle$ to refer to the diagram above (not including the normalisation factor), i.e., an element of the above set of basis states is then denoted $\frac{1}{N_{\vec{p}}}|\vec{p}\rangle$. We want to compute the matrix element that corresponds to the transformation $|\vec{p}\rangle\mapsto|\vec{p}\,'\rangle$:
		\begin{equation}
		\label{eq:plaquettetrafo}
			%
.
		\end{equation}
	In order to calculate the inner product between the two basis vectors, we need to take the dual vector $\langle \vec{p}\,'|$ of the final string diagram, which is
		\begin{equation}
			\langle\vec{p}\,'|=
			%
,
		\end{equation}
	where we have used that a string with label $i^*$ corresponds to a string with label $i$ with the arrow pointing in the other direction. Hence, we can calculate $C(\vec{p},\vec{p}\,')$ via the inner product of the two plaquettes:
		\begin{equation}
			C(\vec{p},\vec{p}\,')=\frac{1}{N_{\vec{p}}N_{\vec{p}\,'}}\langle\vec{p}\,'|\vec{p}\rangle=\frac{1}{N_{\vec{p}}N_{\vec{p}\,'}}
			%
\ .
		\end{equation}
	The calculation of $C(\vec{p}\,',\vec{p})$ works analogously, simply by computing the inner product $\langle\vec{p}\ |\vec{p}\,'\rangle$:
		\begin{equation}
		\label{eq:Ctwo}
			C(\vec{p}\,',\vec{p})=\frac{1}{N_{\vec{p}}N_{\vec{p}\,'}}\langle\vec{p}\ |\vec{p}\,'\rangle=\frac{1}{N_{\vec{p}}N_{\vec{p}\,'}}
			%
\ .
		\end{equation}
	The complex conjugate of $C(\vec{p}\,',\vec{p})$ can then be computed by reflecting the diagram at the horizontal axis (and taking the duals of the objects):
		\begin{equation}
			\overline{C(\vec{p}\,',\vec{p})}=\frac{1}{N_{\vec{p}}N_{\vec{p}\,'}}
			%
=C(\vec{p},\vec{p}\,'),
		\end{equation}	
	which completes the proof.

\end{proof}

\begin{thm}
	The Hamiltonian $H=\sum_\mathbf{v}Q_\mathbf{v}+\sum_\mathbf{p}B_\mathbf{p}$ with $B_\mathbf{p}$ given by \cref{eq:Bop} is a projector. 
\end{thm}

\begin{figure}[t]
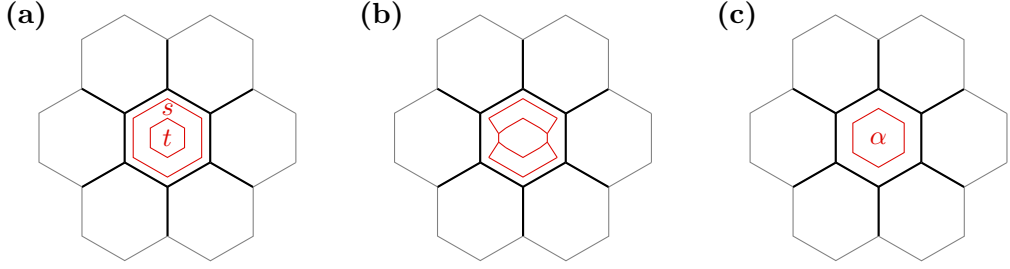

	\centering
	%
	\caption{\small \label{fig:proj}\textbf{Projector.} \textbf{(a)} Applying the operator $B_\mathbf{p}$ twice to the same plaquette inserts two loops, one of type $s$ and one of type $t$ (keep in mind that we sum over $s$ and $t$ which is not depicted in the picture). \textbf{(b)} The two loops are fused together. \textbf{(c)} The result is (a sum over) a single loop of type $\alpha$.}
\end{figure}

\begin{proof}
	The structure of this proof is as follows: Applying the operator $B_\mathbf{p}$ twice corresponds to acting simultaneously with $B_\mathbf{p}=\sum_t a_t B_\mathbf{p}^t$ and $B_\mathbf{p}=\sum_s a_s B_\mathbf{p}^s$ on the plaquette $\mathbf{p}$. The corresponding joint operator is $B_\mathbf{p}^2=\sum_{s,t} a_s a_t B_\mathbf{p}^s B_\mathbf{p}^t$ and for $B_\mathbf{p}$ to be a projector we need to show that this is equal to $\sum_\alpha B_\mathbf{p}^\alpha$. Graphically, $B_\mathbf{p}^2$ acts by adding type $s$ and type $t$ loops to the plaquette and summing over $s,t$ as depicted in \cref{fig:proj}. To keep the following computation as clear as possible we only depict the loops inside the plaquette and not the plaquette itself. Similar to the derivation of the Hamiltonian we have to insert vacuum lines in order to apply $F$-moves. The evaluation of the diagram is then a special case of \cref{eq:bigdiagram} with all outer labels being the vacuum. Since the category is spherical, it holds that $d_{a^*}=d_a$, which we use in the following calculation.
		\begin{align}
			\label{eq:Proj1}
			B_\mathbf{p}^2&=\sum_{s,t} \frac{d_sd_t}{D^4}
\\
			&=B_\mathbf{p}.
		\end{align}
	
	Let us explain some aspects of this computation in more detail. In \cref{eq:Proj2} the multiplicity factor $N_{st}^\alpha N_{s^*t^*}^{\beta^*}$ is included to ensure that the sum only involves terms with valid configurations. In \cref{eq:Proj3} we have used the formulas derived for \itemone\ and \itemfour\ in the special case where the outer labels are the vacuum. To arrive at the terms in \cref{eq:Proj4} we use the fact that $\left(F_{s^*}^{s^*t^*t}\right)_{0\beta^*}\overline{\left(F_0^{\beta^*ts}\right)}_{\alpha s^*}=\overline{\left(F_s^{t^*ts}\right)}_{\alpha 0} \left(F_0^{s^*t^*\alpha}\right)_{s\beta^*}$, which is a consequence of the pentagon equation\footnote{It is also possible to derive this equation by fusing the vacuum string to the right side (i.e., to the string labelled $h$) in the diagram in \itemone, which yields the expression on the right hand side of the equation.}. Furthermore, after carrying out the $\delta_{\alpha,\beta}$ we use the fact that $N_{s^*t^*}^{\alpha^*}=N_{st}^\alpha$, which is true because both correspond to the dimension of the vector space that consists of vertices
		\begin{equation}
			\begin{tikzpicture}
				\draw (90:0) -- (90:0.5);
				\draw (220:0) -- (220:0.5);
				\draw (320:0) -- (320:0.5);
				\node at (90:0.75) {\small $\alpha$};
				\node at (220:0.75) {\small $s$};
				\node at (320:0.75) {\small $t$};
			\end{tikzpicture}
		\end{equation}
	simply with different directions of the arrows on the edges. In \cref{eq:Proj5} we then exploit that we are only working in the multiplicity-free case where $N_{st}^\alpha\in\{0,1\}$, hence $(N_{st}^\alpha)^2=N_{st}^\alpha$. In this step we also used the relations \cref{eq:Fidentities} and \cref{eq:Funitary} since all $F$-symbols are one-dimensional. In \cref{eq:Proj6} we have used the relation \cref{eq:Nidentities} to replace $N_{st}^\alpha$ with $N_{t\alpha^*}^{s^*}$. In \cref{eq:Proj7} we have rearranged the factors and used the fact that $d_s=d_{s^*}$ such that we can apply \cref{thm:qdimrelation} to arrive at the expression in \cref{eq:Proj8}.
\end{proof}

\begin{figure}[t]
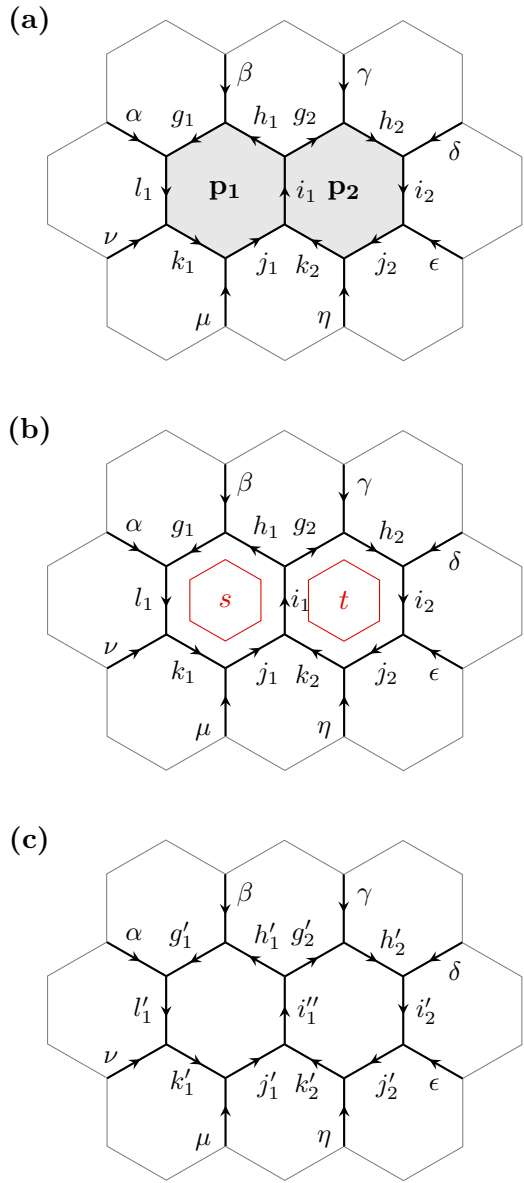

	\centering

	\caption{\small \label{fig:commut}\textbf{Commutativity.} \textbf{(a)} We consider a scenario where the operators $B_\mathbf{p_1}$ and $B_\mathbf{p_2}$ are applied to two neighbouring plaquettes $\mathbf{p_1}$ and $\mathbf{p_2}$. \textbf{(b)} The operator $B^s_\mathbf{p_1}$ puts a loop of type $s$ into plaquette $\mathbf{p_1}$, and $B^t_\mathbf{p_2}$ puts a loop of type $t$ into $\mathbf{p_2}$ (although not depicted in the picture, we take the sum over $s$ and $t$ here). \textbf{(c)} After fusing both loops into the respective plaquettes, the result is a linear combination of string-net configurations which is independent of the order in which the loops have been fused in.}
\end{figure}

\begin{thm}
	The operators $B_\mathbf{p}$ (given by \cref{eq:Bop}) and $Q_\mathbf{v}$ commute with each other when applied to two different plaquettes $\mathbf{p}$ and $\mathbf{p'}$.
\end{thm}

\begin{proof}
	Showing the commutativity of the operator when applying it to two neighbouring plaquettes $\mathbf{p_1}$ and $\mathbf{p_2}$ is a very tedious and lengthy calculation. In order to keep it as simple and clear as possible, we only write out those terms that differ between the different orders of application of $B_\mathbf{p_1}$ and $B_\mathbf{p_2}$ and denote the remaining factors by $(\dots)$. The general idea of the proof is depicted in \cref{fig:commut}. First applying $B_{\mathbf{p_1}}$ to the left plaquette $\mathbf{p_1}$ and applying $B_{\mathbf{p_2}}$ to the right plaquette $\mathbf{p_2}$ afterwards yields
		\begin{align}
		B_\mathbf{p_2}&\Bigg(B_\mathbf{p_1}\Bigg\vert

		\Bigg\rangle.
		\end{align}
	The result is a linear combination of string-net configurations, where in each term we have a sum over the loop values $s$ and $t$ and also over the intermediate label $i'$. Doing the same calculation for the other order of applying the operators, i.e., first applying $B_\mathbf{p_2}$ to the right plaquette $\mathbf{p_2}$ and applying $B_\mathbf{p_1}$ to the left plaquette $\mathbf{p_1}$ afterwards yields
		\begin{align}
		\begin{split}
		\label{eq:commut}
		B_\mathbf{p_1}&\Bigg(B_\mathbf{p_2}\Bigg\vert
		%
		\caption{\label{fig:penta}\textbf{Commuting diagrams.} The fact that these two diagrams commute yields the equations required to prove the commutativity of the magnetic flux operator.}
	\end{figure}	
		
	It remains to be shown that the following equality holds: 
		\begin{align}
			\begin{split}
			\sum_{s,t,i_1'}&\overline{\left(F_{i_1'}^{s h_1 g_2}\right)}_{i_1h_1'} \left(F_{i_1'}^{s j_1 k_2}\right)_{i_1j_1'}\left(F_{i_1''}^{h_1' g_2 t}\right)_{g_2'i_1'}\overline{\left(F_{i_1''}^{j_1'k_2t}\right)}_{k_2'i_1'}\\
			&=\sum_{s,t,\tilde{i}_1}\left(F_{\tilde{i}_1}^{h_1 g_2 t}\right)_{g_2'i_1}\overline{\left(F_{\tilde{i}_1}^{j_1k_2t}\right)}_{k_2'i_1}\overline{\left(F_{i_1''}^{s h_1 g_2'}\right)}_{\tilde{i}_1h_1'} \left(F_{i_1''}^{s j_1 k_2'}\right)_{\tilde{i}_1j_1'}.
			\label{eq:verdict}
			\end{split}
		\end{align}
	In order to achieve this goal, we use the fact that the diagrams in \cref{fig:penta} commute (which is a consequence of Mac Lane's coherence theorem). From the first diagram, we get the relation
		\begin{equation}
			\left(F_{i_1'}^{s h_1 g_2}\right)^\dagger_{h_1'i_1}\left(F_{i_1''}^{h_1' g_2 t}\right)_{g_2'i_1'}=\sum_{\tilde{i}_1}\left(F_{i_1''}^{si_1 t}\right)_{\tilde{i}_1i_1'} \left(F_{\tilde{i}_1}^{h_1g_2t}\right)_{g_2'i_1}\left(F_{i_1''}^{sh_1g_2'}\right)^\dagger_{h_1'\tilde{i}_1}
		\end{equation}
	and the second one yields the identity
		\begin{equation}
			\left(F_{i_1''}^{sj_1k_2'}\right)_{\tilde{i}_1j_1'}\left(F_{\tilde{i}_1}^{j_1k_2t}\right)^\dagger_{i_1k_2'}=\sum_{i_1'} \left(F_{i_1''}^{j_1'k_2t}\right)^\dagger_{i_1'k_2'}\left(F_{i_1'}^{sj_1k_2}\right)_{i_1j_1'}\left(F_{i_1''}^{si_1t}\right)_{\tilde{i}_1i_1'}.
		\end{equation}
	Inserting these into equation \cref{eq:verdict}, and using the fact that in a unitary fusion category it holds that $\left(F_u^{xyz}\right)^\dagger_{\alpha\beta}=\overline{\left(F_u^{xyz}\right)}_{\beta\alpha}$, we arrive at 
		\begin{align}
			\sum_{s,t,i_1',\tilde{i}_1}&\left(F_{i_1'}^{s j_1 k_2}\right)_{i_1j_1'}\overline{\left(F_{i_1''}^{j_1'k_2t}\right)}_{k_2'i_1'}\left(F_{i_1''}^{si_1 t}\right)_{\tilde{i}_1i_1'} \left(F_{\tilde{i}_1}^{h_1g_2t}\right)_{g_2'i_1}\overline{\left(F_{i_1''}^{sh_1g_2'}\right)}_{\tilde{i}_1h_1'}\\
			&=\sum_{s,t,\tilde{i}_1,i_1'}\overline{\left(F_{i_1''}^{j_1'k_2t}\right)}_{k_2'i_1'}\left(F_{i_1'}^{sj_1k_2}\right)_{i_1j_1'}\left(F_{i_1''}^{si_1t}\right)_{\tilde{i}_1 i_1'}\left(F_{\tilde{i}_1}^{h_1 g_2 t}\right)_{g_2'i_1}\overline{\left(F_{i_1''}^{s h_1 g_2'}\right)}_{\tilde{i}_1 h_1'},
		\end{align}
	which is obviously true and therefore
		\begin{equation}
			B_\mathbf{p_2}B_\mathbf{p_1}=B_\mathbf{p_1}B_\mathbf{p_2}.
		\end{equation}
	This calculation can analogously be done for every configuration of neighbouring plaquettes. It always results in exploiting two commuting diagrams such as the ones depicted in \cref{fig:penta}. Since the operator trivially commutes on distant plaquettes, we conclude that it commutes in general.
\end{proof}

\section{Connection to the original Hamiltonian}
\label{sec:originalHam}

It is possible to show that the formula of the Hamiltonian \cref{eq:Bop} that is derived without imposing the conditions \cref{eq:LWcond1,eq:LWcond2,eq:LWcond3,eq:LWcond4} transforms into the original one by Levin and Wen \cref{eq:LWoriginal} when imposing the conditions. Hence, the derivation of the Hamiltonian in the previous section is indeed a true generaliation of the Levin-Wen model to unitary fusion categories that do not fulfil tetrahedral symmetry. In this section, we show that both formulas \cref{eq:LWoriginal} and \cref{eq:Bop} are equal if we assume that the conditions \cref{eq:LWcond1,eq:LWcond2,eq:LWcond3,eq:LWcond4} are fulfilled.

To make this calculation as clear as possible, we manipulate each factor in the formula \cref{eq:Bop} individually. Moreover, we number the individual terms in the tetrahedral symmetry condition to make it easier to indicate which of these identities we are applying:
	\begin{equation}
		\underbrace{\left(F_i^{jkl}\right)_{nm}}_{\equiv\Fone}=\underbrace{\left(F_{l^*}^{kji^*}\right)_{n^*m}}_{\equiv\Ftwo}=\underbrace{\left(F_{j^*}^{i^*lk}\right)_{nm^*}}_{\equiv\Fthree}=\underbrace{\sqrt{\frac{d_md_n}{d_ld_j}}\left(F_{i^*}^{m^*kn^*}\right)_{l^*j^*}}_{\equiv\Ffour}.
	\end{equation}
In the following, we indicate the equality we use by $\Fx\to\Fy$. We can now transform the individual factors of \cref{eq:Bop}:
	\begin{equation}
\begin{array}[c]{r@{}c@{}l} 
		\left(F_{g^*}^{g^*s^* s}\right)_{0g'^*}&{}\overset{\cref{eq:LWcond1}}{=}{}&\sqrt{\frac{d_{g'^*}}{d_s d_g}}\\[1.5em]
		\overline{\left(F_{b^*}^{g'^*sh}\right)}_{h' g^*}&{}\overset{\cref{eq:LWcond4}}{=}{}&\left(F_{b}^{g's^*h^*}\right)_{h'^* g}\overset{\Ftwo\to\Fthree}{=}\left(F_s^{h^*b^*g'}\right)_{h'g^*}\overset{\Fone\to\Ffour}{=}\sqrt{\frac{d_{g^*}d_{h'}}{d_{h^*}d_{g'}}}\left(F_{s^*}^{gb^*h'^*}\right)_{g'^* h}\\[1.5em]
		\overline{\left(F_{i'}^{shc^*}\right)}_{ih'}&{}\overset{\cref{eq:LWcond4}}{=}{}&\left(F_{i'^*}^{s^*h^*c}\right)_{i^*h'^*}\overset{\Ffour\to\Ftwo}{=}\sqrt{\frac{d_{i}d_{h'}}{d_s d_c}}\left(F_{i^*}^{h^*h'i'^*}\right)_{cs}\overset{\Fthree\to\Ffour}{=}\left(F_{h^*}^{si'^*c^*}\right)_{h'^*i^*}\\&{}\overset{\Fone\to\Fthree}{=}{}&\left(F_{s^*}^{hc^*i'^*}\right)_{h'^*i}\\[1.5em]
		\left(F_{i'}^{sjd}\right)_{ij'}&{}\overset{\Ffour\to\Ftwo}{=}{}&\sqrt{\frac{d_id_{j'}}{d_sd_d}}\left(F_{i}^{jj'^*i'}\right)_{ds^*}\overset{\Fthree\to\Ffour}{=}\left(F_j^{s^*i'd^*}\right)_{j'i}\overset{\Fone\to\Fthree}{=}\left(F_{s}^{j^*d^*i'}\right)_{j'i^*}\\
		&{}\overset{\Fone\to\Ffour}{=}{}&\sqrt{\frac{d_{j'}d_i}{d_jd_{i'}}}\left(F_{s^*}^{id^*j'^*}\right)_{i'^*j}\\[1.5em]
		\left(F_j^{s^*sj'}\right)_{j'0}&{}\overset{\Ffour\to\Ftwo}{=}{}&\sqrt{\frac{d_{j'}}{d_sd_j}}\left(F_{j'}^{s0j}\right)_{js}\overset{\Fthree\to\Ffour}{=}\left(F_s^{sjj^*}\right)_{oj'}\overset{\cref{eq:LWcond1}}{=}\sqrt{\frac{d_{j'}}{d_sd_j}}
		\end{array}
	\end{equation}
	\begin{equation}\begin{array}[c]{r@{}c@{}l} 		
		\overline{\left(F_e^{k^*s^*j'}\right)}_{jk'^*}&{}\overset{\cref{eq:LWcond4}}{=}{}&\left(F_{e^*}^{ksj'^*}\right)_{j^*k'}\overset{\Fone\to\Ffour}{=}\sqrt{\frac{d_{k'}d_j}{d_{j'}d_k}}\left(F_e^{k'sj^*}\right)_{j'k^*}\overset{\Ftwo\to\Fthree}{=}\sqrt{\frac{d_{k'}d_j}{d_{j'}d_k}}\left(F_{s^*}^{j^*e^*k'}\right)_{j'^*k}\\[1.5em]
		\overline{\left(F_{l'^*}^{fk^*s^*}\right)}_{k'^*l^*}&{}\overset{\cref{eq:LWcond4}}{=}{}&\left(F_{l'}^{f^*ks}\right)_{k'l}\overset{\Fone\to\Fthree}{=}\left(F_{f}^{l'^*sk}\right)_{k'l^*}\overset{\Ftwo\to\Fthree}{=}\left(F_{s^*}^{kf^*l'^*}\right)_{k'^*l}\\[1.5em]
		\left(F_{l'^*}^{a^*g^*s^*}\right)_{g'^*l^*}&{}\overset{\Fone\to\Fthree}{=}{}&\left(F_a^{l's^*g^*}\right)_{g'^*l}\overset{\Ftwo\to\Fthree}{=}\left(F_s^{g^*a^*l'}\right)_{g'l*}
		\overset{\Fone\to\Ffour}{=}\sqrt{\frac{d_{g'}d_l}{d_{l'}d_g}}\left(F_{s^*}^{la^*g'^*}\right)_{l'^*g}.
		\end{array}
	\end{equation}
Inserting the above relations into \cref{eq:Bop} yields the coefficients of the original Levin-Wen Hamiltonian \cref{eq:LWoriginal}. In this sense, the form of the Hamiltonian \cref{eq:Bop} derived above is a true generalisation of the original Levin-Wen Hamiltonian to fusion categories that do not fulfil the tetrahedral symmetry condition.

\section{Excitations}
\label{sec:excitations}

One of the most interesting aspects of the Levin-Wen construction are the excitations of the model, both physically and mathematically. From a physical perspective, the excitations correspond to topologically non-trivial quasiparticles, or, simply put, anyons\index{anyons}. From a mathematical perspective, they yield the quantum double of the underlying fusion category, hence calculating the excitations of the model is a way to construct a UMTC from a UFC.

Due to the construction of the Hamiltonian, for the ground state string-net configurations it holds that $Q_\mathbf{v}=B_\mathbf{p}=1$ for all $\mathbf{v}$ and all $\mathbf{p}$. Therefore, excited states are those that violate this constraint for some local collection of vertices and plaquettes. These excitations are always created in pairs (see for example \cite{Levin2003}), and the pair creation operator has a string-like structure with the emerging quasiparticles at the end. These particles are independent of the actual form of the string operator since the structure of the string is unobservable. Only the excitations at the endpoints can be observed. From this argumentation it also follows that if the two endpoints coincide such that the string forms a loop then the operator commutes with the Hamiltonian, since it is unobservable. 

In this sense, we can associate with each topologically non-trivial particle\footnote{By topologically non-trivial we mean that it exhibits non-trivial statistics.} a closed string operator that commutes with the Hamiltonian. To find these particles we therefore need to find all closed string operators. An example of such an operator is depicted in \cref{fig:excitations}. We denote a closed string operator that acts along a closed path $P$ by creating a type-$s$ string by $W^s(P)$. In contrast to the operators $B_\mathbf{p}$ and $Q_\mathbf{v}$, $W^s(P)$ is no longer a local operator acting on a single plaquette or vertex. 

\begin{figure}[t]
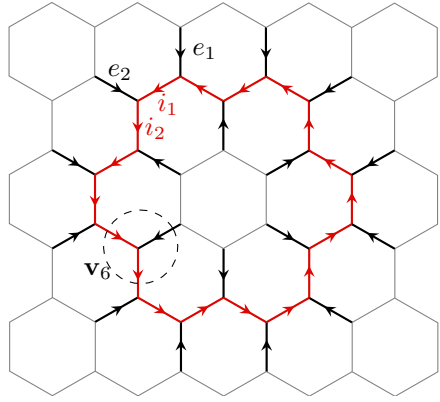

	\centering

	\caption{\small \textbf{Closed string operator on the honeycomb lattice.} A closed string operator $W^s(P)$ only acts non-trivially on spin states along the closed path $P$, depicted as thick red line. It creates a type-$s$ string and fuses it into each vertex $\mathbf{v}_1,\mathbf{v}_2,...\mathbf{v}_N$ along $P$. This action only changes the two sites $i_{k-1},i_k$ of $\mathbf{v}_k$, whereas the third site $e_k$ of the vertex is unaffected. In total, $W^s(P)$ transforms the initial spin state $i_1,i_2,...,i_N$ to the final spin state $i'_1,i'_2,...,i'_N$. The labeling convention is demonstratively shown in the picture for $i_1,i_2,e_1,e_2$ and $\mathbf{v}_6$ of an example for a $W^s(P)$ operator.\label{fig:excitations}}
\end{figure} 

To determine these operators, we follow the approach in \cite{Levin2005}. By assuming that the operator $W^s(P)$ only changes spins along the path $P=\mathbf{v}_1,\dots,\mathbf{v}_N$ (where the $\mathbf{v}_i$ denote the vertices along that path), we can define its action vertex-wise. The ansatz we make is the following: Suppose the operator $W^s(P)$ transforms the initial label configuration $i_1,\dots,i_N$ on the path $P$ to a configuration $i_1',\dots,i_N'$. Its matrix elements are then given by
	\begin{equation}
		W_{i_1 i_2...i_N}^{s,i'_1 i'_2...i'_N}(e_1 e_2...e_N)=\left(\prod_{k=1}^{N} F_k^s\right)\left(\prod_{k=1}^{N} \omega_k\right)\label{eq:StringOperator}.
	\end{equation}
Here, the labels $e_i$ denote the external legs of the path that are not changed by the operator. $F_k^s$ is a combination of $F$-symbols which acts on the vertex $\mathbf{v}_k$ and fuses the type-$s$ string into the respective vertex. The form of $F_k^s$ depends on the form of the vertex $\mathbf{v}_k$. Hence, we can think of it as one of the corners \itemone\ - \itemsix\ of the diagram \cref{eq:bigdiagram} in the calculation of $B_\mathbf{p}^s$. In fact, $B_\mathbf{p}^s$ is a special case of $W^s(P)$ when the path $P$ goes around exactly one plaquette $\mathbf{p}$ (i.e., $P$ turns left at every vertex.). We therefore have to distinguish between six different cases of vertices:
	\begin{align}
		F_k^s=\begin{cases}
		d_s\sqrt{\frac{d_{i_k} d_{i_{k-1}}}{d_{i'_k} d_{i'_{k-1}}}} \left(F_{i^*_k}^{i^*_k s^* s}\right)_{0 i'^*_k} \overline{\left(F_{e^*_k}^{i'^*_k s i_{k-1}}\right)}_{i'_{k-1}i^*_k},
		&\text{if $P$ runs as }
\text{ at $\mathbf{v}_k$ (\itemsix)}
		\end{cases}\label{eq:Excitations}
	\end{align}
	
The additional factors $\omega_k$ can be determined by exploiting the fact that the operator $W^s(P)$ commutes with the Hamiltonian if $P$ is a closed path. However, even though this approach was suitable in the construction by Levin and Wen in \cite{Levin2005}, in the more general approach we presented in this chapter the derivation of these factors is highly complicated, mostly because the matrix elements of the general Hamiltonian lack certain symmetries that are present in the original Hamiltonian.

One solution to this problem is using a different approach to the Levin-Wen model. It is possible to use tensor network states (which we have already seen in \cref{sec:numerics}) to study gapped, topologically ordered systems, see for example \cite{Bultinck2017}. Using this framework, one can construct a tensor network representation of the ground state of the Levin-Wen Hamiltonian, which was done in \cite{Buerschaper2009}, and also study the excitations of the network, see for instance \cite{Sahinoglu2014} and \cite{Williamson2017}. However, although this approach sounds promising, there is one caveat: In all of these constructions it is assumed that the fusion category fulfils the tetrahedral symmetry condition. Hence, in order to build a tensor network to study the excitations of the model it is necessary to first construct the corresponding matrix product operators without imposing tetrahedral symmetry. We go into more detail about the construction of the ground state of the Levin-Wen model for unitary fusion categories without tetrahedral symmetry in the following.

\subsection*{Tensor network representation of string-nets.} \index{tensor network} To find a tensor network representation of the ground state of the Levin-Wen model, we follow the technique described in \cite{Buerschaper2009} and generalise it to fusion categories that do not fulfil the tetrahedral symmetry condition. The goal is to find a tensor network representation of the ground state of the Levin-Wen model in order to efficiently calculate the excitations of the model. In the tensor network representation, plaquettes in the hexagonal lattice are of the form depicted in \cref{fig:TNplaquette}. Hence, we have to construct two types of operators:

\begin{figure}[t]
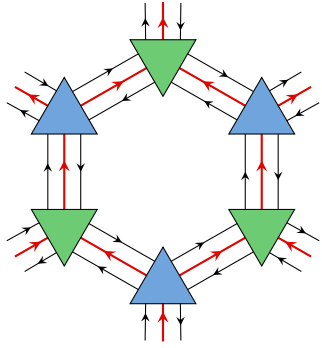

	\centering

	\caption{\small \label{fig:TNplaquette} \textbf{Tensor network representation of the Levin-Wen model.} A plaquette in the Tensor Network (TN) representation of the ground state of the Levin-Wen Hamiltonian consists of two types of tensors, depicted in blue and green. They split the lattice into two sublattices, and even (blue) and an odd (green) sublattice. The black edges (which are virtual bonds in the TN) are contracted while the red edges (that correspond to physical indices) are left uncontracted.}                                        
\end{figure}

\begin{equation}
%
\end{equation}
According to the description of tensor networks in \cref{sec:anyons}, these are rank-$9$ tensors. From these two types of operators we can build the whole hexagonal lattice, hence they split the lattice into an \emph{even} sublattice $\Lambda_\mathrm{even}$ (blue) and an \emph{odd} sublattice $\Lambda_\mathrm{odd}$ (green). For the case of unitary fusion categories that fulfil tetrahedral symmetry, the formulas for these operators are given in \cite{Buerschaper2009}. 

For general unitary fusion categories, we can make use of the calculations we have done to derive the matrix elements of the operator $B_\mathbf{p}^s$. The construction goes as follows: Consider a hexagonal lattice (also referred to as the \emph{physical lattice}). To obtain the ground state $|\Psi_0\rangle$ of the Levin-Wen model, we apply the plaquette operator $B_\mathbf{p}$ to each plaquette of the lattice. As described above, this operator is given by
	\begin{equation}
		B_\mathbf{p}=\sum_{\alpha_\mathbf{p}}\frac{d_{\alpha_\mathbf{p}}}{D^2} B_\mathbf{p}^{\alpha_\mathbf{p}},
	\end{equation}
where $D^2=\sum_s d_s$. Applying the operator $B_\mathbf{p}^{\alpha_\mathbf{p}}$ corresponds to inserting a loop of type $\alpha_\mathbf{p}$ into the plaquette. As a result, we get a lattice with loops inserted at each plaquette as depicted in \cref{fig:TNplaquette}. While a loop in a plaquette $\mathbf{p}$ carries a label $\alpha_\mathbf{p}$, the edges of the physical lattice all carry the vacuum label $0$. The ground state is hence given by
	\begin{equation}
		|\Psi_0\rangle=\frac{1}{D^2}\left(\prod_\mathbf{p}d_{\alpha_\mathbf{p}}\right)|\{\alpha_\mathbf{p}\}\rangle,
	\end{equation}
where $|\{\alpha_\mathbf{p}\}\rangle$ denotes the configuration of the lattice with loops inserted (as in \cref{fig:fatlattice}). The first step in the reduction process is fusing the loops into the lattice by applying the completeness relation at each edge. As a result, the form of the ground state is
	\begin{equation}
		|\Psi_0\rangle=\frac{1}{D^2}\sum_{\{\alpha_\mathbf{p},i_\mathbf{p},j_\mathbf{p},k_\mathbf{p}\}}\left(\prod_\mathbf{p}\frac{\sqrt{d_{i_\mathbf{p}}d_{j_\mathbf{p}}d_{k_\mathbf{p}}}}{d_{\alpha_\mathbf{p}}^2}\right)|\{\alpha_\mathbf{p},i_\mathbf{p},j_\mathbf{p},k_\mathbf{p}\}\rangle,
	\end{equation}
where $\{\alpha_\mathbf{p},i_\mathbf{p},j_\mathbf{p},k_\mathbf{p}\}$ denotes the state of the lattice after having applied the completeness relation to every edge. For further reduction, we have to distinguish between vertices of the even lattice and vertices of the odd lattice.

\begin{figure}[t]
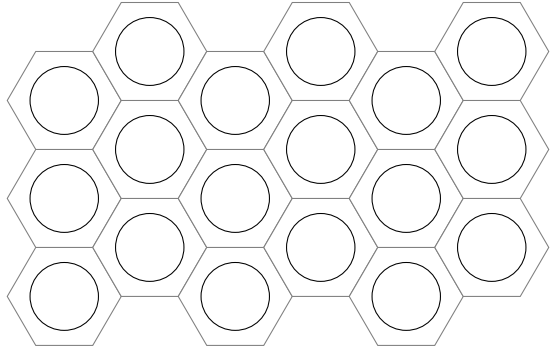

	\centering

	\caption{\small \label{fig:fatlattice}\textbf{Lattice with loops inserted.} An isolated loop at plaquette $\mathbf{p}$ carries a label $\alpha_\mathbf{p}$, while all edges of the hexagonal lattice (grey) carry the vacuum label $0$.}
\end{figure}

For calculations, we use the following convention: At a plaquette $\mathbf{p}$, we label the edges on the right side by $k_\mathbf{p}$, $i_\mathbf{p}$, $j_\mathbf{p}$:
\begin{equation}
%
.
\end{equation}
Note that by using this notation, the sum of configurations $\{i_\mathbf{p},j_\mathbf{p},k_\mathbf{p}\}$ over all plaquettes $\mathbf{p}$ of the lattice includes all edges of the physical lattice. After applying the completeness relation to every edge, the situation at the vertices of the odd sublattice is the following (where the grey lines are labelled by the vacuum):
\begin{equation}
%
.
\end{equation}
The diagram on the right hand side of the equation can be simplified by applying an $F$-move:
\begin{align}
%
.
\end{align}
A similar calculation yields the following expression for the vertices in the even sublattice:
\begin{equation}
%
.
\end{equation}

In summary, we can now write the ground state in terms of the state of the physical lattice (i.e., the state given by labels $i_\mathbf{p}$, $j_\mathbf{p}$, and $k_\mathbf{p}$) only:
\begin{equation}
|\Psi_0\rangle=\frac{1}{D^2}\sum_{\{\alpha_\mathbf{p},i_\mathbf{p},j_\mathbf{p},k_\mathbf{p}\}}\left(\prod_\mathbf{p}\sqrt{d_{i_\mathbf{p}}}\right)\prod_{v\in\Lambda_\mathrm{even}}f(v)\prod_{w\in\Lambda_\mathrm{odd}}g(w)\ |\{i_\mathbf{p},j_\mathbf{p},k_\mathbf{p}\}\rangle
\end{equation}
with
\begin{align}
f(v)&=\left(F_{i_\mathbf{p}}^{k_\mathbf{p}\alpha_\mathbf{q}\alpha_\mathbf{r}^*}\right)_{j_\mathbf{q}\alpha_\mathbf{p}}\\
g(w)&=\overline{\left(F_{i_\mathbf{p}}^{j_\mathbf{p}\alpha_\mathbf{r}\alpha_\mathbf{q}^*}\right)}_{k_\mathbf{r}\alpha_\mathbf{p}}.
\end{align}
This can be further rewritten as
	\begin{equation}
		|\Psi_0\rangle=\frac{1}{D^2}\sum_{\{i_\mathbf{p},j_\mathbf{p},k_\mathbf{p}\}}\lambda_{\{i_\mathbf{p},j_\mathbf{p},k_\mathbf{p}\}}\ |\{i_\mathbf{p},j_\mathbf{p},k_\mathbf{p}\}\rangle,
	\end{equation}
where 
	\begin{equation}
	\label{eq:basiscoeff}
		\lambda_{\{i_\mathbf{p},j_\mathbf{p},k_\mathbf{p}\}}=\left(\prod_\mathbf{p}\sqrt{d_{i_\mathbf{p}}}\right)\sum_{\{\alpha_\mathbf{p}\}}\prod_{v\in\Lambda_\mathrm{even}}f(v)\prod_{w\in\Lambda_\mathrm{odd}}g(w)
	\end{equation}
are the basis coefficients of the string-net ground state. Our goal is to write these coefficients as a contracted tensor network. For this purpose, consider a plaquette $\mathbf{a}$ with neighbouring plaquettes $\mathbf{b},\dots, \mathbf{g}$:
	\begin{equation}
		\begin{tikzpicture}[scale=1,baseline=(current bounding box.center)]
			\node[circle, draw=white, fill=LinkColor3!70!white] at (90:1) {};
			\node[circle, draw=white, fill=LinkColor2!70!white] at (150:1) {};
			\node[circle, draw=white, fill=LinkColor3!70!white] at (210:1) {};
			\node[circle, draw=white, fill=LinkColor2!70!white] at (270:1) {};
			\node[circle, draw=white, fill=LinkColor3!70!white] at (330:1) {};
			\node[circle, draw=white, fill=LinkColor2!70!white] at (30:1) {};
			\draw[->-] (270:1) to node[below,pos=0.7] {\small $k_\mathbf{f}$} (210:1);
			\draw[->-] (210:1) to node[left] {\small $i_\mathbf{g}$} (150:1);
			\draw[->-] (150:1) to node[above,pos=0.3] {\small $j_\mathbf{b}$} (90:1);
			\draw[->-] (270:1) to node[below,pos=0.7] {\small $j_\mathbf{a}$} (330:1);
			\draw[->-] (330:1) to node[right] {\small $i_\mathbf{a}$} (30:1);
			\draw[->-] (30:1) to node[above,pos=0.3] {\small $k_\mathbf{a}$} (90:1);
			\draw[->-] (90:1) -- (90:1.75);
			\draw[->-] (150:1) -- (150:1.75);
			\draw[->-] (30:1) -- (30:1.75);
			\draw[->-] (210:1.75) -- (210:1);
			\draw[->-] (270:1.75) -- (270:1);
			\draw[->-] (330:1.75) -- (330:1);
			\node at (0,0) {\small $\mathbf{a}$};
			\node at (120:1.75) {\small $\mathbf{b}$};
			\node at (60:1.75) {\small $\mathbf{c}$};
			\node at (0:1.75) {\small $\mathbf{d}$};
			\node at (300:1.75) {\small $\mathbf{e}$};
			\node at (240:1.75) {\small $\mathbf{f}$};
			\node at (180:1.75) {\small $\mathbf{g}$};
			\node at (90:2) {\small $i_\mathbf{b}$};
			\node at (30:2) {\small $j_\mathbf{c}$};
			\node at (330:2) {\small $k_\mathbf{e}$};
			\node at (270:2) {\small $i_\mathbf{f}$};
			\node at (210:2) {\small $j_\mathbf{g}$};
			\node at (150:2) {\small $k_\mathbf{g}$};
		\end{tikzpicture}.
	\end{equation}
As indicated in the diagram, we have six vertices that contribute to the coefficient, three of the even sublattice (indicated in blue) and three of the odd sublattice (indicated in green). Hence, the local expression for $\lambda$ at plaquette $\mathbf{a}$ is
	\begin{align}
	\label{eq:lambda}
		\lambda_{\{i_\mathbf{p},j_\mathbf{p},k_\mathbf{p}\}}&\thicksim
		\sqrt{d_\mathbf{a} d_\mathbf{d}}\ \sum_{\alpha_\mathbf{a}}\overline{\left(F_{i_\mathbf{b}}^{j_\mathbf{b}\alpha_\mathbf{a}\alpha_\mathbf{c}}\right)}_{k_\mathbf{a}\alpha_\mathbf{b}}\overline{\left(F_{i_\mathbf{a}}^{j_\mathbf{a}\alpha_\mathbf{e}\alpha_\mathbf{d}}\right)}_{k_\mathbf{e}\alpha_\mathbf{a}}\overline{\left(F_{i_\mathbf{g}}^{j_\mathbf{g}\alpha_\mathbf{f}\alpha_\mathbf{a}}\right)}_{k_\mathbf{f}\alpha_\mathbf{g}}\\
		&\hspace{30pt}\left(F_{i_\mathbf{a}}^{k_\mathbf{a}\alpha_\mathbf{c}\alpha_\mathbf{d}}\right)_{j_\mathbf{c}\alpha_\mathbf{a}}\left(F_{i_\mathbf{f}}^{k_\mathbf{f}\alpha_\mathbf{a}\alpha_\mathbf{e}}\right)_{j_\mathbf{a}\alpha_\mathbf{f}}\left(F_{i_\mathbf{g}}^{k_\mathbf{g}\alpha_\mathbf{b}\alpha_\mathbf{a}}\right)_{j_\mathbf{b}\alpha_\mathbf{g}}.
	\end{align}

This allows us to define vertex tensors in the tensor network built from the plaquettes depicted in \cref{fig:TNplaquette}. They are given by
	\begin{align}
&=\tilde{T}_{\mu\mu'\nu\nu'\lambda\lambda'}^{[ijk]}
	\end{align}
with
\begin{align}
\label{eq:TensorOdd}
T_{\mu\mu'\nu\nu'\lambda\lambda'}^{[ijk]}&=\sqrt{d_{i_\mathbf{p}}}\ \overline{\left(F_i^{j\lambda\mu^*}\right)}_{k\nu}\delta_{\mu\mu'}\delta_{\lambda\lambda'}\delta_{\nu\nu'}\\
\label{eq:TensorEven}
\tilde{T}_{\mu\mu'\nu\nu'\lambda\lambda'}^{[ijk]}&=\left(F_i^{k\lambda\nu^*}\right)_{j\mu}\delta_{\mu\mu'}\delta_{\lambda\lambda'}\delta_{\nu\nu'}.
\end{align}
A quick calculation shows that the formulas for the tensors in \cref{eq:TensorOdd} and \cref{eq:TensorEven} can be transformed into the ones given in \cite{Buerschaper2009} when assuming tetrahedral symmetry \cref{eq:LWcond2}.
With these tensors we can build a regular lattice with the plaquettes depicted in \cref{fig:TNplaquette} and contracting the tensors at the virtual indices $\lambda,\mu,\nu$. If we cut out a single plaquette from the network, one can see that it reproduces the local form of $\lambda$ given in \cref{eq:lambda} (up to some factor which can be absorbed in the summation).

Thus, we have found a tensor network representation of the ground state of the Levin-Wen model for general unitary fusion categories, which includes those that do not fulfil the tetrahedral symmetry condition. The next step in order to obtain the excitations of the model (and hence the UMTC) is to construct the tensors that represent the open string operators (see, for example, \cite{Bultinck2017} and \cite{Williamson2017}). These excitations then yield the UMTC for the UFC from which we built the model, hence this method can be used to construct the Drinfeld centre for the fusion category $\Hd$.

\section{Further generalisations}

The model we described in this chapter is not the most general one possible. As in most of this thesis, we have only worked in a setting where the fusion rules are multiplicity-free. However, a generalization of the above construction to categories with fusion rules that contain multiplicities is straightforward, although somewhat messy: For every vertex that is an element of a higher-dimensional vector space (i.e., a vector space that corresponds to a fusion rule with multiplicities), one has to add a sum over all basis elements of this vector space. As a result, the formulas become a lot more complicated since most of the diagrams consist of many vertices. For this reason we have only considered the multiplicity-free case here.

Another possible direction to further generalize the presented construction is losing the unitarity condition\index{non-unitary fusion category}. This makes the evaluation of the diagram in \cref{eq:bigdiagram} more complicated because some helpful relations we have used have their origin in the unitarity of the category. This affects for example the identity
	\begin{equation}
		\left(F_i^{jkl}\right)^\dagger=\left(F_i^{jkl}\right)^{-1}
	\end{equation}
that is no longer valid in the non-unitary case. Hence, the formula for the operator $G$ given in \cref{eq:Gop} changes to
	\begin{equation}
		\left(G_{ij}^{kl}\right)_{nm}=\sqrt{\frac{d_md_n}{d_jd_k}}\left(F_n^{iml}\right)_{kj}^{-1}.
	\end{equation}
Moreover, the mirror symmetry of the $F$-symbols does not hold any more since a non-unitary category is not automatically spherical. Therefore, the relation between mirrored tree diagrams is not as simple as before. We introduce an additional operator to map between the mirrored diagrams:
	\begin{equation}
		\begin{tikzpicture}[baseline=(current bounding box.center),xscale=0.8,yscale=-0.8]
			\draw (0,0.5) -- (0,1);
			\draw (0,1) -- (-0.5,1.5);
			\draw (-0.5,1.5) -- (-1,2);
			\draw (-0.5,1.5) -- (0,2);
			\draw (0,1) -- (0.5,1.5);
			\draw (0.5,1.5) -- (1,2);
			\node at (0,0.25) {\small $i$};
			\node at (-1,2.25) {\small $j$};
			\node at (0,2.25) {\small $k$};
			\node at (1,2.25) {\small $l$};
			\node at (-0.5,1.1) {\small $m$};
		\end{tikzpicture}=\sum_n\left(E^{i}_{jkl}\right)_{nm}
		\begin{tikzpicture}[baseline=(current bounding box.center),xscale=0.8,yscale=-0.8]
			\draw (0,0.5) -- (0,1);
			\draw (0,1) -- (-0.5,1.5);
			\draw (-0.5,1.5) -- (-1,2);
			\draw (0.5,1.5) -- (0,2);
			\draw (0,1) -- (0.5,1.5);
			\draw (0.5,1.5) -- (1,2);
			\node at (0,0.25) {\small $i$};
			\node at (-1,2.25) {\small $j$};
			\node at (0,2.25) {\small $k$};
			\node at (1,2.25) {\small $l$};
			\node at (0.5,1.1) {\small $n$};
		\end{tikzpicture}
	\end{equation}
with 
	\begin{align}
		\left(E^{i}_{jkl}\right)_{nm}&=B_{ml}^i A_{jk}^m\left(F_k^{j^*il^*}\right)_{nm}^{-1} B_k^{nl^*}A_n^{j^*i}\\
		\left(E^{i}_{jkl}\right)_{nm}^{-1}&=B_{kl}^m A_{jm}^i \left(F_k^{j^*il^*}\right)_{nm} B_n^{il^*} A_k^{j^*n},
	\end{align}
where the $A$ and $B$ operators also slightly differ from the ones derived in \cref{eq:A,eq:As,eq:B,eq:Bs}:
	\begin{align}
		A_{ij}^k&=\sqrt{\frac{d_i d_j}{d_k}} \left(F_j^{i^*ij}\right)_{k0}\label{A_ij}\\
		A_k^{ij}&=\sqrt{\frac{d_i d_j}{d_k}} \left(F_j^{i^*ij}\right)_{0k}^{-1}\\
		B_{ij}^k&=\sqrt{\frac{d_i d_j}{d_k}} \left(F_i^{ijj^*}\right)_{k0}^{-1}\\
		B_k^{ij}&=\sqrt{\frac{d_i d_j}{d_k}} \left(F_i^{ijj^*}\right)_{0k}.\label{B^ij}
	\end{align}
After making these adjustments it is possible to evaluate the six corners of the diagram in \cref{eq:bigdiagram} in this scenario:
	\begin{align}
		\itemone &=\sqrt{\frac{d_{g^*}d_{s^*}}{d_{g'^*}}}\sqrt{\frac{d_s d_h}{d_{h'}}}\left(E_{g^*s^*s}^{g^*}\right)_{g'^*0}^{-1}\left(E_{g'^*sh}^{b^*}\right)_{h'g^*}\\
		\itemtwo &=\sqrt{\frac{d_h d_s}{d_{h'}}}\left(G_{si}^{h'c^*}\right)_{i'h}\\
		\itemthree &=\sqrt{\frac{d_j d_s}{d_{j'}}}\left(H_{j'd}^{si}\right)_{i'j}\\
		\itemfour &=\sqrt{\frac{d_j d_s}{d_j'}}\sqrt{\frac{d_{k^*}d_{s^*}}{d_{k'^*}}}\left(F_j^{s^*sj}\right)_{j'0}\left(F_e^{k^*s^*j'}\right)_{k'^*j}^{-1}\\
		\itemfive &=\sqrt{\frac{d_{l^*}d_{s^*}}{d_{l'^*}}}\left(G_{fk'^*}^{l^*s^*}\right)_{l'^*k^*}\\
		\itemsix &=\sqrt{\frac{d_{g^*}d_{s^*}}{d_{g'^*}}}\left(H_{l^*s^*}^{a^*g'^*}\right)_{l'^*g^*}.
	\end{align}

The main problem with the Levin-Wen construction of a non-unitary fusion category is that the resulting Hamiltonian is not hermitian. This can be easily checked with a simple example, the \emph{Yang-Lee category} \textbf{YL}. This is the Galois conjugate of the Fibonacci category, hence it has the same fusion rules but the non-trivial $F$-symbol $F_\tau^{\tau\tau\tau}$ differs (compare to \cref{ex:FibAnyon}):
	\begin{equation}
		\left(F_\tau^{\tau\tau\tau}\right)_{ef}^{\mathbf{YL}}=\begin{pmatrix}
			-\phi & i \sqrt{\phi}\\
			i \sqrt{\phi} & \phi
		\end{pmatrix},
	\end{equation}
where $\phi=\frac{1+\sqrt{\phi}}{2}$ is the golden ratio. This category is a non-unitary (but still spherical!) fusion category, and an analysis of the corresponding string-net Hamiltonian according to the above construction immediately shows that it is not hermitian and therefore unphysical. 

This problem was also investigated in \cite{Freedman2012}. Here, the authors provided methods to construct a hermitian Hamiltonian from the non-hermitian one that we get from a non-unitary category. However, this comes at the cost of losing the associated stable topological order. Hence, even though a Levin-Wen construction from a non-unitary fusion category is, in principle, possible, the resulting systems are not interesting for our purposes.
\chapter{Conclusion}
\label{ch:conclusion}

Conformal Field Theories (CFTs) are an important area of quantum theory since they can describe systems at a phase transition. In this thesis, we have studied the connection between CFTs and subfactors. Here, we have discussed several approaches to build microscopic models from Unitary Fusion Categories (UFCs) with the aim of finding evidence for a connection between the corresponding subfactor and a CFT. This study is motivated by the concrete example of the Haagerup subfactor, which is the simplest example of a subfactor for which it is not clear whether a corresponding CFT exists. The connection between subfactors and CFTs is mainly drawn by a one-dimensional lattice model, the anyon chain, which is built from the Unitary Modular Tensor Category (UMTC) that can be constructed from the subfactor. Numerically studying the behaviour of the ground state when the system size goes to infinity provides information about the corresponding CFT, given that the system exhibits a continuous quantum phase transition.

In case of the Haagerup subfactor we only know the UFCs but not the UMTC, hence we cannot directly build an anyon chain from it. However, constructing the UMTC from the UFCs is a challenging task on its own. Therefore, the simplest approach to study the connection to CFTs is to build an anyon chain directly from the UFC, even though this is not guaranteed to succeed. The most important result of this thesis, even though it is a negative one, is the insight that it is not sufficient to use the UFC in order to find a CFT via the anyon chain method: in the case of the Haagerup fusion category $\Hd$, we find no evidence for a corresponding CFT in \cref{sec:anyons}.

Therefore, it is inevitable that one needs to build the UMTC, i.e., the Drinfeld centre (or quantum double), of the fusion categories that come from the Haagerup subfactor. Note that it does not matter which of the three categories in the Morita equivalence class we use since their Drinfeld centres are the same. Here, we have discussed two techniques: First, we have introduced the annular category in \cref{ch:defects}, whose representations are connected to the simple objects of the Drinfeld centre of the category. However, this technique involves many tedious calculations, hence it has not yet been pursued for the Haagerup fusion categories. The second way of finding the Drinfeld centre, which is the more promising one, is by constructing the Levin-Wen model and calculating its excitations. Here, the biggest obstacle is to find a generalisation of the model such that it is applicable to fusion categories that do not fulfil the tetrahedral symmetry condition, which was presented in \cref{ch:LW}. However, finding the excitations for the generalised model turned out to be highly complicated due to the lack of symmetries that are present in the original model, but not in the generalised form. 

In summary, we cannot give a conclusive answer to the question we asked in the beginning of this thesis, which was whether there is a CFT that corresponds to the Haagerup subfactor. Nevertheless, through the investigations made in this thesis we have come a long way in answering this question: Firstly, they revealed which methods can \emph{not} be used for the study of this conjecture, namely building anyon chains directly from a UFC. Secondly, they have helped us develop a clear plan for the next steps that are necessary to answer this question:
	\begin{enumerate}
		\item To rule out the possibility of using UFCs instead of UMTCs completely, we can build an anyon chain from the two other fusion categories in the Morita equivalence class of $\Hd$. Since $\Hi_2$ has the same fusion rules, it will not yield any new results. However, it is possible that an anyon chain built from $\Hi_1$ (which has different fusion rules than $\Hi_2$ and $\Hd$) shows critical behaviour and hence yields the desired CFT. Even if this investigation also does not yield any evidence for a CFT, it gives new insights to building an anyon chain for a UFC with multiplicities, since the fusion rules of $\Hi_1$ are not multiplicity-free (in contrast to $\Hi_2$ and $\Hd$). This information would be valuable for the final step of the investigation, which is building an anyon chain from the Drinfeld centre, since this is a UMTC with multiplicities.
		\item To obtain a UMTC that corresponds to the Haagerup subfactor, the Drinfeld centre of the Haagerup fusion categories has to be constructed. The most promising approach here seems to be the construction of a tensor network representation of the ground state of the Levin-Wen model and computing its excitations, as explained in \cref{sec:excitations}. The first step in this approach is then to find a generalisation of the existing methods to UFCs that do not exhibit tetrahedral symmetry. This general method could then be applied to one of the Haagerup fusion categories in order to construct the Drinfeld centre.
		\item Once the Drinfeld centre for the Haagerup fusion categories is constructed, it can be used to build and numerically study an anyon chain in the same way as it is described in \cref{sec:anyons}. This would be the final step in the investigation of the connection between the Haagerup subfactor and a possible corresponding CFT.
	\end{enumerate}

Whilst the techniques presented here have been focused on the specific goal of understanding the possibility of a CFT corresponding to the Haagerup subfactor, they may also be applicable to more complex subfactors. This would help us gain a clearer picture of whether the conjectured general correspondence exists. In particular, the approach via anyon chains in \cref{sec:anyons} is applicable to all subfactors once the UMTC is known, while the UMTC itself can be obtained from the generalisation of the Levin-Wen model presented in \cref{ch:LW}.


\part*{Appendix}
\appendix
\chapter{$F$-symbols for the $\mathcal{H}_3$ fusion category}
\label{app:Fsymbols}

This solution and the code which was used to calculate it can be found under \url{https://github.com/R8monaW/H3Fsymbols}. It is a real solution with two parameters, $p_1,p_2\in\{-1,+1\}$. Additionally, we use the following notation:
	\begin{align}
		A&=\frac{1}{2} \left(\sqrt{13}-3\right),\\
		B&=\frac{1}{3} \left(\sqrt{13}-2\right),\\
		C&=\frac{1}{6} \left(\sqrt{13}+1\right),\\
		D_{\pm}&=\frac{1}{12} \left(5-\sqrt{13}\pm\sqrt{6 \left(\sqrt{13}+1\right)}\right).
	\end{align}
In the following, all $F$-symbols are listed. They are first ordered by their dimension and within that, they are ordered by their labels. The $3$- and $4$-dimensional matrices are depicted as tables where the first row and the first column indicate the admissible labels for the matrix.

Since all occurring values of the $F$-symbols are in the interval $[-1,+1]$, we can visualise this solution as follows: We represent the value $+1$ with a black pixel and the value $-1$ with a white pixel. The values in between are depicted by a green pixel whose darkness depends on where the value lies in the interval $[-1,+1]$, e.g., a value close to $+1$ is depicted by a very dark green pixel, while a value close to $-1$ is depicted by a very light green pixel. For parameter values $p_1=p_2=+1$, this is shown in \cref{fig:pixels}, where the order of the $F$-symbols is chosen randomly.

\begin{figure}[H]
	\includegraphics[width=0.65\textwidth]{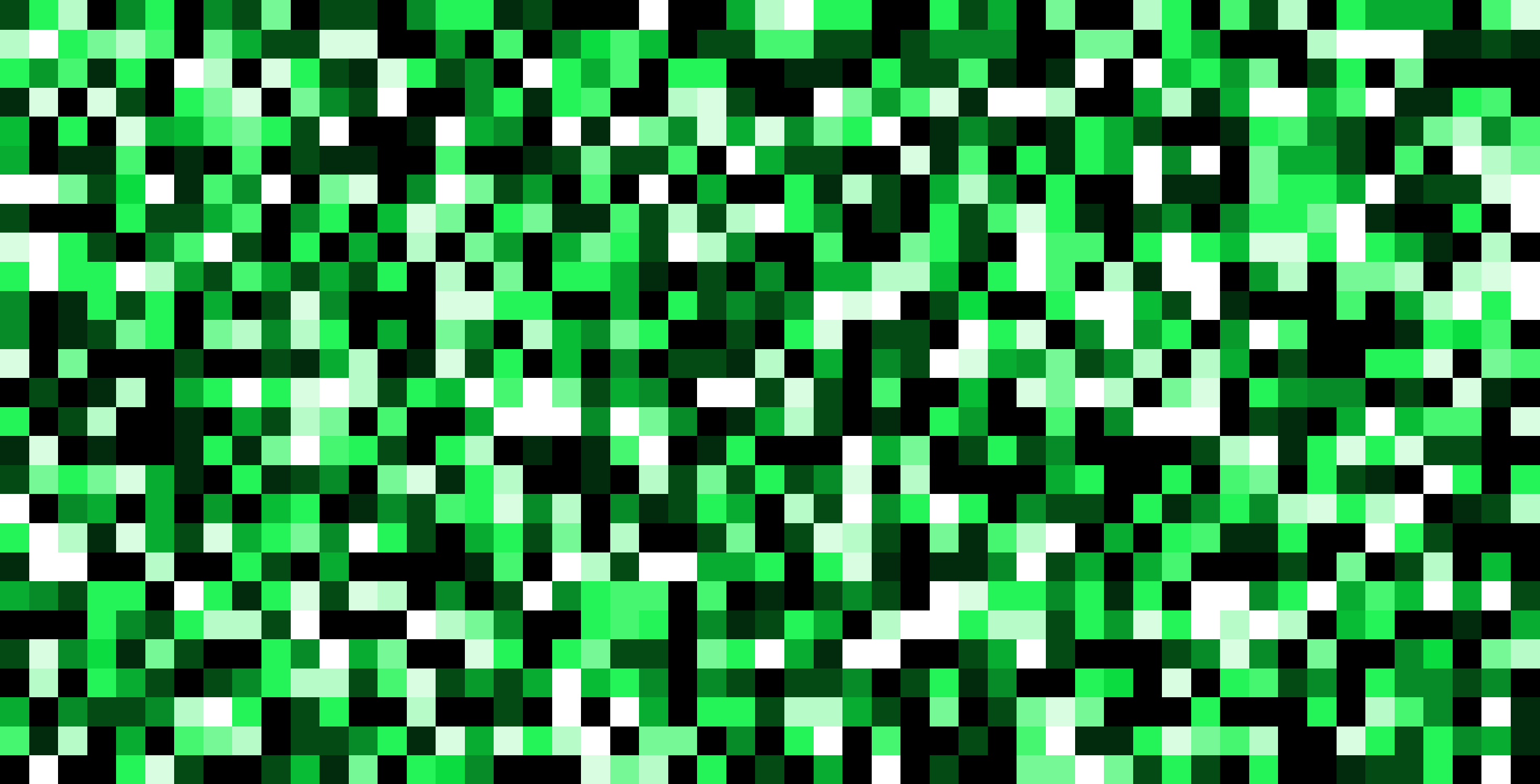}
	\caption{\small \textbf{A visualisation of the $F$-symbols.} To generate this picture, we set the parameters to $p_1=p_2=+1$ and choose a random ordering of the $F$-symbols.}
	\label{fig:pixels}
\end{figure}

\section{$1$-dimensional $F$-symbols}
{\allowdisplaybreaks
}

\section{$3$-dimensional $F$-symbols}
{\allowdisplaybreaks\begin{align}
F_{\rho}^{\rho\rho{}_\alpha\rho}=&{\small\begin{array}{c|ccc}&\rho&{}_\alpha\rho&\alphastarrho\\\hline\rho&D_+&D_-&-C\\{}_\alpha\rho&D_-&C&-D_+\\\alphastarrho&C&D_+&-D_-\end{array}}&F_{\rho}^{\rho\rho\alphastarrho}=&{\small\begin{array}{c|ccc}&\rho&{}_\alpha\rho&\alphastarrho\\\hline\rho&D_-&C&-D_+p_1p_2\\{}_\alpha\rho&Cp_1p_2&D_+p_1p_2&-D_-\\\alphastarrho&D_+&D_-&-Cp_1p_2\end{array}}\\[0.5\baselineskip]F_{\rho}^{\rho{}_\alpha\rho\rho}=&{\small\begin{array}{c|ccc}&\rho&{}_\alpha\rho&\alphastarrho\\\hline\rho&D_+&D_-&C\\{}_\alpha\rho&D_-&C&D_+\\\alphastarrho&-Cp_1&-D_+p_1&-D_-p_1\end{array}}&F_{\rho}^{\rho{}_\alpha\rho\alphastarrho}=&{\small\begin{array}{c|ccc}&\rho&{}_\alpha\rho&\alphastarrho\\\hline\rho&-Cp_1&D_+&D_-p_1p_2\\{}_\alpha\rho&D_+&-D_-p_1&-Cp_2\\\alphastarrho&D_-&-Cp_1&-D_+p_2\end{array}}\\[0.5\baselineskip]F_{\rho}^{\rho\alphastarrho\rho}=&{\small\begin{array}{c|ccc}&\rho&{}_\alpha\rho&\alphastarrho\\\hline\rho&D_-&Cp_1p_2&D_+\\{}_\alpha\rho&Cp_1&D_+p_2&D_-p_1\\\alphastarrho&D_+&D_-p_1p_2&C\end{array}}&F_{\rho}^{\rho\alphastarrho{}_\alpha\rho}=&{\small\begin{array}{c|ccc}&\rho&{}_\alpha\rho&\alphastarrho\\\hline\rho&-Cp_1&D_+p_1p_2&D_-\\{}_\alpha\rho&D_+&-D_-p_2&-Cp_1\\\alphastarrho&D_-&-Cp_2&-D_+p_1\end{array}}\\[0.5\baselineskip]F_{\rho}^{{}_\alpha\rho\rho\rho}=&{\small\begin{array}{c|ccc}&\rho&{}_\alpha\rho&\alphastarrho\\\hline\rho&D_+&D_-&-Cp_1\\{}_\alpha\rho&D_-&C&-D_+p_1\\\alphastarrho&-C&-D_+&D_-p_1\end{array}}&F_{\rho}^{{}_\alpha\rho\rho{}_\alpha\rho}=&{\small\begin{array}{c|ccc}&\rho&{}_\alpha\rho&\alphastarrho\\\hline\rho&D_-&-Cp_1&D_+\\{}_\alpha\rho&-Cp_1&D_+&-D_-p_1\\\alphastarrho&-D_+p_1&D_-&-Cp_1\end{array}}\\[0.5\baselineskip]F_{\rho}^{{}_\alpha\rho{}_\alpha\rho{}_\alpha\rho}=&{\small\begin{array}{c|ccc}&\rho&{}_\alpha\rho&\alphastarrho\\\hline\rho&C&D_+&D_-\\{}_\alpha\rho&D_+&D_-&C\\\alphastarrho&D_-&C&D_+\end{array}}&F_{\rho}^{{}_\alpha\rho{}_\alpha\rho\alphastarrho}=&{\small\begin{array}{c|ccc}&\rho&{}_\alpha\rho&\alphastarrho\\\hline\rho&D_+&D_-&Cp_1p_2\\{}_\alpha\rho&D_-&C&D_+p_1p_2\\\alphastarrho&C&D_+&D_-p_1p_2\end{array}}\\[0.5\baselineskip]F_{\rho}^{{}_\alpha\rho\alphastarrho\rho}=&{\small\begin{array}{c|ccc}&\rho&{}_\alpha\rho&\alphastarrho\\\hline\rho&-Cp_1&D_+&D_-\\{}_\alpha\rho&D_+p_1&-D_-&-C\\\alphastarrho&D_-&-Cp_1&-D_+p_1\end{array}}&F_{\rho}^{{}_\alpha\rho\alphastarrho\alphastarrho}=&{\small\begin{array}{c|ccc}&\rho&{}_\alpha\rho&\alphastarrho\\\hline\rho&D_-p_1p_2&Cp_1p_2&D_+\\{}_\alpha\rho&C&D_+&D_-p_1p_2\\\alphastarrho&D_+&D_-&Cp_1p_2\end{array}}\\[0.5\baselineskip]F_{\rho}^{\alphastarrho\rho\rho}=&{\small\begin{array}{c|ccc}&\rho&{}_\alpha\rho&\alphastarrho\\\hline\rho&D_-&Cp_1&D_+\\{}_\alpha\rho&C&D_+p_1&D_-\\\alphastarrho&-D_+p_1p_2&-D_-p_2&-Cp_1p_2\end{array}}&F_{\rho}^{\alphastarrho\rho\alphastarrho}=&{\small\begin{array}{c|ccc}&\rho&{}_\alpha\rho&\alphastarrho\\\hline\rho&D_+&-D_-&-Cp_2\\{}_\alpha\rho&-D_-p_2&Cp_2&D_+\\\alphastarrho&-Cp_2&D_+p_2&D_-\end{array}}\\[0.5\baselineskip]F_{\rho}^{\alphastarrho{}_\alpha\rho\rho}=&{\small\begin{array}{c|ccc}&\rho&{}_\alpha\rho&\alphastarrho\\\hline\rho&-Cp_1&D_+&D_-\\{}_\alpha\rho&D_+&-D_-p_1&-Cp_1\\\alphastarrho&-D_-p_2&Cp_1p_2&D_+p_1p_2\end{array}}&F_{\rho}^{\alphastarrho{}_\alpha\rho{}_\alpha\rho}=&{\small\begin{array}{c|ccc}&\rho&{}_\alpha\rho&\alphastarrho\\\hline\rho&D_+&D_-&C\\{}_\alpha\rho&D_-&C&D_+\\\alphastarrho&Cp_1p_2&D_+p_1p_2&D_-p_1p_2\end{array}}\\[0.5\baselineskip]F_{\rho}^{\alphastarrho\alphastarrho{}_\alpha\rho}=&{\small\begin{array}{c|ccc}&\rho&{}_\alpha\rho&\alphastarrho\\\hline\rho&D_-p_1p_2&C&D_+\\{}_\alpha\rho&Cp_1p_2&D_+&D_-\\\alphastarrho&D_+&D_-p_1p_2&Cp_1p_2\end{array}}&F_{\rho}^{\alphastarrho\alphastarrho\alphastarrho}=&{\small\begin{array}{c|ccc}&\rho&{}_\alpha\rho&\alphastarrho\\\hline\rho&C&D_+p_1p_2&D_-\\{}_\alpha\rho&D_+p_1p_2&D_-&Cp_1p_2\\\alphastarrho&D_-&Cp_1p_2&D_+\end{array}}\\[0.5\baselineskip]F_{{}_\alpha\rho}^{\rho\rho\rho}=&{\small\begin{array}{c|ccc}&\rho&{}_\alpha\rho&\alphastarrho\\\hline\rho&D_+&D_-&-Cp_1\\{}_\alpha\rho&D_-&C&-D_+p_1\\\alphastarrho&-Cp_1p_2&-D_+p_1p_2&D_-p_2\end{array}}&F_{{}_\alpha\rho}^{\rho\rho\alphastarrho}=&{\small\begin{array}{c|ccc}&\rho&{}_\alpha\rho&\alphastarrho\\\hline\rho&-Cp_2&D_+&-D_-\\{}_\alpha\rho&D_+&-D_-p_2&Cp_2\\\alphastarrho&D_-&-Cp_2&D_+p_2\end{array}}\\[0.5\baselineskip]F_{{}_\alpha\rho}^{\rho{}_\alpha\rho\rho}=&{\small\begin{array}{c|ccc}&\rho&{}_\alpha\rho&\alphastarrho\\\hline\rho&D_-&-Cp_1&-D_+\\{}_\alpha\rho&-Cp_1&D_+&D_-p_1\\\alphastarrho&D_+p_1p_2&-D_-p_2&-Cp_1p_2\end{array}}&F_{{}_\alpha\rho}^{\rho{}_\alpha\rho{}_\alpha\rho}=&{\small\begin{array}{c|ccc}&\rho&{}_\alpha\rho&\alphastarrho\\\hline\rho&C&D_+&D_-\\{}_\alpha\rho&D_+&D_-&C\\\alphastarrho&D_-p_1p_2&Cp_1p_2&D_+p_1p_2\end{array}}\\[0.5\baselineskip]F_{{}_\alpha\rho}^{\rho\alphastarrho{}_\alpha\rho}=&{\small\begin{array}{c|ccc}&\rho&{}_\alpha\rho&\alphastarrho\\\hline\rho&D_+&D_-p_1p_2&C\\{}_\alpha\rho&D_-&Cp_1p_2&D_+\\\alphastarrho&Cp_1p_2&D_+&D_-p_1p_2\end{array}}&F_{{}_\alpha\rho}^{\rho\alphastarrho\alphastarrho}=&{\small\begin{array}{c|ccc}&\rho&{}_\alpha\rho&\alphastarrho\\\hline\rho&D_-&C&D_+p_1p_2\\{}_\alpha\rho&Cp_1p_2&D_+p_1p_2&D_-\\\alphastarrho&D_+&D_-&Cp_1p_2\end{array}}\\[0.5\baselineskip]F_{{}_\alpha\rho}^{{}_\alpha\rho\rho{}_\alpha\rho}=&{\small\begin{array}{c|ccc}&\rho&{}_\alpha\rho&\alphastarrho\\\hline\rho&C&D_+&-D_-p_1\\{}_\alpha\rho&D_+&D_-&-Cp_1\\\alphastarrho&D_-&C&-D_+p_1\end{array}}&F_{{}_\alpha\rho}^{{}_\alpha\rho\rho\alphastarrho}=&{\small\begin{array}{c|ccc}&\rho&{}_\alpha\rho&\alphastarrho\\\hline\rho&D_+p_1p_2&D_-&-Cp_1\\{}_\alpha\rho&D_-&Cp_1p_2&-D_+p_2\\\alphastarrho&Cp_1p_2&D_+&-D_-p_1\end{array}}\\[0.5\baselineskip]F_{{}_\alpha\rho}^{{}_\alpha\rho{}_\alpha\rho\rho}=&{\small\begin{array}{c|ccc}&\rho&{}_\alpha\rho&\alphastarrho\\\hline\rho&C&D_+&D_-p_1\\{}_\alpha\rho&D_+&D_-&Cp_1\\\alphastarrho&-D_-p_1&-Cp_1&-D_+\end{array}}&F_{{}_\alpha\rho}^{{}_\alpha\rho{}_\alpha\rho\alphastarrho}=&{\small\begin{array}{c|ccc}&\rho&{}_\alpha\rho&\alphastarrho\\\hline\rho&D_-p_1p_2&-Cp_1&-D_+p_1\\{}_\alpha\rho&-Cp_2&D_+&D_-\\\alphastarrho&-D_+p_2&D_-&C\end{array}}\\[0.5\baselineskip]F_{{}_\alpha\rho}^{{}_\alpha\rho\alphastarrho\rho}=&{\small\begin{array}{c|ccc}&\rho&{}_\alpha\rho&\alphastarrho\\\hline\rho&D_+&D_-&Cp_1\\{}_\alpha\rho&D_-p_1&Cp_1&D_+\\\alphastarrho&C&D_+&D_-p_1\end{array}}&F_{{}_\alpha\rho}^{{}_\alpha\rho\alphastarrho{}_\alpha\rho}=&{\small\begin{array}{c|ccc}&\rho&{}_\alpha\rho&\alphastarrho\\\hline\rho&D_-&-Cp_1&-D_+p_1\\{}_\alpha\rho&-Cp_1&D_+&D_-\\\alphastarrho&-D_+p_1&D_-&C\end{array}}\\[0.5\baselineskip]F_{{}_\alpha\rho}^{\alphastarrho\rho\rho}=&{\small\begin{array}{c|ccc}&\rho&{}_\alpha\rho&\alphastarrho\\\hline\rho&-C&D_+&D_-\\{}_\alpha\rho&D_+&-D_-&-C\\\alphastarrho&-D_-&C&D_+\end{array}}&F_{{}_\alpha\rho}^{\alphastarrho\rho{}_\alpha\rho}=&{\small\begin{array}{c|ccc}&\rho&{}_\alpha\rho&\alphastarrho\\\hline\rho&D_+p_1&D_-&Cp_1\\{}_\alpha\rho&D_-&Cp_1&D_+\\\alphastarrho&C&D_+p_1&D_-\end{array}}\\[0.5\baselineskip]F_{{}_\alpha\rho}^{\alphastarrho{}_\alpha\rho{}_\alpha\rho}=&{\small\begin{array}{c|ccc}&\rho&{}_\alpha\rho&\alphastarrho\\\hline\rho&D_-p_1&-C&-D_+\\{}_\alpha\rho&-Cp_1&D_+&D_-\\\alphastarrho&-D_+p_1&D_-&C\end{array}}&F_{{}_\alpha\rho}^{\alphastarrho{}_\alpha\rho\alphastarrho}=&{\small\begin{array}{c|ccc}&\rho&{}_\alpha\rho&\alphastarrho\\\hline\rho&Cp_2&-D_+&-D_-\\{}_\alpha\rho&-D_+p_2&D_-&C\\\alphastarrho&-D_-p_2&C&D_+\end{array}}\\[0.5\baselineskip]F_{{}_\alpha\rho}^{\alphastarrho\alphastarrho\rho}=&{\small\begin{array}{c|ccc}&\rho&{}_\alpha\rho&\alphastarrho\\\hline\rho&D_-p_2&Cp_1&D_+\\{}_\alpha\rho&Cp_2&D_+p_1&D_-\\\alphastarrho&D_+p_1p_2&D_-&Cp_1\end{array}}&F_{{}_\alpha\rho}^{\alphastarrho\alphastarrho\alphastarrho}=&{\small\begin{array}{c|ccc}&\rho&{}_\alpha\rho&\alphastarrho\\\hline\rho&D_+p_2&-D_-p_1p_2&-Cp_1p_2\\{}_\alpha\rho&-D_-p_1&C&D_+\\\alphastarrho&-Cp_1&D_+&D_-\end{array}}\\[0.5\baselineskip]F_{\alphastarrho}^{\rho\rho\rho}=&{\small\begin{array}{c|ccc}&\rho&{}_\alpha\rho&\alphastarrho\\\hline\rho&D_-&Cp_1&D_+\\{}_\alpha\rho&-C&-D_+p_1&-D_-\\\alphastarrho&D_+&D_-p_1&C\end{array}}&F_{\alphastarrho}^{\rho\rho{}_\alpha\rho}=&{\small\begin{array}{c|ccc}&\rho&{}_\alpha\rho&\alphastarrho\\\hline\rho&Cp_1&-D_+&D_-\\{}_\alpha\rho&D_+&-D_-p_1&Cp_1\\\alphastarrho&D_-&-Cp_1&D_+p_1\end{array}}\\[0.5\baselineskip]F_{\alphastarrho}^{\rho{}_\alpha\rho{}_\alpha\rho}=&{\small\begin{array}{c|ccc}&\rho&{}_\alpha\rho&\alphastarrho\\\hline\rho&-D_+&-D_-&-C\\{}_\alpha\rho&D_-&C&D_+\\\alphastarrho&C&D_+&D_-\end{array}}&F_{\alphastarrho}^{\rho{}_\alpha\rho\alphastarrho}=&{\small\begin{array}{c|ccc}&\rho&{}_\alpha\rho&\alphastarrho\\\hline\rho&-D_-&-C&-D_+\\{}_\alpha\rho&C&D_+&D_-\\\alphastarrho&D_+&D_-&C\end{array}}\\[0.5\baselineskip]F_{\alphastarrho}^{\rho\alphastarrho\rho}=&{\small\begin{array}{c|ccc}&\rho&{}_\alpha\rho&\alphastarrho\\\hline\rho&D_+&-D_-p_1p_2&Cp_1\\{}_\alpha\rho&D_-&-Cp_1p_2&D_+p_1\\\alphastarrho&Cp_1&-D_+p_2&D_-\end{array}}&F_{\alphastarrho}^{\rho\alphastarrho\alphastarrho}=&{\small\begin{array}{c|ccc}&\rho&{}_\alpha\rho&\alphastarrho\\\hline\rho&-Cp_1p_2&-D_+&-D_-p_1p_2\\{}_\alpha\rho&D_+&D_-p_1p_2&C\\\alphastarrho&D_-&Cp_1p_2&D_+\end{array}}\\[0.5\baselineskip]F_{\alphastarrho}^{{}_\alpha\rho\rho\rho}=&{\small\begin{array}{c|ccc}&\rho&{}_\alpha\rho&\alphastarrho\\\hline\rho&-C&D_+&D_-\\{}_\alpha\rho&-D_+&D_-&C\\\alphastarrho&D_-&-C&-D_+\end{array}}&F_{\alphastarrho}^{{}_\alpha\rho\rho\alphastarrho}=&{\small\begin{array}{c|ccc}&\rho&{}_\alpha\rho&\alphastarrho\\\hline\rho&-D_-p_1&-Cp_1&D_+\\{}_\alpha\rho&Cp_1p_2&D_+p_1p_2&-D_-p_2\\\alphastarrho&D_+&D_-&-Cp_1\end{array}}\\[0.5\baselineskip]F_{\alphastarrho}^{{}_\alpha\rho{}_\alpha\rho\rho}=&{\small\begin{array}{c|ccc}&\rho&{}_\alpha\rho&\alphastarrho\\\hline\rho&D_+p_1&D_-&-Cp_1\\{}_\alpha\rho&-D_-&-Cp_1&D_+\\\alphastarrho&Cp_1&D_+&-D_-p_1\end{array}}&F_{\alphastarrho}^{{}_\alpha\rho{}_\alpha\rho{}_\alpha\rho}=&{\small\begin{array}{c|ccc}&\rho&{}_\alpha\rho&\alphastarrho\\\hline\rho&-D_-p_1&C&D_+\\{}_\alpha\rho&-Cp_1&D_+&D_-\\\alphastarrho&-D_+p_1&D_-&C\end{array}}\\[0.5\baselineskip]F_{\alphastarrho}^{{}_\alpha\rho\alphastarrho{}_\alpha\rho}=&{\small\begin{array}{c|ccc}&\rho&{}_\alpha\rho&\alphastarrho\\\hline\rho&-Cp_1&D_+&D_-\\{}_\alpha\rho&-D_+p_1&D_-&C\\\alphastarrho&-D_-p_1&C&D_+\end{array}}&F_{\alphastarrho}^{{}_\alpha\rho\alphastarrho\alphastarrho}=&{\small\begin{array}{c|ccc}&\rho&{}_\alpha\rho&\alphastarrho\\\hline\rho&-D_+p_2&D_-p_1p_2&Cp_1p_2\\{}_\alpha\rho&-D_-p_1&C&D_+\\\alphastarrho&-Cp_1&D_+&D_-\end{array}}\\[0.5\baselineskip]F_{\alphastarrho}^{\alphastarrho\rho{}_\alpha\rho}=&{\small\begin{array}{c|ccc}&\rho&{}_\alpha\rho&\alphastarrho\\\hline\rho&D_-&Cp_1&D_+\\{}_\alpha\rho&C&D_+p_1&D_-\\\alphastarrho&D_+&D_-p_1&C\end{array}}&F_{\alphastarrho}^{\alphastarrho\rho\alphastarrho}=&{\small\begin{array}{c|ccc}&\rho&{}_\alpha\rho&\alphastarrho\\\hline\rho&C&D_+p_1&D_-\\{}_\alpha\rho&D_+p_1p_2&D_-p_2&Cp_1p_2\\\alphastarrho&D_-&Cp_1&D_+\end{array}}\\[0.5\baselineskip]F_{\alphastarrho}^{\alphastarrho{}_\alpha\rho\rho}=&{\small\begin{array}{c|ccc}&\rho&{}_\alpha\rho&\alphastarrho\\\hline\rho&-D_-&-Cp_1&D_+\\{}_\alpha\rho&-C&-D_+p_1&D_-\\\alphastarrho&D_+p_1&D_-&-Cp_1\end{array}}&F_{\alphastarrho}^{\alphastarrho{}_\alpha\rho\alphastarrho}=&{\small\begin{array}{c|ccc}&\rho&{}_\alpha\rho&\alphastarrho\\\hline\rho&D_+&-D_-p_1&-Cp_1\\{}_\alpha\rho&-D_-p_1&C&D_+\\\alphastarrho&-Cp_1&D_+&D_-\end{array}}\\[0.5\baselineskip]F_{\alphastarrho}^{\alphastarrho\alphastarrho\rho}=&{\small\begin{array}{c|ccc}&\rho&{}_\alpha\rho&\alphastarrho\\\hline\rho&-Cp_1p_2&-D_+p_1&D_-\\{}_\alpha\rho&-D_+p_2&-D_-&Cp_1\\\alphastarrho&-D_-p_1p_2&-Cp_1&D_+\end{array}}&F_{\alphastarrho}^{\alphastarrho\alphastarrho{}_\alpha\rho}=&{\small\begin{array}{c|ccc}&\rho&{}_\alpha\rho&\alphastarrho\\\hline\rho&D_+p_1p_2&-D_-p_1&-Cp_1\\{}_\alpha\rho&-D_-p_2&C&D_+\\\alphastarrho&-Cp_2&D_+&D_-\end{array}}
\end{align}}

\section{$4$-dimensional $F$-symbols}
{\allowdisplaybreaks\begin{align}
F_{\rho}^{\rho\rho\rho}=&{\small\begin{array}{c|cccc}&1&\rho&{}_\alpha\rho&\alphastarrho\\\hline1&A&\sqrt{A}&-p_1\sqrt{A}&p_1\sqrt{A}\\\rho&\sqrt{A}&-B&-D_+p_1&D_-p_1\\{}_\alpha\rho&-p_1\sqrt{A}&-D_+p_1&D_-&B\\\alphastarrho&p_1\sqrt{A}&D_-p_1&B&D_+\end{array}}\\[0.5\baselineskip]F_{\rho}^{\rho{}_\alpha\rho{}_\alpha\rho}=&{\small\begin{array}{c|cccc}&\alpha^*&\rho&{}_\alpha\rho&\alphastarrho\\\hline1&A&-p_1\sqrt{A}&\sqrt{A}&\sqrt{A}\\\rho&\sqrt{A}&-D_-p_1&-B&D_+\\{}_\alpha\rho&-p_1\sqrt{A}&-B&-D_+p_1&-D_-p_1\\\alphastarrho&-p_1\sqrt{A}&D_+&-D_-p_1&Bp_1\end{array}}\\[0.5\baselineskip]F_{\rho}^{\rho\alphastarrho\alphastarrho}=&{\small\begin{array}{c|cccc}&\alpha&\rho&{}_\alpha\rho&\alphastarrho\\\hline1&A&p_1\sqrt{A}&-p_1p_2\sqrt{A}&-p_1p_2\sqrt{A}\\\rho&-p_1p_2\sqrt{A}&-D_+p_2&D_-&-B\\{}_\alpha\rho&p_1\sqrt{A}&D_-&Bp_2&-D_+p_2\\\alphastarrho&p_1\sqrt{A}&-B&-D_+p_2&-D_-p_2\end{array}}\\[0.5\baselineskip]F_{\rho}^{{}_\alpha\rho\rho\alphastarrho}=&{\small\begin{array}{c|cccc}&\alpha&\rho&{}_\alpha\rho&\alphastarrho\\\hline\alpha&-A&p_1\sqrt{A}&-\sqrt{A}&p_2\sqrt{A}\\\rho&-p_1\sqrt{A}&-B&-D_+p_1&D_-p_1p_2\\{}_\alpha\rho&p_1p_2\sqrt{A}&-D_+p_2&D_-p_1p_2&Bp_1\\\alphastarrho&\sqrt{A}&-D_-p_1&-B&-D_+p_2\end{array}}\\[0.5\baselineskip]F_{\rho}^{{}_\alpha\rho{}_\alpha\rho\rho}=&{\small\begin{array}{c|cccc}&1&\rho&{}_\alpha\rho&\alphastarrho\\\hline\alpha&A&\sqrt{A}&-p_1\sqrt{A}&-p_1\sqrt{A}\\\rho&-p_1\sqrt{A}&-D_-p_1&-B&D_+\\{}_\alpha\rho&\sqrt{A}&-B&-D_+p_1&-D_-p_1\\\alphastarrho&-p_1\sqrt{A}&-D_+p_1&D_-&-B\end{array}}\\[0.5\baselineskip]F_{\rho}^{{}_\alpha\rho\alphastarrho{}_\alpha\rho}=&{\small\begin{array}{c|cccc}&\alpha^*&\rho&{}_\alpha\rho&\alphastarrho\\\hline\alpha&-A&-p_1\sqrt{A}&-p_1\sqrt{A}&-p_1\sqrt{A}\\\rho&p_1\sqrt{A}&D_+&D_-&-B\\{}_\alpha\rho&p_1\sqrt{A}&D_-&-B&D_+\\\alphastarrho&p_1\sqrt{A}&-B&D_+&D_-\end{array}}\\[0.5\baselineskip]F_{\rho}^{\alphastarrho\rho{}_\alpha\rho}=&{\small\begin{array}{c|cccc}&\alpha^*&\rho&{}_\alpha\rho&\alphastarrho\\\hline\alpha^*&-A&p_1\sqrt{A}&-p_1\sqrt{A}&-\sqrt{A}\\\rho&-p_1\sqrt{A}&-B&-D_+&-D_-p_1\\{}_\alpha\rho&\sqrt{A}&-D_+p_1&D_-p_1&-B\\\alphastarrho&p_1p_2\sqrt{A}&-D_-p_2&-Bp_2&D_+p_1p_2\end{array}}\\[0.5\baselineskip]F_{\rho}^{\alphastarrho{}_\alpha\rho\alphastarrho}=&{\small\begin{array}{c|cccc}&\alpha&\rho&{}_\alpha\rho&\alphastarrho\\\hline\alpha^*&A&-p_1\sqrt{A}&-p_1\sqrt{A}&-p_2\sqrt{A}\\\rho&-p_1\sqrt{A}&D_-&-B&D_+p_1p_2\\{}_\alpha\rho&-p_1\sqrt{A}&-B&D_+&D_-p_1p_2\\\alphastarrho&-p_2\sqrt{A}&D_+p_1p_2&D_-p_1p_2&-B\end{array}}\\[0.5\baselineskip]F_{\rho}^{\alphastarrho\alphastarrho\rho}=&{\small\begin{array}{c|cccc}&1&\rho&{}_\alpha\rho&\alphastarrho\\\hline\alpha^*&A&-p_1p_2\sqrt{A}&p_1\sqrt{A}&p_1\sqrt{A}\\\rho&p_1\sqrt{A}&-D_+p_2&D_-&-B\\{}_\alpha\rho&-p_1\sqrt{A}&D_-p_2&B&-D_+\\\alphastarrho&-p_1p_2\sqrt{A}&-B&-D_+p_2&-D_-p_2\end{array}}\\[0.5\baselineskip]F_{{}_\alpha\rho}^{\rho\rho{}_\alpha\rho}=&{\small\begin{array}{c|cccc}&1&\rho&{}_\alpha\rho&\alphastarrho\\\hline\alpha^*&A&\sqrt{A}&-p_1\sqrt{A}&p_1\sqrt{A}\\\rho&-p_1\sqrt{A}&-D_-p_1&-B&-D_+\\{}_\alpha\rho&\sqrt{A}&-B&-D_+p_1&D_-p_1\\\alphastarrho&p_1p_2\sqrt{A}&D_+p_1p_2&-D_-p_2&-Bp_2\end{array}}\\[0.5\baselineskip]F_{{}_\alpha\rho}^{\rho{}_\alpha\rho\alphastarrho}=&{\small\begin{array}{c|cccc}&\alpha^*&\rho&{}_\alpha\rho&\alphastarrho\\\hline\alpha^*&Ap_1&p_2\sqrt{A}&p_1\sqrt{A}&p_1\sqrt{A}\\\rho&\sqrt{A}&D_+p_1p_2&D_-&-B\\{}_\alpha\rho&\sqrt{A}&D_-p_1p_2&-B&D_+\\\alphastarrho&p_1p_2\sqrt{A}&-B&D_+p_1p_2&D_-p_1p_2\end{array}}\\[0.5\baselineskip]F_{{}_\alpha\rho}^{\rho\alphastarrho\rho}=&{\small\begin{array}{c|cccc}&\alpha&\rho&{}_\alpha\rho&\alphastarrho\\\hline\alpha^*&-A&-p_1\sqrt{A}&p_1p_2\sqrt{A}&p_1\sqrt{A}\\\rho&p_1\sqrt{A}&-B&-D_+p_2&-D_-\\{}_\alpha\rho&-p_1\sqrt{A}&-D_+&D_-p_2&-B\\\alphastarrho&-p_1p_2\sqrt{A}&-D_-p_2&-B&D_+p_2\end{array}}\\[0.5\baselineskip]F_{{}_\alpha\rho}^{{}_\alpha\rho\rho\rho}=&{\small\begin{array}{c|cccc}&\alpha&\rho&{}_\alpha\rho&\alphastarrho\\\hline1&A&-p_1\sqrt{A}&\sqrt{A}&-\sqrt{A}\\\rho&\sqrt{A}&-D_-p_1&-B&-D_+\\{}_\alpha\rho&-p_1\sqrt{A}&-B&-D_+p_1&D_-p_1\\\alphastarrho&p_1\sqrt{A}&-D_+&D_-p_1&Bp_1\end{array}}\\[0.5\baselineskip]F_{{}_\alpha\rho}^{{}_\alpha\rho{}_\alpha\rho{}_\alpha\rho}=&{\small\begin{array}{c|cccc}&1&\rho&{}_\alpha\rho&\alphastarrho\\\hline1&A&\sqrt{A}&-p_1\sqrt{A}&-p_1\sqrt{A}\\\rho&\sqrt{A}&D_+&-D_-p_1&Bp_1\\{}_\alpha\rho&-p_1\sqrt{A}&-D_-p_1&-B&D_+\\\alphastarrho&-p_1\sqrt{A}&Bp_1&D_+&D_-\end{array}}\\[0.5\baselineskip]F_{{}_\alpha\rho}^{{}_\alpha\rho\alphastarrho\alphastarrho}=&{\small\begin{array}{c|cccc}&\alpha^*&\rho&{}_\alpha\rho&\alphastarrho\\\hline1&A&-p_1p_2\sqrt{A}&p_1\sqrt{A}&p_1\sqrt{A}\\\rho&-p_1p_2\sqrt{A}&-B&-D_+p_2&-D_-p_2\\{}_\alpha\rho&p_1\sqrt{A}&-D_+p_2&D_-&-B\\\alphastarrho&p_1\sqrt{A}&-D_-p_2&-B&D_+\end{array}}\\[0.5\baselineskip]F_{{}_\alpha\rho}^{\alphastarrho\rho\alphastarrho}=&{\small\begin{array}{c|cccc}&\alpha^*&\rho&{}_\alpha\rho&\alphastarrho\\\hline\alpha&A&p_1p_2\sqrt{A}&p_1\sqrt{A}&\sqrt{A}\\\rho&p_1\sqrt{A}&D_-p_2&-B&D_+p_1\\{}_\alpha\rho&p_1p_2\sqrt{A}&-B&D_+p_2&D_-p_1p_2\\\alphastarrho&\sqrt{A}&D_+p_1p_2&D_-p_1&-B\end{array}}\\[0.5\baselineskip]F_{{}_\alpha\rho}^{\alphastarrho{}_\alpha\rho\rho}=&{\small\begin{array}{c|cccc}&\alpha&\rho&{}_\alpha\rho&\alphastarrho\\\hline\alpha&Ap_1&-\sqrt{A}&-\sqrt{A}&-p_1\sqrt{A}\\\rho&-\sqrt{A}&D_+p_1&D_-p_1&-B\\{}_\alpha\rho&-p_1\sqrt{A}&D_-&-B&D_+p_1\\\alphastarrho&\sqrt{A}&Bp_1&-D_+p_1&-D_-\end{array}}\\[0.5\baselineskip]F_{{}_\alpha\rho}^{\alphastarrho\alphastarrho{}_\alpha\rho}=&{\small\begin{array}{c|cccc}&1&\rho&{}_\alpha\rho&\alphastarrho\\\hline\alpha&A&-p_1p_2\sqrt{A}&p_1\sqrt{A}&p_1\sqrt{A}\\\rho&-p_1\sqrt{A}&-Bp_2&-D_+&-D_-\\{}_\alpha\rho&p_1\sqrt{A}&-D_+p_2&D_-&-B\\\alphastarrho&p_1\sqrt{A}&-D_-p_2&-B&D_+\end{array}}\\[0.5\baselineskip]F_{\alphastarrho}^{\rho\rho\alphastarrho}=&{\small\begin{array}{c|cccc}&1&\rho&{}_\alpha\rho&\alphastarrho\\\hline\alpha&A&\sqrt{A}&-p_1\sqrt{A}&p_1\sqrt{A}\\\rho&p_1\sqrt{A}&D_+p_1&-D_-&-B\\{}_\alpha\rho&p_1p_2\sqrt{A}&D_-p_1p_2&Bp_2&D_+p_2\\\alphastarrho&\sqrt{A}&-B&-D_+p_1&D_-p_1\end{array}}\\[0.5\baselineskip]F_{\alphastarrho}^{\rho{}_\alpha\rho\rho}=&{\small\begin{array}{c|cccc}&\alpha^*&\rho&{}_\alpha\rho&\alphastarrho\\\hline\alpha&-A&-p_1\sqrt{A}&p_1\sqrt{A}&-\sqrt{A}\\\rho&p_1\sqrt{A}&-B&-D_+&D_-p_1\\{}_\alpha\rho&\sqrt{A}&D_+p_1&-D_-p_1&-B\\\alphastarrho&-p_1\sqrt{A}&-D_-&-B&-D_+p_1\end{array}}\\[0.5\baselineskip]F_{\alphastarrho}^{\rho\alphastarrho{}_\alpha\rho}=&{\small\begin{array}{c|cccc}&\alpha&\rho&{}_\alpha\rho&\alphastarrho\\\hline\alpha&-Ap_1&p_1\sqrt{A}&p_2\sqrt{A}&p_1\sqrt{A}\\\rho&\sqrt{A}&-D_-&Bp_1p_2&-D_+\\{}_\alpha\rho&-\sqrt{A}&-B&D_+p_1p_2&D_-\\\alphastarrho&-\sqrt{A}&D_+&D_-p_1p_2&-B\end{array}}\\[0.5\baselineskip]F_{\alphastarrho}^{{}_\alpha\rho\rho{}_\alpha\rho}=&{\small\begin{array}{c|cccc}&\alpha&\rho&{}_\alpha\rho&\alphastarrho\\\hline\alpha^*&-A&-\sqrt{A}&-\sqrt{A}&p_1\sqrt{A}\\\rho&-p_1\sqrt{A}&-D_+p_1&-D_-p_1&-B\\{}_\alpha\rho&\sqrt{A}&D_-&-B&-D_+p_1\\\alphastarrho&\sqrt{A}&-B&D_+&-D_-p_1\end{array}}\\[0.5\baselineskip]F_{\alphastarrho}^{{}_\alpha\rho{}_\alpha\rho\alphastarrho}=&{\small\begin{array}{c|cccc}&1&\rho&{}_\alpha\rho&\alphastarrho\\\hline\alpha^*&A&\sqrt{A}&-p_1\sqrt{A}&-p_1\sqrt{A}\\\rho&-p_1\sqrt{A}&Bp_1&D_+&D_-\\{}_\alpha\rho&-p_1\sqrt{A}&-D_+p_1&D_-&-B\\\alphastarrho&-p_1\sqrt{A}&-D_-p_1&-B&D_+\end{array}}\\[0.5\baselineskip]F_{\alphastarrho}^{{}_\alpha\rho\alphastarrho\rho}=&{\small\begin{array}{c|cccc}&\alpha^*&\rho&{}_\alpha\rho&\alphastarrho\\\hline\alpha^*&-Ap_1&\sqrt{A}&p_1\sqrt{A}&-\sqrt{A}\\\rho&-\sqrt{A}&D_-p_1&-B&-D_+p_1\\{}_\alpha\rho&\sqrt{A}&Bp_1&-D_+&D_-p_1\\\alphastarrho&p_1\sqrt{A}&-D_+&-D_-p_1&-B\end{array}}\\[0.5\baselineskip]F_{\alphastarrho}^{\alphastarrho\rho\rho}=&{\small\begin{array}{c|cccc}&\alpha^*&\rho&{}_\alpha\rho&\alphastarrho\\\hline1&A&p_1\sqrt{A}&-\sqrt{A}&\sqrt{A}\\\rho&\sqrt{A}&D_+p_1&-D_-&-B\\{}_\alpha\rho&-p_1\sqrt{A}&-D_-&-Bp_1&-D_+p_1\\\alphastarrho&p_1\sqrt{A}&-B&-D_+p_1&D_-p_1\end{array}}\\[0.5\baselineskip]F_{\alphastarrho}^{\alphastarrho{}_\alpha\rho{}_\alpha\rho}=&{\small\begin{array}{c|cccc}&\alpha&\rho&{}_\alpha\rho&\alphastarrho\\\hline1&A&\sqrt{A}&-p_1\sqrt{A}&-p_1\sqrt{A}\\\rho&\sqrt{A}&-B&-D_+p_1&-D_-p_1\\{}_\alpha\rho&-p_1\sqrt{A}&-D_+p_1&D_-&-B\\\alphastarrho&-p_1\sqrt{A}&-D_-p_1&-B&D_+\end{array}}\\[0.5\baselineskip]F_{\alphastarrho}^{\alphastarrho\alphastarrho\alphastarrho}=&{\small\begin{array}{c|cccc}&1&\rho&{}_\alpha\rho&\alphastarrho\\\hline1&A&-p_1p_2\sqrt{A}&p_1\sqrt{A}&p_1\sqrt{A}\\\rho&-p_1p_2\sqrt{A}&D_-&Bp_2&-D_+p_2\\{}_\alpha\rho&p_1\sqrt{A}&Bp_2&D_+&D_-\\\alphastarrho&p_1\sqrt{A}&-D_+p_2&D_-&-B\end{array}}
\end{align}}
\chapter{Numerical results for the $\Hd$ anyon chain}
\label{app:H3chain}

In the following, we depict the entropy landscape for the Hamiltonian described in \cref{sec:H3chain}:
	\begin{equation}
		H=-\sum_i \cos\psi\ p_i^{(\mathbf{1})}+\sin\psi\cos\theta\  p_i^{(\rho)}+\sin\psi\sin\theta\cos\varphi\ p_i^{(\alpharho)}+\sin\psi\sin\theta\sin\varphi\  p_i^{(\alphastarrho)}.
	\end{equation}
Each diagram corresponds to a fixed value of $\varphi$. The interval $[0,\pi]$ is always divided into $16$ pieces in order to generate data points. The resulting landscape is then visualised by a contour plot, where dark colours correspond to a low entropy and bright colours correspond to a high entropy.

\noindent
\begin{minipage}{0.48\textwidth}
	\centering
	\begin{figure}[H]
\\
		\includegraphics[width=1\textwidth]{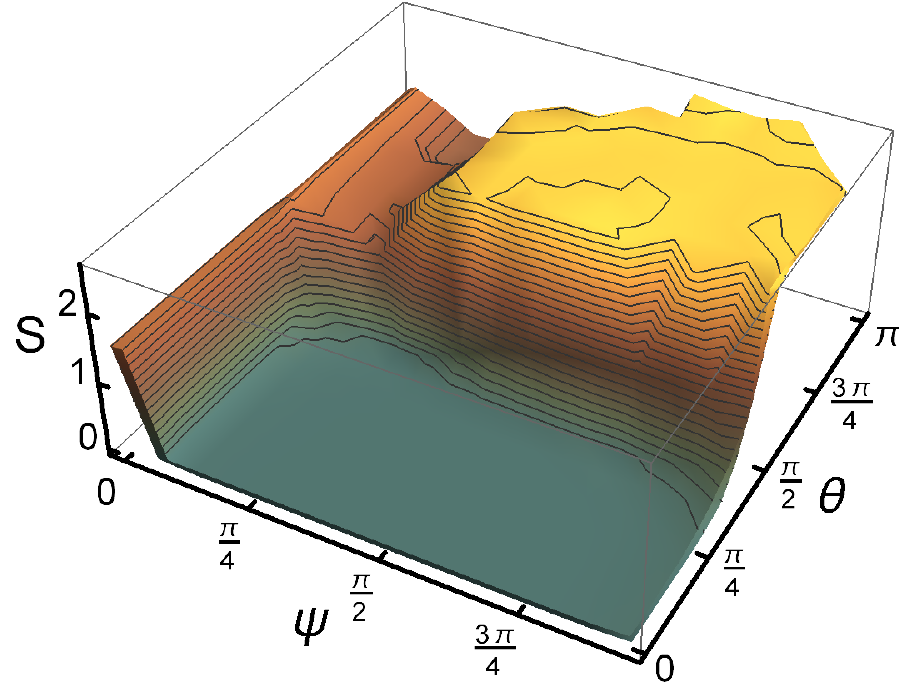}
	\end{figure}
\end{minipage}\hspace{10pt}
\begin{minipage}{0.48\textwidth}
	\centering
	\begin{figure}[H]
		%
\\
		\includegraphics[width=1\textwidth]{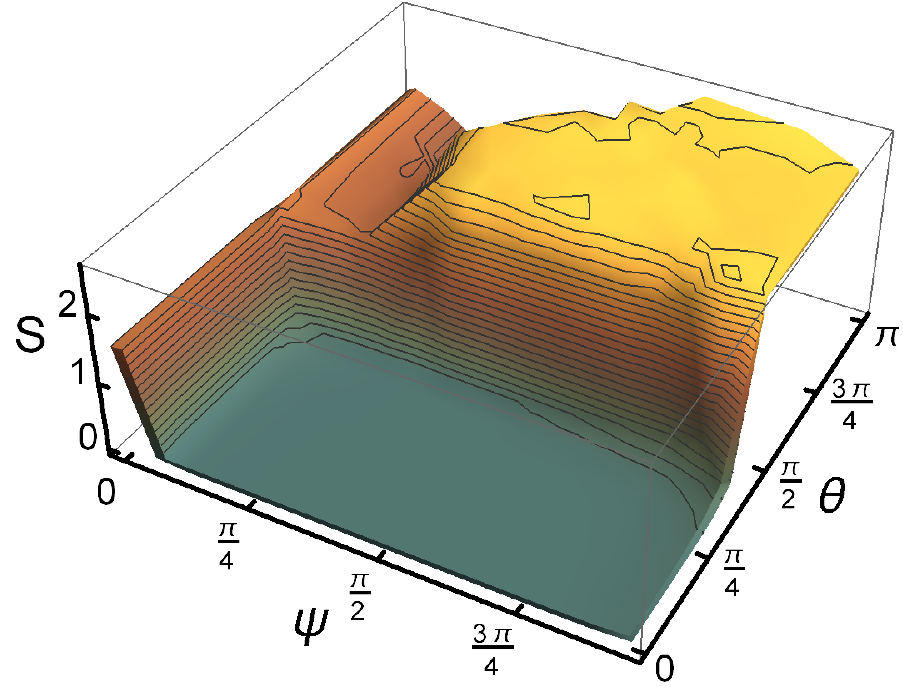}
	\end{figure}
\end{minipage}
\vspace{10pt}

\noindent
\begin{minipage}{0.48\textwidth}
	\centering
	\begin{figure}[H]
		%
\\
		\includegraphics[width=1\textwidth]{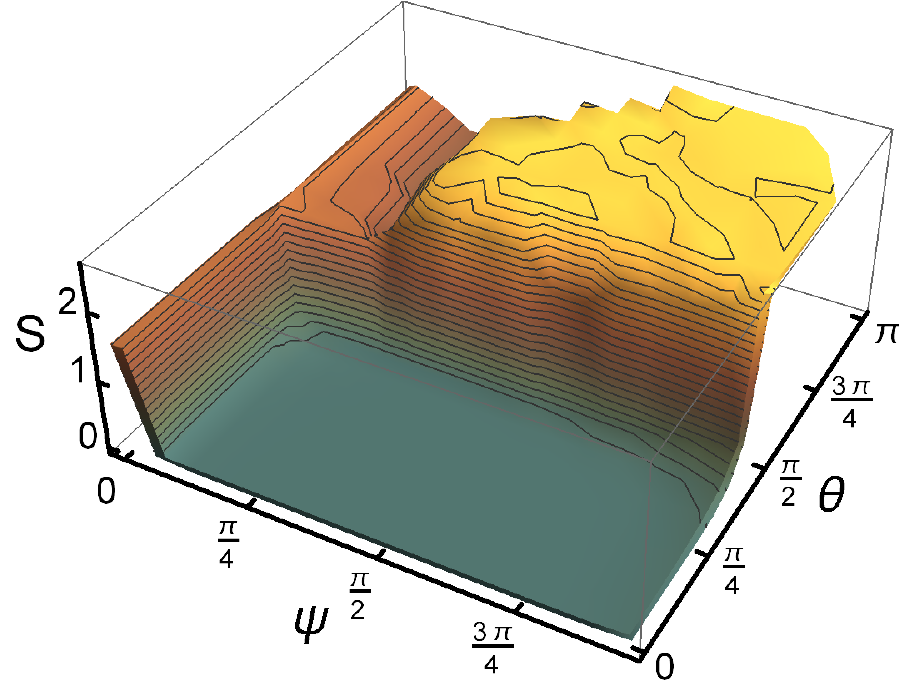}
	\end{figure}
\end{minipage}\hspace{10pt}
\begin{minipage}{0.48\textwidth}
	\centering
	\begin{figure}[H]
		%
\\
		\includegraphics[width=1\textwidth]{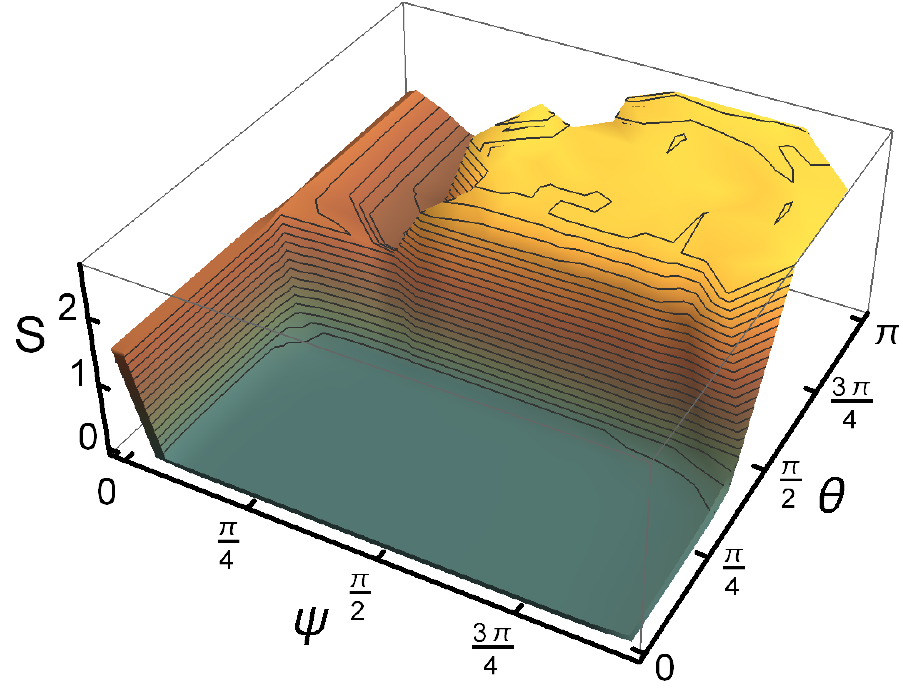}
	\end{figure}
\end{minipage}
\vspace{10pt}

\noindent
\begin{minipage}{0.48\textwidth}
	\centering
	\begin{figure}[H]
		%
\\
		\includegraphics[width=1\textwidth]{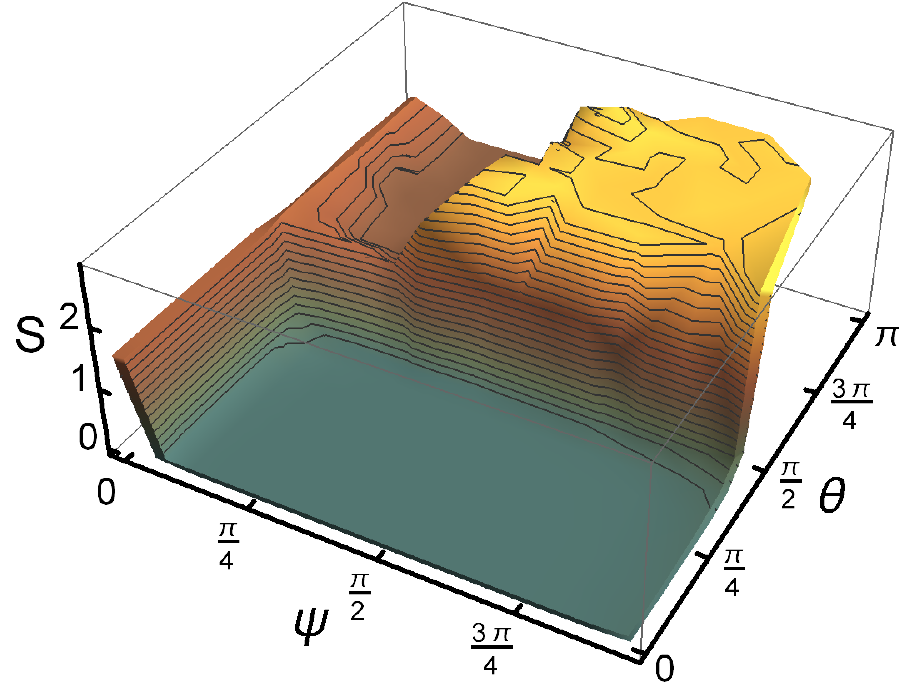}
	\end{figure}
\end{minipage}\hspace{10pt}
\begin{minipage}{0.48\textwidth}
	\centering
	\begin{figure}[H]
		%
\\
		\includegraphics[width=1\textwidth]{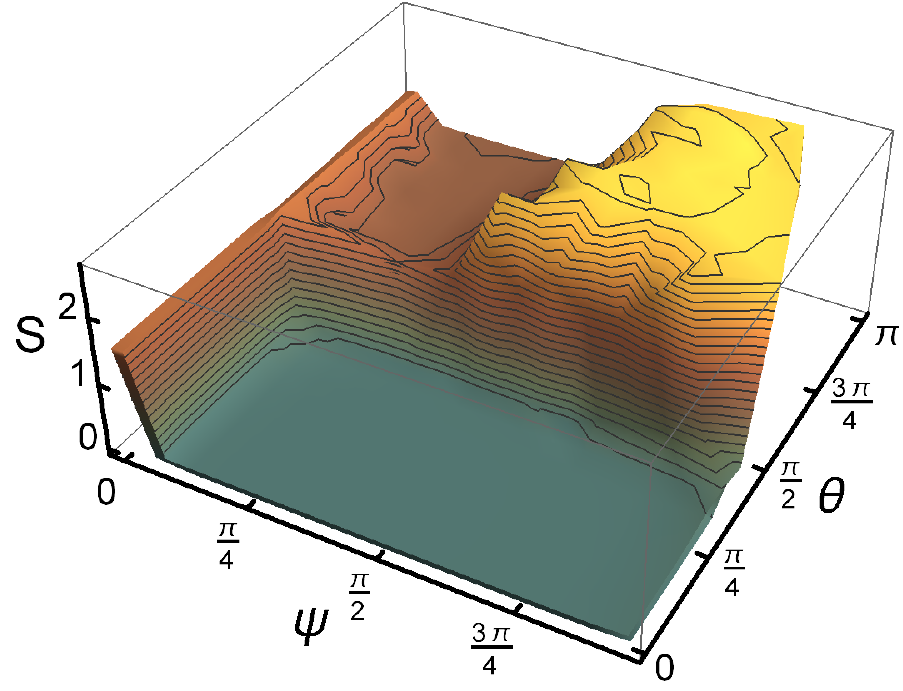}
	\end{figure}
\end{minipage}
\vspace{10pt}

\noindent
\begin{minipage}{0.48\textwidth}
	\centering
	\begin{figure}[H]
		%
\\
		\includegraphics[width=1\textwidth]{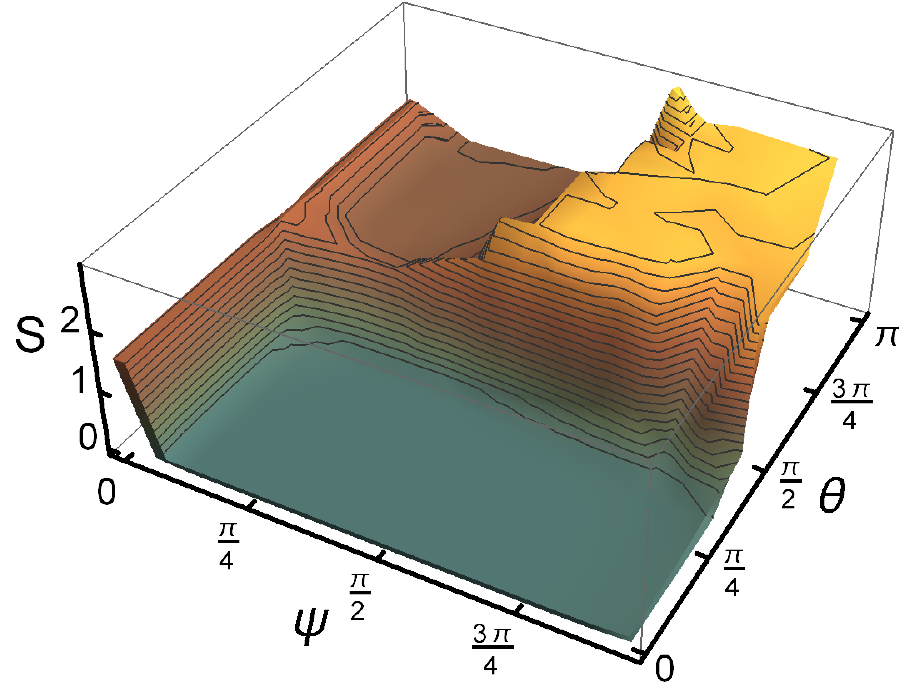}
	\end{figure}
\end{minipage}\hspace{10pt}
\begin{minipage}{0.48\textwidth}
	\centering
	\begin{figure}[H]
		%
\\
		\includegraphics[width=1\textwidth]{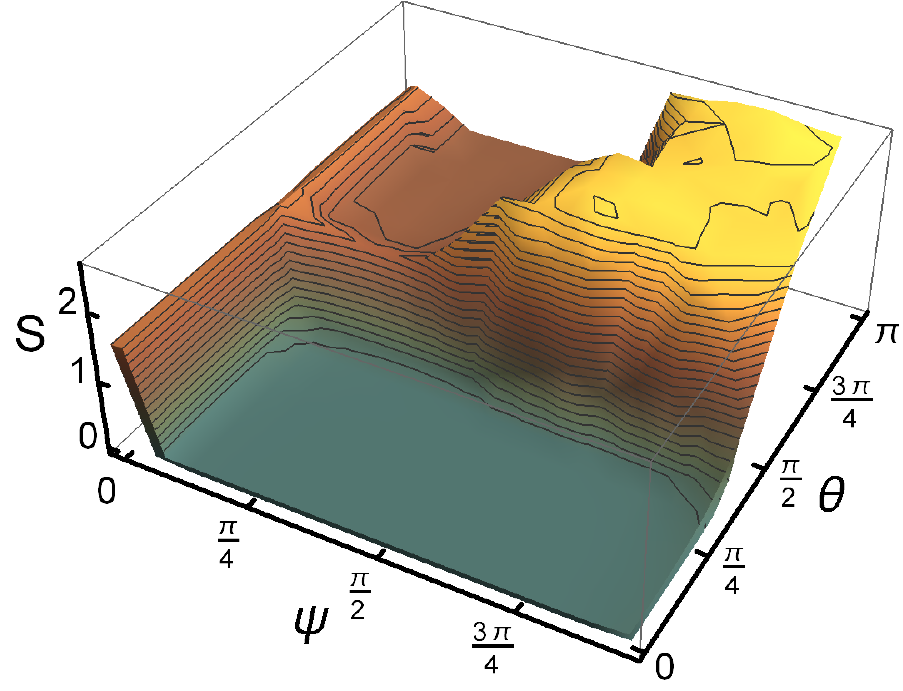}
	\end{figure}
\end{minipage}
\vspace{10pt}

\noindent
\begin{minipage}{0.48\textwidth}
	\centering
	\begin{figure}[H]
		%
\\
		\includegraphics[width=1\textwidth]{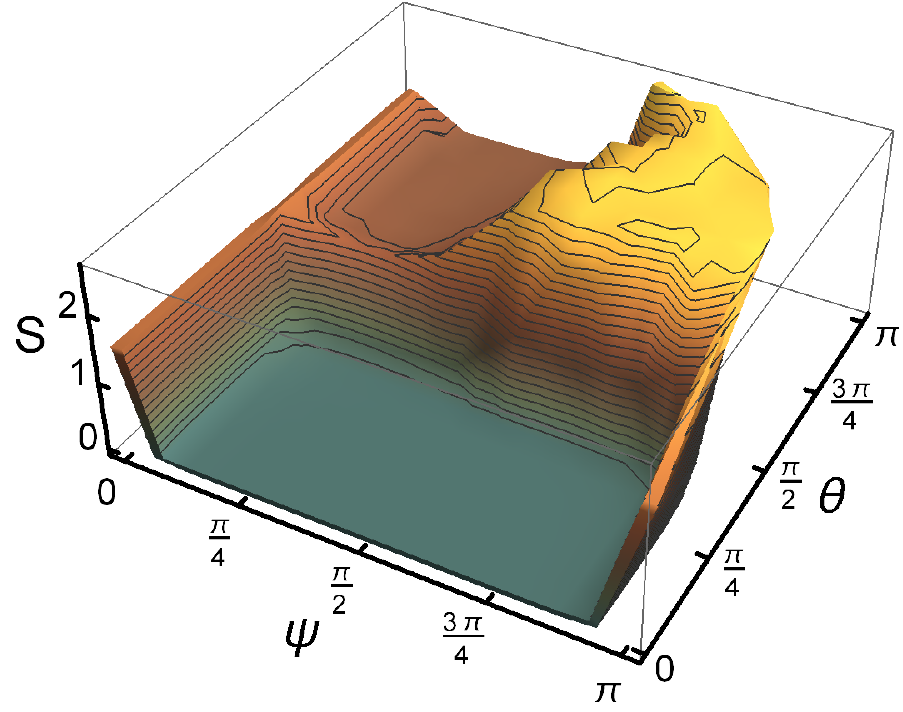}
	\end{figure}
\end{minipage}\hspace{10pt}
\begin{minipage}{0.48\textwidth}
	\centering
	\begin{figure}[H]
		%
\\
		\includegraphics[width=1\textwidth]{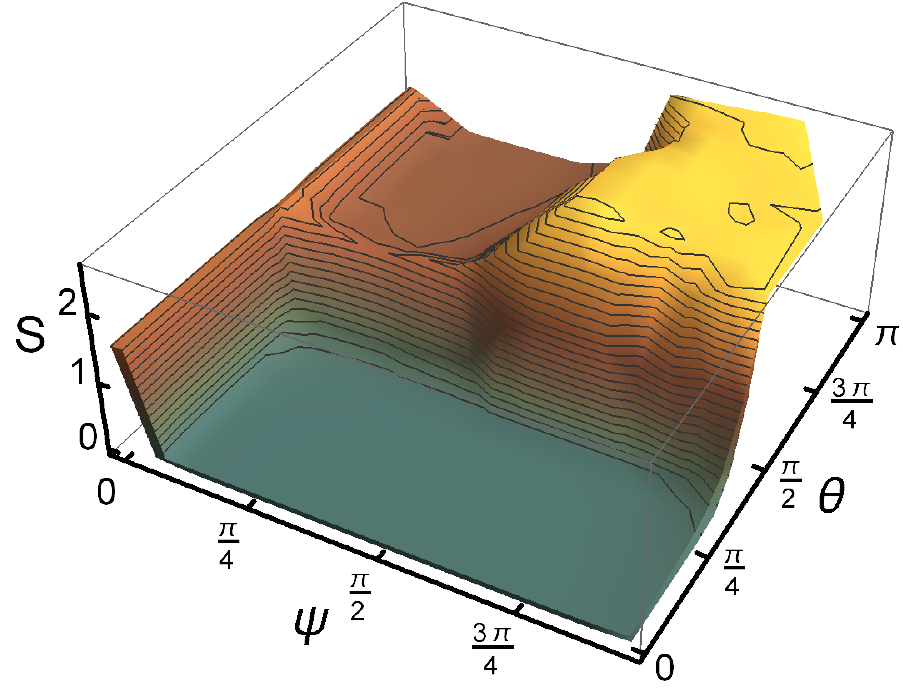}
	\end{figure}
\end{minipage}
\vspace{10pt}

\noindent
\begin{minipage}{0.48\textwidth}
	\centering
	\begin{figure}[H]
		%
\\
		\includegraphics[width=1\textwidth]{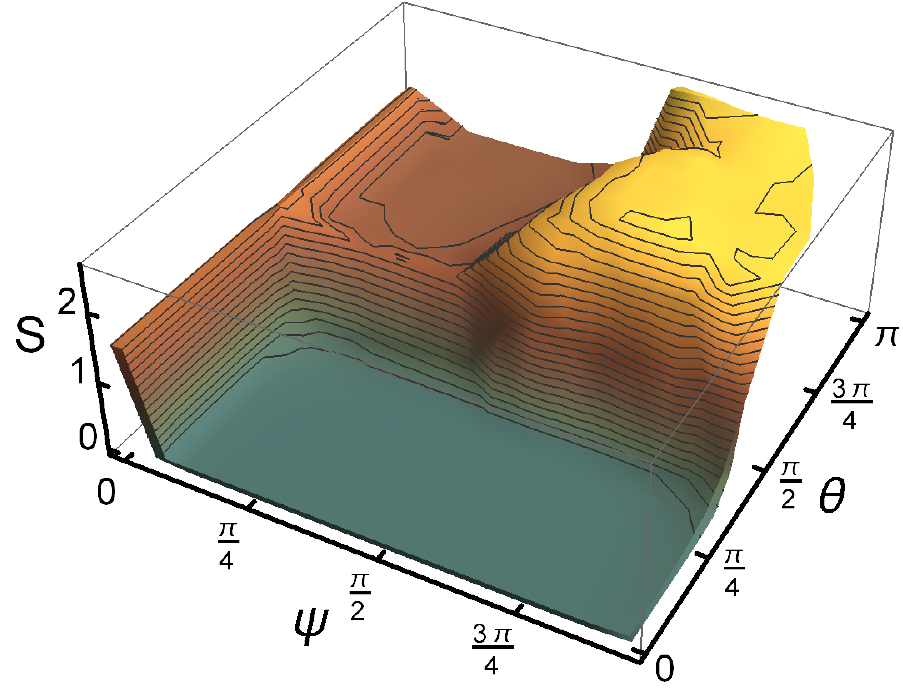}
	\end{figure}
\end{minipage}\hspace{10pt}
\begin{minipage}{0.48\textwidth}
	\centering
	\begin{figure}[H]
		%
\\
		\includegraphics[width=1\textwidth]{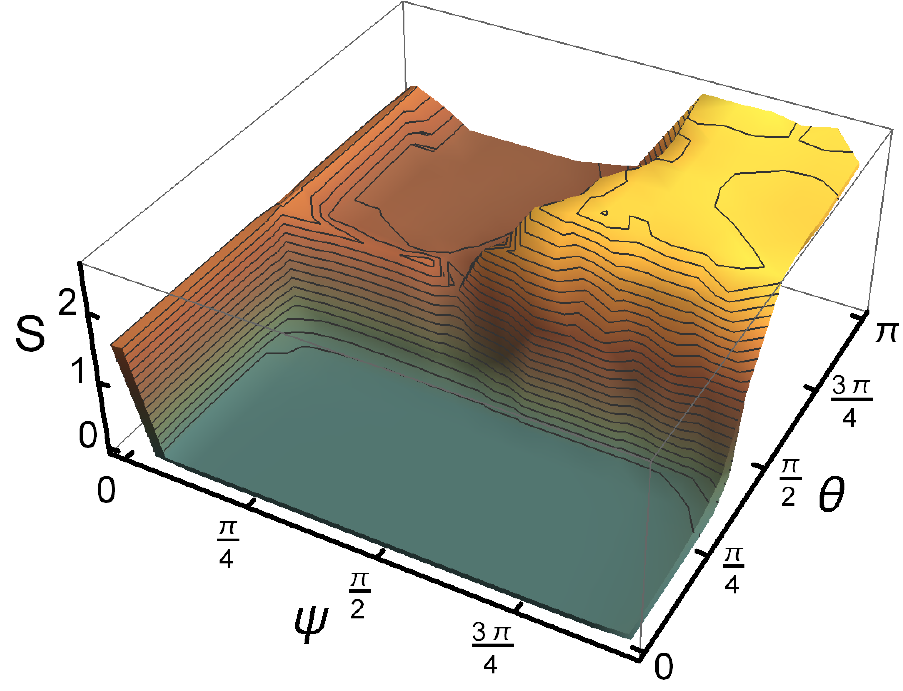}
	\end{figure}
\end{minipage}
\vspace{10pt}

\noindent
\begin{minipage}{0.48\textwidth}
	\centering
	\begin{figure}[H]
		%
\\
		\includegraphics[width=1\textwidth]{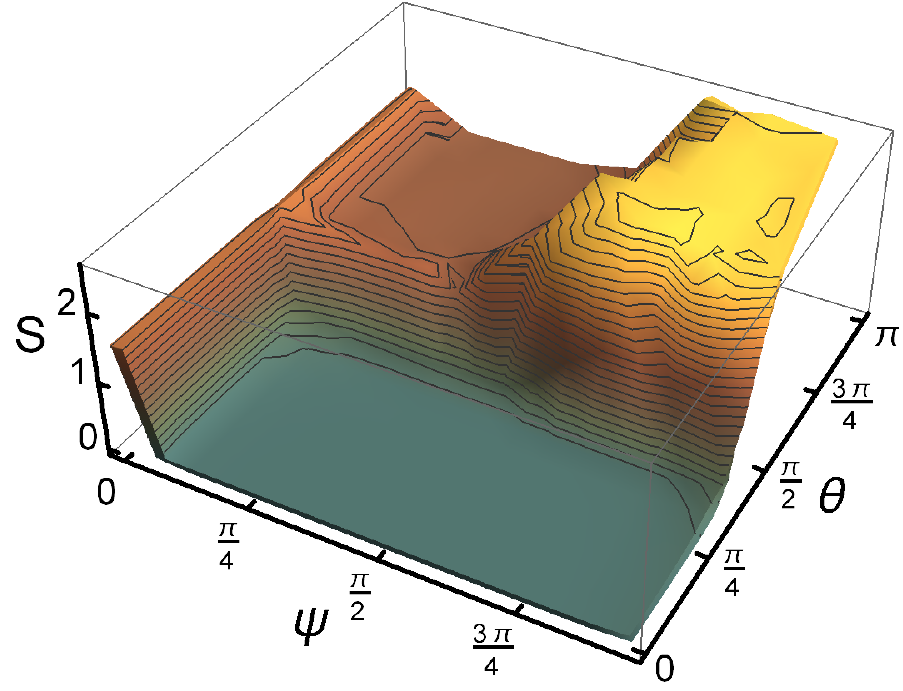}
	\end{figure}
\end{minipage}\hspace{10pt}
\begin{minipage}{0.48\textwidth}
	\centering
	\begin{figure}[H]
		%
\\
		\includegraphics[width=1\textwidth]{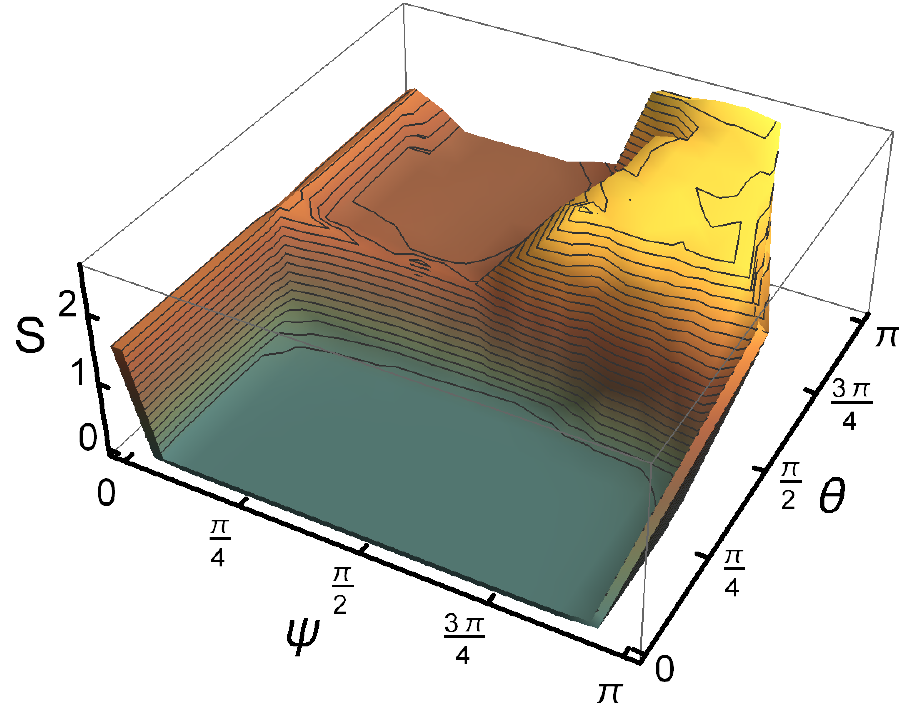}
	\end{figure}
\end{minipage}
\vspace{10pt}

\noindent
\begin{minipage}{0.48\textwidth}
	\centering
	\begin{figure}[H]
		%
\\
		\includegraphics[width=1\textwidth]{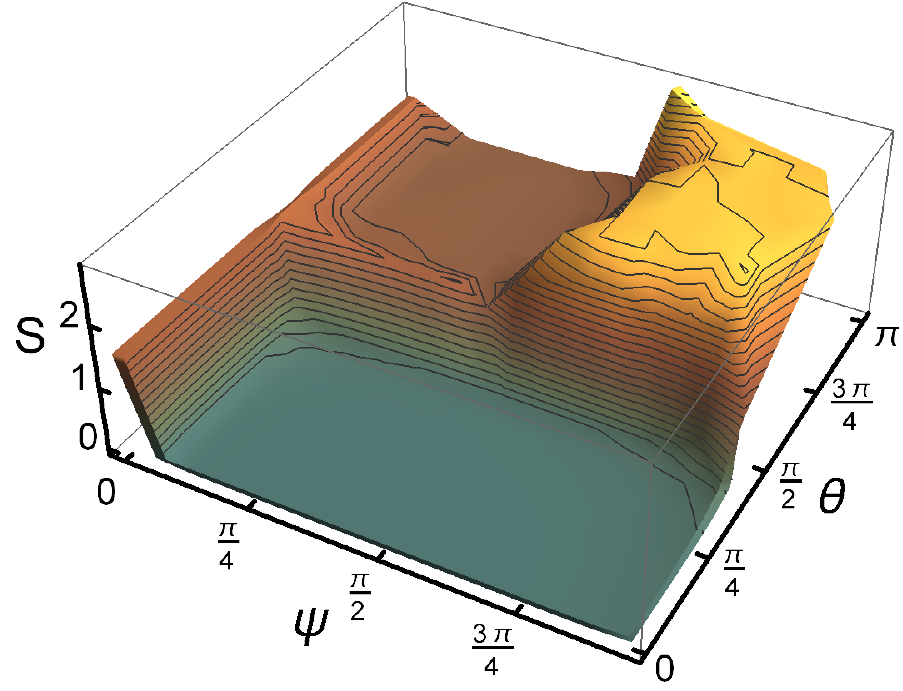}
	\end{figure}
\end{minipage}\hspace{10pt}
\begin{minipage}{0.48\textwidth}
	\centering
	\begin{figure}[H]
		%
\\
		\includegraphics[width=1\textwidth]{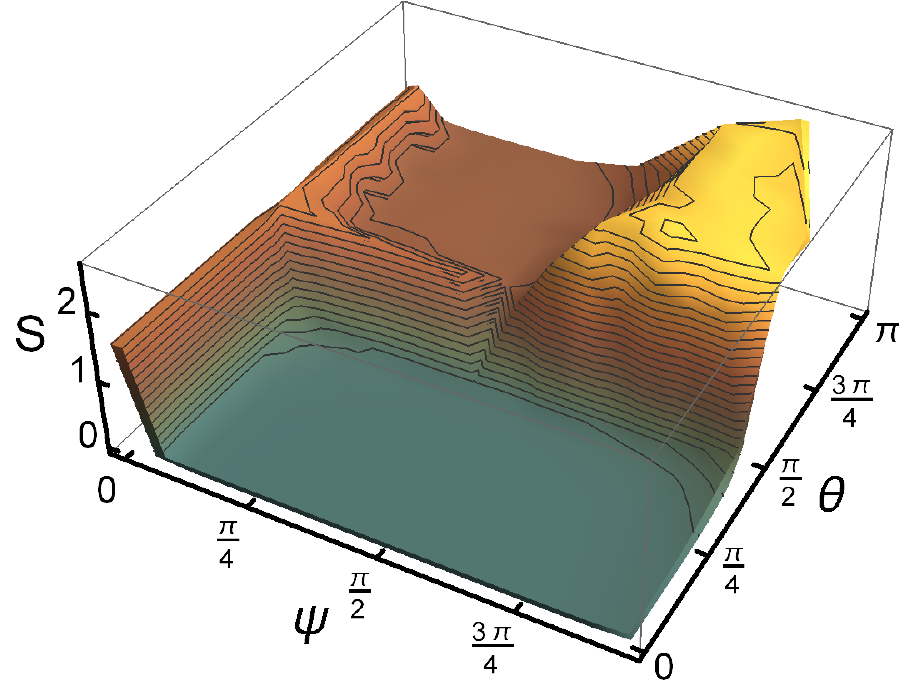}
	\end{figure}
\end{minipage}
\vspace{10pt}

\noindent
\begin{minipage}{0.48\textwidth}
	\centering
	\begin{figure}[H]
		%
\\
		\includegraphics[width=1\textwidth]{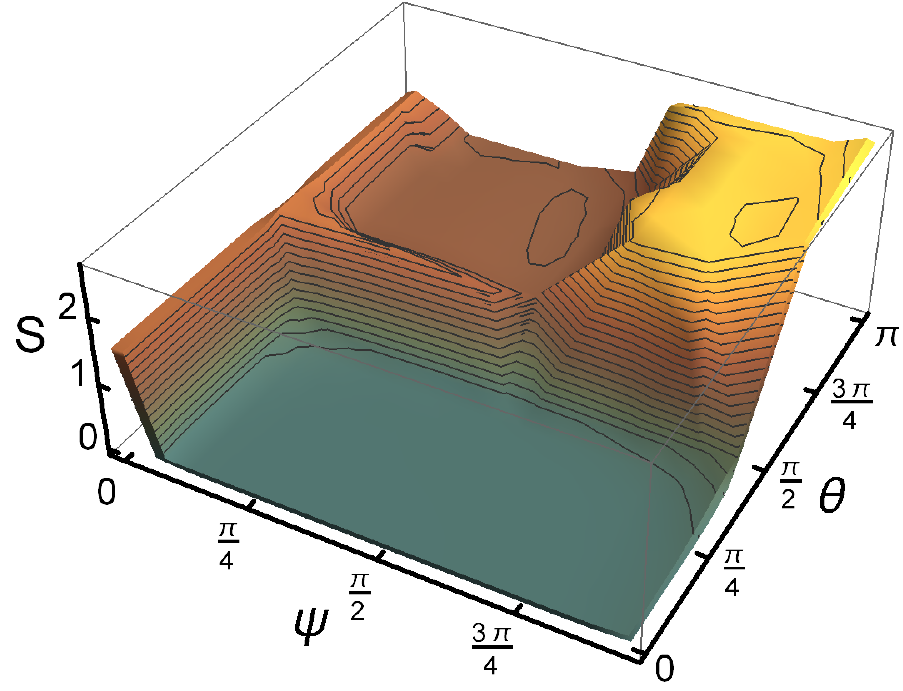}
	\end{figure}
\end{minipage}\hspace{10pt}
\begin{minipage}{0.48\textwidth}
	\centering
	\begin{figure}[H]
		%
\\
		\includegraphics[width=1\textwidth]{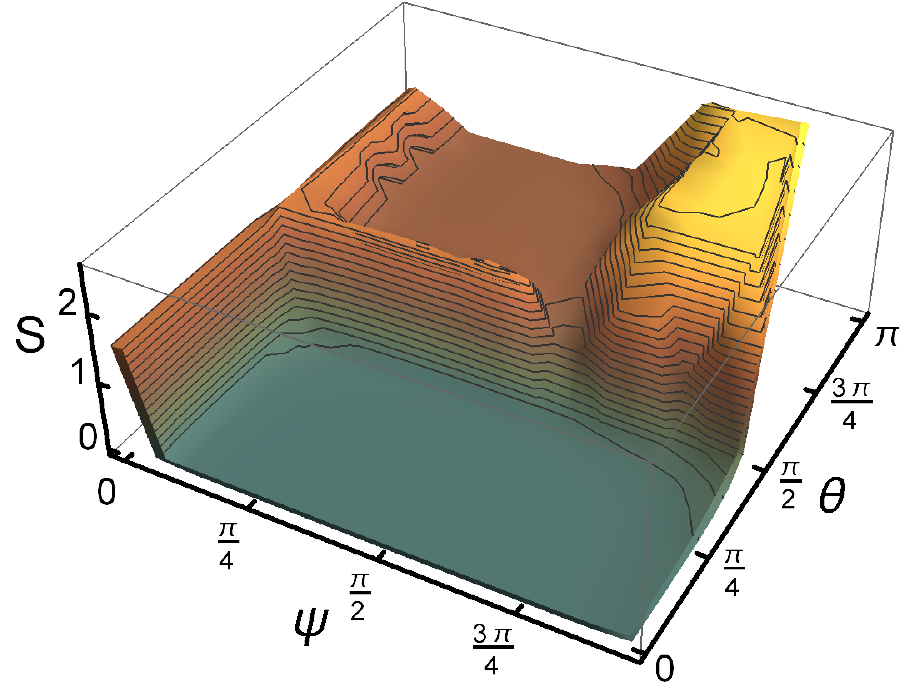}
	\end{figure}
\end{minipage}
\vspace{10pt}

\noindent
\begin{minipage}{0.48\textwidth}
	\centering
	\begin{figure}[H]
		%
\\
		\includegraphics[width=1\textwidth]{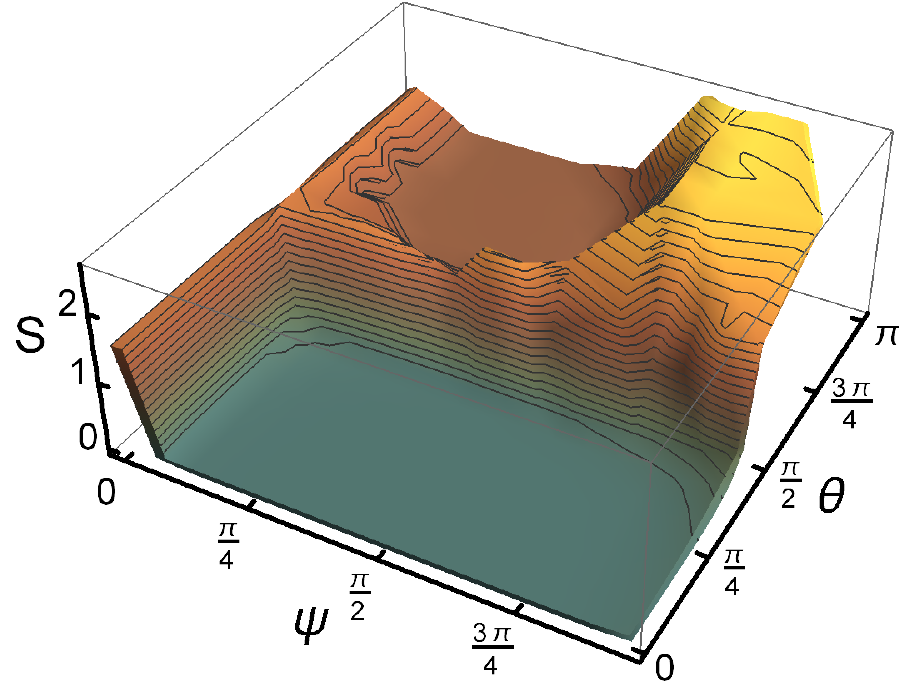}
	\end{figure}
\end{minipage}\hspace{10pt}
\begin{minipage}{0.48\textwidth}
	\centering
	\begin{figure}[H]
		%
\\
		\includegraphics[width=1\textwidth]{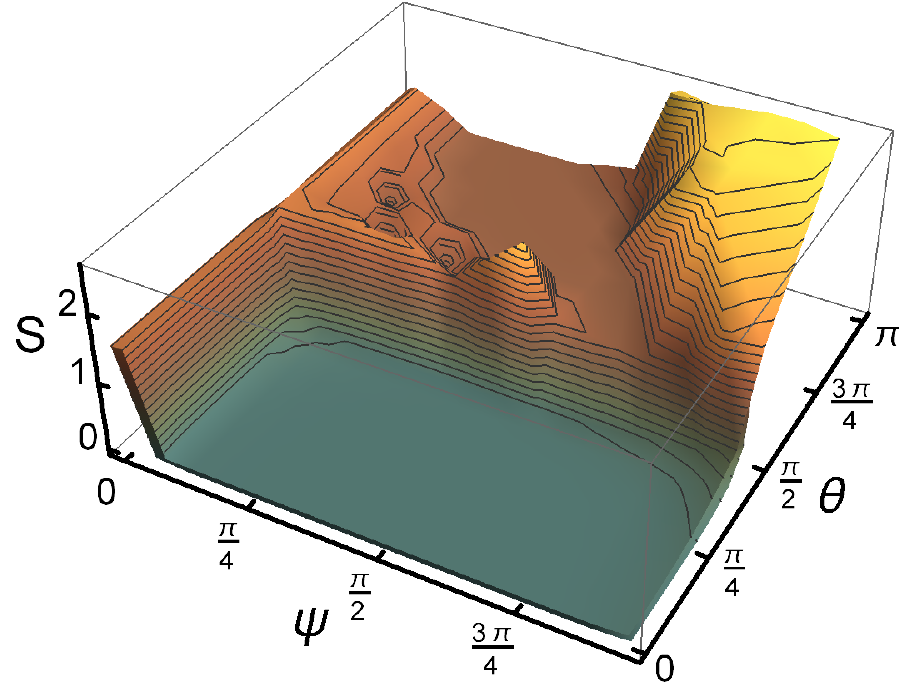}
	\end{figure}
\end{minipage}
\vspace{10pt}

\noindent
\begin{minipage}{0.48\textwidth}
	\centering
	\begin{figure}[H]
		%
\\
		\includegraphics[width=1\textwidth]{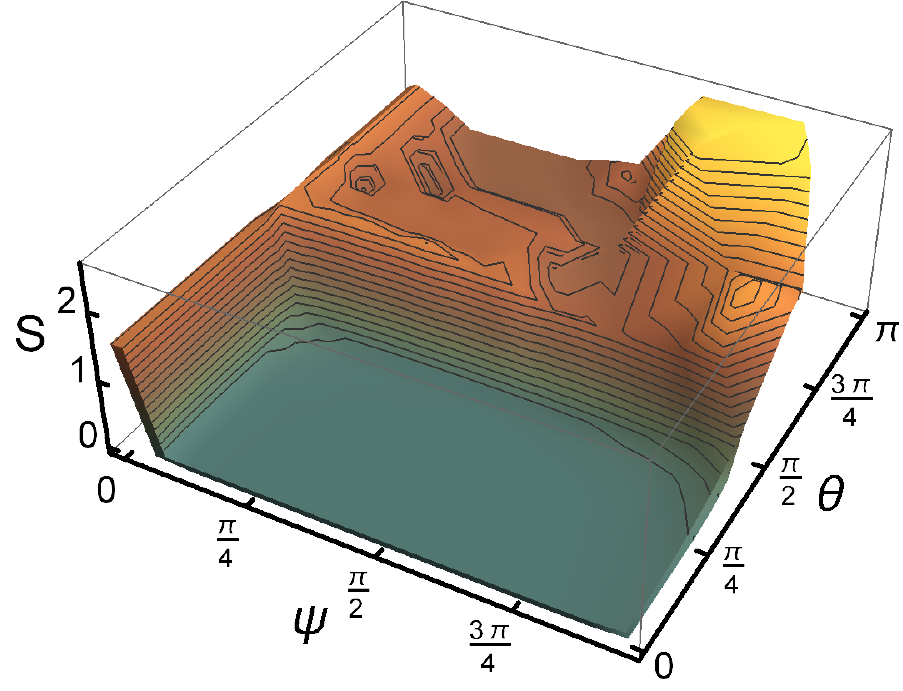}
	\end{figure}
\end{minipage}\hspace{10pt}
\begin{minipage}{0.48\textwidth}
	\centering
	\begin{figure}[H]
		%
\\
		\includegraphics[width=1\textwidth]{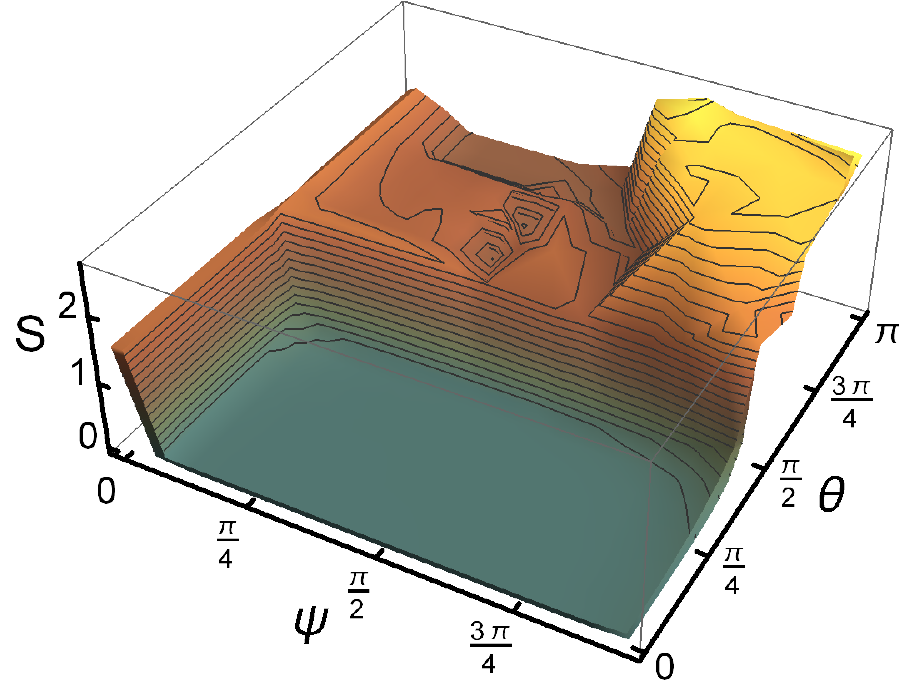}
	\end{figure}
\end{minipage}
\vspace{10pt}

\noindent
\begin{minipage}{0.48\textwidth}
	\centering
	\begin{figure}[H]
		%
\\
		\includegraphics[width=1\textwidth]{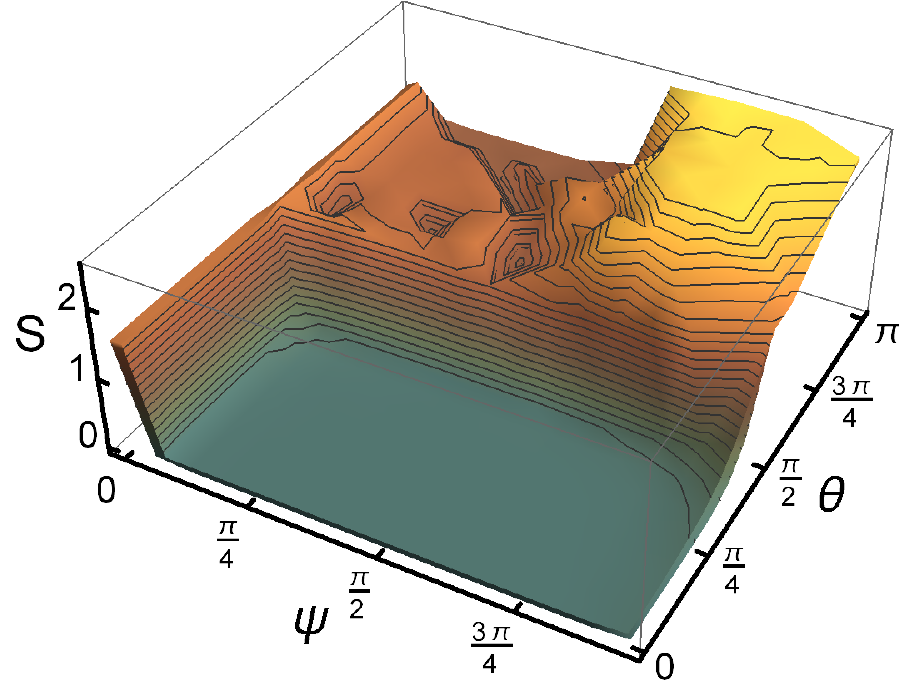}
	\end{figure}
\end{minipage}\hspace{10pt}
\begin{minipage}{0.48\textwidth}
	\centering
	\begin{figure}[H]
		%
\\
		\includegraphics[width=1\textwidth]{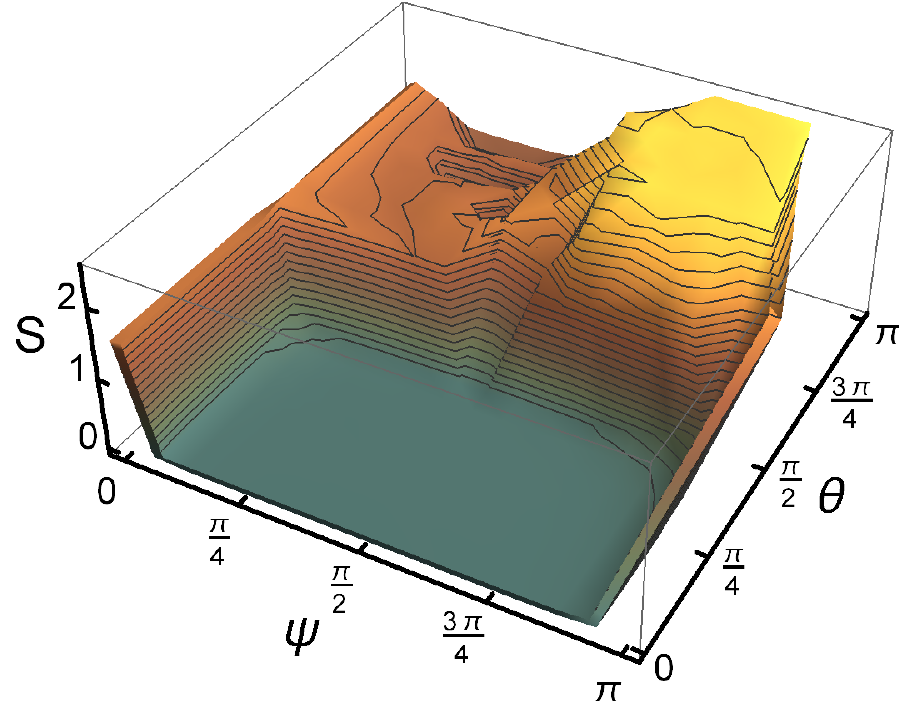}
	\end{figure}
\end{minipage}
\vspace{10pt}

\noindent
\begin{minipage}{0.48\textwidth}
	\centering
	\begin{figure}[H]
		%
\\
		\includegraphics[width=1\textwidth]{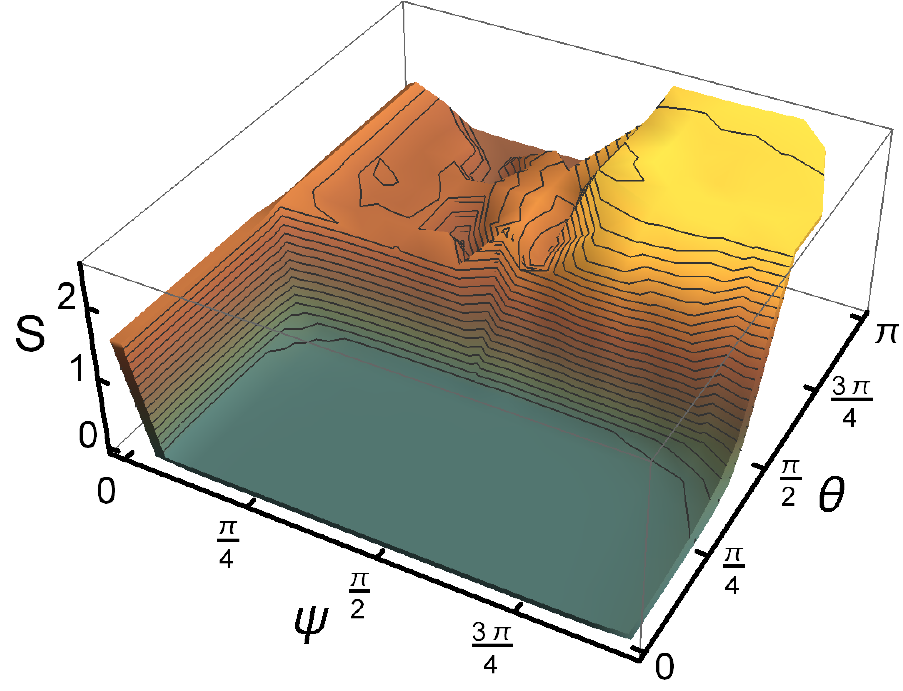}
	\end{figure}
\end{minipage}\hspace{10pt}
\begin{minipage}{0.48\textwidth}
	\centering
	\begin{figure}[H]
		%
\\
		\includegraphics[width=1\textwidth]{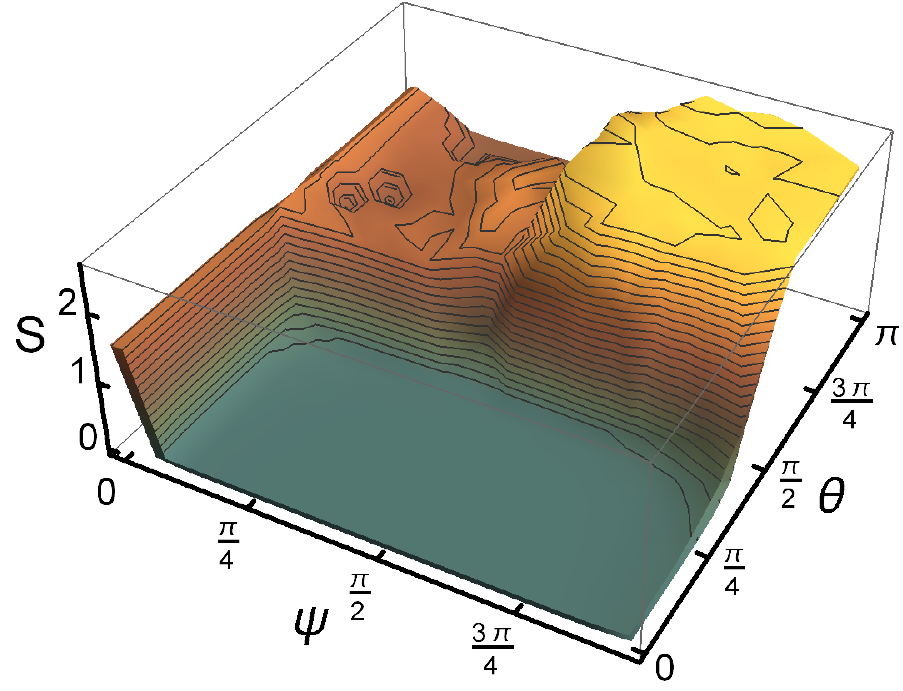}
	\end{figure}
\end{minipage}
\vspace{10pt}

\noindent
\begin{minipage}{0.48\textwidth}
	\centering
	\begin{figure}[H]
		%
\\
		\includegraphics[width=1\textwidth]{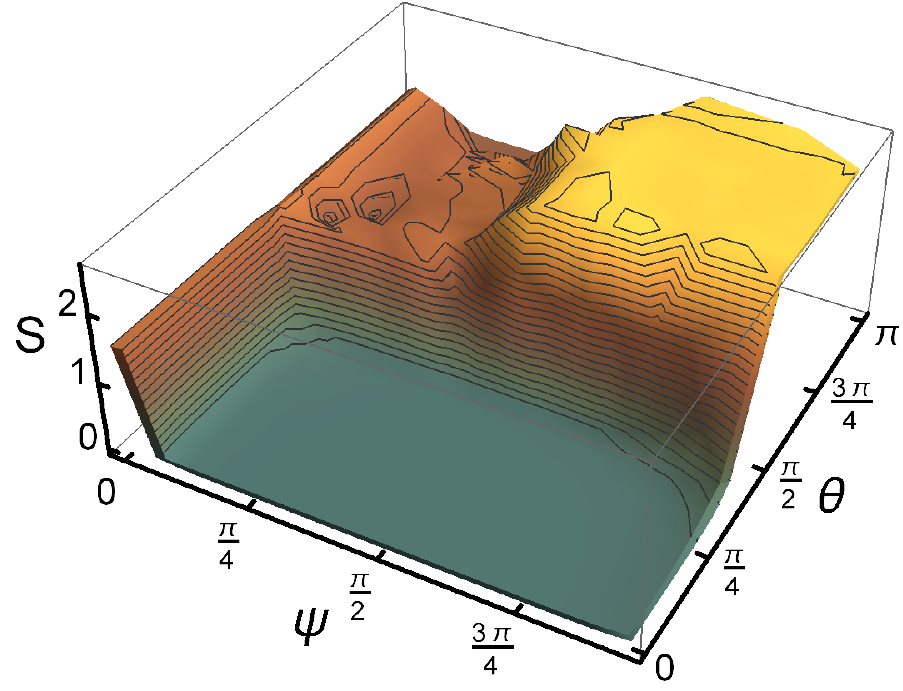}
	\end{figure}
\end{minipage}\hspace{10pt}
\begin{minipage}{0.48\textwidth}
	\centering
	\begin{figure}[H]
		%
\\
		\includegraphics[width=1\textwidth]{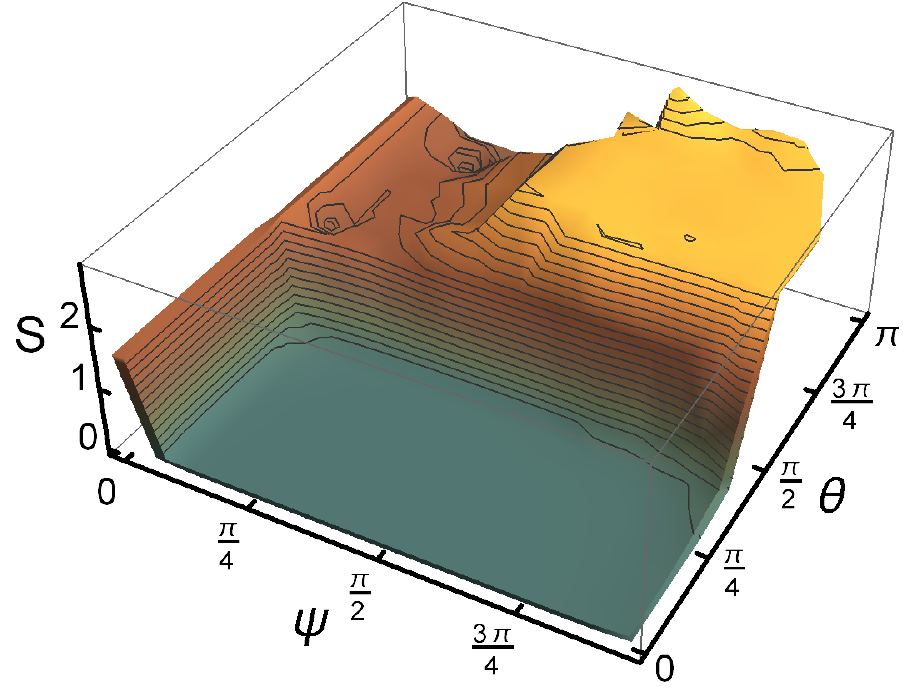}
	\end{figure}
\end{minipage}
\vspace{10pt}

\noindent
\begin{minipage}{0.48\textwidth}
	\centering
	\begin{figure}[H]
		%
\\
		\includegraphics[width=1\textwidth]{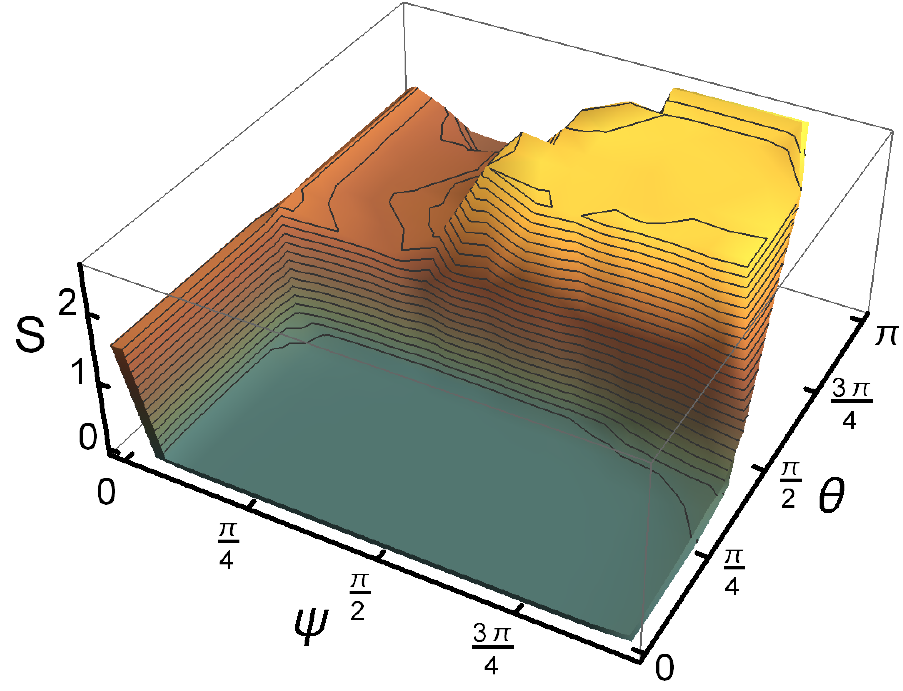}
	\end{figure}
\end{minipage}\hspace{10pt}
\begin{minipage}{0.48\textwidth}
	\centering
	\begin{figure}[H]
		%
\\
		\includegraphics[width=1\textwidth]{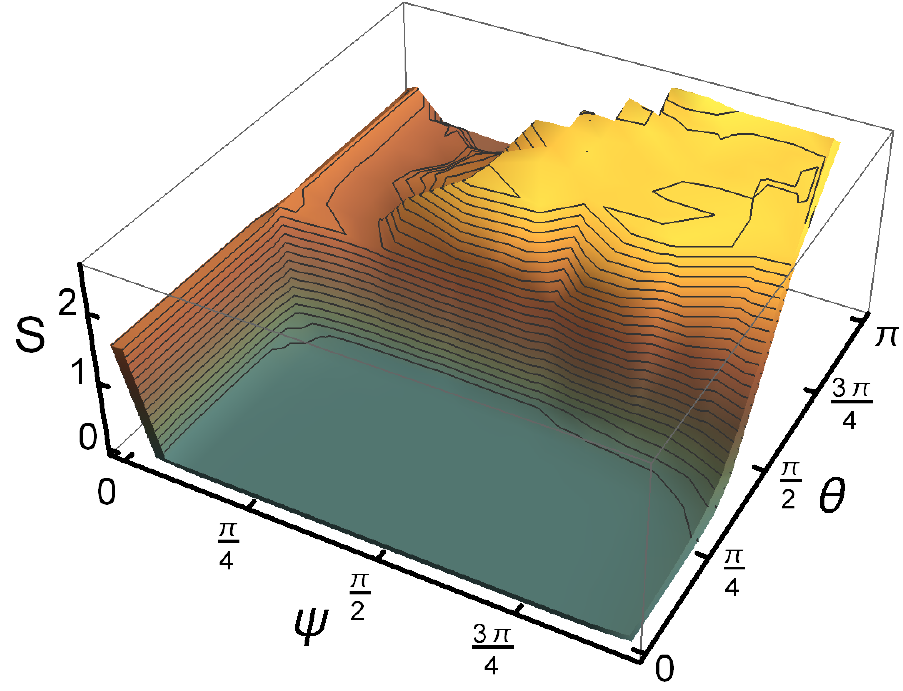}
	\end{figure}
\end{minipage}
\vspace{10pt}

\noindent
\begin{minipage}{0.48\textwidth}
	\centering
	\begin{figure}[H]
		%
\\
		\includegraphics[width=1\textwidth]{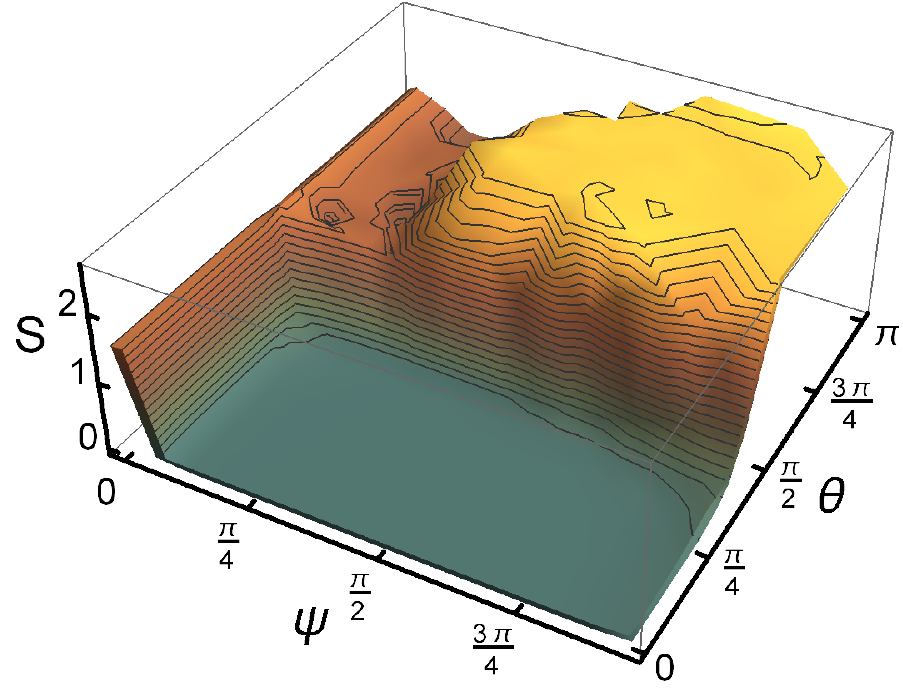}
	\end{figure}
\end{minipage}\hspace{10pt}
\begin{minipage}{0.48\textwidth}
	\centering
	\begin{figure}[H]
		%
\\
		\includegraphics[width=1\textwidth]{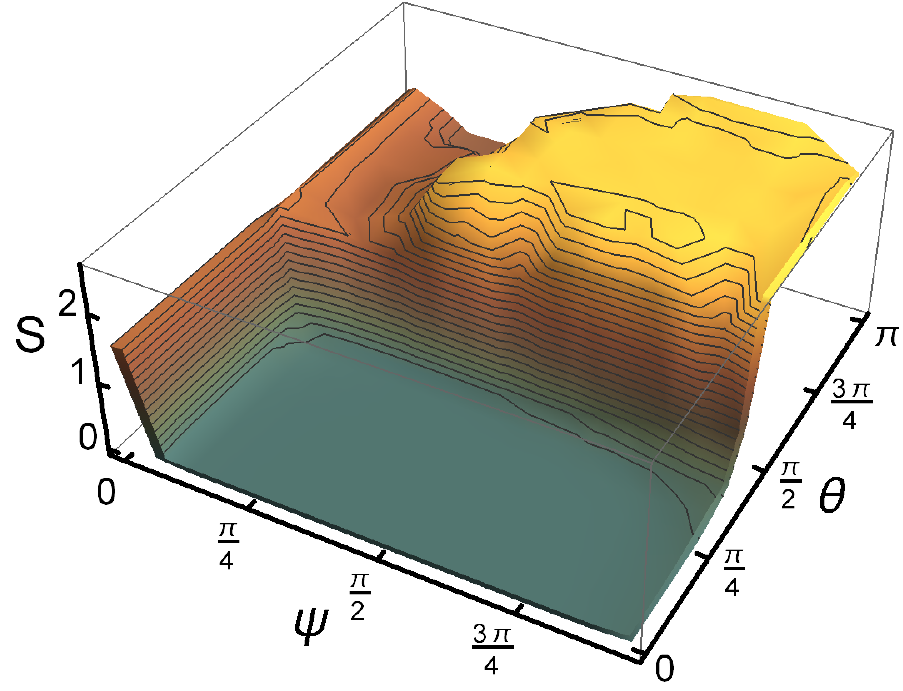}
	\end{figure}
\end{minipage}
\vspace{10pt}

\noindent
\begin{minipage}{0.48\textwidth}
	\centering
	\begin{figure}[H]
		%
\\
		\includegraphics[width=1\textwidth]{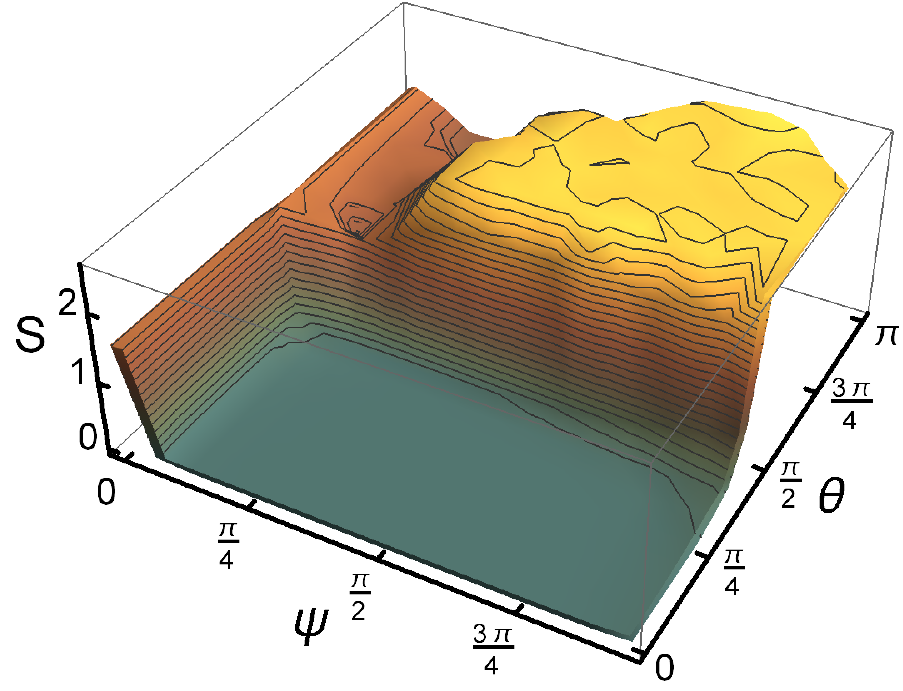}
	\end{figure}
\end{minipage}


\backmatter

\def\backref#1{{\scriptsize $\hookrightarrow$ [#1]}}

\bibliographystyle{amsalpha}
\bibliography{PhDbib}

\newcommand{\etalchar}[1]{$^{#1}$}
\providecommand{\bysame}{\leavevmode\hbox to3em{\hrulefill}\thinspace}
\providecommand{\MR}{\relax\ifhmode\unskip\space\fi MR }
\providecommand{\MRhref}[2]{%
  \href{http://www.ams.org/mathscinet-getitem?mr=#1}{#2}
}
\providecommand{\href}[2]{#2}
\begin{thebibliography}{BDSPV15}

\bibitem[ABF84]{Andrews1984}
George~E. Andrews, Rodney~J. Baxter, and Peter~J. Forrester,
  \emph{{Eight-vertex {SOS} model and generalized Rogers-Ramanujan-type
  identities}}, Journal of Statistical Physics \textbf{35} (1984), no.~3-4,
  193--266.

\bibitem[AGW09]{Ahlbrecht2009}
Andr{\'{e}} Ahlbrecht, Lachezar~S. Georgiev, and Reinhard~F. Werner,
  \emph{Implementation of {Clifford} gates in the {Ising}-anyon topological
  quantum computer}, Physical Review A \textbf{79} (2009), no.~3.

\bibitem[AH99]{asaeda_exotic_1999}
Marta Asaeda and Uffe Haagerup, \emph{Exotic subfactors of finite depth with
  {Jones} indices $(5+\sqrt{13})/2$ and $(5+\sqrt{17})/2$}, Communications in
  Mathematical Physics \textbf{202} (1999), no.~1, 1--63.

\bibitem[AKK94]{araki_subfactors_1994}
Huzihiro Araki, Yasuyuki Kawahigashi, and Hideki Kosaki, \emph{Subfactors},
  Subfactors, World Scientific, 1994, pp.~1--304.

\bibitem[Asa07]{Asaeda2007}
Marta Asaeda, \emph{{Galois Groups and an Obstruction to Principal Graphs of
  Subfactors}}, International Journal of Mathematics \textbf{18} (2007),
  no.~02, 191--202.

\bibitem[ASW84]{Arovas1984}
Daniel Arovas, John~R. Schrieffer, and Frank Wilczek, \emph{{Fractional
  Statistics and the Quantum Hall Effect}}, Physical Review Letters \textbf{53}
  (1984), no.~7, 722--723.

\bibitem[Awo10]{Awodey2010}
Steve Awodey, \emph{Category {Theory}}, Oxford University Press, 2010.

\bibitem[AY08]{Asaeda2008}
Marta Asaeda and Seidai Yasuda, \emph{{On Haagerup's List of Potential
  Principal Graphs of Subfactors}}, Communications in Mathematical Physics
  \textbf{286} (2008), no.~3, 1141--1157.

\bibitem[Bal10]{Balsam2010}
Benjamin Balsam, \emph{{Turaev-Viro invariants as an extended TQFT III}},
  arXiv:1012.0560.

\bibitem[BAV09]{Buerschaper2009}
Oliver Buerschaper, Miguel Aguado, and Guifr{\'{e}} Vidal, \emph{Explicit
  tensor network representation for the ground states of string-net models},
  Physical Review B \textbf{79} (2009), no.~8.

\bibitem[BB20a]{Bridgeman2020b}
Jacob~C. Bridgeman and Daniel Barter, \emph{Computing data for {Levin-Wen} with
  defects}, Quantum \textbf{4} (2020), 277.

\bibitem[BB20b]{Bridgeman2020a}
\bysame, \emph{Computing defects associated to bounded domain wall structures:
  the $\mathbb{Z}/p\mathbb{Z}$ case}, Journal of Physics A: Mathematical and
  Theoretical \textbf{53} (2020), no.~23, 235206.

\bibitem[BBC{\etalchar{+}}19]{Barkeshli2019a}
Maissam Barkeshli, Parsa Bonderson, Meng Cheng, Chao-Ming Jian, and Kevin
  Walker, \emph{{Reflection and Time Reversal Symmetry Enriched Topological
  Phases of Matter: Path Integrals, Non-orientable Manifolds, and Anomalies}},
  Communications in Mathematical Physics \textbf{374} (2019), no.~2,
  1021--1124.

\bibitem[BBCW19]{Barkeshli2019}
Maissam Barkeshli, Parsa Bonderson, Meng Cheng, and Zhenghan Wang,
  \emph{Symmetry fractionalization, defects, and gauging of topological
  phases}, Physical Review B \textbf{100} (2019), no.~11.

\bibitem[BBD17]{Bridgeman2017}
Jacob~C. Bridgeman, Stephen~D. Bartlett, and Andrew~C. Doherty, \emph{Tensor
  networks with a twist: {Anyon}-permuting domain walls and defects in
  projected entangled pair states}, Physical Review B \textbf{96} (2017),
  no.~24.

\bibitem[BBH{\etalchar{+}}18]{Beer2018}
Kerstin Beer, Dmytro Bondarenko, Alexander Hahn, Maria Kalabakov, Nicole Knust,
  Laura Niermann, Tobias~J. Osborne, Christin Schridde, Stefan Seckmeyer,
  Deniz~E. Stiegemann, and Ramona Wolf, \emph{From categories to anyons: a
  travelogue}, arXiv:1811.06670.

\bibitem[BBJ19a]{barter_domain_2019}
Daniel Barter, Jacob~C. Bridgeman, and Corey Jones, \emph{Domain {Walls} in
  {Topological} {Phases} and the {Brauer}–{Picard} {Ring} for
  $\mathbf{Vec}(\mathbb{Z}/p\mathbb{Z})$}, Communications in Mathematical
  Physics \textbf{369} (2019), no.~3, 1167--1185.

\bibitem[BBJ19b]{Bridgeman2019}
Jacob~C. Bridgeman, Daniel Barter, and Corey Jones, \emph{Fusing binary
  interface defects in topological phases: {The} $\mathbb{Z}/p\mathbb{Z}$
  case}, Journal of Mathematical Physics \textbf{60} (2019), no.~12, 121701.

\bibitem[BBS10]{Baraban2010}
Mara Baraban, Nick~E. Bonesteel, and Steven~H. Simon, \emph{Resources required
  for topological quantum factoring}, Physical Review A \textbf{81} (2010),
  no.~6.

\bibitem[BC17]{Bridgeman2017a}
Jacob~C. Bridgeman and Christopher~T. Chubb, \emph{Hand-waving and interpretive
  dance: an introductory course on tensor networks}, Journal of Physics A:
  Mathematical and Theoretical \textbf{50} (2017), no.~22, 223001.

\bibitem[BDSPV15]{Bartlett2015}
Bruce Bartlett, Christopher~L. Douglas, Christopher~J. Schommer-Pries, and
  Jamie Vicary, \emph{Modular categories as representations of the
  3-dimensional bordism 2-category}, arXiv:1509.06811.

\bibitem[Bea99]{Beachy1999}
John~A. Beachy, \emph{{Introductory Lectures on Rings and Modules}}, Cambridge
  University Press, 1999.

\bibitem[BEW19]{Bauer2019}
Andreas Bauer, Jens Eisert, and Carolin Wille, \emph{Towards a mathematical
  formalism for classifying phases of matter}, arXiv:1903.05413.

\bibitem[BHOW20]{Bridgeman2020}
Jacob~C. Bridgeman, Alexander Hahn, Tobias~J. Osborne, and Ramona Wolf,
  \emph{{Gauging defects in quantum spin systems: A case study}}, Physical
  Review B \textbf{101} (2020), no.~13.

\bibitem[BHZS05]{Bonesteel2005}
Nick~E. Bonesteel, Layla Hormozi, Georgios Zikos, and Steven~H. Simon,
  \emph{{Braid Topologies for Quantum Computation}}, Physical Review Letters
  \textbf{95} (2005), no.~14.

\bibitem[Bis98]{Bisch1998}
Dietmar Bisch, \emph{Principal graphs of subfactors with small {Jones} index},
  Mathematische Annalen \textbf{311} (1998), no.~2, 223--231.

\bibitem[Bis15]{Bischoff2015}
Marcel Bischoff, \emph{{The Relation between Subfactors arising from Conformal
  Nets and the Realization of Quantum Doubles}}, arXiv:1511.08931.

\bibitem[Bis16]{Bischoff2016}
\bysame, \emph{{A Remark on CFT Realization of Quantum Doubles of Subfactors:
  Case Index $<4$}}, Letters in Mathematical Physics \textbf{106} (2016),
  no.~3, 341--363.

\bibitem[BJQ13]{Barkeshli2013}
Maissam Barkeshli, Chao-Ming Jian, and Xiao-Liang Qi, \emph{Theory of defects
  in {Abelian} topological states}, Physical Review B \textbf{88} (2013),
  no.~23.

\bibitem[BK01]{bakalov_lectures_2001}
Bojko Bakalov and Alexander {Kirillov Jr.}, \emph{Lectures on {Tensor}
  {Categories} and {Modular} {Functors}}, University {Lecture} {Series},
  vol.~21, American Mathematical Society, 2001.

\bibitem[BKB{\etalchar{+}}20]{Bartolomei2020}
Hugo Bartolomei, Manohar Kumar, Rémi Bisognin, Arthur Marguerite, Jean-Marc
  Berroir, Erwann Bocquillon, Bernard Pla{\c{c}}ais, Antonella Cavanna,
  Q.~Dong, Ulf Gennser, Yong Jin, and Gwendal F{\`{e}}ve, \emph{Fractional
  statistics in anyon collisions}, Science \textbf{368} (2020), no.~6487,
  173--177.

\bibitem[BLP{\etalchar{+}}16]{Brown2016}
Benjamin~J. Brown, Daniel Loss, Jiannis~K. Pachos, Chris~N. Self, and James~R.
  Wootton, \emph{Quantum memories at finite temperature}, Reviews of Modern
  Physics \textbf{88} (2016), no.~4.

\bibitem[BMPS12]{bigelow_constructing_2012}
Stephen Bigelow, Scott Morrison, Emily Peters, and Noah Snyder,
  \emph{Constructing the extended {Haagerup} planar algebra}, Acta Mathematica
  \textbf{209} (2012), no.~1, 29--82.

\bibitem[BMW{\etalchar{+}}17]{Bultinck2017}
Nick Bultinck, Michael Mari\"{e}n, Dominic~J. Williamson, M.~Burak
  {\c{S}}ahino{\u{g}}lu, Jutho Haegeman, and Frank Verstraete, \emph{Anyons and
  matrix product operator algebras}, Annals of Physics \textbf{378} (2017),
  183--233.

\bibitem[Bon07]{Bond2007}
Parsa~Hassan Bonderson, \emph{{Non-Abelian Anyons and Interferometry}}, Ph.D.
  thesis, California Institute of Technology, 2007.

\bibitem[Bor94]{Borceux1994}
Francis Borceux, \emph{Handbook of {Categorical} {Algebra}}, Cambridge
  University Press, 1994.

\bibitem[BPZ84]{Belavin1984}
Alexander~A. Belavin, Alexander~M. Polyakov, and Alexander~B. Zamolodchikov,
  \emph{Infinite conformal symmetry in two-dimensional quantum field theory},
  Nuclear Physics B \textbf{241} (1984), no.~2, 333--380.

\bibitem[BS10]{Baez2010}
John Baez and Mike Stay, \emph{{Physics, Topology, Logic and Computation: A
  Rosetta Stone}}, New Structures for Physics, Springer Berlin Heidelberg,
  2010, pp.~95--172.

\bibitem[CAS13]{CAS13}
David~J. Clarke, Jason Alicea, and Kirill Shtengel, \emph{Exotic non-{Abelian}
  anyons from conventional fractional quantum {Hall} states}, Nature
  Communications \textbf{4} (2013), no.~1.

\bibitem[CC09]{Calabrese2009}
Pasquale Calabrese and John Cardy, \emph{Entanglement entropy and conformal
  field theory}, Journal of Physics A: Mathematical and Theoretical \textbf{42}
  (2009), no.~50, 504005.

\bibitem[CCW16]{Cong2016}
Iris Cong, Meng Cheng, and Zhenghan Wang, \emph{{Topological Quantum
  Computation with Gapped Boundaries}}, arXiv:1609.02037.

\bibitem[CCW17]{Cong2017}
\bysame, \emph{Defects between gapped boundaries in two-dimensional topological
  phases of matter}, Physical Review B \textbf{96} (2017), no.~19.

\bibitem[CMS11]{Calegari2010}
Frank Calegari, Scott Morrison, and Noah Snyder, \emph{{Cyclotomic Integers,
  Fusion Categories, and Subfactors}}, Communications in Mathematical Physics
  \textbf{303} (2011), no.~3, 845--896.

\bibitem[Con76]{Connes1976}
Alain Connes, \emph{{Classification of Injective Factors Cases $II_1$,
  $II_\infty$, $III_\lambda$ , $\lambda\neq 1$}}, The Annals of Mathematics
  \textbf{104} (1976), no.~1, 73.

\bibitem[Con80]{Connes80}
\bysame, \emph{{On the spatial theory of von Neumann algebras}}, Journal of
  Functional Analysis \textbf{35} (1980), no.~2, 153--164.

\bibitem[DKLP02]{Dennis2002}
Eric Dennis, Alexei Kitaev, Andrew Landahl, and John Preskill,
  \emph{Topological quantum memory}, Journal of Mathematical Physics
  \textbf{43} (2002), no.~9, 4452--4505.

\bibitem[DR89]{Doplicher1989}
Sergio Doplicher and John~E. Roberts, \emph{A new duality theory for compact
  groups}, Inventiones Mathematicae \textbf{98} (1989), no.~1, 157--218.

\bibitem[DSPS19]{Douglas2019}
Christopher~L. Douglas, Christopher Schommer-Pries, and Noah Snyder, \emph{The
  balanced tensor product of module categories}, Kyoto Journal of Mathematics
  \textbf{59} (2019), no.~1, 167--179.

\bibitem[EG11]{Evans2011}
David~E. Evans and Terry Gannon, \emph{{The Exoticness and Realisability of
  Twisted Haagerup-Izumi Modular Data}}, Communications in Mathematical Physics
  \textbf{307} (2011), no.~2, 463--512.

\bibitem[EGNO15]{Etingof2015}
Pavel Etingof, Shlomo Gelaki, Dmitri Nikshych, and Victor Ostrik, \emph{Tensor
  {Categories}}, American Mathematical Society, 2015.

\bibitem[EK95]{Evans1995}
David~E. Evans and Yasuyuki Kawahigashi, \emph{On {Ocneanu’s} theory of
  asymptotic inclusions for subfactors, topological quantum field theories and
  quantum doubles}, International Journal of Mathematics \textbf{06} (1995),
  no.~02, 205--228.

\bibitem[EK98]{Evans1998}
\bysame, \emph{Quantum symmetries on operator algebras}, Oxford Mathematical
  Monographs, Clarendon Press Oxford, 1998.

\bibitem[EM45]{Eilenberg1945}
Samuel Eilenberg and Saunders {Mac Lane}, \emph{General theory of natural
  equivalences}, Transactions of the American Mathematical Society \textbf{58}
  (1945), 231--231.

\bibitem[ENO05]{Etingof2005}
Pavel Etingof, Dmitri Nikshych, and Viktor Ostrik, \emph{On fusion categories},
  Annals of Mathematics \textbf{162} (2005), no.~2, 581--642.

\bibitem[FFL{\etalchar{+}}14]{Finch2014}
Peter~E. Finch, Holger Frahm, Marius Lewerenz, Ashley Milsted, and Tobias~J.
  Osborne, \emph{Quantum phases of a chain of strongly interacting anyons},
  Physical Review B \textbf{90} (2014), no.~8.

\bibitem[FGH{\etalchar{+}}12]{Freedman2012}
Michael~H. Freedman, Jan Gukelberger, Matthew~B. Hastings, Simon Trebst,
  Matthias Troyer, and Zhenghan Wang, \emph{Galois conjugates of topological
  phases}, Physical Review B \textbf{85} (2012), no.~4.

\bibitem[FLW02]{Freedman2002}
Michael~H. Freedman, Michael~J. Larsen, and Zhenghan Wang, \emph{{The
  Two-Eigenvalue Problem and Density of Jones Representation of Braid Groups}},
  Communications in Mathematical Physics \textbf{228} (2002), no.~1, 177--199.

\bibitem[FQS84]{Friedan1984}
Daniel Friedan, Zongan Qiu, and Stephen Shenker, \emph{{Conformal Invariance,
  Unitarity, and Critical Exponents in Two Dimensions}}, Physical Review
  Letters \textbf{52} (1984), no.~18, 1575--1578.

\bibitem[FSV13]{Fuchs2013}
Jürgen Fuchs, Christoph Schweigert, and Alessandro Valentino,
  \emph{{Bicategories for Boundary Conditions and for Surface Defects in $3-d$
  TFT}}, Communications in Mathematical Physics \textbf{321} (2013), no.~2,
  543--575.

\bibitem[FTL{\etalchar{+}}07]{feiguin_interacting_2007}
Adrian Feiguin, Simon Trebst, Andreas W.~W. Ludwig, Matthias Troyer, Alexei
  Kitaev, Zhenghan Wang, and Michael~H. Freedman, \emph{Interacting {Anyons} in
  {Topological} {Quantum} {Liquids}: {The} {Golden} {Chain}}, Physical Review
  Letters \textbf{98} (2007), no.~16, 160409.

\bibitem[GI15]{Grossman2015}
Pinhas Grossman and Masaki Izumi, \emph{Drinfeld centers of fusion categories
  arising from generalized {Haagerup} subfactors}, arXiv:1501.07679.

\bibitem[GS12]{grossman_quantum_2012}
Pinhas Grossman and Noah Snyder, \emph{Quantum subgroups of the {Haagerup}
  fusion categories}, Communications in Mathematical Physics \textbf{311}
  (2012), no.~3, 617--643, arXiv: 1102.2631.

\bibitem[Haa94]{Haagerup1994}
Uffe Haagerup, \emph{{Principal graphs of subfactors in the index range
  $4<[M:N]<3+\sqrt{2}$}}, Subfactors (Kyuzeso, 1993), World Sci. Publ., River
  Edge, NJ, 1994, pp.~1--38.

\bibitem[Hal84]{Halperin1984}
Bertrand~I. Halperin, \emph{{Statistics of Quasiparticles and the Hierarchy of
  Fractional Quantized Hall States}}, Physical Review Letters \textbf{52}
  (1984), no.~18, 1583--1586.

\bibitem[HCO{\etalchar{+}}11]{Haegeman2011}
Jutho Haegeman, J.~Ignacio Cirac, Tobias~J. Osborne, Iztok Pi{\v{z}}orn, Henri
  Verschelde, and Frank Verstraete, \emph{{Time-Dependent Variational Principle
  for Quantum Lattices}}, Physical Review Letters \textbf{107} (2011), no.~7.

\bibitem[HLW94]{Holzhey1994}
Christoph Holzhey, Finn Larsen, and Frank Wilczek, \emph{Geometric and
  renormalized entropy in conformal field theory}, Nuclear Physics B
  \textbf{424} (1994), no.~3, 443--467.

\bibitem[Hon09]{Hong2009}
Seung-Moon Hong, \emph{On symmetrization of $6j$-symbols and {Levin-Wen
  Hamiltonian}}, arXiv:0907.2204.

\bibitem[HOV13]{Haegeman2013}
Jutho Haegeman, Tobias~J. Osborne, and Frank Verstraete, \emph{Post-matrix
  product state methods: {To} tangent space and beyond}, Physical Review B
  \textbf{88} (2013), no.~7.

\bibitem[HRW08]{SEUNGMOON2008}
Seung-Moon Hong, Eric Rowell, and Zhenghan Wang, \emph{{On} {Exotic} {Modular}
  {Tensor} {Categories}}, Communications in Contemporary Mathematics
  \textbf{10} (2008), no.~supp01, 1049--1074.

\bibitem[HV19]{HV19}
Chris Heunen and Jamie Vicary, \emph{Categories for {Quantum} {Theory}}, Oxford
  University Press, 2019.

\bibitem[HW12]{Hung2012}
Ling-Yan Hung and Yidun Wan, \emph{String-net models with {$\mathbb{Z}_N$}
  fusion algebra}, Physical Review B \textbf{86} (2012), no.~23.

\bibitem[HW20]{Hahn2020}
Alexander Hahn and Ramona Wolf, \emph{Generalized string-net model for unitary
  fusion categories without tetrahedral symmetry}, Physical Review B
  \textbf{102} (2020), 115154.

\bibitem[HZBS07]{Hormozi2007}
Layla Hormozi, Georgios Zikos, Nick~E. Bonesteel, and Steven~H. Simon,
  \emph{Topological quantum compiling}, Physical Review B \textbf{75} (2007),
  no.~16.

\bibitem[Izu00]{izumi_structure_2000}
Masaki Izumi, \emph{The {Structure} of {Sectors} {Associated} with
  {Longo}–{Rehren} {Inclusions} {I}. {General} {Theory}}, Communications in
  Mathematical Physics \textbf{213} (2000), no.~1, 127--179.

\bibitem[Izu01]{izumi_structure_2001}
\bysame, \emph{The {Structure} of {Sectors} {Associated} with
  {Longo}–{Rehren} {Inclusions} {II}. {Examples}}, Reviews in Mathematical
  Physics \textbf{13} (2001), no.~05, 603--674.

\bibitem[Jac09a]{Jacobson2009a}
Nathan Jacobson, \emph{{Basic Algebra I}}, second ed., Dover Publications Inc.,
  2009.

\bibitem[Jac09b]{Jacobson2009b}
\bysame, \emph{{Basic Algebra II}}, second ed., Dover Publications Inc., 2009.

\bibitem[JGJ03]{Jeon2003}
Gun~Sang Jeon, Kenneth~L. Graham, and Jainendra~K. Jain, \emph{{Fractional
  Statistics in the Fractional Quantum Hall Effect}}, Physical Review Letters
  \textbf{91} (2003), no.~3.

\bibitem[Jim89]{Jimbo1989}
Michio Jimbo, \emph{{Introduction} to the {Yang}-{Baxter} {Equation}},
  International Journal of Modern Physics A \textbf{04} (1989), no.~15,
  3759--3777.

\bibitem[JMS14]{Jones2013}
Vaughan F.~R. Jones, Scott Morrison, and Noah Snyder, \emph{The classification
  of subfactors of index at most 5}, Bulletin of the American Mathematical
  Society \textbf{51} (2014), no.~2, 277--327.

\bibitem[Jon83]{Jones1983}
Vaughan F.~R. Jones, \emph{Index for subfactors}, Inventiones Mathematicae
  \textbf{72} (1983), no.~1, 1--25.

\bibitem[Jon90]{Jones1990}
\bysame, \emph{{Von Neumann Algebras in Mathematics and Physics}}, Proceedings
  of the International Congress of Mathematicians, Kyoto, 1990, pp.~121--138.

\bibitem[Jon14]{Jones2014}
\bysame, \emph{{Some unitary representations of Thompson's groups F and T}},
  arXiv:1412.7740.

\bibitem[Jon15]{JonesvNA}
\bysame, \emph{{Von Neumann Algebras}}, 2015, Lecture notes.

\bibitem[Jon16]{Jones2016}
Corey Jones, \emph{Quantum {$G_2$} categories have property {(T)}},
  International Journal of Mathematics \textbf{27} (2016), no.~02, 1650015.

\bibitem[JS91]{Joyal1991}
Andr{\'{e}} Joyal and Ross Street, \emph{The geometry of tensor calculus, {I}},
  Advances in Mathematics \textbf{88} (1991), no.~1, 55--112.

\bibitem[JS97]{Jones1997}
Vaughan F.~R. Jones and Vaikalathur~S. Sunder, \emph{{Introduction to
  Subfactors}}, Cambridge University Press, 1997.

\bibitem[Kas95]{kassel_quantum_1995}
Christian Kassel, \emph{Quantum {Groups}}, Graduate {Texts} in {Mathematics},
  vol. 155, Springer, 1995.

\bibitem[Kau01]{Kauffman2001}
Louis~H. Kauffman, \emph{Knots and {Physics}}, World Scientific, 2001.

\bibitem[KB10]{Kirillov2010}
Alexander {Kirillov Jr.} and Benjamin Balsam, \emph{{Turaev-Viro invariants as
  an extended TQFT}}, arXiv:1004.1533.

\bibitem[{Kir}11]{Kirillov2011}
Alexander {Kirillov Jr.}, \emph{{String-net model of Turaev-Viro invariants}},
  arXiv:1106.6033.

\bibitem[Kit03]{Kitaev2003}
Alexei~Y. Kitaev, \emph{Fault-tolerant quantum computation by anyons}, Annals
  of Physics \textbf{303} (2003), no.~1, 2--30.

\bibitem[Kit06]{kitaev_anyons_2006}
\bysame, \emph{Anyons in an exactly solved model and beyond}, Annals of Physics
  \textbf{321} (2006), no.~1, 2--111.

\bibitem[KKR10]{Koenig2010}
Robert Koenig, Greg Kuperberg, and Ben~W. Reichardt, \emph{{Quantum computation
  with Turaev-Viro codes}}, Annals of Physics \textbf{325} (2010), no.~12,
  2707--2749.

\bibitem[KMR10]{Kadar2010}
Zolt{\'{a}}n K{\'{a}}d{\'{a}}r, Annalisa Marzuoli, and Mario Rasetti,
  \emph{{Microscopic Description of $2D$ Topological Phases, Duality, and $3D$
  State Sums}}, Advances in Mathematical Physics \textbf{2010} (2010), 1--18.

\bibitem[KPEB18]{Kesselring2018}
Markus~S. Kesselring, Fernando Pastawski, Jens Eisert, and Benjamin~J. Brown,
  \emph{The boundaries and twist defects of the color code and their
  applications to topological quantum computation}, Quantum \textbf{2} (2018),
  101.

\bibitem[Lei09]{Leinster2009}
Tom Leinster, \emph{Basic {Category} {Theory}}, Cambridge University Press,
  2009.

\bibitem[Lew]{PrivateMarius}
Marius Lewerenz, \emph{Private communication.}

\bibitem[LL14]{Lin2014}
Chien-Hung Lin and Michael Levin, \emph{Generalizations and limitations of
  string-net models}, Physical Review B \textbf{89} (2014), no.~19.

\bibitem[LW03]{Levin2003}
Michael Levin and Xiao-Gang Wen, \emph{Fermions, strings, and gauge fields in
  lattice spin models}, Physical Review B \textbf{67} (2003), no.~24.

\bibitem[LW05]{Levin2005}
Michael~A. Levin and Xiao-Gang Wen, \emph{String-net condensation: {A} physical
  mechanism for topological phases}, Physical Review B \textbf{71} (2005),
  no.~4.

\bibitem[LW14]{Lan2014}
Tian Lan and Xiao-Gang Wen, \emph{Topological quasiparticles and the
  holographic bulk-edge relation in $(2+1)$-dimensional string-net models},
  Physical Review B \textbf{90} (2014), no.~11.

\bibitem[Lü02]{Lueck2002}
Wolfgang Lück, \emph{{$L^ 2$-Invariants: Theory and Applications to Geometry
  and $K$-Theory}}, Springer Berlin Heidelberg, 2002.

\bibitem[{Mac}98]{MacLane1998}
Saunders {Mac Lane}, \emph{Categories for the {Working} {Mathematician}},
  Springer-Verlag New York Inc., 1998.

\bibitem[MHOV13]{Milsted2013}
Ashley Milsted, Jutho Haegeman, Tobias~J. Osborne, and Frank Verstraete,
  \emph{Variational matrix product ansatz for nonuniform dynamics in the
  thermodynamic limit}, Physical Review B \textbf{88} (2013), no.~15.

\bibitem[Mila]{evoMPS}
Ashley Milsted, \emph{{evoMPS}}, https://github.com/amilsted/evoMPS.

\bibitem[Milb]{PrivateAsh}
\bysame, \emph{Private communication.}

\bibitem[MPS10]{Morrison2010}
Scott Morrison, Emily Peters, and Noah Snyder, \emph{Skein theory for the
  ${D}_{2n}$ planar algebras}, Journal of Pure and Applied Algebra \textbf{214}
  (2010), no.~2, 117--139.

\bibitem[MPS17]{Morrison2017}
\bysame, \emph{Categories generated by a trivalent vertex}, Selecta Mathematica
  \textbf{23} (2017), no.~2, 817--868.

\bibitem[MS89]{Moore1989}
Gregory Moore and Nathan Seiberg, \emph{Classical and quantum conformal field
  theory}, Communications in Mathematical Physics \textbf{123} (1989), no.~2,
  177--254.

\bibitem[Mv36]{Murray1936}
Francis~J. Murray and John {von Neumann}, \emph{{On Rings of Operators}}, The
  Annals of Mathematics \textbf{37} (1936), no.~1, 116.

\bibitem[Mü03a]{muger_subfactors_2003}
Michael Müger, \emph{From subfactors to categories and topology {I}:
  {Frobenius} algebras in and {Morita} equivalence of tensor categories},
  Journal of Pure and Applied Algebra \textbf{180} (2003), no.~1-2, 81--157.

\bibitem[Mü03b]{muger_subfactors_2003-1}
\bysame, \emph{From subfactors to categories and topology {II}: {The} quantum
  double of tensor categories and subfactors}, Journal of Pure and Applied
  Algebra \textbf{180} (2003), no.~1, 159--219.

\bibitem[Nik13]{Nikshych2013}
Dmitri Nikshych, \emph{Morita equivalence methods in classification of fusion
  categories}, Hopf Algebras and Tensor Categories, vol. 585, Contemporary
  Mathematics, 2013.

\bibitem[NLGM20]{Nakamura2020}
James Nakamura, Shuang Liang, Geoffrey~C. Gardner, and Michael~J. Manfra,
  \emph{Direct observation of anyonic braiding statistics}, Nature Physics
  \textbf{16} (2020), no.~9, 931--936.

\bibitem[NSS{\etalchar{+}}08]{Nayak2008}
Chetan Nayak, Steven~H. Simon, Ady Stern, Michael Freedman, and Sankar~Das
  Sarma, \emph{Non-{Abelian} anyons and topological quantum computation},
  Reviews of Modern Physics \textbf{80} (2008), no.~3, 1083--1159.

\bibitem[Ocn89]{Ocneanu1989}
Adrian Ocneanu, \emph{{Quantized groups, string algebras, and Galois theory for
  algebras}}, Operator Algebras and Applications, Cambridge University Press,
  1989, pp.~119--172.

\bibitem[Ocn94]{Ocneanu1994}
\bysame, \emph{Chirality for operator algebras}, Subfactors (Kyuzeso, 1993)
  (1994), 39--63.

\bibitem[Ost03]{Ostrik2003}
Victor Ostrik, \emph{Module categories, weak {Hopf} algebras and modular
  invariants}, Transformation Groups \textbf{8} (2003), no.~2, 177--206.

\bibitem[OSW19]{Osborne2019}
Tobias~J. Osborne, Deniz~E. Stiegemann, and Ramona Wolf, \emph{{The F-Symbols
  for the H3 Fusion Category}}, arXiv:1906.01322.

\bibitem[Pac09]{Pachos2009}
Jiannis~K. Pachos, \emph{Introduction to {Topological} {Quantum}
  {Computation}}, Cambridge University Press, 2009.

\bibitem[Pet10]{Peters2010}
Emily Peters, \emph{A planar algebra construction of the {Haagerup} subfactor},
  International Journal of Mathematics \textbf{21} (2010), no.~08, 987--1045.

\bibitem[Pop90]{Popa1990}
Sorin Popa, \emph{Classification of subfactors: the reduction to commuting
  squares}, Inventiones Mathematicae \textbf{101} (1990), no.~1, 19--43.

\bibitem[Pop94]{Popa1994}
\bysame, \emph{Classification of amenable subfactors of type {II}}, Acta
  Mathematica \textbf{172} (1994), no.~2, 163--255.

\bibitem[Pre04]{Preskill2004}
John Preskill, \emph{{Lecture Notes for Physics 219: Quantum Computation}}.

\bibitem[PY15]{Pastawski2015}
Fernando Pastawski and Beni Yoshida, \emph{Fault-tolerant logical gates in
  quantum error-correcting codes}, Physical Review A \textbf{91} (2015), no.~1.

\bibitem[Rob99]{Roberts1999}
Justin Roberts, \emph{Classical $6j$-symbols and the tetrahedron}, Geometry
  {\&} Topology \textbf{3} (1999), no.~1, 21--66.

\bibitem[RR99]{Read1999}
Nicholas Read and Edward Rezayi, \emph{{Beyond paired quantum Hall states:
  Parafermions and incompressible states in the first excited Landau level}},
  Physical Review B \textbf{59} (1999), no.~12, 8084--8092.

\bibitem[RW18]{Rowell2018}
Eric~C. Rowell and Zhenghan Wang, \emph{Mathematics of topological quantum
  computing}, Bulletin of the American Mathematical Society \textbf{55} (2018),
  no.~2, 183--238.

\bibitem[Sac17]{Sachdev2017}
Subir Sachdev, \emph{{Quantum Phase Transitions}}, second ed., Cambridge
  University Press, 2017.

\bibitem[Sel10]{Selinger2010}
Peter Selinger, \emph{A {Survey} of {Graphical} {Languages} for {Monoidal}
  {Categories}}, New Structures for Physics, Springer Berlin Heidelberg, 2010,
  pp.~289--355.

\bibitem[SHM{\etalchar{+}}15]{Stojevic2015}
Vid Stojevic, Jutho Haegeman, I.~P. McCulloch, Luca Tagliacozzo, and Frank
  Verstraete, \emph{Conformal data from finite entanglement scaling}, Physical
  Review B \textbf{91} (2015), no.~3.

\bibitem[{\c{S}}WB{\etalchar{+}}14]{Sahinoglu2014}
M.~Burak {\c{S}}ahino{\u{g}}lu, Dominic Williamson, Nick Bultinck, Michael
  Mari\"{e}n, Jutho Haegeman, Norbert Schuch, and Frank Verstraete,
  \emph{{Characterizing Topological Order with Matrix Product Operators}},
  arXiv:1409.2150.

\bibitem[TAF{\etalchar{+}}08]{Trebst2008a}
Simon Trebst, Eddy Ardonne, Adrian Feiguin, David~A. Huse, Andreas W.~W.
  Ludwig, and Matthias Troyer, \emph{{Collective States of Interacting
  Fibonacci Anyons}}, Physical Review Letters \textbf{101} (2008), no.~5.

\bibitem[Tak79]{Takesaki1979}
Masamichi Takesaki, \emph{{Theory of Operator Algebras I}}, Springer New York,
  1979.

\bibitem[TdIL08]{Tagliacozzo2008}
Luca Tagliacozzo, Thiago~R. {de Oliveira}, Sofyan Iblisdir, and José~I.
  Latorre, \emph{Scaling of entanglement support for matrix product states},
  Physical Review B \textbf{78} (2008), no.~2.

\bibitem[Ter15]{Terhal2015}
Barbara~M. Terhal, \emph{Quantum error correction for quantum memories},
  Reviews of Modern Physics \textbf{87} (2015), no.~2, 307--346.

\bibitem[Tho06]{Thom2006}
Andreas Thom, \emph{A remark about the {Connes} fusion tensor product},
  arXiv:math/0601045.

\bibitem[Tit]{PrivateTitsworth}
Matthew Titsworth, \emph{Private communication.}

\bibitem[TTWL08]{Trebst2008}
Simon Trebst, Matthias Troyer, Zhenghan Wang, and Andreas W.~W. Ludwig,
  \emph{{A Short Introduction to Fibonacci Anyon Models}}, Progress of
  Theoretical Physics Supplement \textbf{176} (2008), 384--407.

\bibitem[Tur92]{Turaev1992}
Vladimir~G. Turaev, \emph{{Modular} {Categories} and 3-{Manifold}
  {Invariants}}, International Journal of Modern Physics B \textbf{06} (1992),
  no.~11n12, 1807--1824.

\bibitem[Tur10]{Turaev2010}
\bysame, \emph{Quantum {Invariants} of {Knots} and 3-{Manifolds}}, De Gruyter
  Studies in Mathematics, vol.~18, De Gruyter, 2010.

\bibitem[TV92]{Turaev1992a}
Vladimir Turaev and Oleg~Y. Viro, \emph{State sum invariants of $3$-manifolds
  and quantum $6j$-symbols}, Topology \textbf{31} (1992), no.~4, 865--902.

\bibitem[TV10]{Turaev2010a}
Vladimir Turaev and Alexis Virelizier, \emph{{On two approaches to
  3-dimensional TQFTs}}, arXiv.1006.3501.

\bibitem[Ver88]{Verlinde1988}
Erik Verlinde, \emph{{Fusion rules and modular transformations in $2D$
  conformal field theory}}, Nuclear Physics B \textbf{300} (1988), 360--376.

\bibitem[vSR15]{Keyserlingk2015}
Curt~W. {von Keyserlingk}, Steven~H. Simon, and Bernd Rosenow, \emph{{Enhanced
  Bulk-Edge Coulomb Coupling in Fractional Fabry-Perot Interferometers}},
  Physical Review Letters \textbf{115} (2015), no.~12.

\bibitem[Wan10]{Wang2010}
Zhenghan Wang, \emph{Topological {Quantum} {Computation}}, CBMS Regional
  Conference Series in Mathematics, vol. 112, American Mathematical Society,
  2010.

\bibitem[WBV17]{Williamson2017}
Dominic~J. Williamson, Nick Bultinck, and Frank Verstraete,
  \emph{Symmetry-enriched topological order in tensor networks: {Defects},
  gauging and anyon condensation}, arXiv:1711.07982.

\bibitem[Wen90]{Wen1990a}
Xiao-Gang Wen, \emph{{Topological} {orders} {in} {rigid} {states}},
  International Journal of Modern Physics B \textbf{04} (1990), no.~02,
  239--271.

\bibitem[Wil82]{Wilczek1982}
Frank Wilczek, \emph{{Quantum Mechanics of Fractional-Spin Particles}},
  Physical Review Letters \textbf{49} (1982), no.~14, 957--959.

\bibitem[Wil17]{Wilde2016}
Mark~M. Wilde, \emph{{Quantum Information Theory}}, second ed., Cambridge
  University Press, 2017.

\bibitem[WN90]{Wen1990b}
Xiao-Gang Wen and Qian Niu, \emph{{Ground-state degeneracy of the fractional
  quantum Hall states in the presence of a random potential and on high-genus
  Riemann surfaces}}, Physical Review B \textbf{41} (1990), no.~13, 9377--9396.

\bibitem[WNS{\etalchar{+}}13]{Willett2013}
Robert~L. Willett, Chetan Nayak, Kirill Shtengel, Loren~N. Pfeiffer, and Ken~W.
  West, \emph{{Magnetic-Field-Tuned Aharonov-Bohm Oscillations and Evidence for
  Non-Abelian Anyons at $\nu=5/2$}}, Physical Review Letters \textbf{111}
  (2013), no.~18.

\bibitem[Xu18]{Xu2017}
Feng Xu, \emph{{Examples of Subfactors from Conformal Field Theory}},
  Communications in Mathematical Physics \textbf{357} (2018), no.~1, 61--75.

\end{thebibliography}

\printindex

\chapter*{Curriculum Vitae}

\begin{table}[H]
	\flushleft
	\begin{tabular}{ll}
		Full name: & Ramona Wolf\\
		Date of birth: & 10.03.1993\\
		Place of birth: & Holzminden\\
		Citizenship: & German
	\end{tabular}
\end{table}

\vspace{20pt}

{\large \textbf{Academic career}}

\begin{table}[H]
	\flushleft
	\begin{tabular}{crl}
		\hspace{10pt} & 2017 - 2020 & Research scientist at the Institute for Theoretical Physics,\\
		\hspace{10pt} &  & Leibniz Universität Hannover\vspace{5pt} \\ 
		\hspace{10pt} & 2017 & Master of Science in Physics\\
		\hspace{10pt} &  & \textit{Thesis: Fusion in Tensor Categories} \vspace{5pt} \\
		\hspace{10pt} & 2015 & Bachelor of Science in Physics\\
		\hspace{10pt} & & \textit{Thesis: Quantum Key Distribution in the Non-Asymptotic Regime} \vspace{5pt} \\
		\hspace{10pt} & 2012 - 2017 & Studies in Physics at Leibniz Universität Hannover \vspace{5pt} \\
		\hspace{10pt} & 2003 - 2012 & Abitur at Städtisches Gymnasium Steinheim
	\end{tabular}
\end{table}

\end{document}